\pdfoutput=1 
\documentclass{aa}  
\usepackage{graphicx}
\usepackage{txfonts}
\usepackage{hyperref}
\def\het{He(2$^{3}$S)}

\begin{document}

\title{Connection between planetary He {\sc i} $\lambda$10830 \AA\ absorption and extreme-ultraviolet emission of planet-host stars}

   \author{J.~Sanz-Forcada\inst{1}
     \and
     M.~L\'opez-Puertas\inst{2}
     \and
     M.~Lamp\'on\inst{2}
     \and
     S.~Czesla\inst{3}
     \and
     L.~Nortmann\inst{4}
     \and
     J.\,A.~Caballero\inst{1}
     \and
     M.\,R.~Zapatero~Osorio\inst{1}
     \and
     P.\,J.~Amado\inst{2}
     \and
     F.~Murgas\inst{5,6}
     \and
     J.~Orell-Miquel\inst{5,6}
     \and
     E.~Pall\'e\inst{5,6}
     \and
     A.~Quirrenbach\inst{7}
     \and
     A.~Reiners\inst{4}
     \and
     I.~Ribas\inst{8,9}
     \and
     A.~S\'anchez-L\'opez\inst{2}
     \and
     E.~Solano\inst{1}
     }

   \institute{ 
     Centro de Astrobiolog\'{i}a, CSIC-INTA, Camino bajo del Castillo s/n, 
     E-28692 Villanueva de la Ca\~nada, Madrid, Spain \\ 
     \email{jsanz@cab.inta-csic.es}
     \and 
     Instituto de Astrof\'{i}sica de Andaluc\'{i}a (IAA-CSIC), Glorieta de la Astronom\'{i}a s/n, E-18008 Granada, Spain
     \and 
     Thüringer Landessternwarte Tautenburg, Sternwarte 5, D-07778 Tautenburg, Germany
     \and 
     Institut für Astrophysik und Geophysik, George-August-Universität, Friedrich-Hund-Platz 1, D-37077 Göttingen, Germany
     \and 
     Instituto de Astrofísica de Canarias (IAC), E-38205 La Laguna, Tenerife, Spain
     \and 
     Departamento de Astrofísica, Universidad de La Laguna (ULL), E-38206 La Laguna, Tenerife, Spain
     \and 
     Landessternwarte, Zentrum für Astronomie der Universität Heidelberg, Königstuhl 12, D-69117 Heidelberg, Germany
     \and 
     Institut de Ciències de l’Espai (ICE, CSIC), Campus UAB, Can Magrans s/n, E-08193 Bellaterra, Barcelona, Spain
     \and 
     Institut d’Estudis Espacials de Catalunya (IEEC), E-08860 Castelldefels, Barcelona, Spain
     }

   \date{Received 26 July 2024 / Accepted 12 December 2024}

 
  \abstract
   {The detection of the \ion{He}{i}~$\lambda$10830\,\AA\ triplet in exoplanet atmospheres has opened a new window for probing planetary properties, including atmospheric escape. Unlike Lyman~$\alpha$, the triplet is significantly less affected by  interstellar medium (ISM) absorption. Sufficient X-ray and extreme ultraviolet (XUV) stellar irradiation may trigger the formation of the \ion{He}{i} triplet via photoionization and posterior recombination processes in the planet atmospheres. Only a weak trend between stellar XUV emission and the planetary \ion{He}{i} strength has been observed so far.}
   {We aim to confirm this mechanism for producing near-infrared \ion{He}{i} absorption in exoplanetary atmospheres by examining a substantial sample of planetary systems.}
   {We obtained homogeneous measurements of the planetary \ion{He}{i} line equivalent width and consistently computed the stellar XUV ionizing irradiation. Our first step was to derive new coronal models for the planet-host stars. We used updated data from the X-exoplanets database, archival X-ray spectra of M-type stars (including AU Mic and Proxima Centauri), and new \textit{XMM-Newton} X-ray data recently obtained for the CARMENES project. These data were complemented at longer wavelengths with publicly available HST, FUSE, and EUVE spectra. A total of 75 stars are carefully analyzed to obtain a new calibration between X-ray and extreme ultraviolet (EUV) emission.}
   {Two distinct relationships between stellar X-ray emission (5--100\,\AA) and EUV$_{\rm H}$ (100--920\,\AA) or EUV$_{\rm He}$ (100--504\,\AA) radiation are obtained to scale the emission from late-type (F to M) stellar coronae. A total of 48 systems with reported planetary \ion{He}{i}~$\lambda$10830\,\AA\ studies, including 21 positive detections and 27 upper limits, exhibit a robust relationship between the strength of the planetary \ion{He}{i} feature and the ionizing XUV$_{\rm He}$ received by the planet, corrected by stellar and planetary radii, as well as the planet's gravitational potential. Some outliers could be explained by a different atmospheric composition or the lack of planetary gaseous atmospheres. This relation may serve as a guide to predict the detectability of the \ion{He}{i}~$\lambda$10830\,\AA\ absorption in exoplanet atmospheres.}
   {}
   \keywords{planetary systems -- stars: coronae -- X-rays: stars -- planets and satellites: atmospheres}

   \titlerunning{Connection between planetary He {\sc i} $\lambda$10830 \AA\ absorption and XUV emission of planet-host stars}
   \maketitle
%

\section{Introduction}
The detection of atmospheric features in transiting exoplanets has become the best approach to understanding the composition and evolution of their atmospheres. The earliest detection reports were made for \ion{Na}{i} \citep{cha02} and \ion{H}{i} Lyman $\alpha$ \citep{vid03}, both in \object{HD~209458~b}. 
A number of atomic and molecular species have been detected since then \citep[e.g.,][and references therein]{mad19}. It is remarkable that helium, the second most abundant element in stars and in Solar System planets such as Jupiter or Saturn, was not detected until 2018.
Theoretical works \citep{sea00} have predicted the possibility to detect the \ion{He}{i} infrared triplet at 10830~\AA\ in the atmosphere of transiting exoplanets. An attempt to detect this feature was made by Sasselov \& Sanz-Forcada in 2000 and 2001, using the NASA Infrared Telescope Facility (IRTF), but no feature was detected due to bad weather conditions in both campaigns combined with insufficient instrument sensitivity.
\citet{mou03} used VLT/ISAAC to search for the same feature, but no detection was made either. During the following years the search for atmospheric features focused on other species. The idea of observing exoplanet atmospheres using the \ion{He}{i} 10830 line was resurrected by \citet{okl18}.
The line was soon detected independently by three different groups \citep{spa18,nor18,all18}, and two more detections were reported in the same year \citep{man18,sal18}. \citet{nor18} also identified, as proposed by \citet{sea00}, that in exoplanet atmospheres the line formation is directly related to the incoming stellar irradiation in the XUV band at $\lambda\lambda\sim 5-504$\,\AA, with 504\,\AA\ being the wavelength corresponding to the first ionization energy of He.

The neutral helium atom possesses two types of configurations, namely, orthohelium, and parahelium, corresponding to different sets of energy levels. In orthohelium, the most conspicuous line is actually a triplet, at 10830\,\AA\footnote{Lines at 10829.09, 10830.25, 10830.34\,\AA\ in the air, and 10832.06, 10833.22, 10833.31\,\AA\ in vacuum.}. This triplet is associated with a less intense optical triplet at 5876\,\AA.  The formation of these triplets has long been studied in both massive or low-mass stars. Two mechanisms have been proposed to explain the formation of the triplets: (i) collisional excitation from the singlet (parahelium) levels may populate the ground level of the triplet (orthohelium), allowing for the formation of the triplet lines at 10830 and 5876\,\AA. A high temperature ($\ga 20,000$\,K) is required for this mechanism to operate. Alternatively, in a colder environment, (ii) the \ion{He}{i} atom can be radiatively ionized with EUV photons ($\lambda < 504$~\AA) that would soon recombine into neutral atoms, some of them populating the triplet lower levels in a de-excitation cascade (photoionization-recombination mechanism).
Examples of the first mechanism are present in stellar winds observed in the G2\,I supergiant $\alpha$ Aqr \citep{dup92}. The second mechanism is thought to be responsible for the formation of the line in stellar chromospheres of late-type stars \citep[e.g.,][]{zar86}, which receive EUV photons from their coronae. In this case, a balance of the two mechanisms could occur in dwarfs because of the high temperatures and density reached in their chromospheres \citep{and95,san08}.

In the case of planet atmospheres, the photoionization-recombination mechanism is likely responsible for producing this line, given their cold environment. Hence, we would expect to see a relation between the ionizing XUV irradiation of the planets and the observed \ion{He}{i} triplet in their atmospheres. A small sample of five objects reported by \citet{nor18} seems to follow this trend, but a recent compilation of all published data on \ion{He}{i}~10830 in exoplanets \citep{fos22} could not find such relation. \citet{fos22} used the (H-ionizing) radiation in the range 5--920~\AA\ (XUV$_{\rm H}$) as a proxy of the He-ionizing radiation. Similar non-conclusive results were found by \citet{kir22}, \citet{fos23}, and \citet{all23}.
A further study was carried out by \citet{zha23c},  using a relationship between planetary mass-loss rates estimated from observations and the theoretical energy-limited mass-loss rates. These authors were able to find a rather good correlation between the XUV flux in the same range and the observed \ion{He}{i} triplet absorption.

In this paper, we attempt to probe the correlation between ionizing stellar XUV$_{\rm He}$ (in the range 5--504~\AA) irradiation and the \ion{He}{i} triplet in exoplanets. Since high-energy photons are easily absorbed by the interstellar medium (ISM), no direct observations of spectra in the range 400--504\,\AA\ are available for stars other than the Sun and most stars do not have reliable data in the 100--400\,\AA\ range either. A coronal model is needed to estimate the stellar spectral energy distribution (SED) in the range not covered by the actual data.
It is thus necessary to calculate accurate coronal models of the exoplanet host stars and to calibrate a relation allowing us to easily calculate the broadband XUV$_{\rm He}$ stellar flux.
Details of the X-ray and UV observations used to prepare these models, as well as on the method employed, are described in Sects.~\ref{sec:obs} and \ref{sec:methods}.
The results obtained for the objects in the sample are given in Sect.~\ref{sec:results}, together with a new scaling law to easily calculate the broadband EUV emission in the H- and He-ionizing ranges, provided that the X-ray flux is available. The different parametrizations to relate the XUV$_{\rm He}$ flux and the \ion{He}{i} line are explored, too.
In Sect.~\ref{sec:discussion}, we discuss the results as compared with actual \ion{He}{i} detections observed in exoplanet atmospheres, along with the implications that they have for research on the planet atmospheres. The conclusions are given in Sect.~\ref{sec:conclusions}. We then provide appendices to further discuss the comparison with other XUV scaling laws (Appendix~\ref{sec:comparison}) and the relation between the \ion{He}{i}~10830 equivalent width and the XUV$_{\rm He}$ flux (Appendix~\ref{sec:heliumfit}).
Tables with the observing logs and main results are found in Appendix~\ref{sec:tables}.
In Appendix~\ref{sec:coronalmodels} we include the tables with X-rays and UV line fluxes, along with the figures and tables providing results of the coronal models.

\section{Observations}\label{sec:obs}

We were granted \textit{XMM-Newton} Director Discretionary Time (DDT, prop. IDs \#101704 and \#106917, PI Sanz-Forcada), and Guest Observer time (prop. ID \#092312, PI Sanz-Forcada) to observe a sample of planet-hosting stars suitable for a search of the \ion{He}{i}~10830 triplet with the Calar Alto high-Resolution search for M dwarfs with Exoearths with Near-infrared and optical Echelle Spectrographs \citep[CARMENES][]{qui14}. \textit{XMM-Newton} simultaneously operates two high-spectral resolution detectors \citep[RGS, $\lambda\lambda \sim$6--38\,\AA, $\lambda$/$\Delta\lambda\sim$100--500,][]{denher01} and three European Photon Imaging Camera (EPIC-PN and EPIC-MOS) detectors \citep[sensitivity range 0.1--15 keV and 0.2--10~keV, respectively, $E/\Delta E\sim$20--50,][]{tur01,str01}. \textit{XMM-Newton} also includes an optical monitor (OM). We did not use the OM data in this work because the light in its UV filters is severely contaminated with stellar photospheric emission \citep{ore23}. The data were reduced using the \textit{XMM-Newton} Science Analysis Software (SAS) v20.0, and analyzed following standard procedures within the Interactive Spectral Interpretation System \citep[ISIS,][]{isis} package. Most targets did not have enough statistics to use RGS, so EPIC data were used to obtain a discrete (one to three temperatures) fit to the spectra.

Some objects were also observed with the \textit{Chandra} X-ray observatory \citep{wei02}. The High Energy Transmission Grating Spectrograph (HETGS) contains two gratings, HEG (High Energy Grating, $\lambda\lambda\sim$1.5--15~\AA, $\lambda/\Delta\lambda\sim$120--1200), and
MEG (Medium Energy Grating $\lambda\lambda\sim$3--30~\AA, $\lambda/\Delta\lambda\sim$60--1200), which operate simultaneously,
permitting the further analysis of the data with different spectral resolutions.
The Low Energy Transmission Grating Spectrograph (LETGS, $\lambda\lambda\sim$3--175~\AA, $\lambda/\Delta\lambda\sim$60--1000)
was used in combination with the High Resolution Camera (HRC-S). The positive and negative orders were summed for the flux
measurements. Lines formed in the first dispersion order, but contaminated with contribution from
higher dispersion orders, were not employed in the analysis.
We also used the Advanced CCD Imaging Spectrometer (ACIS, $E/\Delta E\sim$20--50) with no grating.
Standard reduction tasks present in the CIAO v4.14 package were employed in the reduction of data retrieved from the \textit{Chandra}
archive and the extraction of the HEG and MEG spectra. All objects observed in X-rays are listed in Table~\ref{tabobslog}.
We complemented our sample with the objects in the database X-exoplanets\footnote{\url{http://sdc.cab.inta-csic.es/xexoplanets/jsp/homepage.jsp}} \citep[Sanz-Forcada et al. 2011, hereafter][]{san11}, which were reanalyzed for this work as explained in Sect.~\ref{sec:methods}.
New \textit{XMM-Newton} observations were included for a few targets of the X-exoplanets original sample (Table~\ref{tabobslog}): \object{GJ~436}, \object{GJ~674}, \object{HD~27442}, \object{HD~75289}, \object{HD~108147}, \object{HD~189733}, and \object{HD~190360}. In targets with more than one \textit{XMM-Newton} observation, we combined all the data to improve the quality of the EPIC spectra used in the fitting. Two objects formerly detected only with ROSAT, or with lower quality \textit{Chandra} spectra, have now \textit{XMM-Newton} observations (\object{GJ~832}, and \object{$\upsilon$~And}). In the case of \object{$\iota$~Hor} we used the coronal model from \citet{san19}. A few more targets were added to the sample, either because they are interesting M stars hosting exoplanets (e.g., TRAPPIST-1) because they have been surveyed searching for the \ion{He}{i}~10830 triplet or because they can be used to establish a better relation between X-rays and EUV, such as V1298~Tau \citep{mag23}.

%
\begin{figure*}
  \centering
  \includegraphics[width=0.45\textwidth]{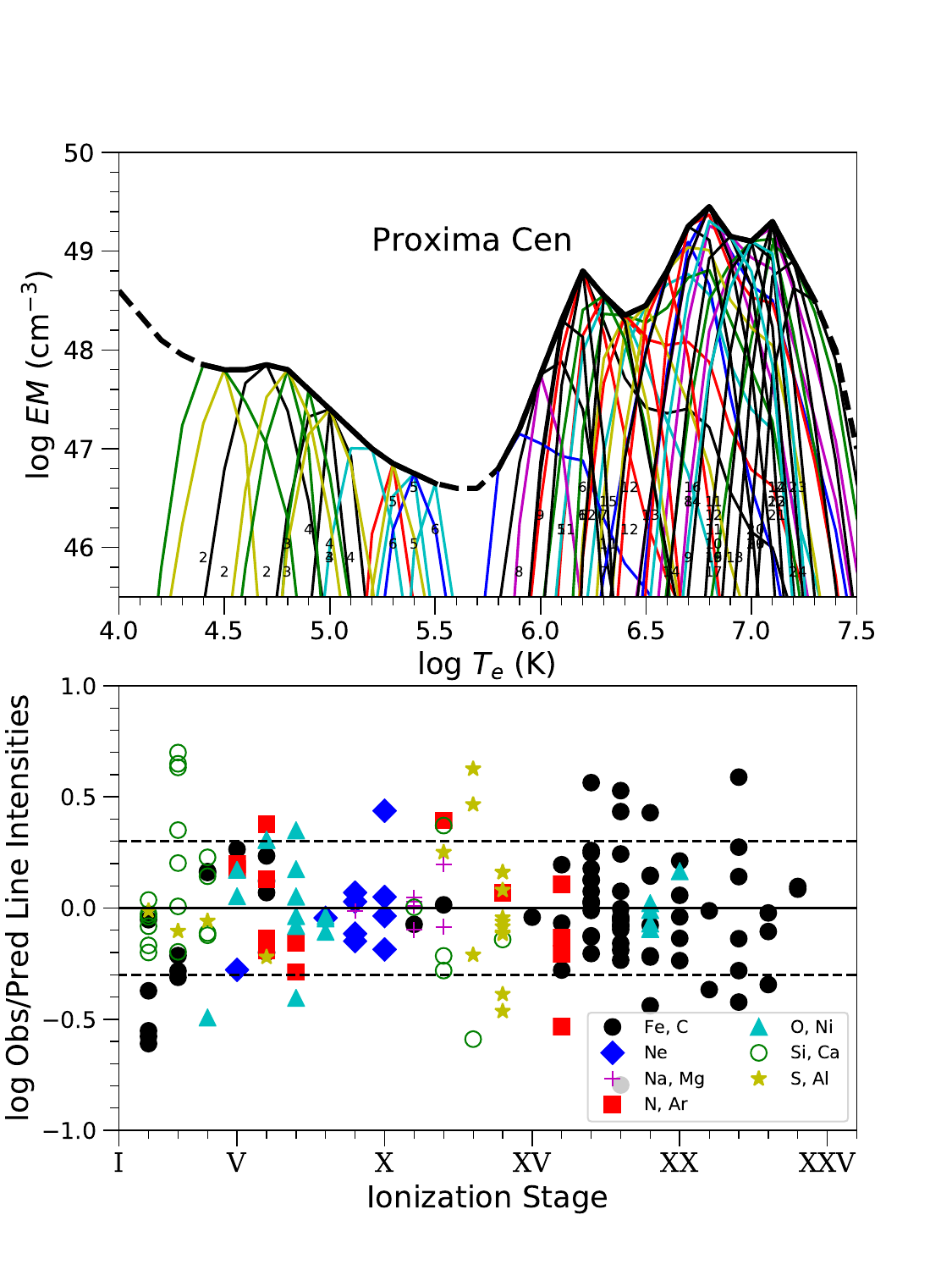}
  \includegraphics[width=0.45\textwidth]{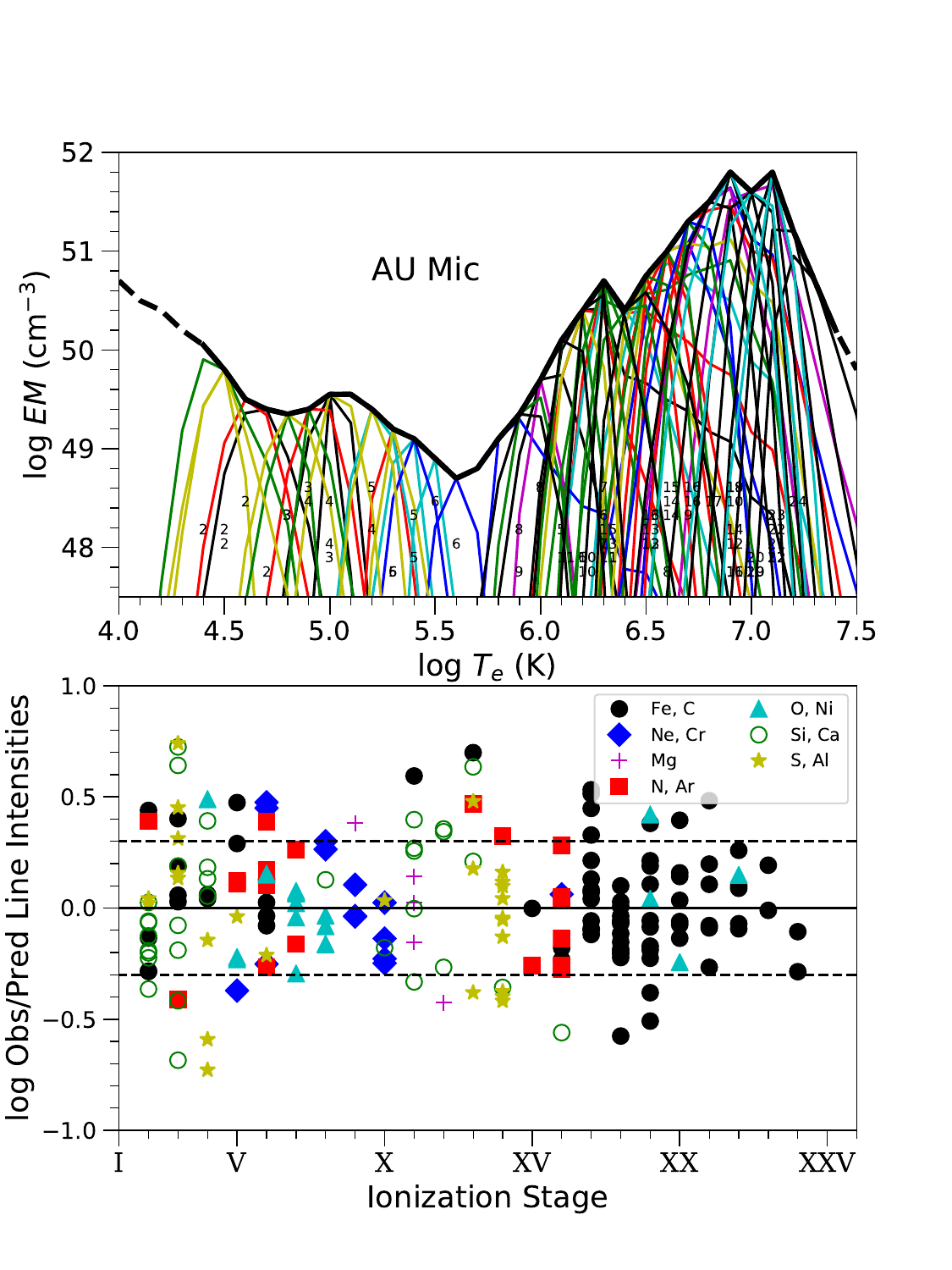}
  \caption{Coronal models of Proxima Cen ({\em left}) and AU~Mic ({\em right}), combining data from \textit{XMM-Newton}, EUVE, HST/STIS, and FUSE.
    {\em Upper panels}: Emission measure distributions (EMDs, thick line).
    The thin lines represent the contribution function for each ion (the emissivity function multiplied by the EMD at each point). The small
    numbers indicate the ionization stages of the species. {\em Lower panels}:  Observed-to-predicted line flux ratios for
    the ion stages in the upper panels. The dashed lines denote a factor of 2. }\label{fig:proxcen} 
\end{figure*}
%

Five special cases of M dwarfs, \object{Proxima~Cen}, \object{AU~Mic}, \object{AD Leo}, Lalande~21185 (\object{GJ~411}, \object{HD 95735}), and \object{GJ 674}, were analyzed in detail using high-resolution X-ray spectra from the \textit{XMM-Newton} or the \textit{Chandra} archives. The use of information from individual spectral lines greatly improves the quality of the coronal models, as shown by \citet{san03}. In these cases, we also included UV spectra (see below). Although the light curves of these targets indicate the possible presence of stellar flares, we preferred to use the combined data of quiescent and flaring states, as we understand that this reflects better the average activity of the star. In the case of Proxima~Cen, it was necessary to correct for the radial velocity of the different datasets to co-add the RGS spectra\footnote{The \textit{XMM-Newton} pointings were not always properly corrected for the proper motion of the star, resulting in small radial velocity shifts between different datasets.}.
Besides these five stars, a few targets for which we have only low-resolution \textit{XMM-Newton}/EPIC or \textit{Chandra}/ACIS spectra were analyzed including high-resolution UV spectra from: \textit{Hubble} Space Telescope (HST) Space Telescope Imaging Spectrograph \citep[STIS,][]{kim98}\footnote{Sensitivity ranges and resolution of the gratings: E140M 1144--1710\,\AA, $\lambda/\Delta \lambda=45,800$; G140L 1150--1730\,\AA,
  $\lambda/\Delta \lambda \sim$960--1440; G140M 1140--1740\,\AA, $\lambda/\Delta \lambda \sim$11,400--17,400} or Cosmic Origins Spectrograph
\citep[COS,][]{ost11}\footnote{Sensitivity ranges and resolution in the gratings: G130M 900--1450\,\AA,
  $\lambda/\Delta \lambda \sim$12,000--17,000; G160M 1360--1775\,\AA, $\lambda/\Delta \lambda \sim$13,000--20,000};
 Far Ultraviolet Spectroscopic Explorer \citep[{\em FUSE}, 905--1187\,\AA, $\lambda/\Delta \lambda \sim$15,000--20,000,][]{moo00};
and Extreme Ultraviolet Explorer \citep[{\em EUVE}, 70--750\,\AA, $\lambda/\Delta \lambda\sim$200--400,][]{hei93}, as listed in
Table~\ref{tabuvlog}. Extracted spectra from HST and FUSE were obtained through the Mikulski Archive for Space Telescopes (MAST), while EUVE spectra were reduced from the raw data to improve data quality, using the Image Reduction and Analysis Facility (IRAF) EUV standard tools.
We also add to this list the X-ray measurements from \object{$\tau$~Boo} \citep{mag11}, now complemented with HST/STIS line fluxes, to extend the coronal model to transition region temperatures (down to $T\sim 10^4$\,K).

The HD~209458 and HD~189733 data were reanalyzed since \citet{san10} to improve the results. In both cases, all the available \textit{XMM-Newton}/EPIC data are combined.
In the case of HD~209458 the data analysis was limited to the spectral range 0.3--8\,keV to improve the statistics of the spectral fitting, which is now characterized by S/N=3.2. The data of both stars were complemented with HST/COS, as detailed by \citet{lam20,lam21}. In the latter case, we also fixed a software bug that led us to overestimate the flux in the EUV range and further made a more accurate analysis of the abundance pattern in the entire temperature range. New updated values of broadband EUV fluxes are made available in Table~\ref{tabresults}. All these new fluxes supersede measurements formerly reported in CARMENES papers \citep[e.g.,][]{nor18,lam21,lam23} or in \citet{cha15} for the cases of AU~Mic and AD~Leo. 

%
\begin{figure*}
  \centering
  \includegraphics[width=0.45\textwidth]{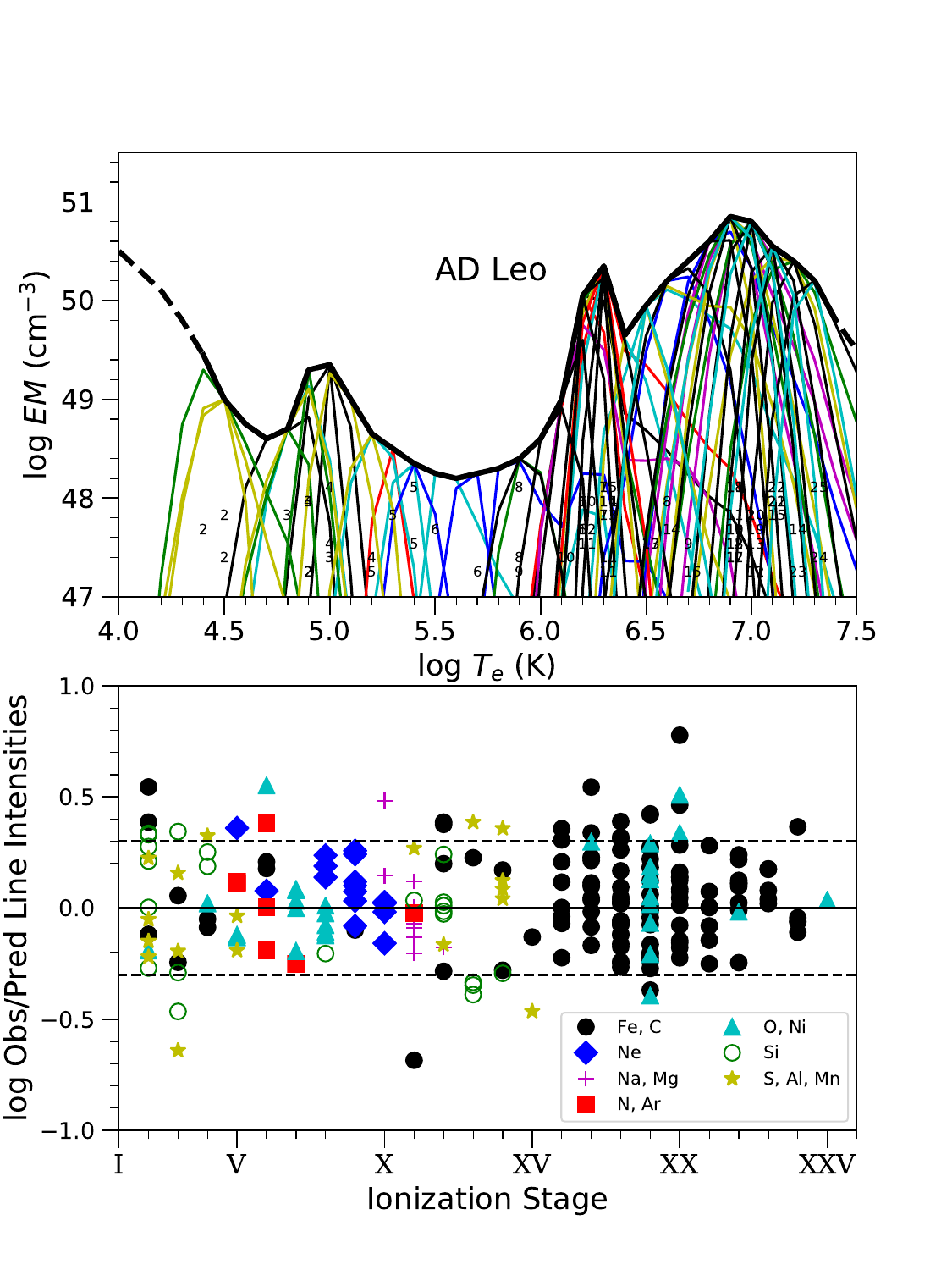}
  \includegraphics[width=0.45\textwidth]{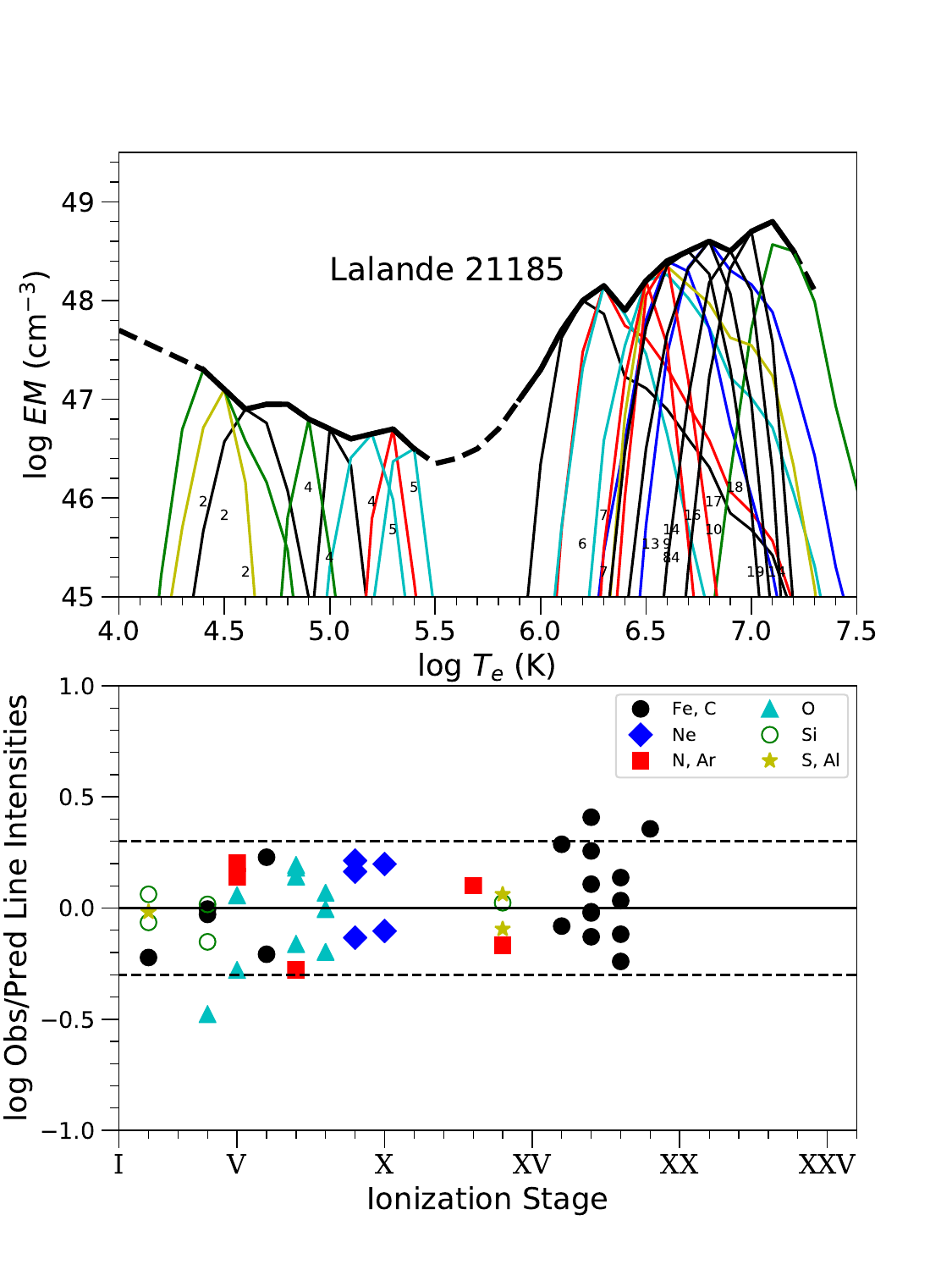}
  \caption{Same as in Fig~\ref{fig:proxcen}, but for AD Leo (GJ 388, {\em left}) and Lalande 21185 ({\em right}), combining data from \textit{Chandra}, EUVE, and HST/STIS (AD~Leo), and \textit{XMM-Newton} and HST/STIS (Lalande 21185).}\label{fig:adleo} 
\end{figure*}
%

\section{Methodology}\label{sec:methods}
We made use of coronal models to produce synthetic XUV spectra. To do so, we first calculated the volume emission measure ($EM$) at different coronal temperatures using the X-ray spectra, with $EM=\int N_{\rm e} \ N_{\rm H} \ {\rm d}V$ as defined by \citet{bri98}, where $N_{\rm e}$ and $N_{\rm H}$ are electron and hydrogen densities, respectively.
The ISIS package and the Astrophysics Plasma Emission Database
\citep[APED,][]{aped} were used to fit the low resolution spectra, and to measure spectral line fluxes in high resolution spectra. We  reanalyzed all the X-ray spectra included in \citet{san11} to account for the updated stellar distances provided by Gaia DR3 \citep{gaia23} and used ATOMDB v3.0.9 in the spectral fitting and further coronal modeling. The flux of the \ion{C}{iii} multiplet at $\sim 1176$ \AA\ is measured as one line, and its theoretical flux is evaluated using \citet{ray88} atomic data instead of ATOMDB\footnote{The fluxes of this multiplet, calculated with current version of ATOMDB, do not match those of the observed counterparts}. 
The X-ray luminosity was calculated in the 0.12--2.48\,keV band ($\sim$5--100\,\AA), similar to the ROSAT/PSPC standard band, by global fitting of the \textit{XMM-Newton}/EPIC or \textit{Chandra}/ACIS low resolution spectra. The spectral fit was used to calculate a coronal model (Table~\ref{tab:2tfits}). The interstellar medium (ISM) hydrogen column density was fixed, using values from ISMTool\footnote{\url{https://heasarc.gsfc.nasa.gov/cgi-bin/Tools/w3nh/w3nh.pl}} adapted to the stellar distances. The X-ray to bolometric luminosities ratio gives an indication of the activity level of the star \citep[e.g.,][]{piz03,wri11}.
The bolometric luminosity of the stars was calculated using the calibration by \citet{pec13} based on $G$ and $K_s$ magnitudes\footnote{\url{https://www.pas.rochester.edu/~emamajek/EEM_dwarf_UBVIJHK_colors_Teff.txt} (accessed in March 2023)}, for stars with spectral types earlier than K5. The calibration from \citet{cif20} was used for later spectral types, based on the $G$ and $J$ magnitudes. 

In  cases with X-ray high-resolution spectra, individual line fluxes were measured considering the point spread function (PSF) of the instrument, as described by \citet{san03}. The EUVE line fluxes were measured using standard IRAF software. The X-ray or EUV line fluxes were used to build the coronal model -- namely, the emission measure distribution (EMD) -- in the range of $\log T$(K)$\sim$5.8--7.4, while the UV line fluxes are employed for the transition region range, $\log T$(K)$\sim$4--5.7, including a few cases in which coronal lines were measured in the UV.  The spectral line fluxes measured in the spectra and the observed-to-predicted line fluxes ratio with the resulting EMD are listed in Tables~\ref{tab:fluxes1}--\ref{tab:uvfluxes4}. The EMDs are listed in Tables~\ref{tabemd1} and \ref{tabemd2} and shown in Figs.~\ref{fig:proxcen}--\ref{fig:adleo} and \ref{fig:tauboo}--\ref{fig:toi836}. In cases where no UV spectra are available, the coronal model is extended to the transition region assuming larger error bars for this temperature range, following \citet{san11}.
The technique employed to build the EMD is described in detail by \citet{san03} and references therein. The basic idea is that an initial EMD is proposed, with emission measure values at each temperature in a grid of 0.1 dex in the $\log T$\,(K)\,$\sim$\,4.0--7.5 range, and an initial set of atomic abundances; the EM is convolved with the emissivity function of each observed spectral line. The same operation is performed for any eventual blends that the observed line could have. The predicted line fluxes are then compared to the observed line fluxes and from this comparison, we changed the EMD to better match the observations in an iterative process. The abundances are calculated by doing this process for the lines of only one element (frequently, Fe) and incorporating the lines of other elements trying to partially overlap the temperature range covered by the new element and the former ones. The solution found through this process is then probed with a Monte Carlo method and modified when needed, to calculate the error bars associated with the EM at each temperature, letting the observed fluxes vary by up to $1\sigma$ in a large number of iterations.

We considered stellar abundances in the corona and transition region in the preparation of the coronal models (Tables~\ref{tab:2tfits}, \ref{tababund1}, \ref{tababund2}). In stars with low statistics X-ray spectra, solar photospheric abundances were used by default, except for the stellar photospheric [Fe/H] when available.
In the case of abundances obtained from the global fit and the EMD, we used Fe from the global fit, and the other elements from the EMD. The level of the EMD was then based on either C or Si. In the case of HD~189733, there is a \ion{Fe}{iii} line formed at transition region temperatures that could be used to fix the EMD level with coronal Fe abundance. As we were not confident in the quality of this line, instead we used  the coronal abundances of Si, Ne, and O to fix the level of the EMD.
Finally, we made use of coronal models to produce synthetic XUV SED, as detailed by \citet{san11}, using the ISIS software.
These SEDs in the 1--2800\,\AA\ range are made available in the X-exoplanets database, and in Table~1 as described in Sect.~\ref{sec:onlinedata}.
Although some individual line fluxes may not be correct given the actual level of accuracy of ATOMDB, we are confident on the overall correctness of these SEDs in the range covered; namely, 5--920\,\AA. The extension to longer wavelengths should be done with care because ATOMDB has not been sufficiently tested for some of the lines at these wavelengths (e.g., the multiplet of \ion{C}{iii} at 1176~\AA). An additional difficulty may arise for stars with substantial photospheric contribution in the UV, which becomes more important for F and G stars. Emission from plasma with photospheric temperatures is not generally covered by ATOMDB.

\section{Results}\label{sec:results}
We calculated the broadband stellar luminosity in two different EUV bands, affecting the H ionization (EUV$_{\rm H}$, $\sim$100--920\,\AA) and He ionization (EUV$_{\rm He}$, $\sim$100--504\,\AA), as listed in Table~\ref{tabresults}. The summed flux in the XUV bands; namely, X-rays (5--100\,\AA) as well as EUV$_{\rm H}$ and EUV$_{\rm He}$, were calculated at the planet orbital separation. We also calculated as a reference the mass loss rates, multiplied by the planet density ($\rho \dot M_p$), assuming the energy-limited approach and ignoring effects related to the mass transfer through the Roche lobe:
\begin{equation}\label{eq:mlr}
\dot M_{\rm XUV} \sim \frac{3 \, F_{\rm XUV,H}}{4\, {\rm G}\, \rho_{\rm p}}\, ,
\end{equation}
where $F_{\rm XUV,H}$ stands for the flux density in the 5--912~\AA\ spectral range, G is the gravitational constant, and $\rho_{\rm p}$ is the planet bulk density, following \citet{san11} and references therein.
Table~\ref{tabresults} includes both the stars in the sample of Table~\ref{tabobslog} and those available in X-exoplanets as listed by \citet{san11}. The latest were updated using new Gaia distances, as well as the new ATOMDB atomic models, as described above.

Details on some individual targets are provided, and the newly calibrated relation between X-rays and EUV flux is described next. We also explored the relation between XUV$_{\rm He}$ stellar irradiation onto the planet atmosphere and the observation of the helium infrared triplet (see Sect.~\ref{sec:heliumresults}).

\subsection{Notes on individual targets}
The sample studied in this work pays special attention to M dwarfs, the main focus of the CARMENES survey. We included some of the brightest M stars in X-rays, such as AU~Mic, Proxima~Cen, and AD~Leo, to obtain the best possible coronal models, so that they can be widely employed in planet atmospheric modeling. The brightest objects have plenty of lines to perform an adequate coronal model and calculate the coronal abundances. Some targets are faint in X-rays and their coronal abundance analysis is more complicated, introducing some uncertainty in the balance between the EM at lower (transition region) and higher (coronal) temperatures. Some details to be considered in individual cases are discussed here. In a few stars (\object{WASP-77}, \object{GJ~357}, \object{GJ~486}, \object{HD~149026}, and \object{TOI-836}), due to the scarcity of bright spectral lines in the COS and STIS ranges, we had to include lines with S/N\,$<$\,3 to prepare the transition region model. We included the candidate planets AD~Leo~b and Barnard's star~b \citep{tuo18,rib18} in the sample, although their detection has been questioned afterwards \citep{car20,lub21,kos22}.

The Barnard's star (\object{GJ 699}) coronal model, in its high temperature component, is based on the \textit{Chandra}/ACIS spectral fit, rather than on the two coronal lines measured with HST/STIS. Ne and N abundances are uncertain because we could not disentangle them from the EM values within the temperature range of line formation. 
Also, no relative abundances relative to Fe were calculated. Thus, the actual level of the EM in the transition region could be different. However, the observed coronal Fe lines are in agreement with the 1-$T$ fit to the \textit{Chandra} value, supporting an Fe abundance that should not deviate substantially from solar photospheric values.
The problem of uncertain abundances affects also \object{GJ~1214} in N and O, in a similar temperature range.

Some of the X-ray fluxes calculated in this paper differ from former publications \citep[e.g.,][]{lou17,kin18}. Sometimes this is due to the assumption of a different ISM absorption or coronal metallicity. We also avoid the fit of X-ray spectra below 0.3~keV, used in the mentioned work, to avoid instrumental problems such as noise in the low-energy end of EPIC spectra.
Our EMD results can also be compared with those by \citet{duv21}, based on a technique that uses smooth polynomials and solar photospheric abundances to fit a differential emission measure (DEM).
Their results are discrepant with ours, as discussed in Appendix~\ref{sec:comparison}. The use of global fits to get the DEM (or EMD) from high-resolution spectra was argued against the analysis of similar data from EUVE spectra \citep{bow00,fav03,gud04} because it tended to produce artificial features not supported by observed line fluxes or continuum \citep[e.g., Fig. 13 of][]{sch95}. This is a variant of the technique used by \citet{duv21}.
  A similar approach was adopted by \citet{lou17}. The use of smooth polynomials imposes constrains to the DEM shape, which may not correspond to the actual DEM of the star.
Another discrepancy arises when the \textit{XMM-Newton}/OM UV signal is used to scale the SED modeled at UV wavelengths \citep{zha23}: the UV filters of this instrument are contaminated by stellar photospheric emission from longer wavelengths. Therefore, they cannot be used to set the UV level of emission of  late-type main sequence stars \citep{kin18,ore23}.

%
\begin{figure*}
  \centering
  \includegraphics[width=0.49\textwidth]{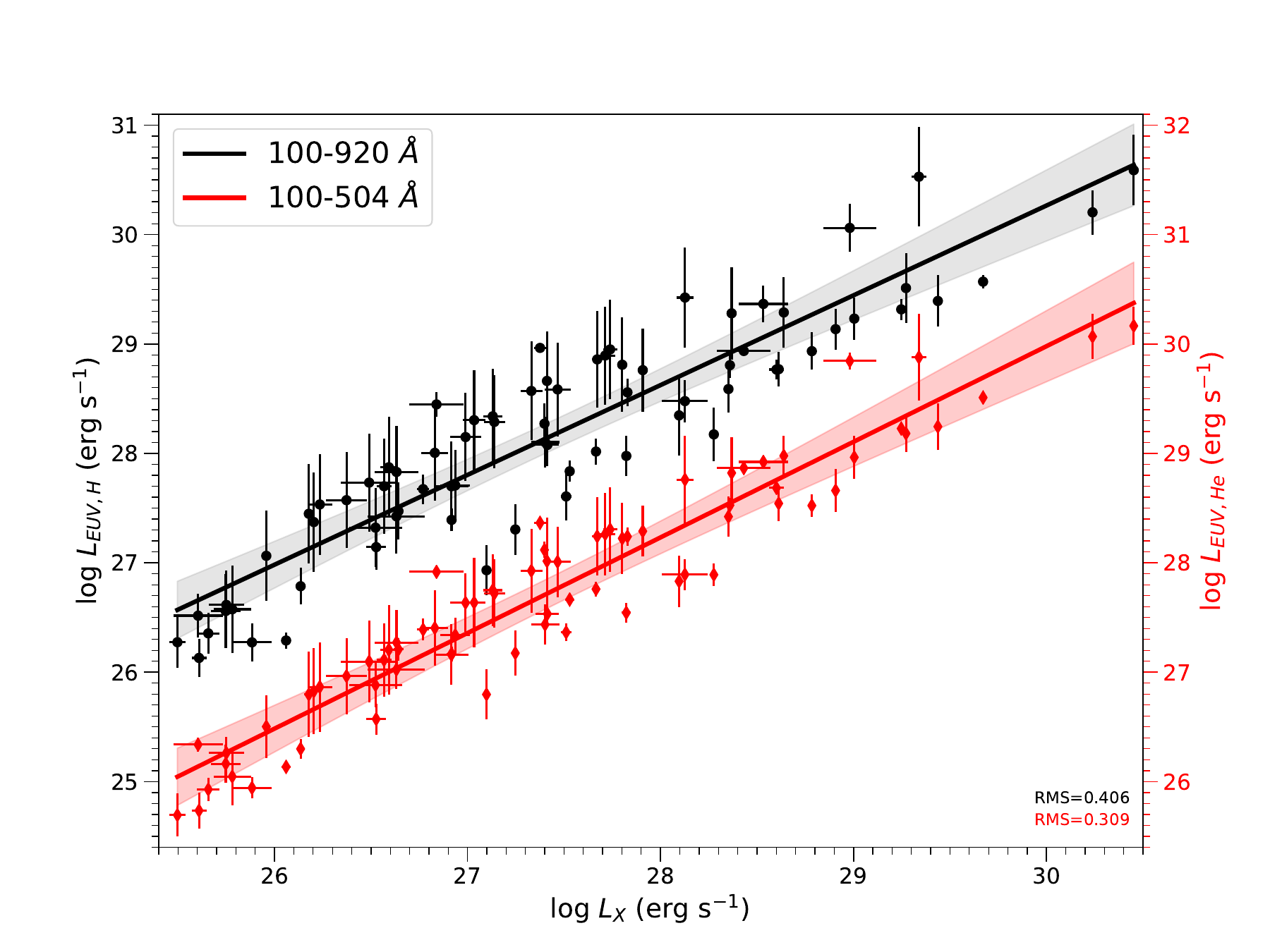}
  \includegraphics[width=0.49\textwidth]{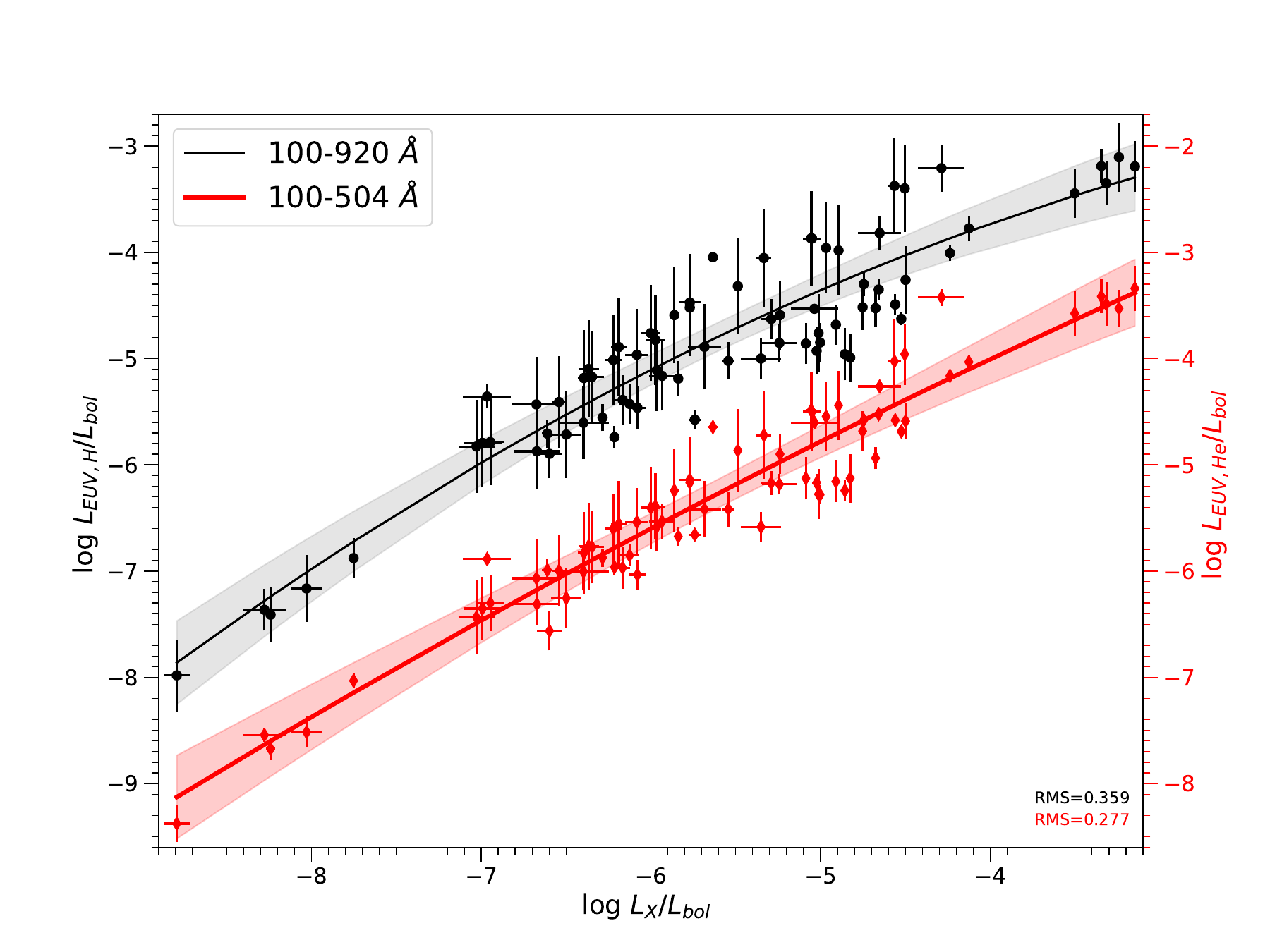}
  \caption{EUV vs X-ray luminosities in two different EUV ranges for the H (100--920\,\AA) and He ionization (100--504\,\AA). There are different scales for the two EUV luminosity ranges.
    {\it Left}: Linear fit over EUV against X-ray luminosity. {\it Right}: Quadratic fit over the luminosity ratios with the bolometric luminosity.}\label{fig:xvseuv} 
\end{figure*}
%

\subsection{X-exoplanets 1.1: New calibration}\label{sec:resultsxuv}
Although it is generally advisable to calculate an individual coronal model for each stellar target, the use of scaling laws is being extensively used in the literature as a first order approach to the problem of planet photoevaporation. We thus consider it important to update the relation of X-exoplanets \citep{san11} with a better coverage of the different stellar activity regimes. 

We calculated a new relation for the EUV$_{\rm H}$ and EUV$_{\rm He}$ against X-ray flux. This relation allows for a fast and reliable calculation of the flux in the EUV spectral ranges, where real data are not available.
The first relation of this kind was provided by \citet{san11}, who fitted a linear relation between $\log L_{\rm X}$ and $\log L_{\rm EUV, H}$, with some deviation from the linear fit at higher values. Such deviation is more evident for higher activity stars \citep{mag23}. When applying a similar linear fit to out current sample, this behavior is confirmed, yielding some data dispersion. The resulting linear fits are:
\begin{eqnarray}
  \log L_{\rm euvH} & = &(0.821 \pm 0.041)\, C_{\rm X} + (28.16 \pm 0.05)
\end{eqnarray}
\begin{eqnarray}
  \log L_{\rm euvHe} & = &(0.874 \pm 0.031)\, C_{\rm X} + (27.74 \pm 0.04)\, ,
\end{eqnarray}
\noindent where $C_{\rm X} = \log L_{\rm X}-27.44$, and they are valid in the range $\log L_{\rm X} \sim 25.5 - 30.5$. These fits have a Pearson's correlation factor $r=0.920$ and 0.957, and a standard deviation of the residuals (RMS) of 0.406 and 0.309, respectively.

The X-ray or EUV luminosities depend on the stellar activity, as well as on the stellar size. To remove the latter dependence, we  may use the stellar surface flux, with the associated uncertainties of the stellar radius (Appendix~\ref{sec:comparison}). Instead, we fit a relation between $\log L_{\rm X}/L_{\rm bol}$ and $\log L_{\rm EUV}/L_{\rm bol}$ (Fig.~\ref{fig:xvseuv}). The use of $L_{\rm bol}$ removes the effect of the stellar size. Linear fits to these data have $r=0.923$ and 0.960, and RMS=0.380 and 0.281, respectively. A substantial bremsstrahlung continuum contribution has an increasing importance with stellar activity, and this continuum is more relevant in X-rays than at EUV wavelengths. A linear fit does not account for the dependence of broadband fluxes on this continuum. Thus a better result is found when applying a quadratic fit: 
\begin{eqnarray}\label{eqxuvh}
  \log \frac{L_{\rm EUV,H}}{L_{\rm bol}} & = & (-0.062 \pm 0.016)\, (U_{\rm X})^{2} + \nonumber\\
  & & (0.763 \pm  0.037)\, U_{\rm X} + (-4.792 \pm 0.051)
,\end{eqnarray}
\begin{eqnarray}\label{eqxuvhe}
  \log \frac{L_{\rm EUV,He}}{L_{\rm bol}}&=& (-0.022\pm 0.016)\, (U_{\rm X})^2 + \nonumber\\
  & & (0.824 \pm 0.029)\, U_{\rm X} + (-5.268 \pm 0.039)\, ,
\end{eqnarray}
\noindent where $U_{\rm X}= \log (L_{\rm X}/L_{\rm bol}) + 5.60$. These results are valid for $\log L_{\rm X}/L_{\rm bol}\sim -3.1$ to $-8.8$, although the application of the equation for values lower than $\sim -7.2$ should be done with caution. Although some dispersion still exists, this approach shows a better fit (RMS=0.359 and 0.277, respectively).
The remaining data dispersion might be related to deficiencies in the coronal models, such as a lack of high-resolution spectral data or problems in the calculations of stellar abundances in the corona and transition region temperature range. These problems can be mitigated by using only stars with good high-resolution spectra, both in the X-ray and UV bands. This will be subject of a future work. An intrinsic problem is also the presence of coronal activity cycles, with an X-ray amplitude that can range from a factor of $\sim 2$ to a factor of $\sim 50$, depending on stellar activity levels. The two extreme cases observed to date are: $\iota$~Hor \citep[$\log L_{\rm X}/L_{\rm bol} \sim -4.9$;][]{san19} and the solar cycle \citep[$\log L_{\rm X}/L_{\rm bol} \sim -5.9$ to $-7.6$;][]{orl01}.
The problem of stellar cycles can only be solved by using the average value of an already known cycle, which is the case in very few stars. The problem is also mitigated if the UV data are taken contemporaneously to the X-ray data.

\subsection{Equivalent widths of the \ion{He}{i} 10830 detections}
To carry out a homogeneous analysis, we needed to revise some of the equivalent widths ($EW$s) of the previously published He detected planets with CARMENES. However, these changes were marginal. The $EW$s of planets HD~209458~b, HD~189733~b, \object{GJ~3470~b}, GJ~1214~b, \object{WASP-69~b}, and \object{HAT-P-32~b} were not changed and, thus, they were taken from Table 3 of \citet{lam23}.
The $EW$s were integrated in the range 10831.0$-$10834.5\,\AA\ (wavelengths in vacuum). 
For the $EW$ of \object{WASP-76~b}, the upper limit of 21.3 m\AA\ reported by \citet{cas21} was adopted instead of the value of 12.4~m\AA\ reported by \citet{lam23}. The latter value was obtained by integrating in a wider range, 10831.0$-$10835.5\,\AA, as the signal of this planet was significantly broadened and red-shifted.
Except as noted, the same procedure was used for the planets with detected signal. 

For \object{WASP-52~b}, the $EW$ was calculated from the model fit to the data performed by \citet{kir22}. That fit shows an offset of 0.1\%, which was subtracted. \citet{kir22} also detected a small and rather noisy signal of \object{WASP-177~b} and used it to fit their model. The $EW$ for the measured spectrum integrated in their spectral range is 5.8~m\AA, while that for the model is 7.5~m\AA. We took the mean value of 6.65~m\AA. 
For \object{HAT-P-18~b}, the $EW$ was obtained from the absorption depth of 0.46\% reported by \citet{para21} integrated over their bandpass of 6.35\,\AA.
The $EW$ of \object{HD\,235088\,b} (TOI-1430\,b) was reported by \citet{ore23} with a value of 9.5$\pm$1.1\,m\AA, while our value obtained by integrating over the usual 10831.0--10834.5\,\AA\ spectral range is slightly larger, 11.1~m\AA. \citet{zha23} reported a value of 6.6$\pm$0.5\,m\AA\ in a different observation. 

\citet{zhan22} reported a value for the $EW$ of 8.6$\pm$0.6~m\AA\ in \object{HD~73583~b} (TOI~560~b). From their figures, by integrating over the spectral range described above, we obtained 7.43~m\AA. 
In the case of \object{TOI-1268~b,} we faced a similar problem. The $EW$ by \citet{mopys} is 19.1~m\AA. The value obtained with the method described above is 17.7~m\AA. We note that in this case the upper limit of the integration was slightly smaller, 10834.2\,\AA, due to a lack of data. The same remarks apply to \object{TOI-2018~b}, where we calculated 6.8~m\AA\ instead of the value of 7.8~m\AA\ from \citet{mopys}. 
 
\object{HAT-P-67\,b} seems to show a rather variable He(2$^{3}$S) absorption (see, e.g., \citealt{bel23} and \citealt{gul24}). The CARMENES measurement, reported by \citet{bel23} in their Fig.~6b, seems to have a problem as the absorption profile shows
an unrealistic emission-like feature at wavelengths shorter than 10832.5~\AA. That feature was likely caused by their lack of a realistic out-of-transit baseline, which prevented them from making an appropriate normalization. That is, it could be due to an increased \ion{He}{i} absorption at bluer wavelengths towards the end of the transit.
This planet was observed with a longer baseline by \citealt{gul24}. Hence, we decided to include these measurements in our analysis. 
By using our method for calculating the EW, we derived from their measurements shown in Fig. 13 (the in-transit mean taken in the visit of May 2020) a value of 147~m\AA. In this case, we consider  the larger value of 25~m\AA\ that covers the significant variability shown by the planet as ``error'' of the EW and not the real
measured uncertainty (see the mid-transit values in  Fig. 9 of the work cited above).
It is worth  noting that \citealt{gul24} measured a rather large EW of $\sim$330~m\AA\ at the mid transit, as well as a significant pre-transit absorption of $\sim$200~m\AA\ (see their Fig. 9). 
They stated, however, that ``only a small fraction of the \ion{He}{i} excess signal would be expected to trace the planet's motion.'' 
Here we focus on the absorption of the \het\ that is being ejected from the atmosphere, not in the \het\ that has already been ejected and forms the cloud around the planet. Then, we can estimate this absorption by taking the difference of the EW at the mid transit, from that of the material already expelled (pre-transit). This results in an absorption of $\sim$130~m\AA, which is very close to the adopted value of 147~m\AA\ for the mid-transit measured in May 2020.

For \object{HAT-P-26\,b}, the $EW$ was obtained from the absorption depth of 0.31\% reported by \citet{vis22} integrated over their full width half maximum (FWHM) bandpass of 6.35\,\AA. We obtained a value of 
19.7$\pm$6.4~m\AA.
The warm super-puffy \object{TOI-1420\,b} planet was observed by \citet{vis24}, who obtained a high (8.5$\sigma$) He triplet signal, with an absorption depth of 0.671\%$\pm$0.079\%. With those values and taking the bandwidth of their filter of 6.4\,\AA, we obtained an $EW$ of 42.9$\pm$5.1~m\AA.

%
\begin{figure*}
  \centering
  \includegraphics[width=\textwidth]{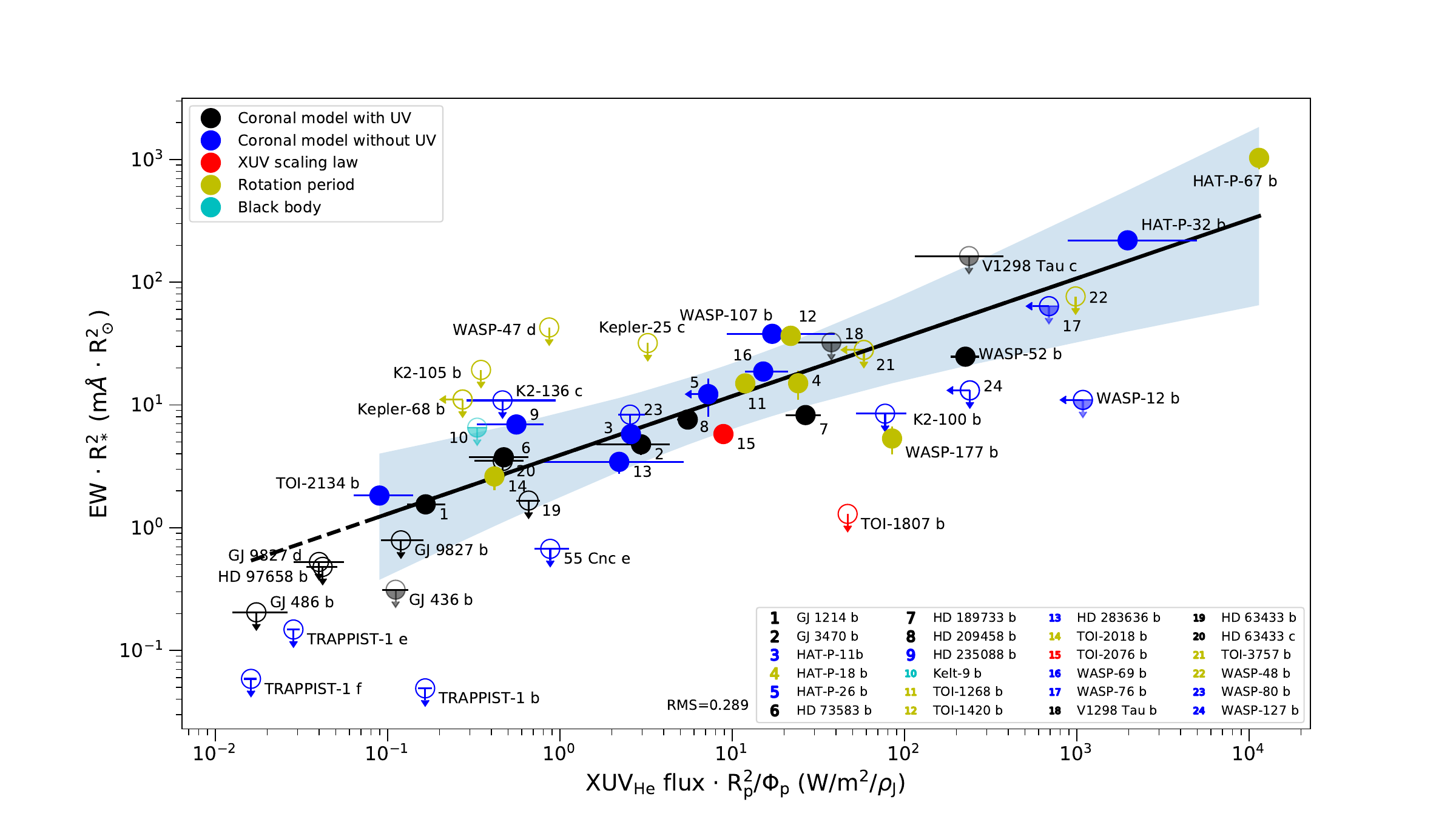}
  \caption{\ion{He}{i}~10830 triplet equivalent width, multiplied by the stellar area, plotted against $F_{\rm XUV,He}$, weighted by $\phi_{\rm p}$/$R_{\rm p}^2$, i.e., by the planet density. The $1\sigma$ error bands to the fit, calculated only with positive detections (filled symbols), are displayed in light blue. Different colors indicate the method used to calculate the XUV flux in the 5--504~\AA\ range (see Sect.~\ref{sec:heliumresults}). Half-filled symbols are used for the equivalent width upper limits calculated by our team, while those calculated by other authors are shown in open symbols.}\label{fig:xuvvsHe} 
\end{figure*}
%

\subsection{The relation between \ion{He}{i} 10830 and XUV ionizing irradiation}\label{sec:heliumresults}
It is expected that the formation of the \ion{He}{i}~10830 triplet in an exoplanet atmosphere is related to the stellar irradiation on the planet by photons capable of ionizing the neutral helium atoms; namely, those with $\lambda \la 504$\,\AA. The actual process of formation of these lines requires also that electrons are available in the plasma for the recombination of He atoms after the radiative ionization, and the main origin of these electrons is the radiative ionization of H atoms by photons with $\lambda \la 912$\,\AA\ (we used 920~\AA\ as ionization edge in our model to cover a slightly wider range).
Furthermore, the stellar irradiation in this wavelength range also affects the He triplet concentration through the total density. As that radiation is responsible for the ionization of the H atmosphere, it changes the mean molecular weight and, hence, through the hydrodynamic equations, it affects the density.
The stellar irradiation also affects the state of the upper atmosphere through the cooling and heating processes, which, in turn, affect the temperature and hence the hydrodynamic and the density.
In addition, a third variable must be considered: the ionization of the ground level of orthohelium atoms, which takes place for $\lambda \la 2600$\,\AA\ \citep{nist} and thus depends on the photospheric emission of the star. Although a combination of these three stellar fluxes should be considered, the right balance between the effects of the fluxes in the three bands is difficult to establish in a general way for all stars. Thus, we took the flux in the XUV range up to 504~\AA\ (XUV$_{\rm He}$) as a proxy of these three processes, as it is expected to be the dominant term and, in a large number of stars, this flux is well correlated with XUV$_{\rm H}$.
Some attempts to derive such dependence have been undertaken\footnote{\url{https://zenodo.org/records/13986513}}.

To further test this hypothesis, we collected information on the detections of the \ion{He}{i}~10830 triplet reported to date (see below) to compare them with the XUV$_{\rm He}$ flux (Table~\ref{tabhelium}).
However, not all the targets in the list have a complete coronal model to calculate the XUV contribution. For those cases we calculated the flux by indirect means. The quality of the XUV$_{\rm He}$ fluxes calculated decreases in this order (as noted in Fig.~\ref{fig:xuvvsHe}): (1) cases with both X-ray and UV spectra, resulting in a complete coronal model; (2) targets with X-ray spectra, but no UV spectra (the coronal model with lower temperature section is extrapolated); (3) X-ray bulk flux available, such as ROSAT fluxes, and the XUV$_{\rm He}$ flux is calculated using the scaling law in Eq.~\ref{eqxuvhe}; (4) X-ray flux determined from the rotation period and the X-ray vs. rotation relation by \citet{wri11}, with subsequent calculation of XUV$_{\rm He}$ flux; and (5) calculation of the XUV$_{\rm He}$ flux based on blackbody emission of the stellar photosphere,
for a star with unlikely or negligible coronal emission (KELT-9)\footnote{We are more confident on our calculation assuming a blackbody emission than on the X-ray flux upper limit provided in Table~\ref{tabresults}. The XUV$_{\rm H}$ flux in this star is more than four orders of magnitude higher than XUV$_{\rm He}$ flux if we assume a blackbody emission.}.
We also included  upper limits for both the He detection and XUV flux.
In case (2) the low temperature part of the coronal model is extrapolated from the high temperature section \citep{san11}; upper and lower boundaries are considered during this extrapolation; thus, the XUV upper limits are based on the upper boundary coronal model,
rather than on the central value used to evaluate the SED in different contexts, such as the case of WASP-12 of \citet{cze24}. The separate fits for the positive detections in the groups (1), (2), and (4) show consistency within the error bars with the general fit.
In addition to the expected dependence of the He absorption on the stellar XUV irradiation as discussed above, we also empirically evaluated  its dependence on other star-planet system parameters. The detailed discussion is found in Appendix~\ref{sec:heliumfit}. Here, we include a summary of that formulation and the results. 

The first point is that we choose as a proxy of the \het\ absorption the $EW$ of the absorption profile, instead of the depth (peak) of such profile, as chosen in some previous studies. The reason is that we are interested in relating the bulk \het\ absorption by the planet atmosphere, and the $EW$ is a better proxy since it is independent of the width of the absorption profile; namely, it is independent of its potential broadening by Doppler temperature and by the radial outflows of the escaping atmosphere.

The second aspect is the star's surface. The transmission (or, equivalently, the $EW$) is measured as a ratio of the equivalent \het\ absorbing area of the planet atmosphere (usually expressed as the surface of a ring) and the area of the stellar disk ($R_*^2$). If we are interested in the properties of the \het\ absorption of a given planet atmosphere independently of the size of its host star, it seems reasonable to de-scale the measured $EW$ by the stellar disk area; that is, to consider $EW \cdot R_*^2$ instead of just $EW$.

An additional parameter that we considered is the planet gravitational potential, $\Phi_{\rm p}$,
as suggested by, for instance, the theoretical model of \citet{Salz2016} and the analysis of the \het\ absorption in diverse planets carried out by \citet{lam23}.
According to those studies the $EW$ is expected to be inversely proportional to the gravitational field because for a planet with a weaker gravitational potential, the atmosphere is escaping more easily, leading to a more extended atmosphere and, hence, a larger absorption.

A further parameter that we considered is a geometric factor related to the planet size (see Appendix~\ref{sec:heliumfit}), included in a general way with $R^{\gamma}_{\rm p}$, where $\gamma$ is likely to vary between 1 and 3 depending on the star-planet system. For planets with very compressed atmospheres as, for example, the case of HD\,189733\,b \citep{sal18, lam23}, the $EW$ is expected to be proportional to $R_{\rm p}$ ($\gamma$=1). However, for those with very extended and flat \het\ distributions, for instance, GJ\,1214\,b  \citep{lam23}, the exponent $\gamma$ could be larger than 2.
Thus, the relationship that we considered generally takes the form (see Eq.\,\ref{eq:ew_total}):\begin{equation}
  EW\cdot {R_{\star}^2} \approx \frac{F_{\rm XUV}\ R_{\rm p}^{\gamma}}{\Phi_{\rm p}}. \label{eq:ew_total2} 
\end{equation}

We performed several tests for different relationships between $EW$ and XUV fluxes, as well as for Eq.\,\ref{eq:ew_total2}, using different values of the exponent $\gamma$ of $R_{\rm p}^{\gamma}$ (Table~\ref{tab:allhefits}).
The $EW$ values collected for all planets with detected \het\ signal and the XUV$_{\rm He}$ fluxes and other parameters listed in Table~\ref{tabhelium} were used.  The stellar spectral type is also listed in the Table to alert for the possible photospheric contribution at wavelengths below 2600\,\AA.

The results for the simple relationship between $EW$ and $F_{\rm XUV}$ (see Fig.\,\ref{fig:xuvvsHe1}) yields a decent Pearson's correlation coefficient of $r=0.579$, but clearly inferior to other tests described below. The correlation coefficient of the fit largely improves to a value of  $r=0.782$ when scaling the $EW$ by the area of the star's disk (see Fig.\,\ref{fig:xuvvsHe2}).
A further test was performed by using Eq.\,\ref{eq:ew_total2} with a value of $\gamma$\,=\,1, namely, $EW$ proportional to $R_{\rm p}$ (see Fig.\,\ref{fig:xuvvsHe2}).
The correlation coefficient obtained in this case is even better, reaching the value of $r=0.891$. A subsequent test was performed by considering $\gamma$\,=\,2, namely, $EW$ proportional to $R^2_{\rm p}$ (see Fig.~\ref{fig:xuvvsHe}), yielding a slightly better correlation than in the previous case, $r=0.898$.
We then proceeded to use the latter fit, which resulted in an empirical equation, excluding upper limits, of
\begin{eqnarray}\label{eq:ewhe}
  \log (EW \cdot R_*^2) &=& \alpha \cdot
  \Bigg(\log \bigg(\frac{F_{\rm XUV,He} \cdot R_{\rm p}^2}{\phi_{\rm p}}\bigg) - 0.96\Bigg) + \beta \, ,
\end{eqnarray}
\noindent where $\alpha = 0.480 \pm 0.054$ and $\beta = 1.052 \pm 0.066$. 
The $EW$ is given in m$\AA$, the stellar radius, $R_*$, in $R_\odot$ units, $F_{\rm XUV,He}$ in W/m$^2$, and $R_{\rm p}^{2}/\Phi_{\rm p}$, which is equivalent to the inverse planet density, $\rho_{\rm p}^{-1}$, in Jovian units. Error bars include the uncertainties in all of these quantities.

\section{Discussion}\label{sec:discussion}
The effects of high-energy stellar irradiation on the exoplanet atmospheres has been subject of increasing interest in recent years. The effects on the planet atmospheres are not limited to the formation of lines such as the \ion{He}{i}~10830 triplet, but they also affect large scale phenomena such as atmospheric mass loss (atmospheric escape).
The actual mass loss rate of a planet depends also on other variables different from stellar irradiation, such as atmospheric composition, stellar winds, or presence of planetary magnetic fields, and it requires an individualized approach to every case \citep[e.g.,][]{kub18}. However Eq.~\ref{eq:mlr} is widely used to roughly compare the expected planet evolution of different cases. Since atmospheric photoevaporation should lead to atmospheric instability, a planet with a current high mass loss rate is unlikely to be habitable.  The approach in Eq.~\ref{eq:mlr} assumes that the atmosphere is based mainly on H, which can be appropriate for Jovian-like planets, but it may not be applicable in the case of rocky planets such as \object{Teegarden's star\,b} or TRAPPIST-1\,b.
In any case, modeling or analyzing planet atmospheres requires (as accurate as possible) knowledge of XUV stellar irradiation. With this work, we are providing new XUV synthetic SEDs for planet-hosting stars, which represents an improvement with respect to the X-exoplanets first release \citep{san11} by updating the stellar coronal models and the atomic data used to model the emission from corona and transition region.
However, the number of planets hosting stars that have accurate X-ray spectra is quite limited. For those cases that do not, we include new XUV scaling laws that can be easily applied even in absence of an X-ray detection (e.g., by using the rotation period of the star, as explained in Sect.~\ref{sec:results}). For more in-depth studies, we advise users to select a star with a similar stellar spectral type and $\log L_{\rm X}/L_{\rm bol}$ as a proxy. A more detailed work providing a grid of models of this kind will be released in the near future.

\subsection{XUV scaling laws}
Early works dealing with the planet photoevaporation problem assumed that the XUV emission from exoplanet host stars would be similar to the solar spectrum, scaled by the size of the star \citep{lec07}. Alternatively, they have made use of the UV flux from the Sun and a young star for their calculations \citep{mur09}. A slight variation was applied by \citet{lin14}, \citet{fra16}, and \citet{sre20}, who used the solar EUV SED scaled by the overall emission in X-rays and UV of the stars. Their models were subsequently used, for instance, in \citet{vis22}, but their XUV fluxes are discrepant with ours by up to a factor of six. This approach is deficient, given that EUV spectra of active stars can be very different from those of quiet stars like the Sun \citep[e.g.,][]{san02b}. \citet{nam23} used the stellar magnetic flux and the solar spectra to extrapolate the XUV flux in active solar-like stars. The relation is calibrated mainly based on solar data, with some data coming from more active stars, along with additional uncertainties as the stellar magnetic flux is required. 

A different approach was introduced by \citet{san11}, including coronal models built with X-rays and UV data to create synthetic SEDs in the non-observed EUV spectral range. A relation between the transition region and the corona was calibrated using coronal models based on EUVE and International Ultraviolet Explorer (IUE) data, then used to complement the coronal models of stars with no UV data available. One of the results of this approach was a scaling law between X-rays and EUV luminosity that could be applied for cases where only a broadband X-ray flux is known. \citet{cha15} tried to apply a more physical relation between X-rays and EUV emission by considering the activity level of the star (using the surface stellar flux, $F_{\rm X}$ and $F_{\rm EUV}$) in six stars, one of them based on actual solar data. \citet{kin18} followed a similar approach (without defining the $F_{\rm X}$ used) using the same stellar data from three stars, along with solar data with some corrections applied with respect to \citet{cha15}. In both cases the sample was small, with most information taken from the Sun and a few very active stars. In addition, the use of stellar radius introduces further uncertainties to the problem.
\citet{jon21} calibrated a relation between X-rays (5--124\,\AA) and the EUV surface flux in the range 100--360\,\AA\ in a sample dominated by active stars, where a new set of solar data was included.
The stellar data in this case came from EUVE spectra, but with some spectra having a low quality above 180~\AA\ \citep{san02b}, and being usually contaminated by geocoronal emission around the \ion{He}{ii} $\lambda$304\,\AA\ line.
As in the other cases, this calibration also propagates the uncertainties of the stellar radii. For the rest of the EUV range, 360--920\,\AA, \citet{jon21} extrapolated the ratio between these two EUV bands in the Sun to the entire range of stellar activity. While plasma in X-rays and in the EUV 100--360\,\AA\ range is mainly made of spectral lines (and continuum) formed above $10^6$\,K, the 360--920\,\AA\ flux has a substantial contribution of lines formed at lower temperatures.
Thus, it is not evident that the solar spectrum paradigm can be used for the active stars. The range of activity levels explored by \citet{jon21} included only a few low-activity stars, which are, nonetheless, still more active than some of our stars. Further comparison with these three relations is shown in Appendix~\ref{sec:comparison}.
More recently \citet{kri24} made use of the relation between age and X-ray luminosity applied by \citet{san11} to calculate the accumulated XUV flux over time on a planet atmosphere.
While the X-ray luminosity can be used, in general terms, to estimate an approximate stellar age, the opposite is not advisable. Obtaining an accurate measurement of a stellar age is a difficult task in astrophysics; thus, using a poorly determined stellar age to calculate the X-ray and  EUV fluxes may end up accumulating large uncertainties. We prefer to calculate the X-rays flux from stellar rotation if no actual measurement is available. 

The new scaling law that we provide in Eq.~\ref{eqxuvh} and Fig.~\ref{fig:xvseuv} solves the limitations of past works by using coronal models for the calibration of the relation. It also uses variables that also depend on the stellar activity level to sample better the different ranges of activity. This new relation overcomes the observed problems with high activity stars \citep{mag23} and covers a wide range of activity levels. We also calculated a new relation in the range that affects the neutral helium ionization, namely, $\lambda < 504$\,\AA\ (Eq.~\ref{eqxuvhe}). This allows us to better estimate the impact of the stellar high-energy irradiation on the line formation of the planetary \ion{He}{i}~10830 triplet. Further comparisons with other recent works in the literature are provided in Appendix~\ref{sec:comparison}.

%
\begin{figure}
  \includegraphics[width=0.52\textwidth]{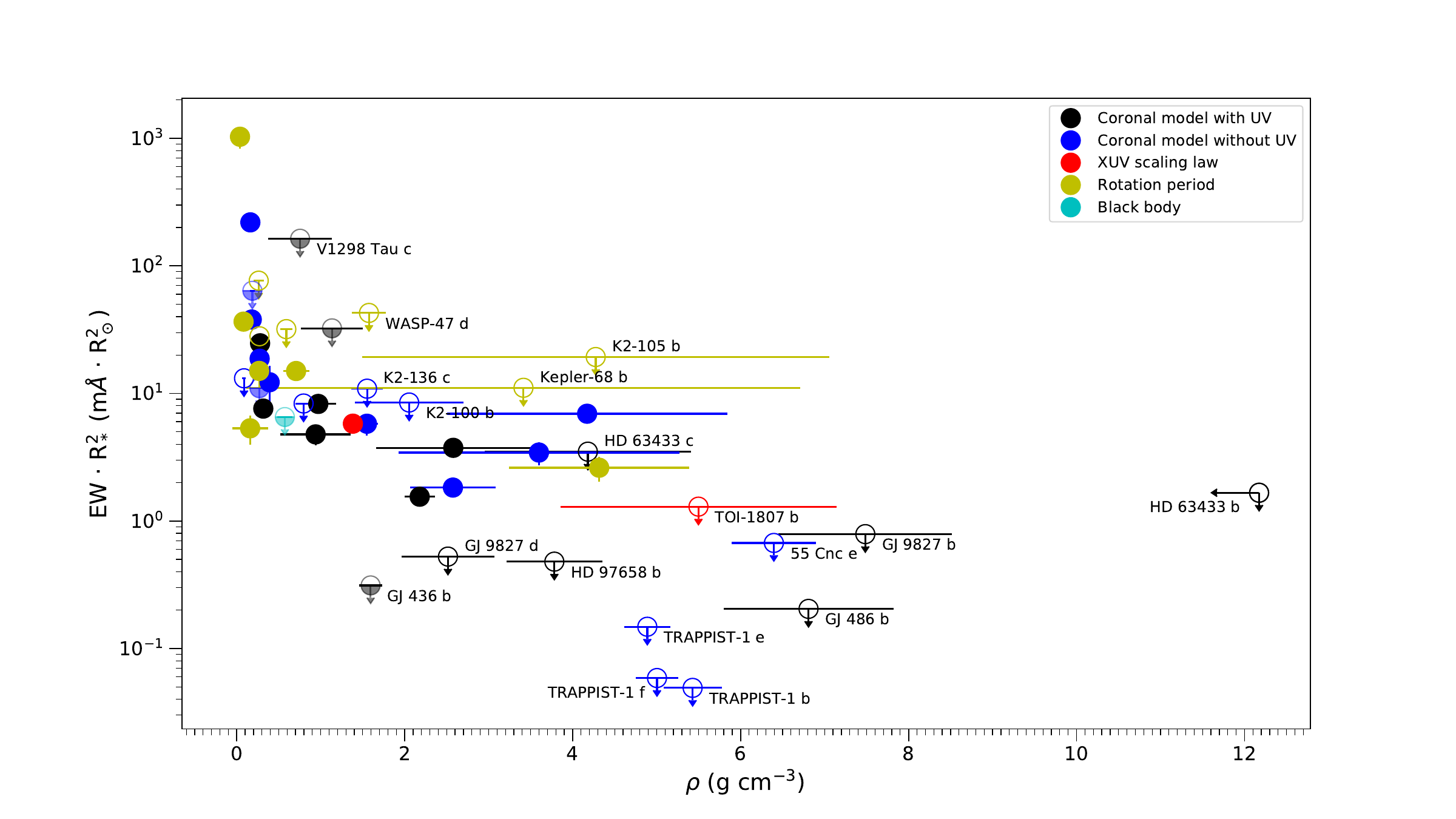}
  \caption{\ion{He}{i}~10830 triplet equivalent width, multiplied by the stellar area, plotted against planet density $1\sigma$ error bars are indicated. Same colors and symbols to Fig.~\ref{fig:xuvvsHe} apply for an easier identification. Labels are included for some planets detected as upper limits (see Sect.~\ref{sec:discusshe}).}\label{fig:density} 
\end{figure}
%

\subsection{The empirical relation between XUV and \ion{He}{i} 10830}\label{sec:discusshe}
Earlier attempts to find an empirical relation between \ion{He}{i}~10830 equivalent width and XUV (either at the H or the He ionization edges) have  only indicated a trend, but not a clear relation \citep{nor18,fos22,kir22,all23}. To evaluate the relation, these authors used $\delta R_{\rm p}/H_{\rm eq}$, which is the amplitude of excess helium absorption observed, in units of the planets' atmospheric scale heights, against the XUV stellar flux at the planet.
A further study by \cite{zha23c} used a different approach. They did not use the atmospheric scale height, which strictly speaking, corresponds to hydrostatic but not to hydrodynamic conditions. Instead, they plotted the mass-loss rate estimated from observations versus the theoretical energy-limited mass-loss rate (i.e., assuming a heating efficiency of 1), where the former were assumed proportional to $R_{\rm star} \cdot EW$ and the latter to $F_{\rm XUV,H}/\rho_{\rm XUV}$, where $\rho_{\rm XUV}$ is the density of the planet calculated with the planet radius of XUV absorption ($R_{\rm XUV}$). In that way, they reached a rather good correlation.

To reduce the data dispersion in those relationships, we sought alternative parametrizations.
The \ion{He}{i}~10830 equivalent width and the XUV$_{\rm He}$ irradiation cannot directly be related with each other, as this relation neglects the impact of the stellar radius on the depth of any absorption signals measured, and the impact of the planet radius on the irradiation energy received by the planet. We therefore used
the $EW$ multiplied by stellar area, and the XUV$_{\rm He}$ flux multiplied by the planetary disk area and divided by the planetary gravitational potential (equivalent to a division by the planet density; see Sect.~\ref{sec:heliumresults}, Fig.~\ref{fig:xuvvsHe}, and Table~\ref{tabhelium}).

Although the observed empirical trend is clear, some dispersion is still present in the data. One reason could be the contribution of photospheric flux with $\lambda < 2600$\,\AA, which ionizes the ground level of the He triplet, diminishing the expected absorption of the line. However, no deviation related to the spectral type is evident. Other relevant factors are likely related to the atmospheric composition, whether there is a primordial atmosphere, mainly composed of H and He, or the current He content is lower.
Some outliers (the \object{TRAPPIST-1} planets, \object{55 Cnc~e}, \object{TOI-1807~b}, GJ~486~b, \object{GJ~9827~b}) lie below the He level that would be expected for their XUV irradiation: because of their Earth- and super-Earth size and high densities (Fig.~\ref{fig:density}, Table~\ref{tabhelium}), they may be fully rocky without an atmosphere or they may have an atmosphere that is rather tenuous; otherwise, it may be thick, but with a high mean molecular weight. Some upper limits correspond to sub-Neptunian planets (GJ~9827~d, GJ~436~b, \object{HD~97658~b}, \object{K2-100~b}), which might have atmospheric conditions and chemical compositions that complicate the helium detection with current instrumentation.
From Figs.~\ref{fig:xuvvsHe} and \ref{fig:density}, we identify that among planets with density lower than 2~g\,cm$^{-3}$, 15 out of 28 planets were detected ($54\pm 4$\%).
Yet another physical reason that could explain the dispersion observed is stellar variability. The ionization levels of \ion{He}{i} depend on the stellar XUV emission, which is highly variable, especially during flares. However, the XUV level is based on X-ray observations that are not simultaneous to the \ion{He}{i}~10830 observations.
An issue that is more related to the data acquisition and analysis is the difficulties in the measurements of exoplanets orbiting distant (i.e., faint) stars. In the particular case of WASP-177~b, \citet{kir22} had some problems with systematics in the measurement of the He triplet.
Another source of dispersion could be the actual dependence on the planet radius or surface, as explained in Appendix~\ref{sec:heliumfit}.

\section{Conclusions}\label{sec:conclusions}
Planet atmospheric photoevaporation takes place due to XUV ($\lambda \la 920 \AA$) stellar irradiation. The \ion{He}{i}~10830~\AA\ triplet is used to study this phenomenon. The formation of the line in a planet atmospheric environment follows a process of ionization by photons with $\lambda$<504~\AA, followed by a recombination with electrons that are mainly the result of hydrogen ionization (by photons with $\lambda$<912~\AA). The main problem to evaluate the stellar flux at those wavelengths is the absorption of these photons by the ISM. To overcome this problem, we constructed coronal (and transition region) models of the star. We then calculated a synthetic SED in the whole XUV range ($\lambda\lambda \sim 5-920$~\AA).

We analyzed new \textit{XMM-Newton}, \textit{Chandra}, and EUV spectra of 50 stars, either from proprietary or archival observations. We also used high-resolution spectra to measure individual lines formed at these temperatures in 5 stars. In the case of 26 stars, we extended the analysis to the transition region temperatures by measuring UV lines in high-resolution spectra from HST and FUSE. We built detailed emission measure distributions (EMDs) for this group.  We also updated the analysis of the spectra in X-exoplanets \citep{san11} using the latest Gaia distances, and the latest version of ATOMDB. The overall sample includes 100 stars hosting 163 planets, of which 75 stars have S/N$>3$, and the rest are considered upper limits.
The whole sample, excluding upper limits, was then used to calculate new scaling laws for an easier calculation of the broadband fluxes that matter for the He and H ionization. A new approach to the problem was introduced by using the X-ray and EUV luminosity weighted by the bolometric luminosity. This allows us to remove effects related to stellar size, while closely considering  the behavior of the scaling laws with the stellar activity.
Future improvements will include more coronal high quality EMDs to reduce the dispersion observed in the data when the low-resolution X-ray spectra alone are used to model the coronae.

The newly calculated scaling laws are then used to evaluate the stellar He-ionizing irradiation in 48 exoplanets for which the \ion{He}{i}~10830~\AA\ triplet has been measured, including upper limits. In the cases with no X-rays measurements, we used an X-ray luminosity based on the stellar rotation period. We then checked the expected trend of the formation of this line in exoplanet atmospheres with the XUV$_{\rm He}$ stellar irradiation. A clear relation is observed, once the planet gravitational potential, along with the stellar and planetary surface are included. The remaining dispersion observed in the data can be attributed to the He content in the planet atmospheres.
Some outliers, such as the TRAPPIST-1 b, e, and f planets, as well as 55 Cnc e, could be explained by the lack of planetary gaseous atmospheres.
The stellar variability and the difficulties to measure the \ion{He}{i}~10830~\AA\ triplet in the transmission spectrum of the planet also contribute to the data dispersion.
The observed relation can be used to predict the detectability of the \ion{He}{i}~10830~\AA\ triplet in a transiting planet.

\section*{Data availability}\label{sec:onlinedata}
The SEDs modeled are listed in Table~1, only available in electronic form at the CDS via anonymous ftp to cdsarc.u-strasbg.fr (130.79.128.5) or via \href{http://cdsweb.u-strasbg.fr/cgi-bin/qcat?J/A+A/}{http://cdsweb.u-strasbg.fr/cgi-bin/qcat?J/A+A/}.

\begin{acknowledgements}
  We thank the anonymous referee for the useful comments that helped improve the manuscript.
  We acknowledge financial support from the Agencia Estatal de Investigaci\'on (AEI/10.13039/501100011033) of the Ministerio de Ciencia e Innovaci\'on and the ERDF ``A way of making Europe'' through projects 
  PID2022-137241NB-C4[1:4],     
  PID2022-141216NB-I00,    
  PID2021-125627OB-C31,         
  and the Centre of Excellence ``Severo Ochoa'' and ``Mar\'ia de Maeztu'' awards to the Instituto de Astrof\'isica de Canarias (CEX2019-000920-S), Instituto de Astrof\'isica de Andaluc\'ia (CEX2021-001131-S) and Institut de Ci\`encies de l'Espai (CEX2020-001058-M).
We also acknowledge the support of the DFG priority program SPP 1992 “Exploring the Diversity of Extrasolar Planets" (CZ 222/5-1).
We acknowledge Norbert Schartel for the observations granted as \textit{XMM-Newton} Director Discretionary Time (DDT).
This research has made use of the NASA's High Energy Astrophysics Science Archive Research Center (HEASARC), and it is based on observations made with HST, FUSE, EUVE, \textit{XMM-Newton} and \textit{Chandra}, and obtained from the MAST data archive at the Space Telescope Science Institute, and the public archives of \textit{XMM-Newton} and \textit{Chandra}. 
CARMENES is an instrument at the Centro Astron\'omico Hispano en Andaluc\'ia (CAHA) at Calar Alto (Almer\'{\i}a, Spain), operated jointly by the Junta de Andaluc\'ia and the Instituto de Astrof\'isica de Andaluc\'ia (CSIC).
  CARMENES was funded by the Max-Planck-Gesellschaft (MPG), the Consejo Superior de Investigaciones Cient\'{\i}ficas (CSIC),
  the Ministerio de Econom\'ia y Competitividad (MINECO) and the European Regional Development Fund (ERDF) through projects FICTS-2011-02, ICTS-2017-07-CAHA-4, and CAHA16-CE-3978, and the members of the CARMENES Consortium (Max-Planck-Institut f\"ur Astronomie,
  Instituto de Astrof\'{\i}sica de Andaluc\'{\i}a, Landessternwarte K\"onigstuhl, Institut de Ci\`encies de l'Espai, Institut f\"ur Astrophysik G\"ottingen, Universidad Complutense de Madrid,
  Th\"uringer Landessternwarte Tautenburg, Instituto de Astrof\'{\i}sica de Canarias, Hamburger Sternwarte, Centro de Astrobiolog\'{\i}a and
  Centro Astron\'omico Hispano-Alem\'an), with additional contributions by the MINECO, 
  the Deutsche Forschungsgemeinschaft (DFG) through the Major Research Instrumentation Programme and Research Unit FOR2544 ``Blue Planets around Red Stars'', the Klaus Tschira Stiftung, the states of Baden-W\"urttemberg and Niedersachsen, and by the Junta de Andaluc\'{\i}a.
\end{acknowledgements}

\newpage

\begin{appendix}

  \section{Comparison with other XUV determination approaches}\label{sec:comparison} 

  In this section, we show further comparisons between our XUV modeling approach and others employed in the literature, as outlined in Sect.~\ref{sec:results}. Here, we also discuss in more detail the use of DEM polynomial fits as an alternative technique to our EMD. The results from our scaling laws are also compared against other works in the literature.
  
    \subsection{Comparison with polynomial fits of the differential emission measure}
\begin{figure}
  \centering
  \includegraphics[width=0.49\textwidth]{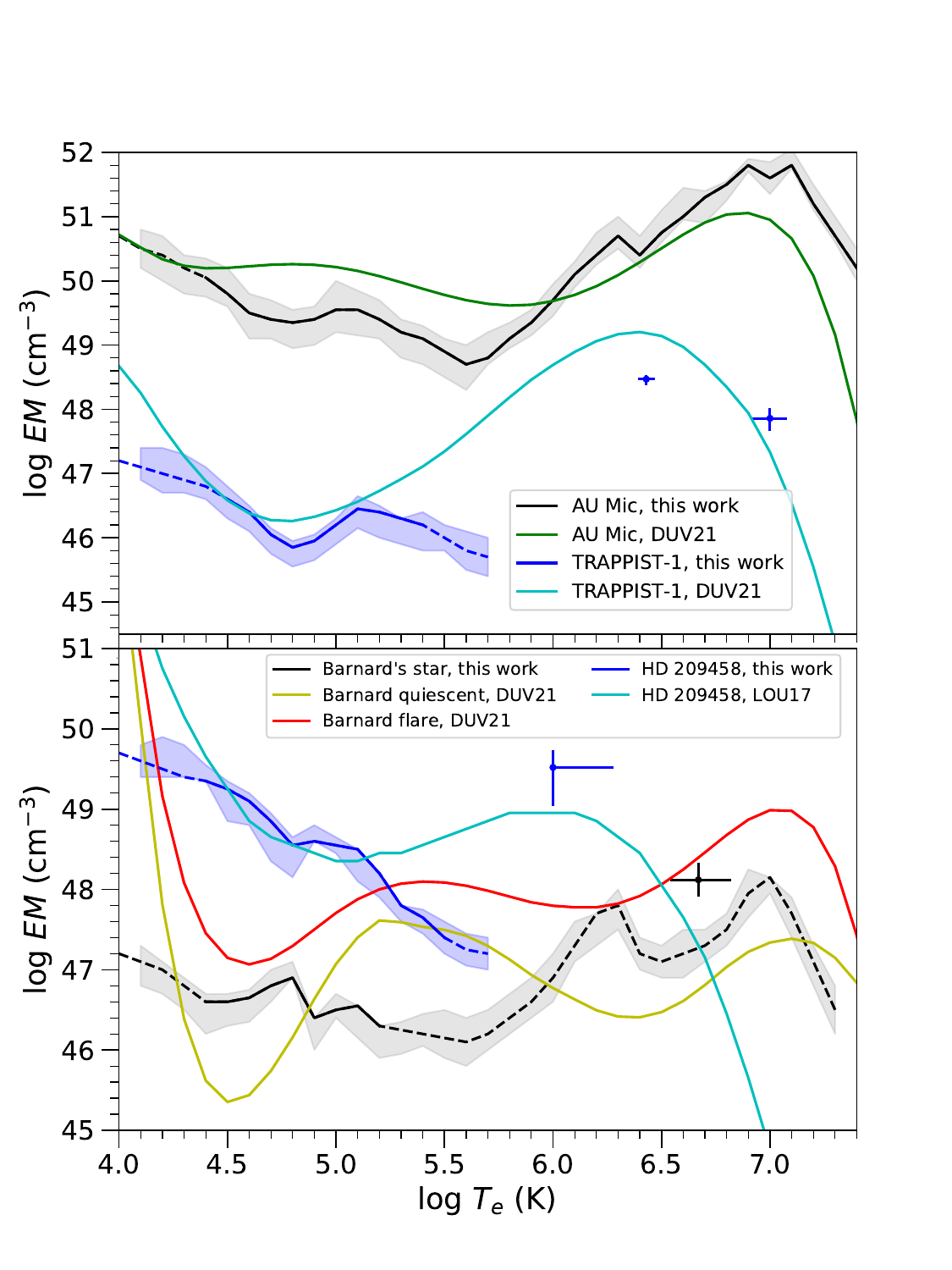}
  \vspace{-8mm}
  \caption{EMD calculated in our work, with shaded 1$\sigma$ error bands, as compared with those of LOU17 and DUV21. Dashed lines indicate an uncertain EMD despite formal error bands. The results from our global fit to the X-ray spectra are indicated with error bars.}\label{fig:dememd} 
\end{figure}

\begin{figure*}
  \centering
  \includegraphics[width=0.33\textwidth]{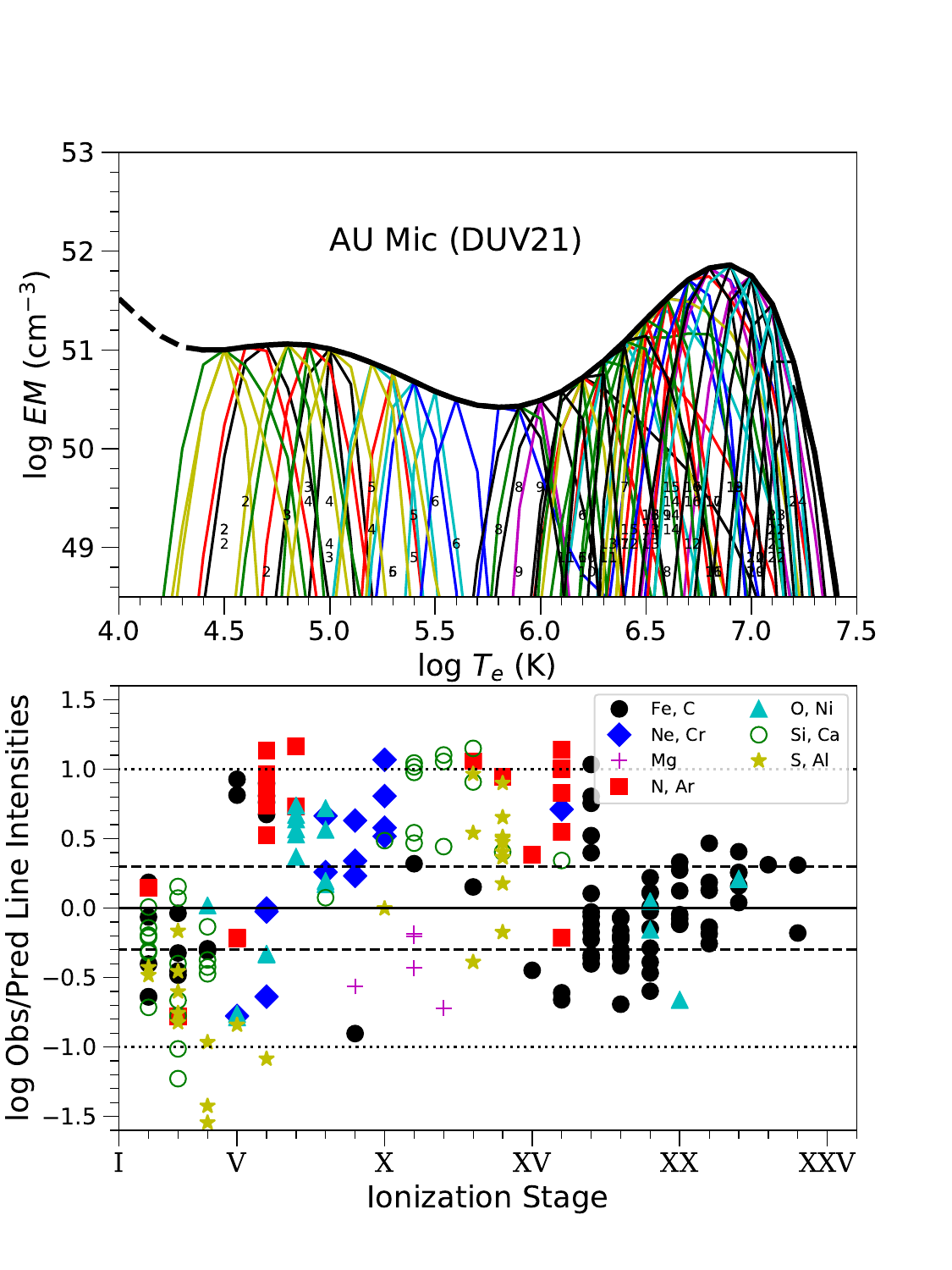}
  \includegraphics[width=0.33\textwidth]{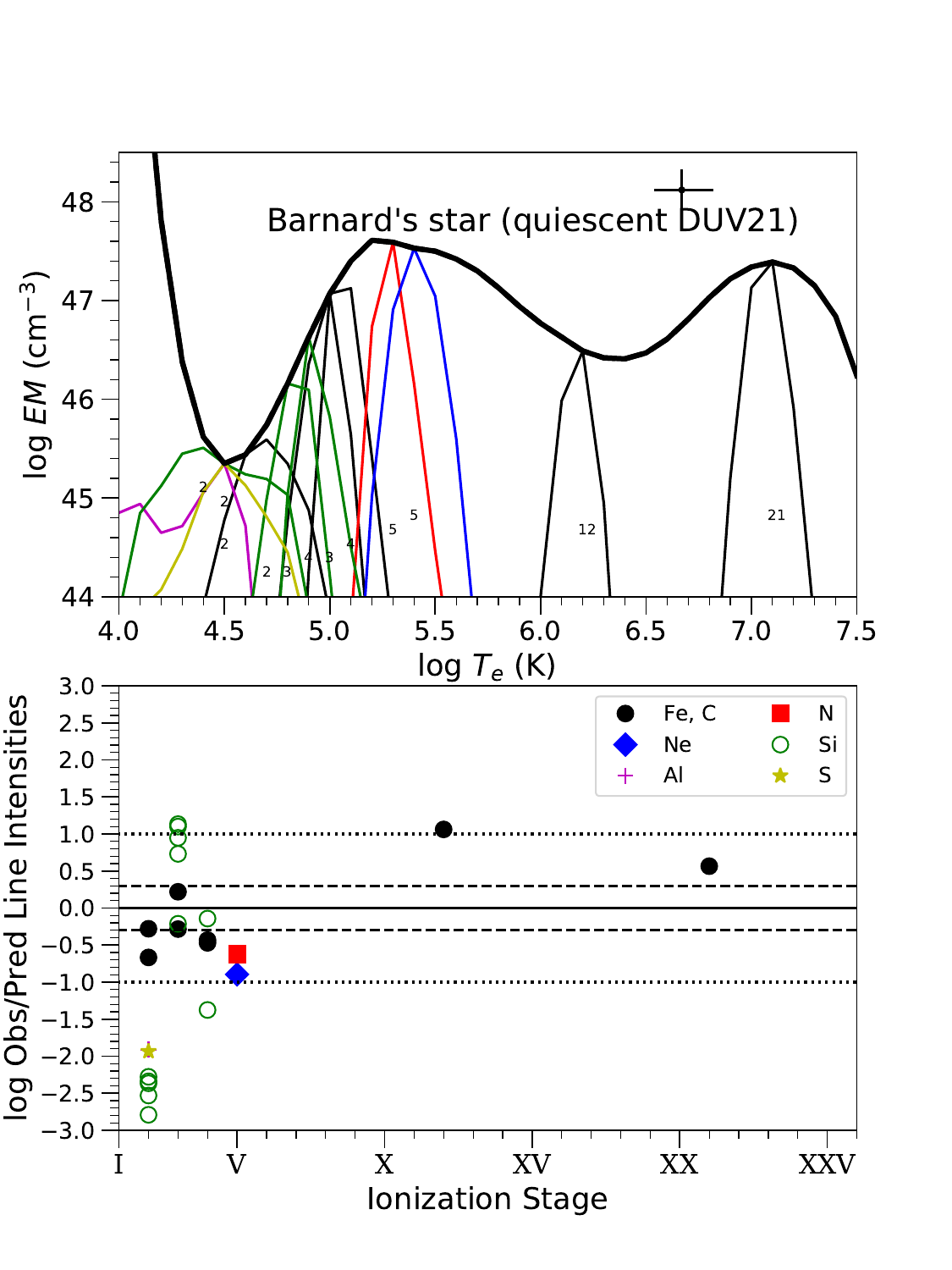}
  \includegraphics[width=0.33\textwidth]{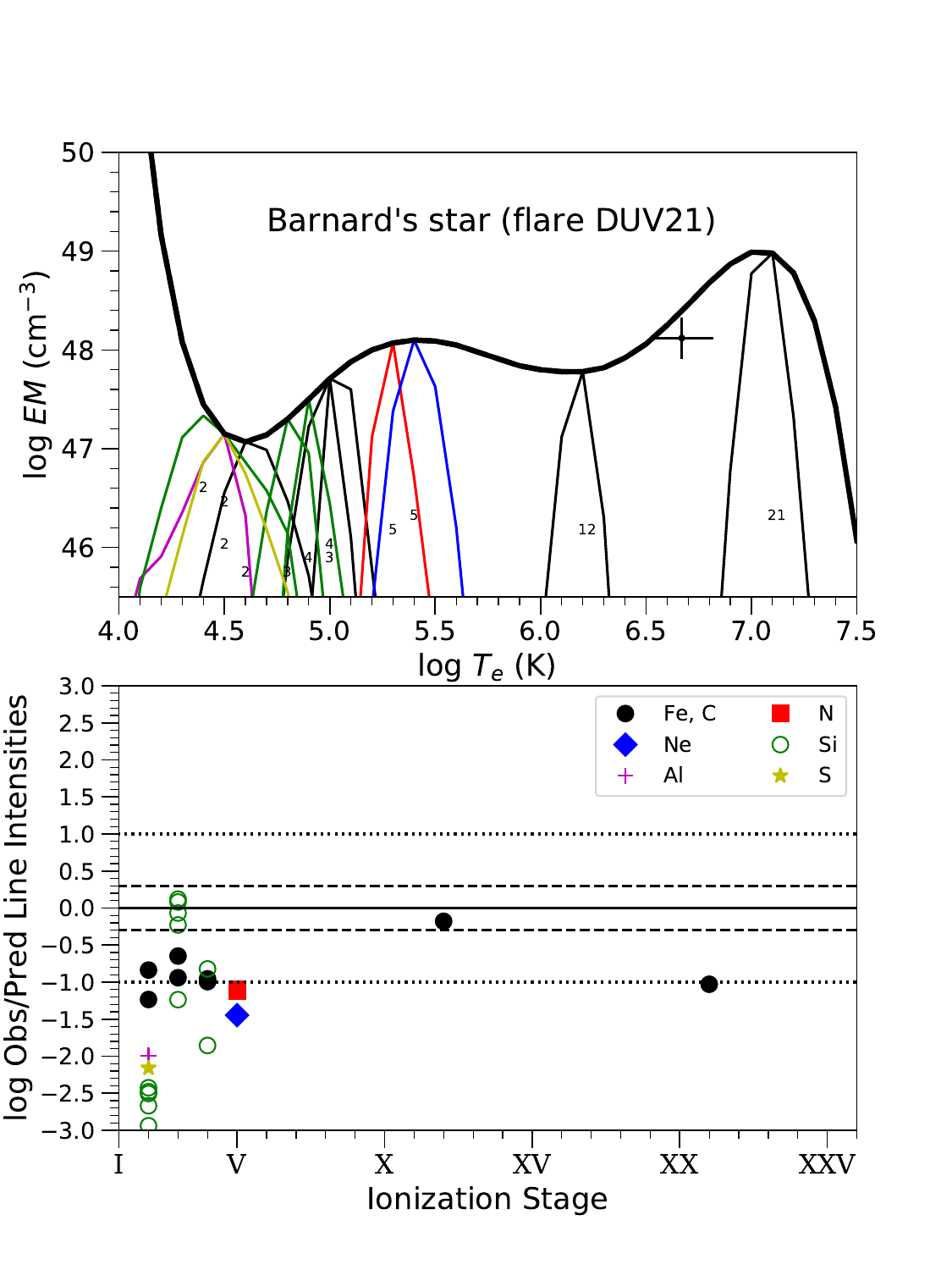}
  \includegraphics[width=0.33\textwidth]{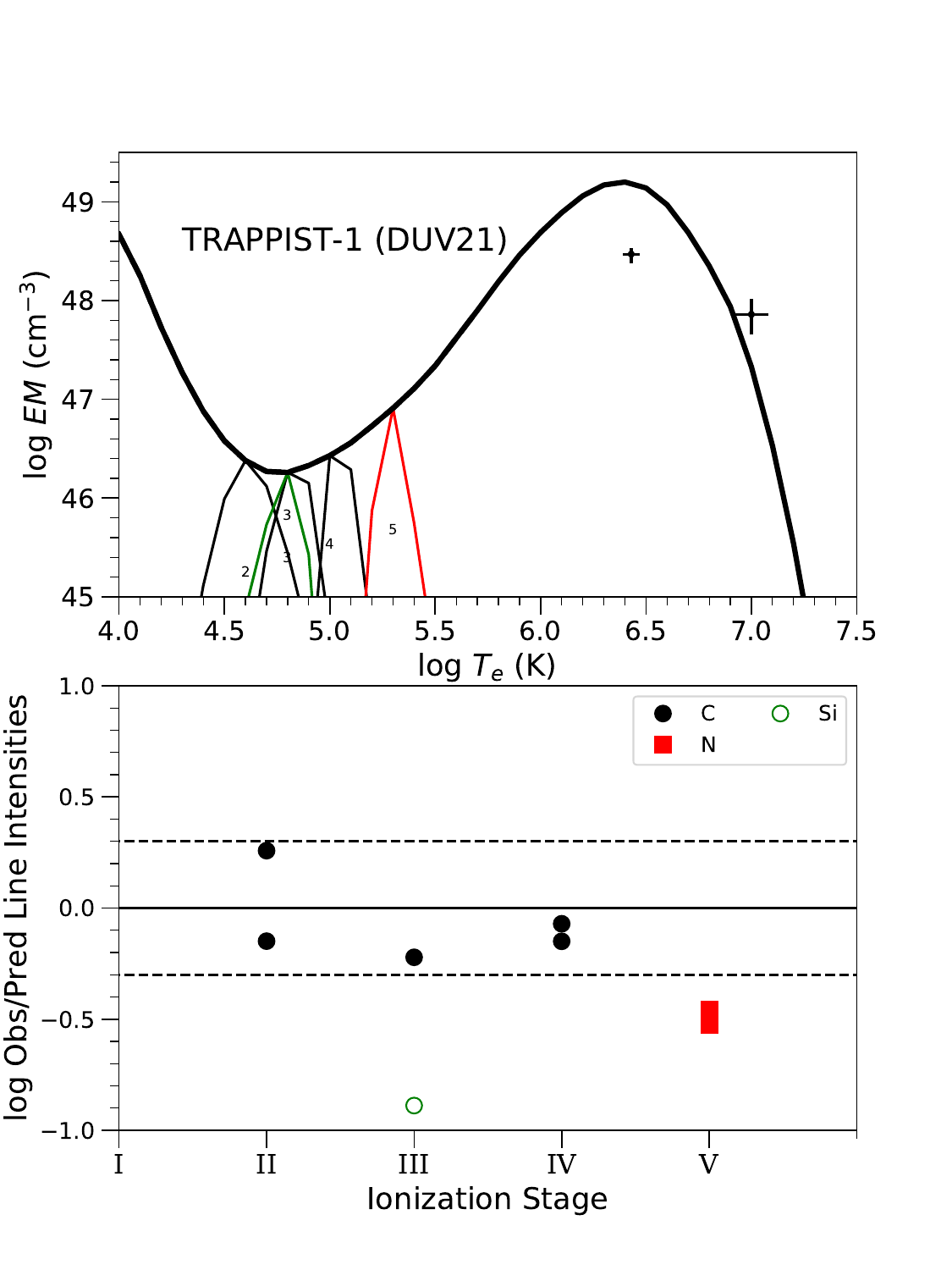}
  \includegraphics[width=0.33\textwidth]{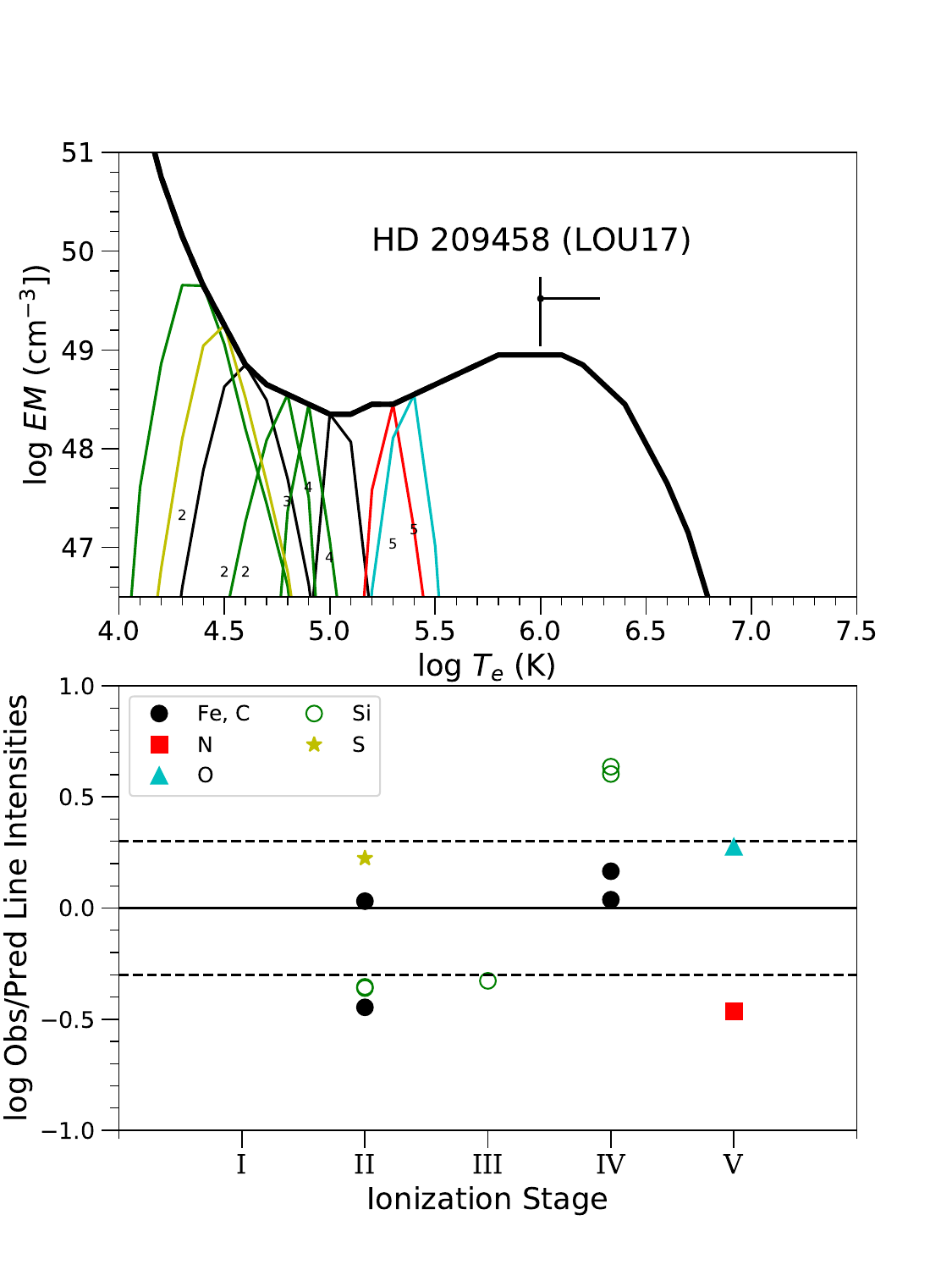}
  \caption{EMDs calculated from the DEM published in LOU17 and DUV21, using our line fluxes measurements to test the quality of the fit. Symbols and lines as in Fig.~\ref{fig:proxcen}. In the lower panels the axis range has been adapted to show all available line fluxes ratios. An offset of one order of magnitude is indicated with dotted lines.}\label{fig:emdothers} 
\end{figure*}

    Some recent publications have developed a technique to calculate the DEM of a star by fitting smooth polynomials functions \citep[][hereafter LOU17 and DUV21 respectively]{lou17,duv21}. Here we compare our results with those of LOU17 and DUV21. G.~Duvvuri kindly provided the DEM solutions for three targets in common with ours: AU Mic, Barnard's Star, and Trappist-1. A similar approach was applied by LOU17 to determine the DEM of HD 209458. To transform their DEM into volume DEM at the distances ($d'$) used in our work, we corrected by the stellar radius ($R$) and distance ($d$) used by them.
    Although the basic concept of DEM and EMD are similar, there are differences in the way they relate to the temperature.  Therefore we needed to transform the DEM into EMD.
    The definition of the DEM and EMD and formulae needed to transform them were given by, e.g., \citet{bow00}, and \citet{delz02} and references therein.
We use the definition of the volume emission measure, evaluated in a temperature interval 2$\delta$ around $T_a$ as in \citet{bri98}, namely, $EM=\int_{T_a-\delta}^{T_a+\delta} N_{\rm e} \ N_{\rm H} \ {\rm d}V$. The DEM is defined as $DEM=N_{\rm e} \ N_{\rm H} \ \frac{{\rm d}V}{{\rm d}T}$.
We included another term to account for the temperature grid in a $\log T$ scale: $\frac{{\rm d}(\log \ T)}{{\rm d} T} = \frac{1}{\ln(10) \ T}$. We integrated the DEM in a temperature interval $\Delta (\log T)$ around a temperature $T_a$, assuming that the emission measure is constant in that interval. Then the emission measure (in the EMD definition) relates to DEM as
\begin{eqnarray}
  EM & = & DEM \cdot 4 \pi \, \Big(\frac{R d'}{d}\Big)^2 \cdot T_a \cdot \ln 10 \cdot \Delta (\log T)\, ,
\end{eqnarray}
\noindent where $\Delta (\log T)$\,(K)\,=\,0.1 is the temperature resolution of the ATOMDB models used in our work. Once the DEMs were transformed into EMDs (Fig.~\ref{fig:dememd}), we calculated the predicted line fluxes using the solar coronal abundances assumed in LO17 and DUV21 to test how good are the models provided by those authors. The results of this comparison are shown in Fig.~\ref{fig:emdothers}.

The smooth shape imposed by polynomial fits yields large differences with our results, in special in the temperature ranges with fewer spectral lines, such as the regions with $\log T$(K)$<$4.5, and $\sim$5.2--6.0.
DUV21 used solar coronal abundances in all cases, which might explain some of the discrepancies, as acknowledged by DUV21. While this assumption could be a good approximation for a solar-like quiet corona, like that of HD~209458, it is inadequate for a very active star like AU~Mic\footnote{DUV21 DEM was based on a different dataset from ours.}, as shown in Fig.~\ref{fig:emdothers} \citep[see e.g.,][]{san03}.
In the case of Barnard's star, DUV21 made a different DEM for flaring and quiescent stages, while we used only one overall EMD. However none of the DUV21 functions match the minimum of the EMD as sampled by us.
Finally, TRAPPIST-1 model also over-interprets the information available in the X-rays spectra by providing a continuous DEM along the high temperature range.
In all cases, there are obvious discrepancies, of up to three orders of magnitude, between predicted and observed line fluxes when the EMDs of DUV21 and LOU17 are employed. Their EUV broadband fluxes differ by up to $\sim$1\,dex from ours (Table~\ref{tab:leuvcompared}).

%
%

%
%

\begin{table}
  \caption[]{Comparison of modeled EUV broadband luminosities\tablefootmark{a}.}\label{tab:leuvcompared}
  \tabcolsep 3.pt
  \vspace{-8mm}
  \begin{center}
  \begin{scriptsize}
    \begin{tabular}{lccccc}
\hline \hline
Reference & AU Mic & TRAPPIST-1 & Barnard's Star & Barnard's Star & HD 209458 \\
     &        &            & (quiescence) & (flare) & \\
\hline
DUV21, LOU17 & 28.93 & 27.33 & 25.71 & 26.61 & 28.20 \\ 
This work    & 29.39 & 26.29 & \multicolumn{2}{c}{26.13} & 28.45 \\ 
\hline
\end{tabular}
  \tablefoot{
   \tablefoottext{a}{$\log L_{\rm EUV,H}$ (erg\,s$^{-1}$) in the 100--912~\AA\ range as reported by DUV21 and LOU17, and compared with our model.}}
  \end{scriptsize}
  \end{center}
\end{table}

  \subsection{Comparison with other scaling laws}

\begin{figure}
  \centering
  \includegraphics[width=0.49\textwidth]{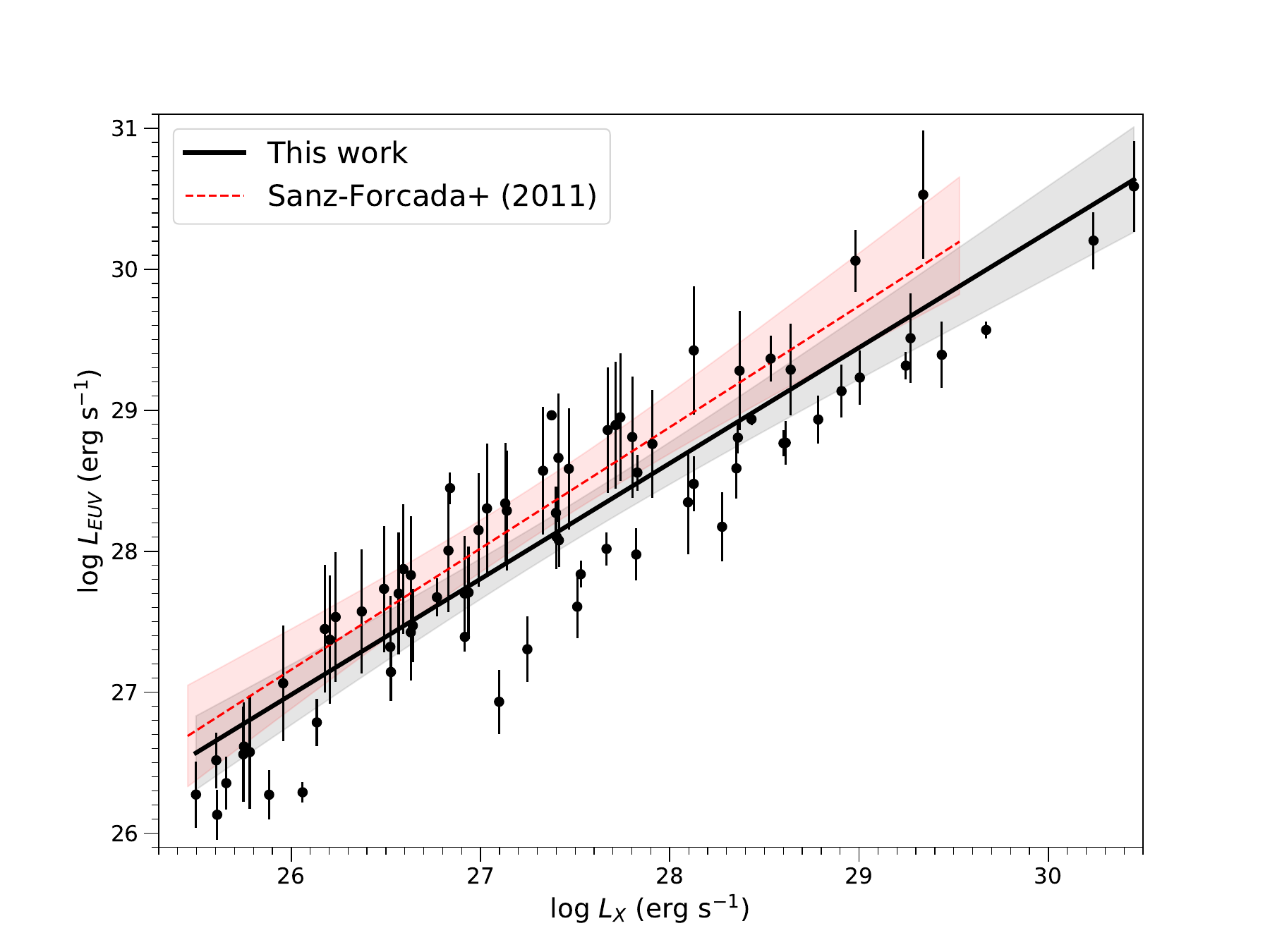}
  \caption{EUV vs. X-ray flux with our data and the fit by \citet{san11}. $1\sigma$ error bands to both fits, and their original datasets, are displayed in light gray and red respectively.}\label{fig:LxLeuv} 
\end{figure}

In this section, we compare the results from our scaling law described in Sect.~\ref{sec:resultsxuv} and other works in the literature. Our X-ray vs. EUV$_{\rm H}$ luminosity linear fit has an RMS=0.406. Figure~\ref{fig:LxLeuv} includes a comparison with the \citet{san11} linear fit (RMS=0.411 with the current dataset) to $L_{\rm X}$ and $L_{\rm EUV,H}$. The addition of UV high-spectral resolution data and values with larger X-ray luminosity seems to lower the EUV modeled luminosity, but both fits are consistent.

\begin{figure}
  \centering
  \includegraphics[width=0.49\textwidth]{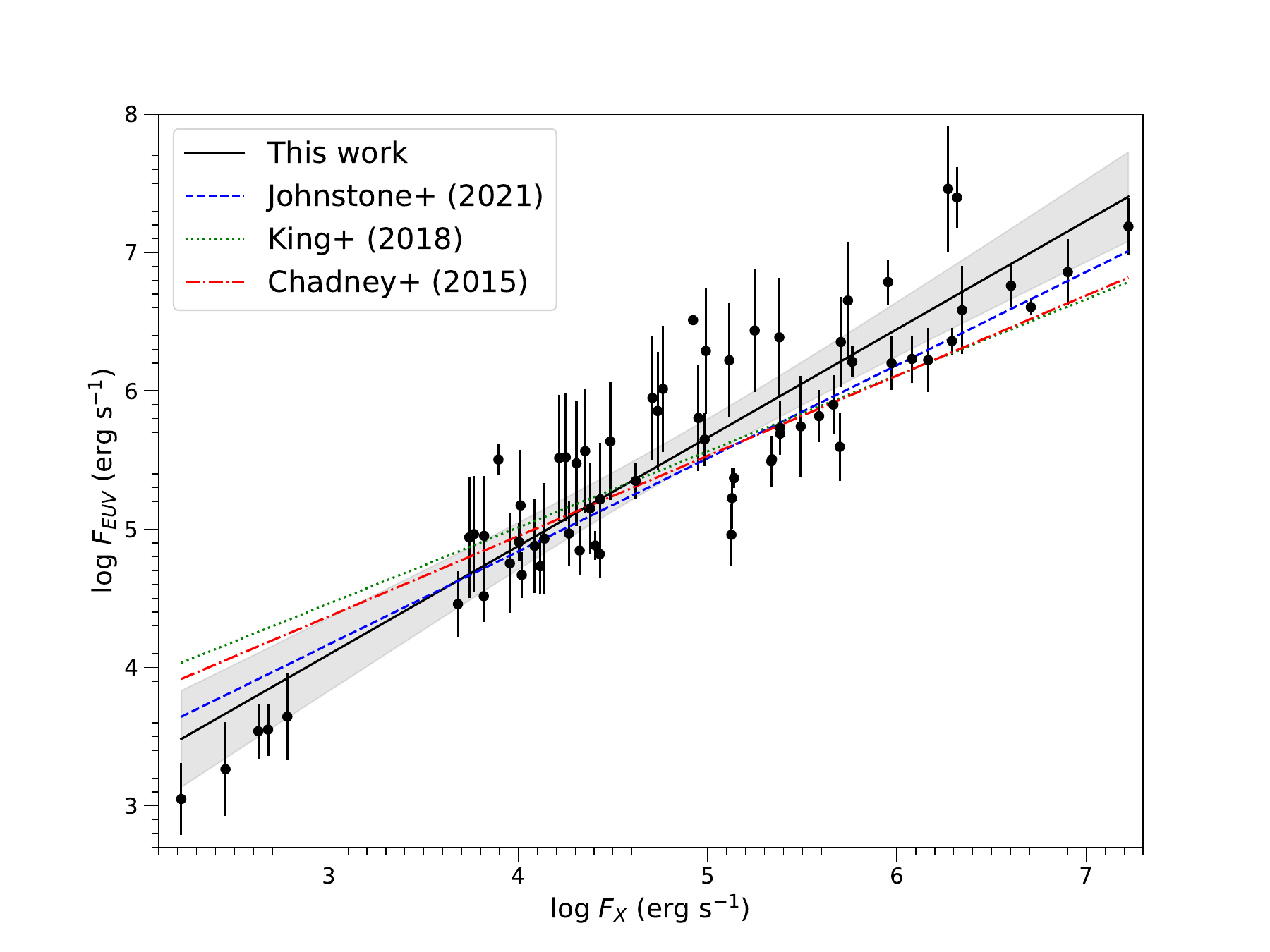}
  \caption{EUV vs. X-ray surface flux with our data and some literature fits. $1\sigma$ error bands to our fit is displayed in light gray.}\label{fig:fxfeuv} 
\end{figure}

Scaling laws based on the surface stellar flux are shown in Fig.~\ref{fig:fxfeuv} together with our own fit to the data. Our fit is
\begin{eqnarray}
  \log F_{\rm EUV,H} & = &(0.784 \pm 0.044)\, C_{\rm X} + (5.54 \pm 0.05)\, ,
\end{eqnarray}
\noindent where $C_{\rm X} = \log F_{\rm X}-4.88$, all in c.g.s units.
The plot includes only the stars of our sample with a stellar radius available in the Extrasolar Planet Encyclopaedia\footnote{\url{https://exoplanet.eu}} database, 65 stars. Our linear fit to the data shows a Pearson's correlation factor $r=0.911$, and an RMS of 0.386. The \citet{jon21} relation has an RMS of 0.405 with this dataset, \citet{cha15} an RMS=0.445, and \citet{kin18} an RMS=0.462.  The literature fits have a worse RMS than our linear fit between $\log L_{\rm X}$ and $\log L_{\rm EUV,H}$ (Fig.~\ref{fig:xvseuv}). They also have a dependence on the stellar radius, introducing a new source of uncertainty. We display the literature fits extrapolated to cover our range of values. Although all of the fits are roughly consistent with ours, discrepancies arise for the most active stars, and especially for the least active stars.

%
%
%
%
%
\begin{figure}
  \centering
  \includegraphics[width=0.49\textwidth]{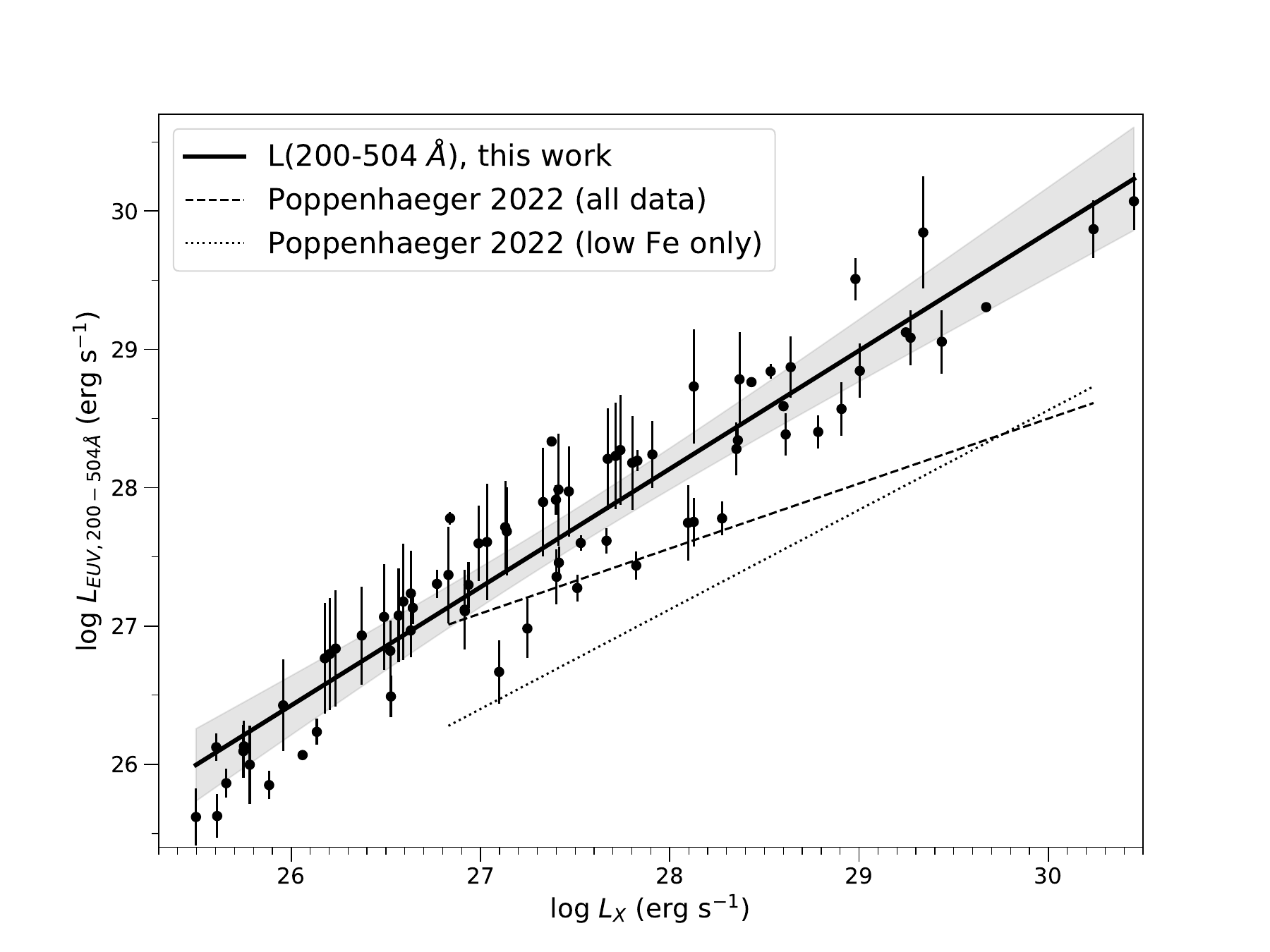}
  \caption{X-ray vs. EUV (in the range 200--504~\AA) luminosity with our data and the fits calculated by \citet{pop22}. $1\sigma$ error bands to our fit are displayed in light gray.}\label{fig:poppenrel} 
\end{figure}
%

\citet{pop22} calculated the EUV flux in the range 200--504~\AA\ based on EUVE observations in a number of targets. This is a spectral range severely affected by ISM absorption \citep[see, e.g.,][]{san02b} and the spectra are 
contaminated by geocoronal emission around the \ion{He}{ii}~304~\AA\ line, as those in \citet{jon21}. Since \citet{pop22} calculated the absorption based on two lines that are absent in most of the spectra, there may be systematic under- or over-estimation in this calculation.
Moreover, most lines above $\sim 370~\AA$ are absent in EUVE spectra, and some of the observed lines are actually the second order of stronger bluewards lines. An example is the EUVE spectrum of AD~Leo \citep{san02}, a star at only $\sim$5~pc from us. The low quality spectrum of AD~Leo shown in Fig. 9 of \citet{pop22} came from an automatic reduction available in HEASARC, quite below the quality level that can be achieved by a customized data reduction, which would anyway be insufficient to accurately calculate the flux in the 200--504\,\AA\ spectral range.
In Fig.~\ref{fig:poppenrel} we show the fit that we did to our calculated EUV fluxes in this range, as compared with the \citet{pop22} fits \citep[their relation was also applied by][]{fos23}. They are obviously discrepant.
Our linear fit is
\begin{equation}
  \log L_{\rm EUV,200-504 \AA} = (0.855 \pm 0.033)\, C_{\rm X} + (27.66 \pm 0.04)\, ,
\end{equation}
\noindent where $C_{\rm X} = \log L_{\rm X}-27.44$ in c.g.s. units. This is valid in the range $\log L_{\rm X} \sim 25.5 - 30.5$. This fit has a Pearson's correlation factor $r=0.951$ and an RMS=0.326.
The spectral range used to calculate $L_{\rm X}$ by \citet{pop22} was 0.2--5~keV, instead of our 0.12--2.4~keV range. We tested the impact of this difference by measuring the flux in an active (AD~Leo) star and a more quiet (GJ~357) star. The flux in the 0.2--5~keV range is $\sim$~4\% lower for AD~Leo, and $\sim 0.3$~dex lower for GJ~357, thus increasing the differences between the two relations.


\section{Relations between He {\sc i} $\lambda$10830 \AA\ equivalent width and XUV flux}\label{sec:heliumfit}

\begin{figure}
  \centering
  \includegraphics[width=0.4\textwidth]{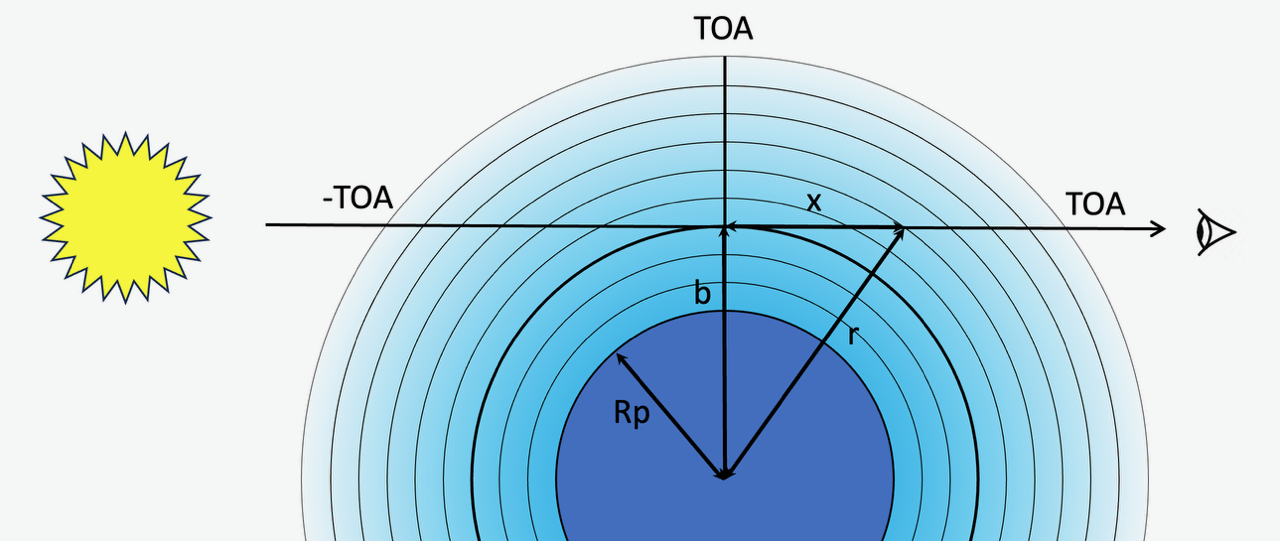}
  \caption{He(2$^{3}$S) column along the line of sight, $x$, at the impact parameter, $b$, and distance from the center of the planet, $r$.}
  \label{limb1}
\end{figure}

In this section, we seek  an empirical relationship between the stellar
XUV irradiation, measured in the 5--504~\AA\ range, and the equivalent
width ($EW$) of the \ion{He}{i} $\lambda$10830~\AA\ line in the planet atmosphere.
The equivalent width, $EW$, of an absorbing line over the frequency interval $\Delta \nu$ in a planetary
atmosphere during the primary transit is given by
\begin{equation}\label{eq:eqw}
 EW \simeq \frac{\int_{\Delta \nu}{\cal A}_{\nu}\,{\rm d} \nu}{\pi\, R_{\star}^2}\, , 
\end{equation}
where $\nu$ is the frequency, ${\cal A}_{\nu}$ is the absorption of the entire planetary
atmosphere, $R_{\star}$ is the stellar radius, and we assume that $R_{\star}^2 >> R_{p}^2$. The absorption ${\cal A}_{\nu}$ is
obtained by integrating the partial absorption of the areas covered by
infinitesimal spherical rings over all impact parameter $b$ values
(see, e.g., Sect. 3.3 in \citealt{lam20} and Fig.\,\ref{limb1}) as:\ 
\begin{equation}\label{eq:abs}
{\cal A}_{\nu} = 2 \pi \int_{R_p}^{\rm R_*} b\ [1-{\cal T}_{\nu}(b)]\  {\rm d} b\, ,    
\end{equation}
where we assume spherical symmetry, $R_{\rm p}$ is the radius of the
planet, TOA is the top of the atmosphere in front of the star, and the transmission ${\cal T}_{\nu}$ is given by
\begin{equation}\label{eq:tau}
{\cal T}_{\nu}(b) = \exp\left[-\int_{\rm -TOA}^{\rm TOA} k_{\nu}(x)\ n(x)\ {\rm d} x\ \right]\, , 
\end{equation}
where $k_{\nu}$ is the absorption coefficient (also denoted by $\sigma_{\nu}(x)$, the absorption cross section) of the He(2$^{3}$S) lines
and $n(x)$ is the He(2$^{3}$S) concentration.

To estimate the dependence of $EW$ on the different parameters, we
considered the two extreme conditions of (a) optically thick, where the transmission near the line center, 
${\cal T}_{\nu={\rm He\,I}}(b)$, is close to zero, and (b) optically thin, where ${\cal T}_{\nu={\rm He\,I}}(b)$ is close to unity.
This approach in two extreme situations seems reasonable as we know that most gas giant atmospheres have, in
general, a narrow (compressed) region at low altitudes with large
He(2$^{3}$S) concentrations, and an extended and slowly decreasing He(2$^{3}$S) concentration at medium to larger radii (see, e.g.,
Fig.\,A.1 in \citealt{lam23}). The first region can be considered in
optically thick conditions while the latter in the optically thin regime.

\subsection{Optically thick}
In this case, ${\cal T}(b)\simeq 0$, the $EW$ is given by that produced
by an opaque ring of height $H_{\rm He,0}$. From Eqs.~\ref{eq:eqw} and \ref{eq:abs}, with ${\cal T}(b)\simeq 0$, we obtain
\begin{equation}
 EW_{\rm thick} \propto  \frac{2\,\ R_{\rm p}\  H_{\rm He,0}}{R_{\star}^2}.
\end{equation}
To estimate the extension of the opaque region, $H_{\rm He,0}$, it is reasonable to assume that it is proportional to the
irradiating XUV flux $F_{\rm XUV}$, as this flux controls the He(2$^{3}$S) concentration that enters into Eq.\,\ref{eq:tau}. In fact,
the planets with larger XUV fluxes have larger peaks of He(2$^{3}$S) concentrations (see Fig.\,A.1 in \citealt{lam23}). In addition, for a
planet with a stronger gravitational potential, $\Phi_{\rm p}$, it is expected that its atmosphere (and the He(2$^{3}$S) concentration) is
more compressed and hence $H_{\rm He,0}$ should be smaller. It is then reasonable that in this optically thick limit we have
\begin{equation}
 EW_{\rm thick} \propto   \frac{R_{\rm p}}{R_{\star}^2} \frac{F_{\rm XUV}}{\Phi_{\rm p}}. \label{eq:thick}
\end{equation}

%
%

\subsection{Optically thin}

In this case the transmission over the interval $\Delta\nu$ can be
approximated by
\begin{equation}\label{eq:tau2}
{\cal T}(b) \simeq 1-\int_{\rm -TOA}^{\rm TOA} k(x)\ n(x)\ {\rm d} x\ =1-k \ m(b)\, , 
\end{equation}
where $m(b) = \int_{\rm -TOA}^{\rm TOA} n(x)\ {\rm d} x$ is the
  column density along the line of sight at impact parameter $b$ (see Fig.~\ref{limb1}) and we have assumed that the absorption coefficient does not depend on $x$.
Assuming spherical symmetry and changing the integration variable to $r$, the distance from the center of the planet, e.g., $x=\sqrt{r^2-b^2}$ and ${\rm d}x= \frac{r}{\sqrt{r^2-b^2}}\ {\rm d}r$, we obtain
\begin{equation}
m(b) = 2\ \int_{b}^{\infty} \frac{n(r)\ r}{\sqrt{r^2-b^2}}\ {\rm d}r.
\end{equation}

To further evaluate $m(b)$ and the $EW$ let us make some assumptions on $n(r)$.
The \het\ concentration can be well approximated by
\begin{equation}\label{eq:nr} 
n(r)=n_0\ (r_0/r)^p\, , 
\end{equation} 
where the exponent $p$ usually takes values between 2 and 5 and $n_0$ is the \het\ concentration at $r_0$ \cite[see,
  e.g., Fig.\,A.1(c) in][]{lam23}. Note that the \het\ peak concentration, $n_0$, usually occurs very close to the lower boundary of the atmosphere; hence $r_0 \simeq R_P$. Then, we obtain
\begin{equation}\label{eq:mb}
m(b) \simeq 2\ n_0\ R_P^p \int_{b}^{\infty} \frac{r^{1-p}}{\sqrt{r^2-b^2}}\ {\rm d}r 
,\end{equation}
and, by performing the integral, we obtain
\begin{equation}\label{eq:mb2} 
m(b) \simeq \sqrt{\pi}\ n_0\ R_P^p\ \frac{\Gamma((p-1)/2)}{\Gamma(p/2)}\ b^{1-p}\, ,   
\end{equation}
valid for $p>1$, which is amply met in our case.

%
\begin{figure}
  \includegraphics[width=0.54\textwidth]{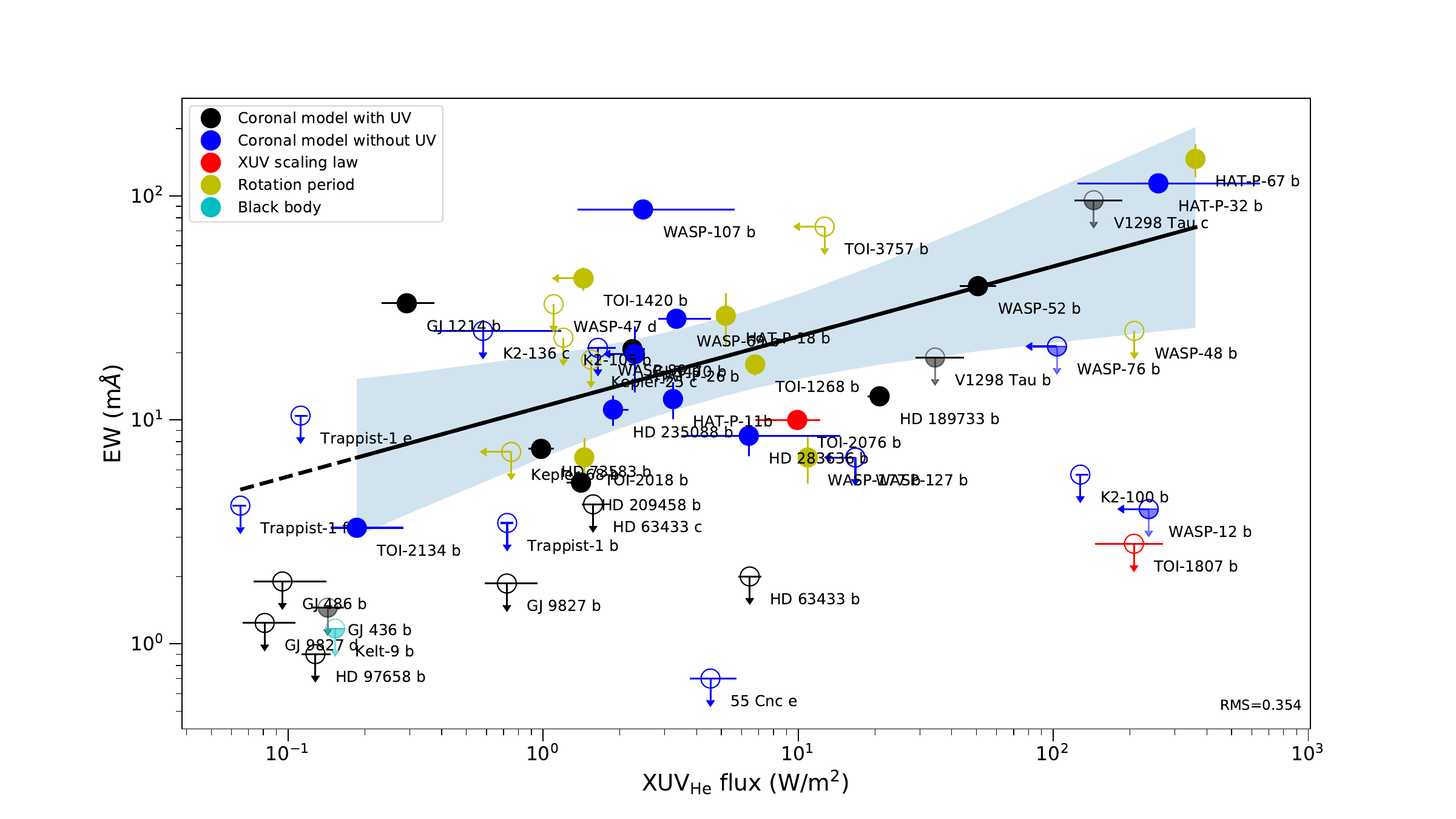}
  \caption{\ion{He}{i}~10830 triplet equivalent width, plotted against $F_{\rm XUV,He}$. $1\sigma$ error bands to the fit are displayed in light blue. Colors and symbols as in Fig.~\ref{fig:xuvvsHe}. Pearson's coefficient is $r=0.579$}\label{fig:xuvvsHe1} 
\end{figure}
%

%
\begin{figure}
  \includegraphics[width=0.54\textwidth]{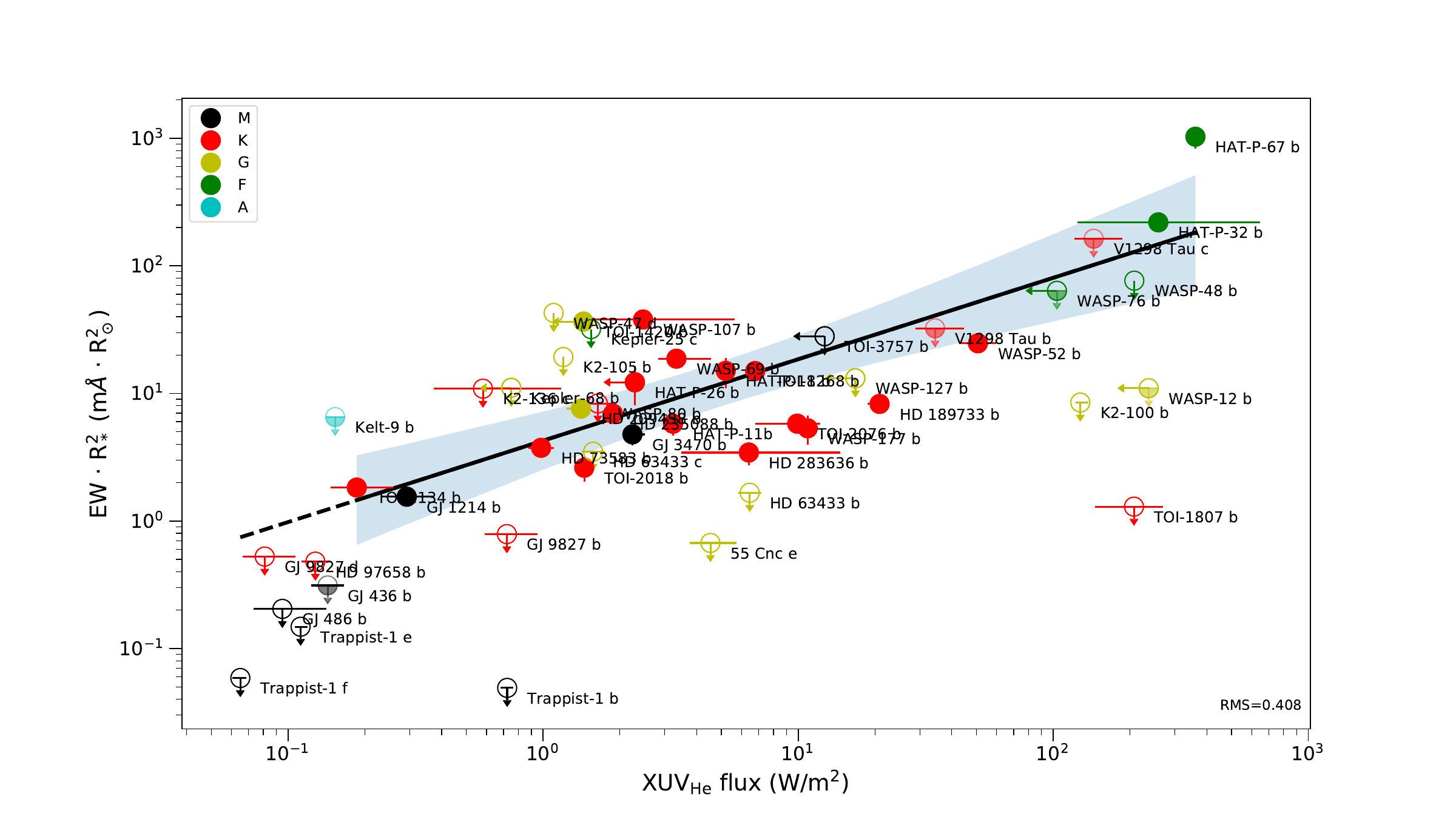}
  \caption{\ion{He}{i}~10830 triplet equivalent width, multiplied by the stellar area, plotted against $F_{\rm XUV,He}$. The color scheme indicates the stellar spectral types, with symbols as in Fig.~\ref{fig:xuvvsHe}. The fit gives $r=0.782$.}\label{fig:xuvvsHe2} 
\end{figure}
%

Including this expression into Eq. \ref{eq:tau2}, and considering
Eqs.\,\ref{eq:eqw} and \ref{eq:abs}, we obtain
\begin{equation}
EW \simeq  2\ \sqrt{\pi}\ \frac{\Gamma((p-1)/2)}{\Gamma(p/2)}\ k\ n_0\ \frac{R_{\rm P}^p}{R_{\star}^2} \int_{R_P}^{R_*}  b^{2-p} {\rm d} b. \label{eq:ew_thin_int}
\end{equation}Performing the integral in Eq.\,\ref{eq:ew_thin_int} and considering that $R_* \gg R_P$, we find that the integral is independent of $R_P$ for $1<p<3$, it is proportional to $\ln(R_*/R_P)$ for $p$\,=\,3, and is proportional to $R_P^{3-p}$ for $p>3$. Hence, as the $\Gamma$ functions are constants, 
\begin{equation}
EW_{\rm thin} \propto \frac{1}{R_{\star}^2} \frac{F_{\rm XUV}}{\Phi_{\rm p}} \times 
\begin{cases}
   R_P^{p}  & \text{for  $1<p<3$ }, \\
   R_P^{3}\, \ln(R_*/R_P) & \text{for  $p$=3, and} \\
   R_P^{3}    & \text{for  $p>$3 }, 
\end{cases}
\end{equation}
where we also considered that $n_0 \propto F_{\rm XUV}$ and the $EW$ is, as in
the case of the optically thick, proportional to $1/{\Phi_{\rm p}}$.
Then, for a given planet with an \het\ concentration $n(r)$ profile
with a slope $p$ as in Eq.\,\ref{eq:nr}, the $EW$ in the optically thin limit would be proportional to $R_{p}^\gamma$ with $\gamma$ ranging from 1 to 3.
For the optically thick limit we obtain a dependence directly
proportional to $R_{\rm p}$.
Hence, overall, we expect a dependence of $EW$ of the form of\begin{equation}\label{eq:ew_total}
EW \propto  \frac{1}{{R_{\star}^2}} \frac{F_{\rm XUV}}{\Phi_{\rm p}}\ \ R_{\rm p}^{\gamma}\, , 
\end{equation}
with $\gamma$ varying between 1 and 3 depending on the specific conditions of
the planets.

%
\begin{figure}
  \includegraphics[width=0.54\textwidth]{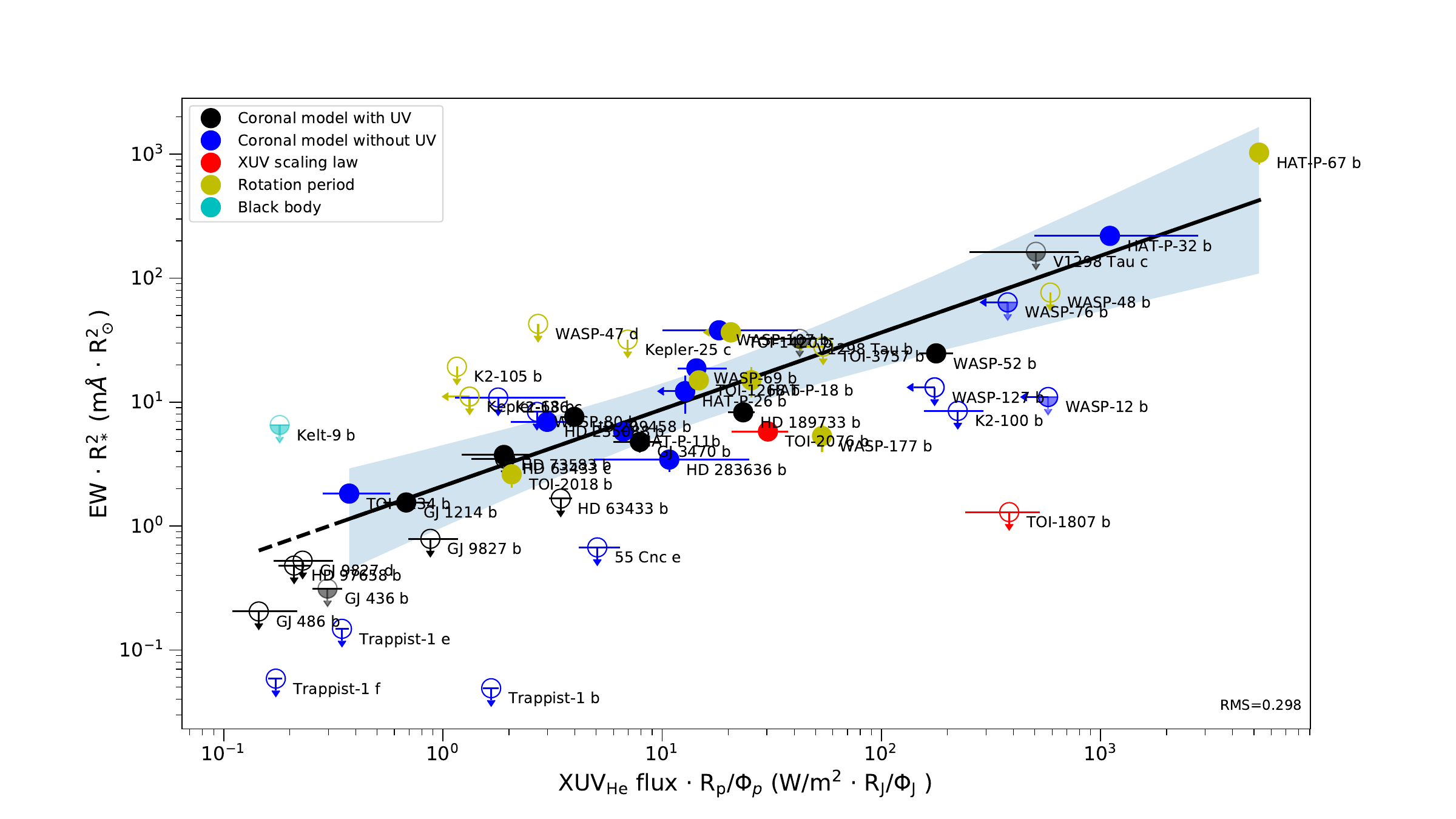}
  \caption{\ion{He}{i}~10830 triplet equivalent width, multiplied by the stellar area, plotted against $F_{\rm XUV,He}$ multiplied by planet radius $R_{\rm p}$, divided by the planet gravitational potential $\Phi_p$ (in Jovian units). Colors and symbols are similar to Fig.~\ref{fig:xuvvsHe}. The fit gives $r=0.891$.}\label{fig:xuvvsHe5} 
\end{figure}
%

We probed the dependence on the stellar area in Fig.~\ref{fig:xuvvsHe2}, with a substantial improvement with respect to Fig.~\ref{fig:xuvvsHe1}. If we consider the relation between $EW \cdot R_\star^2$ and $F_{\rm XUV,He} / \Phi_{\rm p}$ we get a Pearson's correlation factor of $r=0.803$.
For the relation between $EW \cdot R_\star^2$ and $F_{\rm XUV,He} \cdot R_{\rm p}^2$ we get a correlation factor of $r=0.843$, and for that against $F_{\rm XUV,He} \cdot R_{\rm p} / \Phi_{\rm p}$ we reach $r=0.891$ (Fig.~\ref{fig:xuvvsHe5}). Finally, a dependence on  $F_{\rm XUV,He} \cdot R_{\rm p}^2 / \Phi_{\rm p}$ results in a further improvement (Fig.~\ref{fig:xuvvsHe}, $r=0.898$). Table~\ref{tab:allhefits} shows the $r$ and RMS of the fits with different exponents of $R_{\rm p}$.
If we use $EW \cdot R_\star$ against $F_{\rm XUV,He} \cdot R_{\rm p}^2 / \Phi_{\rm p} \sim \rho_p$  as in \citet{zha23c} assuming that $R_{\rm XUV}$=$R_{\rm p}$, our data improve their log-log fit, with a correlation coefficient of $r=0.857$, but it is still worse than the fit using $EW \cdot R_\star^2$.

\begin{table}
  \caption[]{{Fits using different exponents on planet radii\tablefootmark{a}.}}\label{tab:allhefits}
  \vspace{-8mm}
  \begin{center}
    \begin{tabular}{lccccc}
\hline \hline
$\gamma$ & $r$ & RMS & $\alpha$ & $\beta$ & $X_0$\\
\hline
1    & 0.891 & 0.298 & $0.620 \pm 0.073$ & $1.053 \pm 0.068$ & 1.18 \\
1.5  & 0.899 & 0.287 & $0.546 \pm 0.061$ & $1.053 \pm 0.006$ & 1.07 \\
2    & 0.898 & 0.289 & $0.480 \pm 0.054$ & $1.052 \pm 0.066$ & 0.96 \\
2.5  & 0.893 & 0.295 & $0.424 \pm 0.049$ & $1.056 \pm 0.068$ & 0.86 \\
3    & 0.886 & 0.304 & $0.378 \pm 0.045$ & $1.056 \pm 0.070$ & 0.75 \\
\hline
\end{tabular}
  \tablefoot{
   \tablefoottext{a}{Pearson's $r$ coefficient, RMS, and parameters of equation $\log \big(EW\, R_*^2\big) = \alpha \Bigg(\log \bigg(\frac{F_{\rm XUV,He} \cdot R_{\rm p}^\gamma}{\phi_{\rm p}}\bigg) - X_0\Bigg) + \beta$, with different $R_{\rm p}$ exponent ($\gamma$).}}
\end{center}
\end{table}

\section{Main tables}\label{sec:tables} 

\begin{table}
\caption[]{X-rays observation log of stars with exoplanets.}\label{tabobslog}
\vspace{-5mm}
\tabcolsep 2.5pt
\begin{center}
\begin{scriptsize}
  \begin{tabular}{lcrrcccrc}
\hline \hline
{Star name} & SpT & $\alpha$ (J2000.0) & $\delta$ (J2000.0) & {Date} & {Instr.\tablefootmark{a}} & {t (ks)} & {S/N} \\ 
\hline
\multicolumn{7}{l}{New \textit{XMM-Newton} observations} \\
\hline
WASP-76     & F7IV-V& 01:46:31.9 & $+02$:42:02 & 2020-01-01 & EPIC &  9.0 & 1.2 \\
HAT-P-32A   & F8V:  & 02:04:10.5 & $+46$:41:15 & 2019-08-30 & EPIC & 15.3 & 11.3 \\
\object{KELT-7}     & F3IV-V& 05:13:11.1 & $+33$:19:05 & 2019-08-29 & EPIC &  7.2 & 11.7 \\
WASP-12     & G0V   & 06:30:32.8 & $+29$:40:20 & 2019-09-14 & EPIC &  8.8 & 1.3 \\
WASP-13     & G1V-IV& 09:20:24.7 & $+33$:52:56 & 2019-10-18 & EPIC &  8.9 & 3.2 \\
WASP-127    & G5V   & 10:42:14.7 & $-03$:50:06 & 2019-11-24 & EPIC &  5.9 & 0.4 \\
WASP-107    & K7V   & 12:33:32.7 & $-10$:08:46 & 2018-06-22 & EPIC & 57.5 & 16.7 \\
\object{HAT-P-12}   & K4V   & 13:57:33.1 & $+43$:29:35 & 2019-11-25 & EPIC &  7.4 & 2.2 \\
\object{WASP-39}    & G9V   & 14:29:18.2 & $-03$:26:40 & 2020-02-08 & EPIC & 13.4 & 1.5 \\
\object{KELT-8}     & G2V-IV& 18:53:13.3 & $+24$:07:37 & 2019-09-14 & EPIC &  7.8 & 1.1 \\
\object{GJ 806}     & M1.5V & 20:45:04.1 & $+44$:29:57 & 2023-12-07 & EPIC & 17.5 & 6.9 \\
\object{HD 201585}  & A8V   & 21:10:12.4 & $+10$:44:21 & 2019-10-29 & EPIC &  5.7 & 2.4 \\
\object{HAT-P-1}    & G0V   & 22:57:46.9 & $+38$:40:29 & 2019-12-02 & EPIC & 16.8 & 2.1 \\
\hline
\multicolumn{7}{l}{Other objects from \textit{XMM-Newton} and \textit{Chandra} archives} \\
\hline
\object{GJ 12}    & M3.0V & 00:15:49.2 & $+13$:33:22 & 2018-06-16 & EPIC & 17.8 & 4.5 \\
WASP-77A    & G8V & 02:28:37.2 & $-07$:03:39 & 2013-11-09 & ACIS & 9.9  & 3.7 \\
Teegarden's star   & M7.0V & 02:53:00.9 & $+16$:52:53 & 2021-08-03 & EPIC & 29.1 & 27.3 \\
HD 283636   & K0V & 04:23:55.1 & $+27$:49:21 & 2021-08-23 & EPIC & 4.8 & 8.1 \\ 
K2-136      & K5V & 04:29:39.0 & $+22$:52:58 & 2018-09-11 & EPIC & 37.2 & 21.9 \\
HD 63433    & G5V & 07:49:55.1 & $+27$:21:47 & 2021-03-26 & EPIC & 4.5 & 57.4 \\  
GJ 3470     & M1.5V & 07:59:05.8 & $+15$:23:29 & 2015-04-15 & EPIC & 15.2 & 18.0 \\
K2-100      & G0V & 08:38:24.3 & $+20$:06:22 & 2013-10-30 & EPIC & 66.3 & 34.6 \\
K2-100      &     &            &             & 2015-05-05 & EPIC & 54.0 & \dots\\
HD 73583    & K4V & 08:38:45.3 & $-13$:15:24 & 2021-04-21 & EPIC & 9.4 & 19.2 \\ 
GJ 357      & M2.5V & 09:36:01.6 & $-21$:39:39 & 2019-05-19 & EPIC & 29.1 & 7.5 \\
AD Leo      & M3.0V & 10:19:36.3 & $+19$:52:12 & 2000-01-22 & LETG & 10 & 168  \\
AD Leo      &     &            &             & 2000-10-24 & LETG & 48 & \dots \\
AD Leo      &     &            &             & 2002-06-01 & HETG & 46 & 112  \\
Lalande 21185 & M1.5V & 11:03:20.2 & $+35$:58:12 & 2001-05-15 & RGS & 19.7 & 60.3 \\
Lalande 21185 &     &            &             & 2004-05-12 & RGS & 44.3 & \dots\\
HD 97658    & K1V & 11:14:33.2 & $+25$:42:37 & 2015-06-04 & EPIC & 29.9 & 16.6 \\
GJ 486      & M3.5V&12:47:56.6 & $+09$:45:05 & 2021-12-23 & EPIC & 31.5 & 10.3 \\
Proxima Cen  & M5.5V & 14:29:42.9 & $-62$:40:46 & 2001--2018 & RGS & 506  & 360 \\
TOI-836     & K7V & 15:00:19.4 & $-24$:27:15 & 2022-07-25 & EPIC & 10.1 & 7.2 \\
HD 149026   & G0V & 16:30:29.6 & $+38$:20:50 & 2015-08-14 & EPIC & 17.1 & 6.0 \\
GJ 1214     &M4.5V& 17:15:18.9 & $+04$:57:50 & 2013-09-27 & EPIC & 31.7 & 4.3 \\ 
Barnard's star & M3.5V & 17:57:48.5 & $+04$:41:36 & 2019-06-17 & ACIS & 26.7 & 11.4 \\
TOI-2134    & K5V & 18:07:44.4 & $+39$:04:27 & 2022-09-12 & EPIC & 17.7 & 5.7 \\
HAT-P-11    & K4V & 19:50:50.2 & $+48$:04:51 & 2015-05-19 & EPIC & 25.0 & 13.4 \\
HD 235088   & K2V & 20:02:27.4 & $+53$:22:37 & 2021-07-07 & EPIC & 11.4 & 26.6 \\
WASP-80     & K7V & 20:12:40.2 & $-02$:08:39 & 2014-05-15 & EPIC & 15.3 & 15.6 \\
WASP-80     &     &            &             & 2015-05-13 & EPIC & 29.2 & \dots \\
KELT-9      & A0V & 20:31:26.4 & $+39$:56:20 & 2017-11-09 & EPIC & 50.0 & 2.4 \\
AU Mic      & M0.5V & 20:45:09.5 & $-31$:20:27 & 2000-10-13 & RGS & 52.7 & 670 \\
WASP-69     & K4V   & 21:00:06.2 & $-05$:05:40 & 2016-10-21 & EPIC & 27.4 & 30.6 \\
TRAPPIST-1  & M8.0V & 23:06:29.4 & $-05$:02:29 & 2014-12-17 & EPIC & 27.2 & 19.0 \\
TRAPPIST-1  &     &            &             & 2017-11-23 & EPIC & 49.8 & \dots\\
TRAPPIST-1  &     &            &             & 2017-11-29 & EPIC & 39.8 & \dots\\
TRAPPIST-1  &     &            &             & 2017-12-12 & EPIC & 35.3 & \dots\\
TRAPPIST-1  &     &            &             & 2018-12-10 & EPIC & 15.6 & \dots\\
WASP-52     & K2V & 23:13:58.8 & $+08$:45:41 & 2014-01-03 & ACIS & 9.8  & 3.5 \\
GJ 9827     & K6V & 23:27:04.8 & $-01$:17:11 & 2018-05-27 & EPIC & 19.0 & 8.7 \\
GJ 9827     &     &            &             & 2018-06-23 & EPIC & 14.1 & \dots\\
GJ 9827     &     &            &             & 2018-11-27 & EPIC & 11.3 & \dots\\
\hline
\multicolumn{7}{l}{New archival data of X-exoplanets objects included in this work} \\
\hline
$\upsilon$ And & F8V & 01:36:47.8 & $+41$:24:20 & 2013-08-04 & EPIC & 16.7 & 49.4 \\
HD 27442    &K2IV & 04:16:29.0 & $-59$:18:08 & 2009-02-10 & EPIC & 7.50 & 16.6 \\
HD 27442    &     &            &             & 2016-09-30 & EPIC & 35.7 & \dots\\
HD 75289     &G0V  & 08:47:40.4 & $-41$:44:12 & 2005-04-28 & EPIC & 8.14 & 3.5  \\
HD 75289     &     &            &             & 2013-11-27 & EPIC & 15.1 &\dots \\
GJ 436      & M2.5V & 11:42:11.1 & $+26$:42:24 & 2008-12-10 & EPIC & 30.3 & 21.7 \\
GJ 436      &     &            &             & 2015-11-21 & EPIC & 22.5 & \dots\\
HD 108147    &F9V  & 12:25:46.3 & $-64$:01:20 & 2002-08-10 & EPIC & 5.74 & 35.8 \\
HD 108147    &     &            &             & 2010-08-08 & EPIC & 52.5 &\dots \\
GJ 674      &M2.5V& 17:28:39.9 & $-46$:53:43 & 2008-09-05 & EPIC & 20.0 &  223 \\
GJ 674      &     &            &             & 2018-04-03 & EPIC & 24.5 & \dots\\
HD 189733    &K2V  & 20:00:43.7 & $+22$:42:39 & 2007-04-17 & EPIC & 38.2 & 460  \\
HD 189733    &     &            &             & 2009-05-18 & EPIC & 30.1 &\dots \\
HD 189733    &     &            &             & 2011-04-30 & EPIC & 28.1 &\dots \\
HD 189733    &     &            &             & 2012-05-08 & EPIC & 56.9 &\dots \\
HD 189733    &     &            &             & 2013-05-09 & EPIC & 29.3 &\dots \\
HD 189733    &     &            &             & 2013-11-03 & EPIC & 28.0 &\dots \\
HD 189733    &     &            &             & 2014-04-05 & EPIC & 26.7 &\dots \\
HD 189733    &     &            &             & 2014-10-18 & EPIC & 26.1 &\dots \\
HD 189733    &     &            &             & 2014-11-08 & EPIC & 25.6 &\dots \\
HD 189733    &     &            &             & 2015-04-03 & EPIC & 28.8 &\dots \\
HD 190360    & G6IV & 20:03:37.4 & $+29$:53:48 & 2005-04-25 & EPIC & 4.21 & 3.0  \\
HD 190360    &     &            &             & 2005-11-18 & EPIC & 3.64 &\dots \\
GJ 832       & M1.5V & 21:33:34.0 & $-49$:00:32 & 2014-10-11 & EPIC & 10.2 & 17.6 \\ 
\hline
\end{tabular}
\end{scriptsize}
\end{center}
\vspace{-5mm}
 \tablefoot{
 \tablefoottext{a}{\textit{XMM-Newton} or \textit{Chandra} instrument used to measure the X-ray flux. EPIC exposure time is the average of all EPIC detectors. S/N is the signal-to-noise ratio of the summed observations of the same star.}}
\end{table}

\begin{table}
\caption[]{UV observations of stars in the sample$^a$.}\label{tabuvlog}
\vspace{-0.5cm}
\begin{center}
\begin{scriptsize}
  \begin{tabular}{lcrrcccrc}
\hline \hline
{Star name} & {Date} & {Instrument} \\ 
\hline
55 Cnc      & 2016-04-04--2017-02-14  & COS/G130M \\
AD Leo      &  2000-03-10, 2002-06-01 & STIS/E140M \\
AD Leo      &  1993--2000             & EUVE\tablefootmark{a} \\
AU Mic      &  1998-09-06, 2020-07-02 & STIS/E140M \\
AU Mic      &  2000-08-26, 2001-10-10 & FUSE       \\
AU Mic      &  1993--1996             & EUVE       \\
Barnard's star &  2019-03-04 & COS/G130M+STIS/G140L\tablefootmark{b} \\
GJ 357      & 2021-04-21 & COS/G130M+G160M \\
GJ 436      & 2012-06-23, 2015-06-25  & COS/G130M+G160M \\
GJ 486      &  2022-03-15             & STIS/G140L+G140M\\
GJ 674      & 2018-04-02              & COS/G130M+STIS/G140L & \\
GJ 1214     &  2012-08-04, 2015-08-20 & COS/G130M+G160M \\
GJ 3470     &  2017-12-27--2018-12-23 & COS/G130M       \\
GJ 9827     &  2018-08-23--2020-10-10 & STIS/G140M \\
\object{HD 63443}    & 2020-10-29--2022-02-17  & STIS/G140M \\
HD 73583    & 2021-12-24--2023-03-02  & STIS/G140M \\  
HD 97658    & 2015-02-13              & COS/G130M+G160M \\
HD 149026   &  2021-07-26, 2022-10-05 & STIS/G140L      \\
HD 189733   & 2009-0916--2017-07-03   & COS/G130M\tablefootmark{c} \\
HD 209458   & 2009-09-15              & COS/G130M+G160M\tablefootmark{d} \\
Lalande 21185 & 2020-01-14            & STIS/E140M \\
Proxima Cen &  2019-04-28--2019-07-01 & STIS/E140M \\
Proxima Cen &  2003-04-05             & FUSE       \\
Proxima Cen &  1993-05-21             & EUVE       \\
$\tau$ Boo  & 1999-06-09, 1999-06-26  & STIS/G140M \\
TOI-836     & 2022-07-28              & STIS/G140M+G140L \\
TRAPPIST-1  &  2016-09-26--2019-06-07 & COS/G130M+G160M \\
$\upsilon$ And & 2011-11-09           & STIS/E140M \\
\object{WASP-13}     &  2015-04-21             & COS/G140L \\
WASP-52     &  2021-10-05, 2021-10-07 & STIS/G140M+G140L \\ 
WASP-77A    &  2022-02-02             & STIS/G140L \\ 
\hline
\end{tabular}
\end{scriptsize}
\end{center}
\vspace{-3mm}
 \tablefoot{
   \tablefoottext{a}{Fluxes from \citet{san02}.}
   \tablefoottext{b}{STIS fluxes were corrected to match the level of the mean of the fluxes of the \ion{C}{iii} 1176~\AA, \ion{Si}{iii} 1206~\AA, and \ion{C}{ii} 1335 lines in the COS spectrum.}
   \tablefoottext{c}{Summed spectrum from HSLA (\textit{Hubble} Spectroscopic Legacy Archive), available at \url{https://archive.stsci.edu/hst/spectral_legacy/}.}
   \tablefoottext{d}{Fluxes from \citet{fra10}.}
 }
\end{table}

\begin{table*}[ht!]
\caption[]{X-ray (5--100~\AA), EUV$_{\rm H}$ (100--920~\AA), and EUV$_{\rm He}$ (100--504~\AA) luminosities of stars in the sample$^a$.}\label{tabresults}
\renewcommand{\arraystretch}{1.3} 
\vspace{-5mm}
\begin{center}
\begin{scriptsize}
\begin{tabular}{llcccccccccc}
\hline \hline
Planet name & SpT & Stellar distance & $\log L_{\rm bol}$ & $\log L_{\rm X}$ & $\log L_{\rm EUV,H}$ & $\log L_{\rm EUV,He}$ & $M_p \sin i$ & $a_p$ & $\log F_{\rm XUV,H}$ & $\log F_{\rm XUV,He}$ & $\rho \dot M_p$ \\
& & (pc) & {(erg s$^{-1}$)} & {(erg s$^{-1}$)} & {(erg s$^{-1}$)} & {(erg s$^{-1}$)} & {(M$_{\rm J}$)} & {(au)} &
\multicolumn{2}{c}{(erg s$^{-1}$cm$^{-2}$)} & {(g$^2$s$^{-1}$cm$^{-3}$)\tablefootmark{b}} \\
\hline
14 Her b & K0V & 17.899$\pm$0.010 & 33.41 & 26.92 & 27.70$^{+0.44}_{-0.38}$ & 27.16$^{+0.33}_{-0.22}$ & 5.215 & 2.820 & $-$0.58 & $-$0.99 & 2.9e+06 \\
14 Her c &  &  &  &  &  &  & 7.031 & 27.000 & $-$2.55 & $-$2.96 & 3.2e+04 \\
16 Cyg B b & G2V & 21.128$\pm$0.009 & 33.68 & $<$26.78 & $<$27.98 & $<$27.45 & 1.640 & 1.680 & < 0.11 & <-0.36 & (1.4e+07) \\
2M1207 A b & M8V & 64.68$\pm$0.50 & 31.08 & $<$26.16 & $<$27.36 & $<$26.83 & 5.500 & 42.000 & <-3.29 & <-3.78 & (5.8e+03) \\
30 Ari B b\tablefootmark{c} & F6V & 44.924$\pm$0.061 & 34.20 & 29.67 & 29.57$^{+0.09}_{-0.03}$ & 29.51$^{+0.03}_{-0.01}$ & 9.880 & 0.995 & 2.47 & 2.45 & 3.3e+09 \\
47 Uma b & G0V & 13.887$\pm$0.019 & 33.78 & 25.75 & 26.62$^{+0.37}_{-0.26}$ & 26.26$^{+0.20}_{-0.10}$ & 2.530 & 2.100 & $-$1.42 & $-$1.71 & 4.2e+05 \\
47 Uma c &  &  &  &  &  &  & 0.540 & 3.600 & $-$1.89 & $-$2.18 & 1.4e+05 \\
47 Uma d &  &  &  &  &  &  & 1.640 & 11.600 & $-$2.91 & $-$3.20 & 1.4e+04 \\
51 Peg b & G2IV & 15.528$\pm$0.019 & 33.79 & $<$26.66 & $<$27.86 & $<$27.33 & 0.460 & 0.052 & < 3.01 & < 2.54 & (1.1e+10) \\
55 Cnc b & G8V & 12.587$\pm$0.006 & 33.38 & 26.77 & 27.67$^{+0.15}_{-0.12}$ & 27.39$^{+0.11}_{-0.09}$ & 0.840 & 0.113 & 2.17 & 1.92 & 1.6e+09 \\
55 Cnc c &  &  &  &  &  &  & 0.178 & 0.237 & 1.52 & 1.28 & 3.8e+08 \\
55 Cnc d &  &  &  &  &  &  & 3.840 & 5.446 & $-$1.20 & $-$1.44 & 7.1e+05 \\
55 Cnc e &  &  &  &  &  &  & 0.027 & 0.015 & 3.90 & 3.66 & 8.9e+10 \\
55 Cnc f &  &  &  &  &  &  & 0.148 & 0.773 & 0.50 & 0.26 & 3.5e+07 \\
AD Leo b\tablefootmark{c} & M3.0V & 4.965$\pm$0.001 & 31.96 & 28.61 & 28.77$^{+0.16}_{-0.15}$ & 28.54$^{+0.17}_{-0.15}$ & 0.073 & 0.025 & 4.75 & 4.64 & 6.4e+11 \\
AU Mic b & M1.5V & 9.714$\pm$0.002 & 32.58 & 29.44 & 29.39$^{+0.26}_{-0.21}$ & 29.25$^{+0.26}_{-0.18}$ & 0.028 & 0.070 & 4.58 & 4.51 & 4.2e+11 \\
AU Mic c &  &  &  &  &  &  & 0.0456 & 0.149 & 4.12 & 4.05 & 1.5e+11 \\
AU Mic d &  &  &  &  &  &  & 0.0032 & 0.085 & 4.41 & 4.35 & 2.9e+11 \\
Barnard's star b & M3.5V & 1.828$\pm$0.0001 & 31.15 & 25.61 & 26.13$^{+0.16}_{-0.20}$ & 25.74$^{+0.19}_{-0.15}$ & 0.0012 & 0.0229 & 2.07 & 1.81 & 1.3e+09 \\
CoRoT-7 b & K0V & 159.49$\pm$0.25 & 33.27 & 28.98 & 30.06$^{+0.28}_{-0.16}$ & 29.84$^{+0.11}_{-0.05}$ & 0.019 & 0.017 & 6.16 & 5.98 & 1.6e+13 \\
CoRoT-7 c &  &  &  &  &  &  & 0.042 & 0.046 & 5.30 & 5.13 & 2.3e+12 \\
CoRoT-7 d &  &  &  &  &  &  & 17.142 & 0.080 & 4.82 & 4.65 & 7.4e+11 \\
GJ 12 b & M3.0V & 12.167$\pm$0.004 & 31.46 & 25.78 & 26.58$^{+0.43}_{-0.37}$ & 26.05$^{+0.32}_{-0.21}$ & 0.0049 & 0.066 & 1.55 & 1.15 & 4.0e+08 \\
GJ 86 b & K1V & 10.761$\pm$0.006 & 33.18 & $<$27.41 & $<$28.13 & $<$28.44 & 4.270 & 0.118 & $<$3.10 & $<$2.52 & (1.4e+10) \\
GJ 317 b & M3V & 15.179$\pm$0.009 & 31.93 & 26.59 & 27.87$^{+0.47}_{-0.45}$ & 27.21$^{+0.44}_{-0.39}$ & 1.753 & 1.151 & 0.32 & $-$0.27 & 2.4e+07 \\
GJ 317 c &  &  &  &  &  &  & 1.644 & 5.230 & $-$0.99 & $-$1.59 & 1.1e+06 \\
GJ 357 b & M2.5V & 9.436$\pm$0.002 & 31.78 & 25.66 & 26.36$^{+0.12}_{-0.25}$ & 25.93$^{+0.08}_{-0.12}$ & 0.0066 & 0.033 & 1.95 & 1.63 & 1.0e+09 \\
GJ 357 c &  &  &  &  &  &  & 0.012 & 0.061 & 1.42 & 1.10 & 2.9e+08 \\
GJ 357 d &  &  &  &  &  &  & 0.023 & 0.204 & 0.37 & 0.05 & 2.6e+07 \\
GJ 436 b & M2.5V & 9.775$\pm$0.003 & 31.97 & 26.14 & 26.79$^{+0.14}_{-0.20}$ & 26.30$^{+0.08}_{-0.10}$ & 0.067 & 0.029 & 2.50 & 2.16 & 3.6e+09 \\
GJ 486 b & M3.5V & 8.079$\pm$0.002 & 31.66 & 25.50 & 26.27$^{+0.25}_{-0.22}$ & 25.69$^{+0.23}_{-0.16}$ & 0.0089 & 0.017 & 2.41 & 1.98 & 2.9e+09 \\
GJ 674 b & M2.5V & 4.553$\pm$0.001 & 31.79 & 27.67 & 28.02$^{+0.15}_{-0.08}$ & 27.76$^{+0.10}_{-0.03}$ & 0.040 & 0.039 & 3.55 & 3.39 & 3.9e+10 \\
GJ 806 b & M1.5V & 12.064$\pm$0.003 & 32.01 & 26.23 & 27.53$^{+0.47}_{-0.45}$ & 26.87$^{+0.43}_{-0.38}$ & 0.0060 & 0.014 & 3.81 & 3.21 & 7.2e+10 \\
GJ 806 c &  &  &  &  &  &  & 0.018 & 0.052 & 2.67 & 2.07 & 5.2e+09 \\
GJ 832 b & M1.5V & 4.9671$\pm$0.0006 & 32.04 & 26.18 & 27.45$^{+0.46}_{-0.44}$ & 26.80$^{+0.42}_{-0.36}$ & 0.740 & 3.600 & $-$1.09 & $-$1.67 & 9.1e+05 \\
GJ 876 b & M4V & 4.672$\pm$0.001 & 31.69 & 26.20 & 27.37$^{+0.47}_{-0.44}$ & 26.83$^{+0.42}_{-0.36}$ & 3.534 & 0.214 & 1.29 & 0.81 & 2.2e+08 \\
GJ 876 c &  &  &  &  &  &  & 0.698 & 0.134 & 1.70 & 1.22 & 5.6e+08 \\
GJ 876 d &  &  &  &  &  &  & 0.019 & 0.021 & 3.31 & 2.83 & 2.3e+10 \\
GJ 876 e &  &  &  &  &  &  & 0.042 & 0.345 & 0.88 & 0.39 & 8.5e+07 \\
GJ 1214 b & M4.5V & 14.641$\pm$0.015 & 31.13 & 25.88 & 26.27$^{+0.21}_{-0.14}$ & 25.94$^{+0.12}_{-0.08}$ & 0.026 & 0.014 & 2.67 & 2.47 & 5.3e+09 \\
GJ 3470 b & M1.5V & 29.394$\pm$0.026 & 32.19 & 27.53 & 27.84$^{+0.10}_{-0.09}$ & 27.66$^{+0.06}_{-0.04}$ & 0.040 & 0.036 & 3.46 & 3.35 & 3.2e+10 \\
GJ 9827 b & K6V & 29.656$\pm$0.018 & 32.61 & 26.53 & 27.14$^{+0.24}_{-0.17}$ & 26.57$^{+0.17}_{-0.11}$ & 0.015 & 0.019 & 3.25 & 2.86 & 2.0e+10 \\
GJ 9827 c &  &  &  &  &  &  & 0.0060 & 0.039 & 2.61 & 2.22 & 4.5e+09 \\
GJ 9827 d &  &  &  &  &  &  & 0.011 & 0.056 & 2.30 & 1.91 & 2.2e+09 \\
GQ Lup b & K7V & 154.08$\pm$0.71 & 33.69 & 30.45 & 30.59$^{+0.37}_{-0.28}$ & 30.17$^{+0.22}_{-0.13}$ & 31.000 & 32.000 & 0.17 & 0.17 & 1.7e+07 \\
HAT-P-1 b & G0V & 160.26$\pm$0.26 & 33.79 & $<$28.17 & $<$29.34 & $<$28.90 & 0.523 & 0.056 & < 4.40 & < 4.03 & (2.8e+11) \\
HAT-P-11 b & K4V & 37.836$\pm$0.014 & 33.01 & 27.38 & 28.96$^{+0.00}_{-0.00}$ & 28.37$^{+0.00}_{-0.00}$ & 0.074 & 0.053 & 4.08 & 3.51 & 1.3e+11 \\
HAT-P-11 c &  &  &  &  &  &  & 1.595 & 4.130 & 0.29 & $-$0.27 & 2.2e+07 \\
HAT-P-12 b & K4V & 142.05$\pm$0.20 & 32.88 & $<$27.95 & $<$29.16 & $<$28.62 & 0.210 & 0.038 & < 4.56 & < 4.09 & (4.1e+11) \\
HAT-P-32 b & F8V & 286.5$\pm$1.6 & 33.90 & 29.34 & 30.53$^{+0.47}_{-0.45}$ & 29.88$^{+0.43}_{-0.37}$ & 0.750 & 0.034 & 6.02 & 5.47 & 1.2e+13 \\
HD 4308 b & G5V & 22.046$\pm$0.010 & 33.59 & $<$26.15 & $<$27.35 & $<$26.82 & 0.041 & 0.118 & < 1.78 & < 1.31 & (6.8e+08) \\
HD 20367 b & G0V & 26.055$\pm$0.020 & 33.77 & 29.27 & 29.51$^{+0.38}_{-0.26}$ & 29.18$^{+0.22}_{-0.11}$ & 1.070 & 1.250 & 2.06 & 1.89 & 1.3e+09 \\
HD 27442 A b & K2IV & 18.440$\pm$0.034 & 34.88 & 26.64 & 27.47$^{+0.32}_{-0.20}$ & 27.21$^{+0.15}_{-0.06}$ & 1.350 & 1.160 & $-$0.05 & $-$0.26 & 1.0e+07 \\
HD 46375 A b & K1IV & 29.516$\pm$0.017 & 33.40 & 27.03 & 28.30$^{+0.47}_{-0.45}$ & 27.64$^{+0.44}_{-0.39}$ & 0.230 & 0.041 & 3.65 & 3.06 & 5.0e+10 \\
HD 49674 b & G5V & 42.882$\pm$0.037 & 33.55 & 27.47 & 28.58$^{+0.45}_{-0.41}$ & 28.01$^{+0.37}_{-0.27}$ & 0.100 & 0.058 & 3.64 & 3.14 & 4.9e+10 \\
HD 50554 b & F8V & 31.066$\pm$0.019 & 33.74 & $<$26.36 & $<$27.54 & $<$27.07 & 5.160 & 2.410 & <-0.64 & <-1.07 & (2.6e+06) \\
HD 52265 b & G0V & 29.922$\pm$0.018 & 33.94 & 26.99 & 28.15$^{+0.43}_{-0.37}$ & 27.63$^{+0.32}_{-0.21}$ & 1.210 & 0.520 & 1.30 & 0.84 & 2.2e+08 \\
HD 63433 b & G5V & 22.381$\pm$0.010 & 33.46 & 28.78 & 28.93$^{+0.14}_{-0.21}$ & 28.52$^{+0.10}_{-0.11}$ & $<$0.068 & 0.072 & 4.00 & 3.81 & 1.1e+11 \\
HD 63433 c &  &  &  &  &  &  & 0.048 & 0.146 & 3.39 & 3.20 & 2.8e+10 \\
\hline
\end{tabular}
\end{scriptsize}
\end{center}
\end{table*}
\setcounter{table}{2}
\begin{table*}
\caption{Continued$^a$.}
\begin{center}
\begin{scriptsize}
\begin{tabular}{llcccccccccc}
  \hline \hline
  \noalign{\smallskip}
Planet name & SpT & Stellar distance & $\log L_{\rm bol}$ & $\log L_{\rm X}$ & $\log L_{\rm EUV,H}$ & $\log L_{\rm EUV,He}$ & $M_p \sin i$ & $a_p$ & $\log F_{\rm XUV,H}$ & $\log F_{\rm XUV,He}$ & $\rho \dot M_p$ \\
& & (pc) & {(erg s$^{-1}$)} & {(erg s$^{-1}$)} & {(erg s$^{-1}$)} & {(erg s$^{-1}$)} & {(M$_{\rm J}$)} & {(au)} &
\multicolumn{2}{c}{(erg s$^{-1}$cm$^{-2}$)} & {(g$^2$s$^{-1}$cm$^{-3}$)\tablefootmark{b}} \\
\hline 
HD 63433 d &  &  &  &  &  &  & \dots & 0.050 & 4.31 & 4.12 & 2.3e+11 \\
HD 70642 b & G5IV & 29.283$\pm$0.017 & 33.40 & 26.37 & 27.57$^{+0.46}_{-0.42}$ & 26.96$^{+0.39}_{-0.30}$ & 2.000 & 3.300 & $-$0.89 & $-$1.42 & 1.5e+06 \\
HD 73583 b & K4V & 31.586$\pm$0.020 & 32.83 & 27.82 & 27.98$^{+0.15}_{-0.22}$ & 27.54$^{+0.09}_{-0.09}$ & 0.032 & 0.060 & 3.19 & 3.00 & 1.8e+10 \\
HD 73583 c &  &  &  &  &  &  & 0.030 & 0.123 & 2.57 & 2.38 & 4.2e+09 \\
HD 75289 b & G0V & 29.028$\pm$0.017 & 33.88 & 25.60 & 26.52$^{+0.26}_{-0.14}$ & 26.34$^{+0.09}_{-0.04}$ & 0.470 & 0.046 & 1.79 & 1.64 & 6.9e+08 \\
HD 93083 b & K3V & 28.482$\pm$0.016 & 33.18 & 26.83 & 28.01$^{+0.46}_{-0.42}$ & 27.40$^{+0.39}_{-0.30}$ & 0.370 & 0.477 & 1.23 & 0.70 & 1.9e+08 \\
HD 95089 b & K0IV & 135.50$\pm$0.37 & 34.71 & $<$26.60 & $<$27.80 & $<$27.27 & 1.200 & 1.510 & < 0.01 & $<-$0.45 & (1.1e+07) \\
HD 95089 c &  &  &  &  &  &  & 3.970 & 3.447 & $<-$0.71 & $<-$1.17 & (2.2e+06) \\
HD 97658 b & K1V & 21.561$\pm$0.009 & 33.13 & 26.92 & 27.39$^{+0.11}_{-0.10}$ & 27.17$^{+0.08}_{-0.06}$ & 0.025 & 0.080 & 2.26 & 2.11 & 2.1e+09 \\
HD 99492 b & K2V & 18.162$\pm$0.010 & 33.11 & 26.57 & 27.70$^{+0.45}_{-0.41}$ & 27.11$^{+0.38}_{-0.29}$ & 0.080 & 0.123 & 2.10 & 1.59 & 1.4e+09 \\
HD 99492 c &  &  &  &  &  &  & 0.056 & 5.400 & $-$1.18 & $-$1.69 & 7.4e+05 \\
HD 101930 b & K1V & 29.958$\pm$0.018 & 33.24 & $<$26.00 & $<$27.20 & $<$26.67 & 0.300 & 0.302 & < 0.93 & < 0.35 & (9.5e+07) \\
HD 102195 b & K0V & 29.360$\pm$0.017 & 33.26 & 28.37 & 29.28$^{+0.45}_{-0.40}$ & 28.82$^{+0.38}_{-0.28}$ & 0.490 & 0.049 & 4.50 & 4.12 & 3.6e+11 \\
HD 108147 b & F9V & 38.820$\pm$0.030 & 33.87 & 27.91 & 28.76$^{+0.42}_{-0.34}$ & 28.29$^{+0.29}_{-0.17}$ & 0.261 & 0.102 & 3.35 & 2.97 & 2.5e+10 \\
HD 111232 b & G8V & 28.893$\pm$0.017 & 33.41 & $<$26.50 & $<$27.71 & $<$27.18 & 6.800 & 1.970 & $<-$0.31 & $<-$0.77 & (5.5e+06) \\
HD 114386 b & K3V & 27.980$\pm$0.016 & 33.03 & 26.63 & 27.43$^{+0.39}_{-0.29}$ & 27.02$^{+0.23}_{-0.12}$ & 1.240 & 1.020 & 0.02 & $-$0.29 & 1.2e+07 \\
HD 114386 c &  &  &  &  &  &  & 1.190 & 1.832 & $-$0.49 & $-$0.80 & 3.7e+06 \\
HD 114762 b & F9V & 38.17$\pm$0.16 & 33.75 & $<$26.71 & $<$27.91 & $<$27.38 & 10.980 & 0.353 & < 1.39 & < 0.92 & (2.8e+08) \\
HD 114783 b & K0V & 21.030$\pm$0.013 & 33.19 & 26.52 & 27.32$^{+0.41}_{-0.31}$ & 26.88$^{+0.26}_{-0.14}$ & 1.100 & 1.160 & $-$0.19 & $-$0.54 & 7.2e+06 \\
HD 114783 c &  &  &  &  &  &  & 0.610 & 5.047 & $-$1.47 & $-$1.82 & 3.8e+05 \\
HD 130322 b & K0V & 31.908$\pm$0.031 & 33.33 & 27.33 & 28.57$^{+0.46}_{-0.44}$ & 27.93$^{+0.42}_{-0.35}$ & 1.020 & 0.090 & 3.24 & 2.67 & 1.9e+10 \\
HD 149026 b & G0V & 76.22$\pm$0.12 & 34.00 & 27.40 & 28.10$^{+0.21}_{-0.25}$ & 27.44$^{+0.15}_{-0.22}$ & 0.356 & 0.043 & 3.46 & 3.01 & 3.3e+10 \\
HD 154345 b & G8V & 18.268$\pm$0.007 & 33.35 & 27.13 & 28.34$^{+0.45}_{-0.41}$ & 27.75$^{+0.38}_{-0.28}$ & 1.000 & 4.300 & $-$0.35 & $-$0.87 & 5.0e+06 \\
HD 164922 b & K0V & 21.978$\pm$0.010 & 33.43 & $<$25.51 & $<$26.69 & $<$26.22 & 0.365 & 2.160 & $<-$1.40 & $<-$1.82 & (4.5e+05) \\
HD 164922 c &  &  &  &  &  &  & 0.041 & 0.341 & < 0.21 & $<-$0.22 & (1.8e+07) \\
HD 164922 d &  &  &  &  &  &  & 0.033 & 0.123 & < 1.09 & < 0.67 & (1.4e+08) \\
HD 164922 e &  &  &  &  &  &  & 0.033 & 0.229 & < 0.55 & < 0.13 & (4.0e+07) \\
HD 179949 b & F8V & 27.541$\pm$0.030 & 33.88 & 28.64 & 29.29$^{+0.38}_{-0.27}$ & 28.98$^{+0.24}_{-0.13}$ & 0.920 & 0.045 & 4.62 & 4.39 & 4.7e+11 \\
HD 187123 b & G2V & 46.041$\pm$0.042 & 33.70 & $<$27.28 & $<$28.46 & $<$27.99 & 0.520 & 0.043 & < 3.78 & < 3.36 & (6.8e+10) \\
HD 187123 c &  &  &  &  &  &  & 1.990 & 4.890 & $<-$0.34 & $<-$0.76 & (5.2e+06) \\
HD 189733 b & K2V & 19.775$\pm$0.008 & 33.10 & 28.36 & 28.81$^{+0.08}_{-0.15}$ & 28.53$^{+0.04}_{-0.07}$ & 1.154 & 0.031 & 4.51 & 4.32 & 3.6e+11 \\
HD 190360 b & G6IV & 16.003$\pm$0.010 & 33.69 & $<$26.54 & $<$27.72 & $<$27.25 & 1.495 & 3.920 & $<-$0.88 & $<-$1.31 & (1.5e+06) \\
HD 190360 c &  &  &  &  &  &  & 0.064 & 0.128 & < 2.09 & < 1.66 & (1.4e+09) \\
HD 195019 b & G3IV & 37.523$\pm$0.028 & 33.87 & $<$26.34 & $<$27.51 & $<$27.04 & 3.700 & 0.139 & < 1.81 & < 1.38 & (7.2e+08) \\
HD 201585 b & A8V & 182.15$\pm$0.66 & 34.68 & $<$28.62 & $<$29.80 & $<$29.33 & 3.695 & 0.043 & < 5.10 & < 4.69 & (1.4e+12) \\
HD 209458 b & G0V & 48.146$\pm$0.070 & 33.80 & 26.84 & 28.45$^{+0.06}_{-0.16}$ & 27.92$^{+0.03}_{-0.04}$ & 0.685 & 0.047 & 3.66 & 3.15 & 5.1e+10 \\
HD 216435 b & K6V & 33.124$\pm$0.033 & 34.13 & 27.74 & 28.95$^{+0.46}_{-0.44}$ & 28.30$^{+0.42}_{-0.36}$ & 1.260 & 2.560 & 0.71 & 0.14 & 5.8e+07 \\
HD 216437 b & G4IV & 26.695$\pm$0.021 & 33.62 & 26.63 & 27.83$^{+0.44}_{-0.39}$ & 27.27$^{+0.35}_{-0.24}$ & 1.820 & 2.320 & $-$0.32 & $-$0.82 & 5.3e+06 \\
HD 217107 b & G8IV & 20.088$\pm$0.012 & 33.62 & $<$25.76 & $<$26.93 & $<$26.46 & 1.394 & 0.075 & < 1.77 & < 1.34 & (6.6e+08) \\
HD 217107 c &  &  &  &  &  &  & 4.090 & 5.940 & $<-$2.04 & $<-$2.46 & (1.0e+05) \\
HD 218566 b & K3V & 28.818$\pm$0.025 & 33.11 & 27.14 & 28.29$^{+0.45}_{-0.40}$ & 27.72$^{+0.36}_{-0.26}$ & 0.210 & 0.687 & 1.19 & 0.70 & 1.8e+08 \\
HD 235088 b & K2V & 41.237$\pm$0.017 & 33.13 & 28.28 & 28.17$^{+0.30}_{-0.19}$ & 27.89$^{+0.14}_{-0.07}$ & 0.022 & 0.070 & 3.38 & 3.28 & 2.7e+10 \\
HD 283636 b & K0V & 50.942$\pm$0.052 & 32.76 & 27.71 & 28.89$^{+0.46}_{-0.44}$ & 28.26$^{+0.41}_{-0.34}$ & 0.025 & 0.036 & 4.36 & 3.81 & 2.6e+11 \\
HD 330075 b & G5V & 45.455$\pm$0.021 & 33.16 & 26.49 & 27.73$^{+0.46}_{-0.43}$ & 27.10$^{+0.41}_{-0.34}$ & 0.620 & 0.039 & 3.13 & 2.56 & 1.5e+10 \\
HR 8799 b & A5V & 40.883$\pm$0.084 & 34.32 & 28.13 & 29.42$^{+0.47}_{-0.45}$ & 28.76$^{+0.43}_{-0.38}$ & 3.286 & 68.000 & $-$1.67 & $-$2.26 & 2.4e+05 \\
HR 8799 c &  &  &  &  &  &  & 3.897 & 42.900 & $-$1.27 & $-$1.86 & 6.1e+05 \\
HR 8799 d &  &  &  &  &  &  & 3.897 & 27.000 & $-$0.87 & $-$1.46 & 1.5e+06 \\
HR 8799 e &  &  &  &  &  &  & 4.057 & 16.400 & $-$0.43 & $-$1.03 & 4.1e+06 \\
K2-100 b & G0V & 182.82$\pm$0.67 & 33.81 & 29.25 & 29.32$^{+0.13}_{-0.06}$ & 29.23$^{+0.04}_{-0.02}$ & 0.069 & 0.030 & 5.16 & 5.15 & 1.6e+12 \\
K2-136 b & K5V & 58.893$\pm$0.069 & 32.77 & 27.80 & 28.81$^{+0.45}_{-0.41}$ & 28.22$^{+0.37}_{-0.27}$ & \dots & 0.071 & 3.70 & 3.22 & 5.7e+10 \\
K2-136 c &  &  &  &  &  &  & \dots & 0.118 & 3.25 & 2.77 & 2.0e+10 \\
K2-136 d &  &  &  &  &  &  & \dots & 0.154 & 3.03 & 2.54 & 1.2e+10 \\
KELT-7 b & F3IV-V & 135.50$\pm$0.37 & 34.34 & 28.60 & 28.77$^{+0.13}_{-0.06}$ & 28.68$^{+0.03}_{-0.01}$ & 1.233 & 0.045 & 4.21 & 4.20 & 1.8e+11 \\
KELT-8 b & G2V-IV & 197.24$\pm$0.39 & 34.02 & $<$29.26 & $<$30.40 & $<$30.01 & 0.867 & 0.046 & < 5.64 & < 5.31 & (4.9e+12) \\
KELT-9 b & A0V & 207.04$\pm$0.86 & 35.12 & $<$29.50 & $<$30.68 & $<$30.21 & 2.880 & 0.034 & < 6.18 & < 5.78 & (1.7e+13) \\
Lalande 21185 b & M1.5V & 2.5461$\pm$0.0002 & 31.92 & 27.10 & 26.93$^{+0.25}_{-0.21}$ & 26.80$^{+0.27}_{-0.19}$ & 0.0085 & 0.079 & 2.08 & 2.03 & 1.4e+09 \\
Lalande 21185 c &  &  &  &  &  &  & 13.600 & 2.940 & $-$1.06 & $-$1.11 & 9.7e+05 \\
Lalande 21185 d &  &  &  &  &  &  & 3.890 & 0.514 & 0.45 & 0.40 & 3.2e+07 \\
NGC 2423 3 b & G5IV & 934.6$\pm$8.7 & 35.51 & $<$29.90 & $<$30.44 & $<$29.85 & 10.600 & 2.100 & < 2.39 & < 2.09 & (2.8e+09) \\
Pollux b & K0III & 10.358$\pm$0.029 & 35.15 & 27.40 & 28.27$^{+0.25}_{-0.13}$ & 28.12$^{+0.10}_{-0.04}$ & 2.900 & 1.690 & 0.42 & 0.29 & 3.0e+07 \\
Proxima Cen b & M5.5V & 1.3020$\pm$0.0001 & 30.75 & 27.25 & 27.30$^{+0.27}_{-0.20}$ & 27.17$^{+0.25}_{-0.17}$ & 0.00337 & 0.049 & 2.76 & 2.69 & 6.4e+09 \\
Proxima Cen d &  &  &  &  &  &  & 0.00082 & 0.029 & 3.21 & 3.14 & 1.8e+10 \\
TOI-836 b & K7V & 27.510$\pm$0.015 & 32.70 & 27.41 & 28.08$^{+0.16}_{-0.23}$ & 27.53$^{+0.10}_{-0.13}$ & 0.014 & 0.042 & 3.46 & 3.08 & 3.3e+10 \\
TOI-836 c &  &  &  &  &  &  & 0.030 & 0.075 & 2.96 & 2.58 & 1.0e+10 \\
TOI-2134 b & K5V & 22.671$\pm$0.005 & 32.87 & 26.94 & 27.71$^{+0.38}_{-0.27}$ & 27.34$^{+0.22}_{-0.11}$ & 0.029 & 0.078 & 2.54 & 2.25 & 3.9e+09 \\
TOI-2134 c &  &  &  &  &  &  & 0.132 & 0.371 & 1.19 & 0.90 & 1.7e+08 \\
TRAPPIST-1 b & M8.0V & 12.467$\pm$0.011 & 30.30 & 26.06 & 26.29$^{+0.05}_{-0.09}$ & 26.14$^{+0.02}_{-0.02}$ & 0.0043 & 0.011 & 2.95 & 2.86 & 1.0e+10 \\
TRAPPIST-1 c &  &  &  &  &  &  & 0.0041 & 0.015 & 2.68 & 2.59 & 5.3e+09 \\
TRAPPIST-1 d &  &  &  &  &  &  & 0.0012 & 0.021 & 2.38 & 2.29 & 2.7e+09 \\
TRAPPIST-1 e &  &  &  &  &  &  & 0.0022 & 0.028 & 2.14 & 2.05 & 1.6e+09 \\
TRAPPIST-1 f &  &  &  &  &  &  & 0.0033 & 0.037 & 1.90 & 1.81 & 9.0e+08 \\
TRAPPIST-1 g &  &  &  &  &  &  & 0.0042 & 0.045 & 1.73 & 1.64 & 6.1e+08 \\
TRAPPIST-1 h &  &  &  &  &  &  & 0.0010 & 0.063 & 1.44 & 1.35 & 3.1e+08 \\
Teegarden's b & M7.0V & 3.832$\pm$0.001 & 30.46 & 25.96 & 27.06$^{+0.44}_{-0.38}$ & 26.50$^{+0.35}_{-0.23}$ & 0.0036 & 0.026 & 2.82 & 2.34 & 7.4e+09 \\
Teegarden's c &  &  &  &  &  &  & 0.0033 & 0.045 & 2.33 & 1.85 & 2.4e+09 \\
Teegarden's d &  &  &  &  &  &  & 0.0026 & 0.079 & 1.85 & 1.37 & 8.0e+08 \\
V1298 Tau b & K1IV & 107.99$\pm$0.23 & 33.55 & 30.24 & 30.20$^{+0.20}_{-0.20}$ & 30.07$^{+0.22}_{-0.19}$ & 0.64 & 0.172 & 4.59 & 4.54 & 4.3e+11 \\
V1298 Tau c &  &  &  &  &  &  & 0.062 & 0.084 & 5.21 & 5.17 & 1.8e+12 \\
V1298 Tau d &  &  &  &  &  &  & $<$0.114 & 0.110 & 4.97 & 4.93 & 1.1e+12 \\
V1298 Tau e &  &  &  &  &  &  & 0.66 & 0.267 & 4.20 & 4.16 & 1.8e+11 \\
WASP-12 b & G0V & 413.2$\pm$3.4 & 34.08 & $<$28.80 & $<$29.94 & $<$29.56 & 1.460 & 0.023 & < 5.75 & < 5.44 & (6.4e+12) \\
WASP-13 b & G1V-IV & 230.9$\pm$1.1 & 33.47 & 28.43 & 28.94$^{+0.05}_{-0.03}$ & 28.87$^{+0.01}_{-0.01}$ & 0.484 & 0.054 & 4.10 & 4.09 & 1.4e+11 \\
\hline
\end{tabular}
\end{scriptsize}
\end{center}
\end{table*}
\setcounter{table}{2}
\begin{table*}
\caption{Continued$^a$.}
\begin{center}
\begin{scriptsize}
\begin{tabular}{llcccccccccc}
\hline \hline
  \noalign{\smallskip}
Planet name & SpT & Stellar distance & $\log L_{\rm bol}$ & $\log L_{\rm X}$ & $\log L_{\rm EUV,H}$ & $\log L_{\rm EUV,He}$ & $M_p \sin i$ & $a_p$ & $\log F_{\rm XUV,H}$ & $\log F_{\rm XUV,He}$ & $\rho \dot M_p$ \\
& & (pc) & {(erg s$^{-1}$)} & {(erg s$^{-1}$)} & {(erg s$^{-1}$)} & {(erg s$^{-1}$)} & {(M$_{\rm J}$)} & {(au)} &
\multicolumn{2}{c}{(erg s$^{-1}$cm$^{-2}$)} & {(g$^2$s$^{-1}$cm$^{-3}$)\tablefootmark{b}} \\
\hline
WASP-39 b & G9V & 215.52$\pm$0.46 & 33.38 & $<$28.26 & $<$29.46 & $<$28.95 & 0.280 & 0.049 & < 4.65 & < 4.21 & (5.0e+11) \\
WASP-52 b & K2V & 174.52$\pm$0.30 & 33.18 & 28.53 & 29.37$^{+0.13}_{-0.20}$ & 28.92$^{+0.06}_{-0.05}$ & 0.460 & 0.027 & 5.09 & 4.75 & 1.4e+12 \\
WASP-69 b & K4V & 50.277$\pm$0.051 & 33.11 & 28.10 & 28.35$^{+0.40}_{-0.33}$ & 27.83$^{+0.29}_{-0.18}$ & 0.260 & 0.045 & 3.78 & 3.52 & 6.8e+10 \\
WASP-76 b & F7IV & 189.0$\pm$2.9 & 34.20 & $<$28.72 & $<$29.88 & $<$29.47 & 0.920 & 0.033 & < 5.40 & < 5.05 & (2.8e+12) \\
WASP-77 A b & G8V & 105.71$\pm$0.22 & 33.48 & 28.13 & 28.48$^{+0.14}_{-0.25}$ & 27.89$^{+0.12}_{-0.16}$ & 1.760 & 0.024 & 4.40 & 4.12 & 2.8e+11 \\
WASP-80 b & K7V & 49.727$\pm$0.049 & 32.53 & 27.51 & 27.61$^{+0.28}_{-0.17}$ & 27.36$^{+0.11}_{-0.05}$ & 0.554 & 0.035 & 3.33 & 3.22 & 2.4e+10 \\
WASP-107 b & K7V & 64.392$\pm$0.124 & 32.73 & 27.67 & 28.86$^{+0.46}_{-0.43}$ & 28.24$^{+0.40}_{-0.32}$ & 0.120 & 0.055 & 3.95 & 3.42 & 1.0e+11 \\
WASP-107 c &  &  &  &  &  &  & 0.360 & 1.829 & 0.91 & 0.37 & 9.1e+07 \\
WASP-127 b & G5V & 160.772$\pm$0.517 & 33.84 & $<$28.37 & $<$29.55 & $<$29.08 & 0.180 & 0.052 & < 4.68 & < 4.27 & (5.4e+11) \\
$\beta$ Pic b & A6V & 19.635$\pm$0.058 & 34.54 & 25.75 & 26.56$^{+0.39}_{-0.29}$ & 26.16$^{+0.23}_{-0.12}$ & 11.898 & 9.930 & $-$2.82 & $-$3.14 & 1.7e+04 \\
$\beta$ Pic c &  &  &  &  &  &  & 8.500 & 2.680 & $-$1.68 & $-$2.00 & 2.3e+05 \\
$\epsilon$ Eri b & K2V & 3.220$\pm$0.001 & 33.10 & 28.35 & 28.59$^{+0.21}_{-0.22}$ & 28.42$^{+0.16}_{-0.21}$ & 0.770 & 3.530 & 0.24 & 0.14 & 2.0e+07 \\
$\iota$ Hor b & G0V & 17.358$\pm$0.012 & 33.82 & 28.91 & 29.14$^{+0.19}_{-0.18}$ & 28.66$^{+0.23}_{-0.16}$ & 2.260 & 0.925 & 1.95 & 1.72 & 1.0e+09 \\
$\mu$ Ara b & G3IV & 15.603$\pm$0.022 & 33.75 & $<$26.22 & $<$27.39 & $<$26.94 & 1.676 & 1.500 & $<-$0.38 & $<-$0.79 & (4.6e+06) \\
$\mu$ Ara c &  &  &  &  &  &  & 0.033 & 0.091 & < 2.05 & < 1.65 & (1.3e+09) \\
$\mu$ Ara d &  &  &  &  &  &  & 0.522 & 0.921 & < 0.04 & $<-$0.36 & (1.2e+07) \\
$\mu$ Ara e &  &  &  &  &  &  & 1.814 & 5.235 & $<-$1.47 & $<-$1.87 & (3.8e+05) \\
$\tau$ Boo A b & F7V & 15.613$\pm$0.027 & 34.09 & 29.00 & 29.23$^{+0.14}_{-0.25}$ & 28.96$^{+0.15}_{-0.25}$ & 4.130 & 0.046 & 4.66 & 4.51 & 5.1e+11 \\
$\upsilon$ And b & F8V & 13.479$\pm$0.038 & 34.11 & 27.83 & 28.56$^{+0.13}_{-0.13}$ & 28.24$^{+0.10}_{-0.07}$ & 0.620 & 0.059 & 3.64 & 3.39 & 4.9e+10 \\
$\upsilon$ And c &  &  &  &  &  &  & 1.800 & 0.861 & 1.31 & 1.06 & 2.3e+08 \\
$\upsilon$ And d &  &  &  &  &  &  & 10.190 & 2.550 & 0.37 & 0.12 & 2.6e+07 \\
$\upsilon$ And e &  &  &  &  &  &  & 1.059 & 5.246 & $-$0.26 & $-$0.51 & 6.2e+06 \\
\hline
\end{tabular}
\end{scriptsize}
\end{center}
\vspace{-3mm}
 \tablefoot{
   \tablefoottext{a}{XUV (5--920 or 5--504~\AA) fluxes at the planets' orbits, and mass loss rates assuming the energy-limited approach.}
   \tablefoottext{b}{1~M$_{\rm J}$\,Gyr$^{-1}$=$6.02\times 10^{13}$~g\,s$^{-1}$, 1~M$_{\oplus}$\,Gyr$^{-1}$=$1.89\times 10^{11}$~g\,s$^{-1}$. Mass loss rates based on $F_{\rm XUV,H}$ upper limits are expressed in parenthesis.}   
   \tablefoottext{c}{Controversial or retracted planet detection.}
   }
\renewcommand{\arraystretch}{1.}
\end{table*}

\begin{table*}
\caption[]{\ion{He}{i} 10830 \AA\ triplet equivalent width and XUV (5--504~\AA) flux at planet distance.}\label{tabhelium}
\tabcolsep 3.0pt
\vspace{-8mm}
\renewcommand{\arraystretch}{1.3}  
\begin{center}
\begin{scriptsize}
\begin{tabular}{lccccccccl}
\hline \hline
Planet name & SpT & $R_*$ & EW\tablefootmark{a} & $F_{\rm XUV,He}$ & Model\tablefootmark{b} & $M_{\rm p}$ & $R_{\rm p}$ & $\rho_{\rm p}$ & References (planet data and He observation) \\
 &  & ($R_\odot$) & (m\AA) & (W\,m$^{-2}$) &  & (M$_{\rm J}$) & (R$_{\rm J}$) & (g cm$^{-3}$) &  \\
\hline
GJ 436 b & M2.5V & 0.464$\pm$0.011 & $<$1.4 & 0.14$\pm$0.02 & 1 & 0.067$\pm$0.001 & 0.374$\pm$0.010 & 1.595$\pm$0.125 & \citet{tri18,nor18} \\
GJ 1214 b & M4.5V & 0.216$\pm$0.012 & 33.2$\pm$2.8 & 0.29$^{+0.08}_{-0.06}$ & 1 & 0.026$\pm$0.001 & 0.245$\pm$0.005 & 2.179$\pm$0.170 & \citet{clo21,ore22} \\
GJ 3470 b & M1.5V & 0.48$\pm$0.04 & 20.7$\pm$1.3 & 2.24$^{+0.25}_{-0.17}$ & 1 & 0.040$\pm$0.004 & 0.374$\pm$0.053 & 0.940$\pm$0.408 & \citet{kos19,pal20} \\
\object{HAT-P-11 b} & K4V & 0.683$\pm$0.009 & 12.4$\pm$2.4 & 3.23$\pm$0.02 & 2 & 0.074$\pm$0.005 & 0.389$\pm$0.005 & 1.551$\pm$0.118 & \citet{yee18,all18} \\
HAT-P-18 b & K2V & 0.717$\pm$0.026 & 29.2$\pm$7.6 & 5.2: & 4 & 0.183$\pm$0.034 & 0.947$\pm$0.044 & 0.267$\pm$0.062 & \citet{knu14,para21} \\
HAT-P-26 b & K1V & 0.788$\pm$0.043 & 19.7$\pm$6.4 & $<$2.3 & 2 & 0.059$\pm$0.007 & 0.570$\pm$0.010 & 0.392$\pm$0.052 & \citet{har11,vis22} \\
HAT-P-32 b & F8V & 1.387$\pm$0.067 & 114.0$\pm$4.0 & 258.22$^{+383.18}_{-133.02}$ & 2 & 0.750$\pm$0.130 & 1.789$\pm$0.025 & 0.162$\pm$0.029 & \citet{bon17,cze22} \\
HAT-P-67 b & F5IV & 2.65$\pm$0.12 & 147$\pm$25 & 361.0: & 4 & 0.320$\pm$0.180 & 2.164$\pm$0.010 & 0.039$\pm$0.022 & \citet{gul24} \\
HD 73583 b & K4V & 0.71$\pm$0.02 & 7.4$\pm$0.6 & 0.98$^{+0.11}_{-0.10}$ & 1 & 0.032$\pm$0.011 & 0.249$\pm$0.009 & 2.579$\pm$0.904 & \citet{bar22,zhan22} \\
HD 189733 b & K2V & 0.805$\pm$0.016 & 12.8$\pm$0.4 & 20.86$^{+1.20}_{-1.98}$ & 1 & 1.154$\pm$0.025 & 1.138$\pm$0.077 & 0.971$\pm$0.198 & \citet{pare21,sal18} \\
HD 209458 b & G0V & 1.203$\pm$0.061 & 5.3$\pm$0.5 & 1.41$^{+0.11}_{-0.16}$ & 1 & 0.685$\pm$0.017 & 1.390$\pm$0.018 & 0.316$\pm$0.015 & \citet{wan11,alo19} \\
HD 235088 b & K2V & 0.789$\pm$0.022 & 11.1$\pm$1.7 & 1.88$^{+0.26}_{-0.13}$ & 2 & 0.022$\pm$0.006 & 0.187$\pm$0.018 & 4.172$\pm$1.657 & \citet{zha23,ore23} \\
\object{HD 283636 b} & K0V & 0.636$\pm$0.024 & 8.5$\pm$1.6 & 6.41$^{+8.09}_{-2.89}$ & 2 & 0.025$\pm$0.006 & 0.205$\pm$0.027 & 3.599$\pm$1.664 & \citet{zha23} \\
\object{KELT-9 b} & A0V & 2.362$\pm$0.07 & $<$1.2 & 0.2: & 5 & 2.880$\pm$0.350 & 1.840$\pm$0.040 & 0.573$\pm$0.079 & \citet{bor19,nor18} \\
TOI-1268 b & K2V & 0.92$\pm$0.06 & 17.7$\pm$1.9 & 6.8: & 4 & 0.303$\pm$0.026 & 0.810$\pm$0.050 & 0.707$\pm$0.144 & \citet{sub22,mopys} \\
TOI-1420 b & G6V & 0.923$\pm$0.024 & 42.9$\pm$5.1 & $<$ 1.4 & 4 & 0.079$\pm$0.012 & 1.062$\pm$0.027 & 0.082$\pm$0.014 & \citet{yos23,vis24} \\
TOI-2018 b & K4V & 0.62$\pm$0.01 & 6.8$\pm$1.5 & 1.4: & 4 & 0.029$\pm$0.007 & 0.203$\pm$0.006 & 4.316$\pm$1.062 & \citet{dai23,mopys} \\
\object{TOI-2076 b} & K0 & 0.761$\pm$0.016 & 10.0$\pm$0.7 & 9.91$^{+2.17}_{-3.07}$ & 3 & 0.028:\tablefootmark{c} & 0.293$\pm$0.004 & 1.383: & \citet{hed21,zha23} \\
\object{TOI-2134 b} & K5V & 0.744$\pm$0.027 & 3.3$\pm$0.3 & 0.19$^{+0.10}_{-0.04}$ & 2 & 0.029$\pm$0.002 & 0.240$\pm$0.014 & 2.575$\pm$0.499 & \citet{res24,zha23c} \\
\object{WASP-12 b} & G0V & 1.657$\pm$0.045 & $<$4.0 & $<$236.8 & 2 & 1.460$\pm$0.270 & 1.884$\pm$0.057 & 0.271$\pm$0.056 & \citet{tur21,cze24} \\
WASP-52 b & K2V & 0.79$\pm$0.02 & 39.6$\pm$1.4 & 50.67$^{+8.63}_{-7.29}$ & 1 & 0.460$\pm$0.020 & 1.270$\pm$0.030 & 0.278$\pm$0.023 & \citet{heb13,kir22}\\
WASP-69 b & K4V & 0.813$\pm$0.028 & 28.3$\pm$0.9 & 3.33$^{+1.17}_{-0.48}$ & 2 & 0.260$\pm$0.017 & 1.057$\pm$0.047 & 0.273$\pm$0.041 & \citet{and14,nor18} \\
WASP-76 b & F7IV-V & 1.73$\pm$0.04 & $<$21.3 & $<$103.4 & 2 & 0.920$\pm$0.030 & 1.830$\pm$0.060 & 0.186$\pm$0.019 & \citet{wes16,cas21} \\
\object{WASP-107 b} & K7V & 0.66$\pm$0.02 & 87.2$\pm$7.6 & 2.46$^{+3.12}_{-1.09}$ & 2 & 0.120$\pm$0.001 & 0.940$\pm$0.020 & 0.179$\pm$0.012 & \citet{and17,all19} \\
WASP-177 b & K2V & 0.885$\pm$0.046 & 6.8$\pm$1.6 & 10.9: & 4 & 0.508$\pm$0.038 & 1.580$\pm$0.660 & 0.160$\pm$0.201 & \citet{tur19,kir22} \\
\object{V1298 Tau} b & K1IV & 1.305$\pm$0.07 & $<$19.0 & 34.42$^{+9.84}_{-5.28}$ & 1 & 0.64$\pm$0.19 & 0.888$\pm$0.032 & 0.886$\pm$0.096 & \citet{sua22,mopys} \\
V1298 Tau c & K1IV & 1.305$\pm$0.07 & $<$95.8 & 144.08$^{+41.20}_{-22.09}$ & 1 & 0.062$\pm$0.029 & 0.467$\pm$0.021 & 0.755$\pm$0.368 & \citet{sik23,mopys} \\
\hline
55 Cnc e & G8V & 0.98$\pm$0.016 & $<$0.7 & 4.53$^{+1.15}_{-0.73}$ & 2 & 0.027$\pm$0.001 & 0.174$\pm$0.003 & 6.396$\pm$0.492 & \citet{cri18,zha21} \\
GJ 486 b & M3.5V & 0.328$\pm$0.011 & $<$1.9 & 0.10$^{+0.05}_{-0.02}$ & 1 & (944$\pm$41)e-5 & 0.120$\pm$0.006 & 6.809$\pm$1.000 & \citet{cab22,mas24} \\
GJ 9827 b & K6V & 0.651$\pm$0.065 & $<$1.9 & 0.72$^{+0.22}_{-0.13}$ & 1 & 0.015$\pm$0.001 & 0.136$\pm$0.005 & 7.485$\pm$1.022 & \citet{kos21,car21} \\
GJ 9827 d & K6V & 0.651$\pm$0.065 & $<$1.2 & 0.08$^{+0.02}_{-0.01}$ & 1 & 0.011$\pm$0.002 & 0.174$\pm$0.007 & 2.515$\pm$0.540 & \citet{kos21,car21} \\
HD 63433 b & G5V & 0.912$\pm$0.034 & $<$2.0 & 6.46$^{+0.65}_{-0.59}$ & 1 & $<$0.068:\tablefootmark{c} & 0.191$\pm$0.007 & $<$12.174 & \citet{mall23,zhan22b} \\
HD 63433 c & G5V & 0.912$\pm$0.034 & $<$4.2 & 1.57$^{+0.16}_{-0.14}$ & 1 & 0.048$\pm$0.013 & 0.243$\pm$0.009 & 4.181$\pm$1.220 & \citet{mall23,zhan22b} \\
HD 97658 b & K1V & 0.73$\pm$0.02 & $<$0.9 & 0.13$^{+0.02}_{-0.01}$ & 1 & 0.025$\pm$0.002 & 0.200$\pm$0.009 & 3.782$\pm$0.558 & \citet{vangro14,kas20} \\
K2-100 b & G0V & 1.22$\pm$0.02 & $<$5.7 & 127.72$^{+8.54}_{-4.41}$ & 2 & 0.069$\pm$0.019 & 0.346$\pm$0.014 & 2.054$\pm$0.635 & \citet{bar19,gai20} \\
\object{K2-105 b} & G8V & 0.91$\pm$0.01 & $<$23.3 & 1.2: & 4 & 0.094$\pm$0.060 & 0.301$\pm$0.011 & 4.275$\pm$2.768 & \citet{nar17,zha23} \\
\object{K2-136 c} & K5V & 0.66$\pm$0.02 & $<$25.0 & 0.58$^{+0.59}_{-0.21}$ & 2 & 0.022:\tablefootmark{c} & 0.260$\pm$0.010 & 1.552: & \citet{mann18,gai21} \\
\object{Kepler-25 c} & F6V & 1.31$\pm$0.02 & $<$18.6 & 1.5: & 4 & 0.048$\pm$0.004 & 0.465$\pm$0.006 & 0.591$\pm$0.060 & \citet{mil19,zha23} \\
\object{Kepler-68 b} & G1V & 1.24$\pm$0.02 & $<$7.2 & $<$  0.8 & 4 & 0.024$\pm$0.004 & 0.206$\pm$0.065 & 3.415$\pm$3.287 & \citet{mil19,zha23} \\
TOI-1807 b & K3V & 0.68$\pm$0.015 & $<$2.8 & 207.30$\pm$59.90 & 3 & 0.008$\pm$0.002 & 0.122$\pm$0.009 & 5.498$\pm$1.630 & \citet{nar22,mas24} \\
\object{TOI-3757 b} & M0V & 0.62$\pm$0.01 & $<$73.0 & $<$ 12.7 & 4 & 0.268$\pm$0.028 & 1.071$\pm$0.040 & 0.271$\pm$0.041 & \citet{kan22} \\
TRAPPIST-1 b & M8V & 0.119$\pm$0.001 & $<$3.5 & 0.72$^{+0.03}_{-0.04}$ & 2 & (43.2$\pm$2.2)e-4 & 0.100$\pm$0.001 & 5.429$\pm$0.336 & \citet{agol21,kri21} \\
TRAPPIST-1 e & M8V & 0.119$\pm$0.001 & $<$10.5 & 0.11$\pm$0.01 & 2 & (218$\pm$7)e-5 & 0.082$\pm$0.001 & 4.889$\pm$0.264 & \citet{agol21,kri21} \\
TRAPPIST-1 f & M8V & 0.119$\pm$0.001 & $<$4.1 & 0.07$\pm$0.00 & 2 & (327$\pm$10)e-5 & 0.093$\pm$0.001 & 5.004$\pm$0.244 & \citet{agol21,kri21} \\
\object{WASP-47 d} & G9V & 1.14$\pm$0.01 & $<$32.9 & 1.1: & 4 & 0.041$\pm$0.005 & 0.319$\pm$0.004 & 1.575$\pm$0.190 & \citet{van17,zha23} \\
\object{WASP-48 b} & F8IV & 1.75$\pm$0.09 & $<$25 & 207.9: & 4 & 0.984$\pm$0.085 & 1.670$\pm$0.100 & 0.262$\pm$0.052 & \citet{eno11,ben23} \\
\object{WASP-80 b} & K7V & 0.63$\pm$0.15 & $<$21 & 1.64$^{+0.27}_{-0.14}$ & 2 & 0.554$\pm$0.040 & 0.952$\pm$0.027 & 0.796$\pm$0.089 & \citet{tri13,fos22} \\
\object{WASP-127 b} & G5V & 1.39$\pm$0.03 & $<$6.8 & $<$ 16.8 & 2 & 0.180$\pm$0.020 & 1.370$\pm$0.040 & 0.087$\pm$0.012 & \citet{lam17,dos20} \\
\hline
\end{tabular}
\end{scriptsize}
\end{center}
\renewcommand{\arraystretch}{1.}
\vspace{-3mm}
 \tablefoot{
   \tablefoottext{a}{Equivalent width of the net transmission planetary signal.}
   \tablefoottext{b}{Model employed to calculate XUV flux (see text): coronal model with (1) and without (2) UV data; X vs. EUV relation (3); X-rays vs $P_{\rm rot}$ relation (4); and blackbody emission (5).}
   \tablefoottext{c}{No error bars associated with the mass are used in the plot of these objects.}
 }
\end{table*}

\newpage

\onecolumn

\section{Spectral fits, line fluxes and emission measure distribution}\label{sec:coronalmodels} 

\begin{table*}[ht!]
\caption[]{Low-resolution X-ray spectral fits\tablefootmark{a}.}\label{tab:2tfits}
\vspace{-5mm}
\renewcommand{\arraystretch}{1.3}
\begin{center}
\begin{scriptsize}
  \begin{tabular}{lccccc}
\hline \hline
Star & $\log N_{\rm H}$ & $\log T$ & $\log EM$ & Elements & [X/H] \\
     &  (cm$^{-3}$)     &   (K)   &(cm$^{-3}$)&   (X)    &       \\
\hline
14 Her & 18.3 & 6.50$^{+0.05}_{-0.06}$ & 49.52$^{+0.09}_{-0.13}$ & Fe & $-$0.22$^{+0.12}_{-0.00}$ \\
16 Cyg B & 18.3 & 6.30$^{+0.00}_{-0.00}$ & 49.32$^{+0.23}_{-0.54}$ & Fe & $-$0.20 \\
2M 1207 A & 19.0 & 6.30$^{+0.00}_{-0.00}$ & 48.70$^{+0.28}_{-0.00}$ & Fe & $-$0.20 \\
30 Ari B & 18.7 & 6.67$^{+0.02}_{-0.02}$,6.23$^{+0.16}_{-0.08}$,6.93$^{+0.01}_{-0.01}$ & 51.91$^{+0.04}_{-0.04}$,51.33$^{+0.14}_{-0.18}$,51.79$^{+0.03}_{-0.03}$ & Fe, O, Ne & $-$0.06$^{+0.03}_{-0.03}$,$-$0.34$^{+0.06}_{-0.08}$, 0.03$^{+0.06}_{-0.07}$ \\
47 UMa & 18.3 & 6.25$^{+0.14}_{-0.25}$ & 48.32$^{+0.21}_{-0.47}$ & Fe & $-$0.30$^{+0.20}_{-0.00}$ \\
51 Peg & 18.3 & 6.30$^{+0.00}_{-0.00}$ & 49.20$^{+0.24}_{-0.59}$ & Fe & $-$0.20 \\
55 Cnc & 18.3 & 6.56$^{+0.04}_{-0.05}$ & 49.61$^{+0.08}_{-0.10}$ & Fe, O & $-$0.53$^{+0.28}_{-0.33}$,$-$0.80$^{+0.32}_{-1.05}$ \\
Barnard's star & 18.0 & 6.67$^{+0.15}_{-0.13}$ & 48.12$^{+0.21}_{-0.21}$ & Fe & 0.00 \\
Corot 7 & 20.0 & 6.03$^{+0.45}_{-0.03}$ & 51.60$^{+0.37}_{-1.33}$ & Fe & 0.03 \\
GJ 12 & 18.5 & 6.44$^{+0.12}_{-0.11}$ & 48.39$^{+0.14}_{-0.19}$ & Fe & $-$0.50 \\
GJ 86 & 18.3 & 6.56$^{+0.03}_{-0.03}$ & 50.04$^{+0.07}_{-0.09}$ & Fe, O, Ne & $-$0.27$^{+0.15}_{-0.14}$,$-$0.03$^{+0.10}_{-0.10}$,$-$0.04$^{+0.18}_{-0.20}$ \\
GJ 317 & 18.0 & 6.60$^{+0.10}_{-0.06}$ & 49.24$^{+0.13}_{-0.14}$ & Fe & $-$0.34$^{+0.36}_{-0.49}$ \\
GJ 357 & 18.3 & 6.47$^{+0.04}_{-0.05}$ & 48.22$^{+0.08}_{-0.10}$ & Fe & $-$0.10 \\
GJ 436 & 18.0 & 6.41$^{+0.08}_{-0.08}$,6.94$^{+0.22}_{-0.14}$ & 48.55$^{+0.11}_{-0.11}$,48.23$^{+0.15}_{-0.40}$ & Fe & $-$0.29$^{+0.19}_{-0.01}$ \\
GJ 486 & 18.0 & 6.67$^{+0.40}_{-0.09}$ & 48.07$^{+0.07}_{-0.16}$ & Fe & $-$0.15 \\
GJ 806 & 18.5 & 6.58$^{+0.10}_{-0.09}$ & 48.90$^{+0.10}_{-0.12}$ & Fe & $-$0.45 \\
GJ 832 & 18.0 & 6.52$^{+0.04}_{-0.05}$ & 48.81$^{+0.07}_{-0.10}$ & Fe & $-$0.36$^{+0.37}_{-0.38}$ \\
GJ 876 & 18.0 & 6.90$^{+0.77}_{-0.19}$,6.53$^{+0.05}_{-0.04}$ & 48.19$^{+0.38}_{-0.94}$,48.89$^{+0.11}_{-0.11}$ & Fe, O, Ne & $-$0.50$^{+0.15}_{-0.15}$,$-$0.34$^{+0.12}_{-0.13}$,$-$0.09$^{+0.19}_{-0.22}$ \\
GJ 1214 & 18.5 & 6.61$^{+1.14}_{-0.15}$ & 48.20$^{+0.19}_{-0.34}$ & Fe & 0.39 \\
GJ 3470 & 18.5 & 7.05$^{+0.62}_{-0.11}$,6.47$^{+0.06}_{-0.07}$ & 49.29$^{+0.13}_{-0.18}$,49.89$^{+0.07}_{-0.10}$ & Fe & 0.20 \\
GJ 9827 & 18.5 & 6.60$^{+0.08}_{-0.06}$ & 49.16$^{+0.08}_{-0.10}$ & Fe & $-$0.28 \\
GQ Lup & 21.8 & 6.57$^{+0.12}_{-0.11}$,7.52$^{+0.10}_{-0.08}$ & 52.98$^{+0.55}_{-0.54}$,52.89$^{+0.06}_{-0.07}$ & Fe, Ne & $-$0.77$^{+0.36}_{-1.51}$, 0.16$^{+0.29}_{-0.24}$ \\
HAT-P 1 & 21.0 & 6.30$^{+0.00}_{-0.00}$ & 50.63$^{+0.36}_{- N/A}$ & Fe & 0.13 \\
HAT-P 11 & 18.7 & 6.41$^{+0.05}_{-0.03}$,7.06$^{+0.22}_{-0.20}$ & 49.75$^{+0.08}_{-0.10}$,48.80$^{+0.27}_{-0.88}$ & Fe & 0.30 \\
HAT-P 12 & 20.0 & 6.30$^{+0.00}_{-0.00}$ & 50.51$^{+0.36}_{- N/A}$ & Fe & $-$0.29 \\
HAT-P 32 & 21.1 & 6.61$^{+0.09}_{-0.07}$ & 51.89$^{+0.10}_{-0.11}$ & Fe & $-$0.04 \\
HD 4308 & 18.3 & 6.30$^{+0.00}_{-0.00}$ & 48.69$^{+0.23}_{-0.54}$ & Fe & $-$0.20 \\
HD 20367 & 18.7 & 6.67$^{+0.06}_{-0.02}$,6.30$^{+0.17}_{-0.15}$,6.98$^{+0.02}_{-0.02}$ & 51.61$^{+0.05}_{-0.12}$,50.85$^{+0.36}_{-0.21}$,51.36$^{+0.05}_{-0.07}$ & Fe, O, Ne & $-$0.09$^{+0.04}_{-0.04}$,$-$0.26$^{+0.09}_{-0.07}$,$-$0.22$^{+0.16}_{-0.27}$ \\
HD 27442 & 18.3 & 6.33$^{+0.04}_{-0.04}$ & 49.09$^{+0.08}_{-0.09}$ & Fe & 0.20 \\
HD 46375 & 18.7 & 6.61$^{+0.14}_{-0.11}$ & 49.67$^{+0.09}_{-0.17}$ & Fe & $-$0.30$^{+0.20}_{-0.00}$ \\
HD 49674 & 18.8 & 6.50$^{+0.06}_{-0.05}$ & 49.92$^{+0.09}_{-0.11}$ & Fe & 0.25 \\
HD 50554 & 18.5 & 6.30$^{+0.00}_{-0.00}$ & 48.86$^{+0.20}_{-0.38}$ & Fe & 0.00 \\
HD 52265 & 18.3 & 6.40$^{+0.06}_{-0.04}$ & 49.46$^{+0.13}_{-0.16}$ & Fe & 0.20 \\
HD 63433 & 18.3 & 6.93$^{+0.05}_{-0.06}$,6.64$^{+0.05}_{-0.06}$ & 51.03$^{+0.18}_{-0.16}$,51.11$^{+0.12}_{-0.21}$ & Fe, O & $-$0.18$^{+0.09}_{-0.08}$,$-$0.14$^{+0.14}_{-0.14}$ \\
HD 70642 & 18.5 & 6.49$^{+0.14}_{-0.10}$ & 48.92$^{+0.17}_{-0.27}$ & Fe & 0.00 \\
HD 73583 & 18.7 & 6.64$^{+0.05}_{-0.06}$ & 50.54$^{+0.07}_{-0.08}$ & Fe & $-$0.66$^{+0.24}_{-0.33}$ \\
HD 75289 & 18.7 & 6.23$^{+0.21}_{-0.23}$ & 48.00$^{+0.42}_{-0.81}$ & Fe & 0.30 \\
HD 93083 & 18.5 & 6.50$^{+0.07}_{-0.05}$ & 49.35$^{+0.10}_{-0.12}$ & Fe & 0.10 \\
HD 95089 & 20.5 & 6.30$^{+0.00}_{-0.00}$ & 49.14$^{+1.57}_{- N/A}$ & Fe & $-$0.20 \\
HD 97658 & 18.0 & 6.92$^{+0.26}_{-0.19}$,6.40$^{+0.04}_{-0.07}$ & 48.76$^{+0.24}_{-0.48}$,49.39$^{+0.08}_{-0.09}$ & Fe & $-$0.23 \\
HD 99492 & 18.3 & 6.51$^{+0.05}_{-0.03}$ & 49.04$^{+0.07}_{-0.09}$ & Fe & 0.20 \\
HD 101930 & 18.7 & 6.30$^{+0.00}_{-0.00}$ & 48.54$^{+0.43}_{- N/A}$ & Fe & $-$0.20 \\
HD 102195 & 18.5 & 6.45$^{+0.05}_{-0.06}$,6.84$^{+0.04}_{-0.04}$ & 50.80$^{+0.07}_{-0.08}$,50.62$^{+0.10}_{-0.13}$ & Fe, O & $-$0.17$^{+0.10}_{-0.10}$,$-$0.45$^{+0.10}_{-0.11}$ \\
HD 108147 & 18.7 & 6.67$^{+0.15}_{-0.10}$,6.38$^{+0.10}_{-0.11}$ & 50.02$^{+0.26}_{-0.65}$,50.05$^{+0.21}_{-0.46}$ & Fe & 0.19$^{+0.20}_{-0.14}$ \\
HD 111232 & 18.7 & 6.30$^{+0.00}_{-0.00}$ & 49.05$^{+0.08}_{-0.00}$ & Fe & $-$0.20 \\
HD 114386 & 18.7 & 6.38$^{+0.24}_{-0.23}$ & 49.17$^{+0.21}_{-0.35}$ & Fe & $-$0.10$^{+0.00}_{-0.20}$ \\
HD 114762 & 18.7 & 6.30$^{+0.00}_{-0.00}$ & 49.25$^{+0.34}_{- N/A}$ & Fe & $-$0.20 \\
HD 114783 & 18.3 & 6.37$^{+0.13}_{-0.15}$ & 49.09$^{+0.21}_{-0.45}$ & Fe & $-$0.30$^{+0.20}_{-0.00}$ \\
HD 130322 & 18.7 & 6.54$^{+0.09}_{-0.07}$ & 49.93$^{+0.11}_{-0.13}$ & Fe & $-$0.18$^{+0.09}_{-0.12}$ \\
HD 149026 & 20.0 & 6.94$^{+0.10}_{-0.44}$ & 49.59$^{+0.13}_{-0.18}$ & Fe & 0.36 \\
HD 154345 & 18.3 & 6.43$^{+0.05}_{-0.05}$ & 49.68$^{+0.08}_{-0.10}$ & Fe & $-$0.10 \\
HD 164922 & 18.3 & 6.30$^{+0.00}_{-0.00}$ & 48.01$^{+0.55}_{- N/A}$ & Fe & 0.00 \\
HD 179949 & 18.7 & 6.53$^{+0.04}_{-0.05}$,6.25$^{+0.10}_{-0.08}$,6.82$^{+0.02}_{-0.01}$ & 50.79$^{+0.06}_{-0.21}$,50.70$^{+0.09}_{-0.12}$,50.79$^{+0.05}_{-0.06}$ & Fe, O, Ne & $-$0.08$^{+0.02}_{-0.02}$,$-$0.46$^{+0.07}_{-0.05}$,$-$0.15$^{+0.06}_{-0.07}$ \\
HD 187123 & 18.8 & 6.30$^{+0.00}_{-0.00}$ & 49.78$^{+0.00}_{-0.00}$ & Fe & 0.00 \\
HD 189733 & 18.3 & 7.30$^{+0.00}_{-0.00}$,6.44$^{+0.01}_{-0.01}$,6.88$^{+0.01}_{-0.01}$ & 50.58$^{+0.01}_{-0.01}$,50.61$^{+0.03}_{-0.03}$,50.80$^{+0.04}_{-0.04}$ & Fe, O, Ne, Mg, Si & $-$0.51$^{+0.03}_{-0.03}$,$-$0.27$^{+0.01}_{-0.01}$,$-$0.28$^{+0.06}_{-0.07}$,$-$0.27$^{+0.05}_{-0.06}$,$-$0.19$^{+0.05}_{-0.05}$ \\
HD 190360 & 18.3 & 6.30$^{+0.00}_{-0.00}$ & 49.04$^{+0.08}_{-0.19}$ & Fe & 0.00 \\
HD 195019 & 18.3 & 6.30$^{+0.00}_{-0.00}$ & 48.83$^{+0.24}_{-0.56}$ & Fe & 0.00 \\
HD 201585 & 20.8 & 6.30$^{+0.00}_{-0.00}$ & 51.12$^{+0.25}_{-0.63}$ & Fe & 0.00 \\
\hline
\end{tabular}
\end{scriptsize}
\end{center}
\end{table*}
\setcounter{table}{0}
\begin{table*}
\caption{(continued) Low-resolution X-ray spectral fits\tablefootmark{a}.}
\begin{center}
\begin{scriptsize}
  \begin{tabular}{lccccc}
\hline \hline
Star & $\log N_{\rm H}$ & $\log T$ & $\log EM$ & Elements & [X/H] \\
     &  (cm$^{-3}$)     &   (K)   &(cm$^{-3}$)&   (X)    &       \\
\hline
HD 209458 & 18.7 & 6.00$^{+0.28}_{-0.00}$ & 49.52$^{+0.22}_{-0.48}$ & Fe & $-$0.20 \\
HD 216435 & 18.5 & 6.58$^{+0.05}_{-0.05}$ & 50.31$^{+0.10}_{-0.09}$ & Fe & $-$0.08$^{+0.08}_{-0.22}$ \\
HD 216437 & 18.3 & 6.40$^{+0.15}_{-0.11}$ & 49.16$^{+0.20}_{-0.38}$ & Fe & 0.00 \\
HD 217107 & 18.7 & 6.30$^{+0.00}_{-0.00}$ & 48.25$^{+0.32}_{- N/A}$ & Fe & 0.00 \\
HD 218566 & 18.7 & 6.48$^{+0.09}_{-0.08}$ & 49.62$^{+0.12}_{-0.17}$ & Fe & 0.20 \\
HD 235088 & 19.0 & 6.87$^{+0.07}_{-0.09}$,6.53$^{+0.14}_{-0.13}$ & 50.64$^{+0.18}_{-0.40}$,50.44$^{+0.32}_{-0.31}$ & Fe, Ne & $-$0.24$^{+0.17}_{-0.17}$, 0.17$^{+0.29}_{-0.69}$ \\
HD 283636 & 19.0 & 6.56$^{+0.09}_{-0.07}$ & 50.25$^{+0.09}_{-0.12}$ & Fe & 0.00 \\
HD 330075 & 18.7 & 6.50$^{+0.00}_{-0.00}$ & 49.09$^{+0.27}_{-0.87}$ & Fe & $-$0.20 \\
HR 8799 & 18.5 & 6.55$^{+0.06}_{-0.06}$ & 50.79$^{+0.07}_{-0.09}$ & Fe & $-$0.50 \\
K2 100 & 20.0 & 6.84$^{+0.03}_{-0.04}$,6.33$^{+0.12}_{-0.18}$ & 51.35$^{+0.19}_{-0.20}$,51.11$^{+0.11}_{-0.16}$ & Fe, Ne & 0.22$^{+0.19}_{-0.19}$, 0.60$^{+0.17}_{-0.31}$ \\
K2 136 & 19.0 & 7.01$^{+0.08}_{-0.06}$,6.46$^{+0.09}_{-0.07}$ & 49.87$^{+0.22}_{-0.23}$,50.15$^{+0.11}_{-0.11}$ & Fe & $-$0.04$^{+0.27}_{-0.29}$ \\
Kelt 7 & 20.0 & 6.80$^{+0.22}_{-0.17}$,6.33$^{+0.27}_{-0.27}$ & 50.70$^{+0.21}_{-0.85}$,50.57$^{+0.37}_{-0.50}$ & Fe & 0.24 \\
Kelt 8 & 21.1 & 6.30$^{+0.00}_{-0.00}$ & 51.67$^{+0.30}_{-2.24}$ & Fe & 0.27 \\
Kelt 9 & 22.1 & 6.30$^{+0.00}_{-0.00}$ & 52.00$^{+0.80}_{- N/A}$ & Fe & 0.00 \\
NGC 2423 3 & 21.6 & 7.10$^{+0.00}_{-0.00}$ & 52.48$^{+0.38}_{- N/A}$ & Fe & 0.00 \\
Pollux & 18.0 & 6.23$^{+0.05}_{-0.11}$,6.64$^{+0.17}_{-0.21}$ & 49.86$^{+0.06}_{-0.08}$,48.60$^{+0.59}_{-0.90}$ & Fe, O & 0.20$^{+0.43}_{-0.30}$,$-$0.29$^{+0.23}_{-0.13}$ \\
Teegarden's star & 18.0 & 6.77$^{+0.19}_{-0.06}$,6.10$^{+0.11}_{-0.08}$ & 48.37$^{+0.07}_{-0.19}$,48.39$^{+0.24}_{-0.21}$ & Fe & $-$0.89$^{+0.20}_{-0.26}$ \\
TOI-836 & 18.7 & 6.57$^{+0.07}_{-0.07}$ & 50.04$^{+0.08}_{-0.10}$ & Fe & $-$0.30 \\
TOI-2134 & 18.5 & 6.40$^{+0.05}_{-0.04}$ & 49.43$^{+0.09}_{-0.11}$ & Fe & 0.12 \\
TRAPPIST-1 & 18.0 & 6.43$^{+0.04}_{-0.04}$,7.00$^{+0.08}_{-0.08}$ & 48.47$^{+0.06}_{-0.09}$,47.86$^{+0.16}_{-0.20}$ & Fe & 0.04 \\
WASP 12 & 21.3 & 6.30$^{+0.00}_{-0.00}$ & 51.20$^{+1.04}_{- N/A}$ & Fe & 0.30 \\
WASP 13 & 20.1 & 6.30$^{+0.00}_{-0.00}$ & 50.93$^{+0.22}_{-0.49}$ & Fe & 0.00 \\
WASP 39 & 20.7 & 6.30$^{+0.00}_{-0.00}$ & 50.79$^{+0.50}_{- N/A}$ & Fe & $-$0.12 \\
WASP 52 & 20.6 & 6.55$^{+1.06}_{-0.29}$ & 51.13$^{+0.33}_{-0.63}$ & Fe & $-$0.19 \\
WASP 69 & 18.5 & 6.73$^{+0.03}_{-0.04}$ & 50.80$^{+0.04}_{-0.05}$ & Fe & $-$0.56$^{+0.11}_{-0.12}$ \\
WASP 76 & 20.3 & 6.30$^{+0.00}_{-0.00}$ & 51.15$^{+0.29}_{-1.52}$ & Fe & 0.23 \\
WASP 77A & 20.5 & 6.94$^{+0.12}_{-0.47}$ & 50.57$^{+0.16}_{-0.25}$ & Fe & 0.00 \\
WASP 80 & 18.6 & 6.47$^{+0.08}_{-0.09}$,6.96$^{+0.08}_{-0.08}$ & 49.83$^{+0.15}_{-0.21}$,49.70$^{+0.15}_{-0.25}$ & Fe & $-$0.14 \\
WASP 107 & 20.3 & 6.52$^{+0.04}_{-0.05}$ & 50.21$^{+0.10}_{-0.08}$ & Fe & 0.02 \\
WASP 127 & 20.7 & 6.30$^{+0.00}_{-0.00}$ & 50.87$^{+0.33}_{- N/A}$ & Fe & 0.00 \\
$\beta$ Pic & 18.7 & 6.35$^{+0.07}_{-0.08}$ & 48.30$^{+0.12}_{-0.16}$ & Fe & $-$0.20 \\
$\mu$ Ara & 18.3 & 6.30$^{+0.00}_{-0.00}$ & 48.69$^{+0.20}_{-0.39}$ & Fe & 0.10 \\
$\upsilon$ And & 18.7 & 6.80$^{+0.07}_{-0.19}$,6.46$^{+0.03}_{-0.06}$ & 49.23$^{+1.08}_{-0.52}$,50.22$^{+0.10}_{-0.14}$ & Fe, O & 0.24$^{+0.12}_{-0.12}$, 0.03$^{+0.12}_{-0.11}$ \\
\hline
\end{tabular}
  \tablefoot{
   \tablefoottext{a}{Spectral fits of stars with no X-rays coronal lines in the sample \citep[including stars from][]{san11}. Element abundances and FIP are based on solar photospheric values by \citet{and89}. Abundances with no error bars are set based on photospheric abundances.}}
\end{scriptsize}
\end{center}
\renewcommand{\arraystretch}{1.}
\end{table*}

%
\begin{figure*}
  \centering
  \vspace{-0.5cm}
  \includegraphics[width=0.33\textwidth]{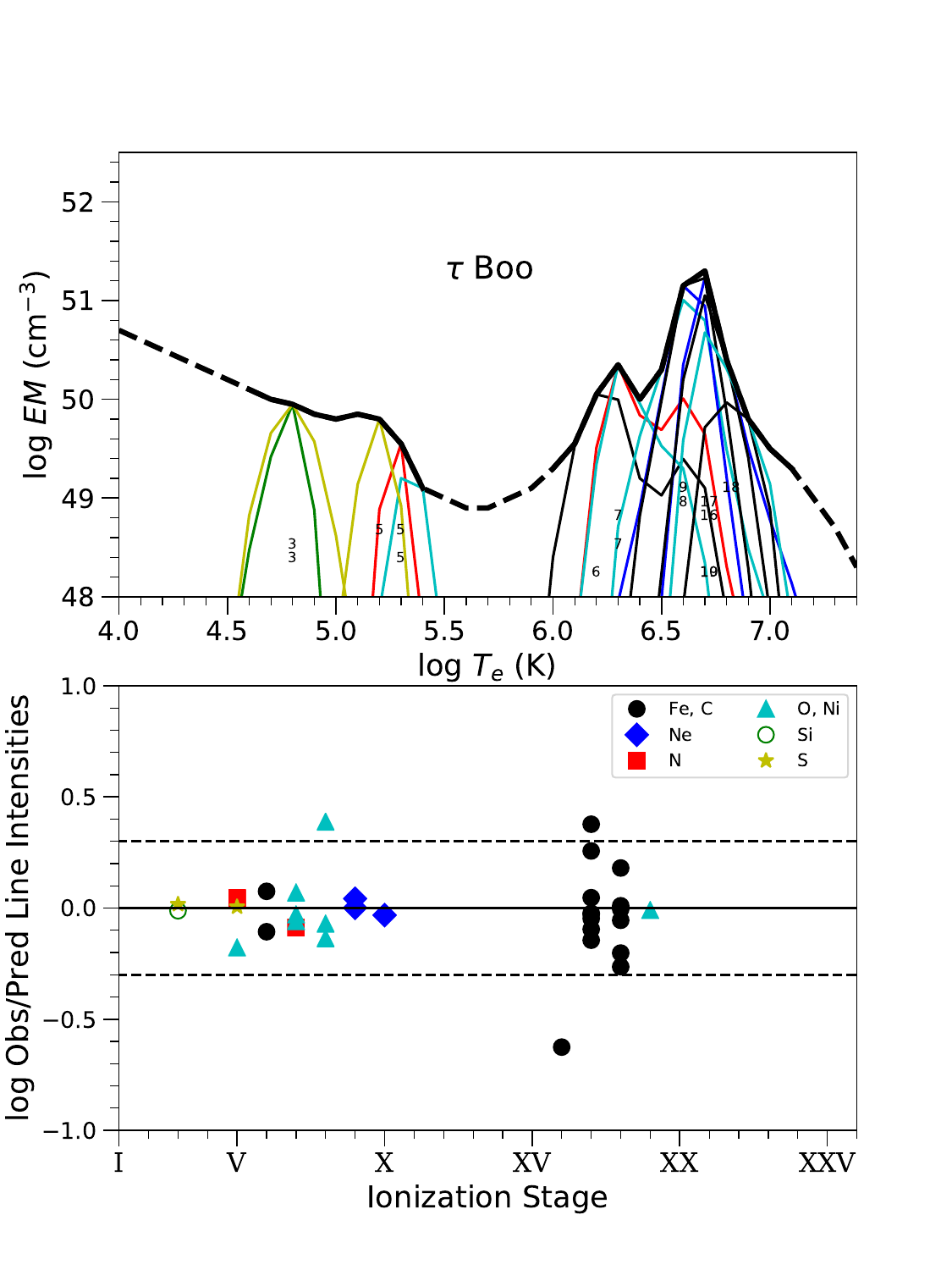}
  \includegraphics[width=0.33\textwidth]{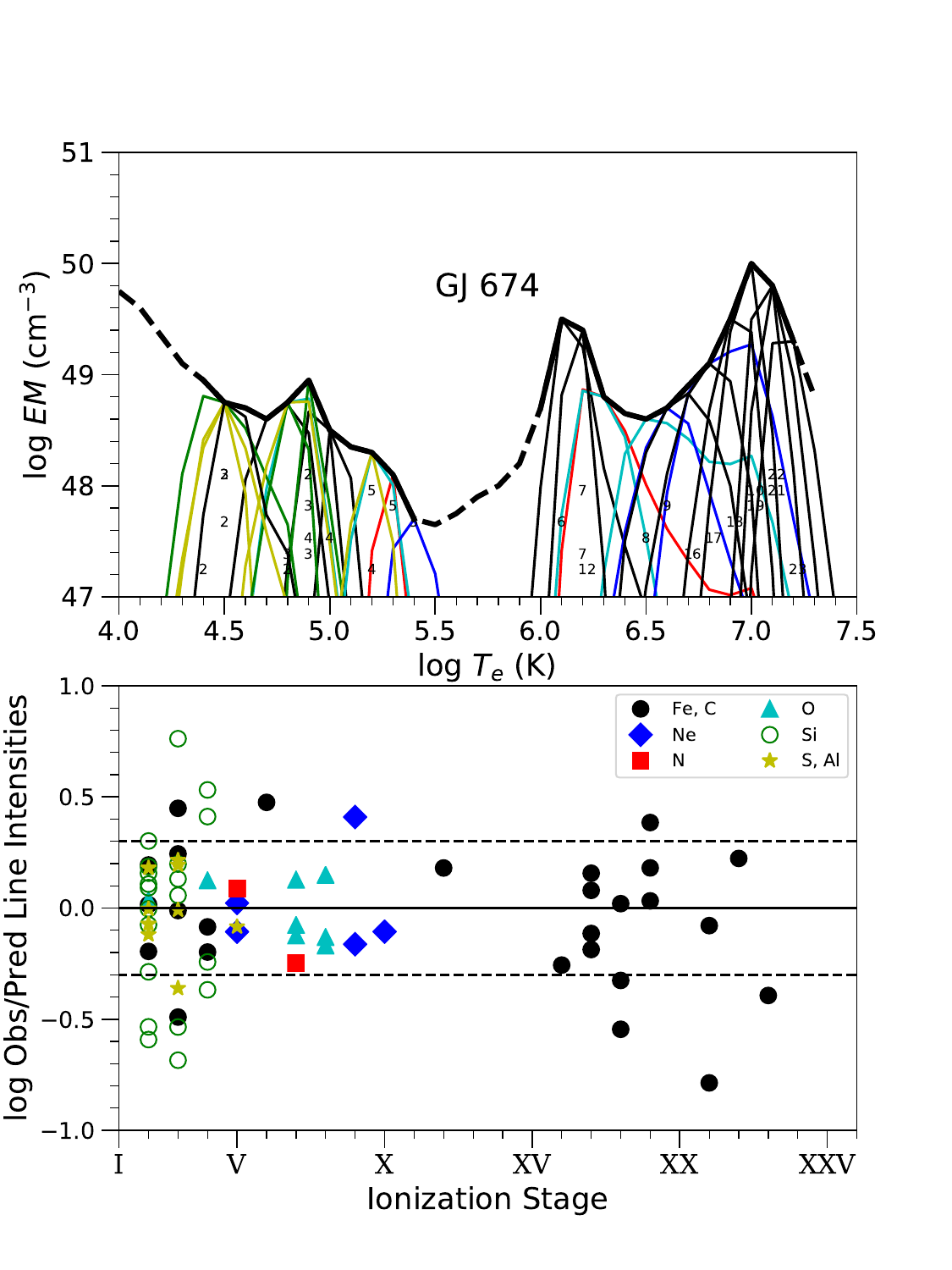}
  \includegraphics[width=0.33\textwidth]{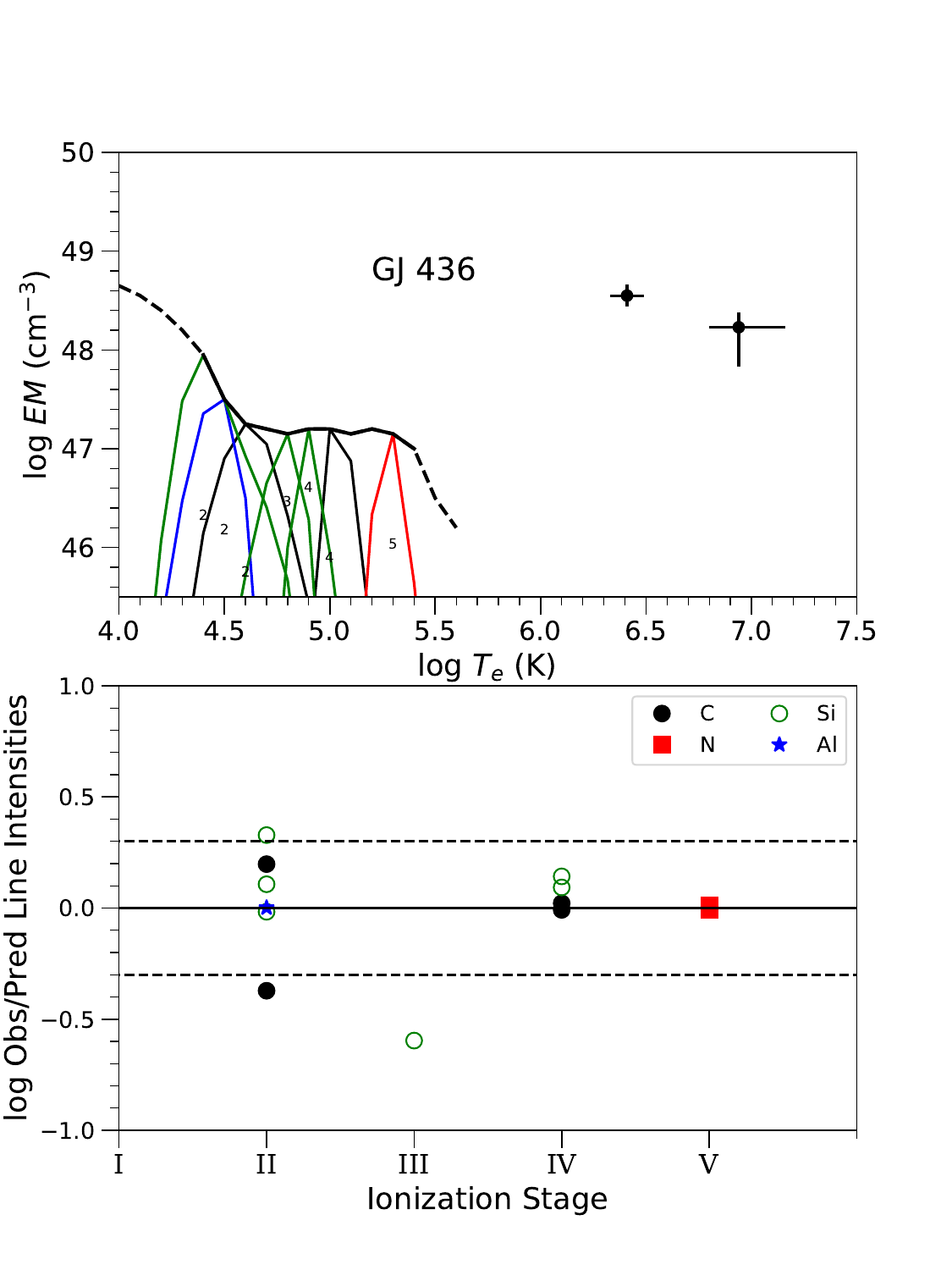}
  \caption{Emission Measure Distributions as in Fig.~\ref{fig:proxcen}. {\it left:} $\tau$ Boo EMD combining data from XMM-Newton \citep{mag11}, and HST/STIS. {\it center:} GJ 674 combining data from XMM-Newton and HST/STIS and COS. {\it right:} GJ 436 using HST/COS data. The 2-$T$ global fit to XMM-Newton/EPIC spectra is also indicated with error bars. }\label{fig:tauboo} 
\end{figure*}
%

%
\begin{table*}
\caption[]{{\em XMM-Newton}/RGS, EUVE, FUSE, and HST/STIS line fluxes of Proxima Centauri\tablefootmark{a}.}\label{tab:fluxes1} 
\tabcolsep 3 pt
\begin{scriptsize}
  \begin{tabular}{lrccrcl}
\hline \hline
Ion & $\lambda_{\rm model}$ (\AA) & $\log T_{\rm max}$ & $F_{\rm obs}$ & $S/N$ & Ratio & Blends \\
\hline
\ion{Si}{xiv} & 6.1804 & 7.3 & 5.30e$-$14 & 10.8 & $-$0.14 & \ion{Si}{xiv}  6.1858 \\
\ion{Si}{xiii} & 6.6479 & 7.1 & 6.43e$-$14 & 16.1 & $-$0.59 & \ion{Si}{xiii}  6.6882,  6.7403 \\
\ion{Mg}{xii} & 7.1058 & 7.1 & 2.85e$-$14 & 12.4 & 0.20 & \ion{Mg}{xii}  7.1069, \ion{Al}{xiii}  7.1710,  7.1764 \\
\ion{Fe}{xxii} & 8.0904 & 7.2 & 5.64e$-$15 & 6.7 & 0.59 & \ion{Mg}{x}  8.0696,  8.0701 \\
\ion{Fe}{xxiv} & 8.2326 & 7.4 & 2.08e$-$15 & 4.1 & 0.10 & \ion{Fe}{xxii}  8.1684, 8.1684, \ion{Ni}{xxi}  8.2329, \ion{Ni}{xxiii}  8.2685, \ion{Fe}{xxiv}  8.2850 \\
\ion{Mg}{xii} & 8.4192 & 7.1 & 4.75e$-$14 & 20.0 & $-$0.09 & \ion{Mg}{xii}  8.4246 \\
\ion{Mg}{xi} & 9.1687 & 6.9 & 6.22e$-$14 & 25.0 & 0.05 &  \\
\ion{Mg}{xi} & 9.2312 & 6.9 & 1.43e$-$14 & 12.1 & $-$0.10 & \ion{Mg}{xi}  9.2282, \ion{Ne}{x}  9.2461,  9.2912, 9.2913, \ion{Ni}{xix}  9.2540, \\
\ion{Mg}{xi} & 9.3143 & 6.9 & 3.51e$-$14 & 19.2 & 0.03 & \ion{Ne}{x}  9.3616 \\
\ion{Ne}{x} & 9.4807 & 6.9 & 2.42e$-$14 & 16.3 & 0.44 & \ion{Ne}{x}  9.4809 \\
\ion{Ne}{x} & 9.7080 & 6.9 & 2.55e$-$14 & 17.3 & 0.05 & \ion{Ne}{x}  9.7085 \\
\ion{Na}{xi} & 10.0232 & 7.0 & 3.04e$-$14 & 19.4 & 0.01 & \ion{Na}{xi} 10.0286 \\
\ion{Fe}{xvii} & 10.1210 & 6.9 & 9.71e$-$15 & 11.1 & 0.26 & \ion{Ni}{xix} 10.1100, \ion{Fe}{xix} 10.1195, 10.1419 \\
\ion{Ne}{x} & 10.2385 & 6.9 & 5.21e$-$14 & 26.0 & $-$0.04 & \ion{Ne}{x} 10.2396 \\
\ion{Fe}{xvii} & 10.5040 & 6.9 & 1.75e$-$14 & 15.6 & 0.56 &  \\
\ion{Fe}{xviii} & 10.5364 & 7.0 & 3.43e$-$15 & 6.9 & $-$0.00 &  \\
\ion{Fe}{xix} & 10.6193 & 7.1 & 3.71e$-$14 & 22.9 & 0.15 & \ion{Fe}{xix} 10.6001, 10.6116,  10.6840, \ion{Fe}{xxiv} 10.6190, 10.6630, \ion{Ne}{ix} 10.6426, \ion{Fe}{xvii} 10.6570 \\
\ion{Fe}{xvii} & 10.7700 & 6.9 & 1.93e$-$14 & 16.7 & 0.18 & \ion{Ne}{ix} 10.7650 \\
\ion{Fe}{xix} & 10.8160 & 7.1 & 7.97e$-$15 & 10.8 & 0.14 & \ion{Fe}{xix} 10.8083 \\
\ion{Fe}{xxiii} & 10.9810 & 7.3 & 1.60e$-$14 & 15.4 & $-$0.02 & \\
\ion{Ne}{ix} & 11.0010 & 6.7 & 2.96e$-$14 & 21.1 & $-$0.15 & \ion{Fe}{xxiii} 10.9810, 10.9810, 11.0190, \ion{Fe}{xvii} 11.0260, \ion{Fe}{xxiv} 11.0290 \\
\ion{Fe}{xvii} & 11.1310 & 6.9 & 1.21e$-$14 & 13.6 & $-$0.01 &  \\
\ion{Fe}{xxiv} & 11.1760 & 7.4 & 9.15e$-$15 & 11.9 & 0.09 & \ion{Fe}{xxiv} 11.1870 \\
\ion{Fe}{xvii} & 11.2540 & 6.9 & 2.27e$-$14 & 18.8 & 0.07 & \ion{Fe}{xxiv} 11.2680 \\
\ion{Fe}{xviii} & 11.4230 & 7.0 & 1.54e$-$14 & 15.7 & $-$0.16 & \ion{Fe}{xxii} 11.4270, \ion{Fe}{xviii} 11.4274, \ion{Fe}{xxiv} 11.4320, \ion{Fe}{xxiii} 11.4580 \\
\ion{Ne}{ix} & 11.5440 & 6.7 & 3.76e$-$14 & 24.7 & $-$0.12 & \ion{Fe}{xviii} 11.5270, 11.5270 \\
\ion{Fe}{xxiii} & 11.7360 & 7.3 & 2.74e$-$14 & 21.2 & $-$0.11 &  \\
\ion{Fe}{xxii} & 11.7700 & 7.2 & 2.16e$-$14 & 18.9 & $-$0.42 & \ion{Fe}{xxiii} 11.7360, 11.7360 \\
\ion{Ni}{xx} & 11.8320 & 7.1 & 1.57e$-$14 & 16.2 & 0.16 & \ion{Fe}{xxii} 11.8020, \ion{Ni}{xx} 11.8460 \\
\ion{Fe}{xxii} & 11.9770 & 7.2 & 1.24e$-$14 & 14.6 & $-$0.14 & \ion{Fe}{xxii} 11.8810, 11.9320, 11.9474, \ion{Fe}{xxi} 11.9023, 11.9750, \ion{Ni}{xx} 11.9617 \\
\ion{Fe}{xviii} & 12.0827 & 7.0 & 8.54e$-$15 & 12.2 & 0.53 & \ion{Ni}{xx} 12.0787, 12.0879, \ion{Fe}{xxii} 12.0928, \ion{Ne}{ix} 12.0960 \\
\ion{Ne}{x} & 12.1321 & 6.9 & 2.67e$-$13 & 68.6 & $-$0.19 & \ion{Fe}{xvii} 12.1240, \ion{Ne}{x} 12.1375 \\
\ion{Fe}{xxii} & 12.2100 & 7.2 & 9.26e$-$15 & 12.8 & 0.27 & \ion{Fe}{xxi} 12.2040, \ion{Ni}{xxi} 12.2080, \ion{Fe}{xix} 12.2120 \\
\ion{Fe}{xxi} & 12.2840 & 7.2 & 6.88e$-$14 & 35.1 & $-$0.01 & \ion{Fe}{xvii} 12.2660 \\
\ion{Ni}{xix} & 12.4350 & 7.0 & 3.16e$-$14 & 24.0 & $-$0.09 & \ion{Fe}{xxi} 12.3930, \ion{Fe}{xx} 12.4234, 12.4310 \\
\ion{Fe}{xvi} & 12.5399 & 6.8 & 3.00e$-$14 & 23.5 & 0.20 & \ion{Fe}{xxi} 12.4990, 12.5048, \ion{Fe}{xx} 12.5260, 12.5760, 12.5760, \ion{Fe}{xvii} 12.5391 \\
\ion{Ni}{xix} & 12.6560 & 7.0 & 1.23e$-$14 & 15.1 & $-$0.07 & \ion{Fe}{xx} 12.6210, \ion{Fe}{xxi} 12.6490, \ion{Fe}{xvii} 12.6950 \\
\ion{Fe}{xxii} & 12.7540 & 7.2 & 1.74e$-$14 & 18.0 & 0.14 & \ion{Fe}{xxi} 12.7200, 12.7209,12.7675 \\
\ion{Fe}{xx} & 12.8460 & 7.1 & 5.96e$-$14 & 33.3 & 0.06 & \ion{Fe}{xxi} 12.8220, \ion{Fe}{xx} 12.8240, 12.8640 \\
\ion{Fe}{xx} & 12.9120 & 7.1 & 3.91e$-$14 & 26.6 & $-$0.04 & \ion{Fe}{xix} 12.9033, 12.9330, 13.0220, 13.0220, \ion{Fe}{xviii} 12.9494, \ion{Fe}{xxii} 12.9530, \ion{Fe}{xx} 12.9650 \\
\ion{Fe}{xix} & 13.0220 & 7.1 & 3.81e$-$15 & 7.1 & $-$0.22 & \\
\ion{Fe}{xvii} & 13.1530 & 6.9 & 1.82e$-$14 & 18.0 & 0.25 & \ion{Fe}{xx} 13.1530, 13.1530, 13.1370, 13.1370 \\
\ion{Fe}{xx} & 13.2740 & 7.1 & 2.16e$-$14 & 20.4 & 0.21 & \ion{Fe}{xx} 13.2453, \ion{Fe}{xix} 13.2261, 13.2933, \ion{Fe}{xxii} 13.2360, \ion{Ni}{xx} 13.2560 \\
\ion{Fe}{xviii} & 13.3230 & 7.0 & 1.28e$-$14 & 15.9 & 0.24 & \ion{Ni}{xx} 13.3090, \ion{Fe}{xviii} 13.3312 \\
\ion{Ne}{ix} & 13.4473 & 6.7 & 3.10e$-$13 & 79.2 & 0.03 & \ion{Fe}{xix} 13.4970, 13.5180, \ion{Ne}{ix} 13.5531 \\
\ion{Ne}{ix} & 13.6990 & 6.7 & 1.23e$-$13 & 50.4 & 0.07 & \ion{Fe}{xix} 13.6450, 13.7054 \\
\ion{Fe}{xvii} & 13.8250 & 6.9 & 9.15e$-$14 & 43.7 & 0.13 & \ion{Ni}{xix} 13.7790, \ion{Fe}{xix} 13.7950, \ion{Fe}{xvii} 13.8920 \\
\ion{Fe}{xviii} & 13.9530 & 7.0 & 2.23e$-$14 & 21.6 & 0.43 & \ion{Fe}{xx} 13.9620 \\
\ion{Ni}{xix} & 14.0430 & 7.0 & 2.60e$-$14 & 23.4 & $-$0.01 & \ion{Fe}{xxi} 14.0080, \ion{Fe}{xix} 14.0179, 14.0340, 14.0388 \\
\ion{Ni}{xix} & 14.0770 & 7.0 & 1.45e$-$14 & 17.5 & 0.02 & \ion{Fe}{xviii} 14.0549, \ion{Fe}{xx} 14.0620, 14.0784 \\
\ion{Fe}{xviii} & 14.2080 & 7.0 & 1.26e$-$13 & 51.5 & $-$0.05 & \ion{Fe}{xviii} 14.2560 \\
\ion{Fe}{xviii} & 14.3730 & 7.0 & 6.33e$-$14 & 36.3 & $-$0.05 & \ion{Fe}{xviii} 14.3430, 14.3990,14.4250, 14.4555 \\
\ion{Fe}{xviii} & 14.5340 & 7.0 & 3.95e$-$14 & 28.9 & $-$0.05 & \ion{O}{viii} 14.5242, \ion{Fe}{xviii} 14.5608, 14.5710 \\
\ion{Fe}{xix} & 14.6640 & 7.1 & 2.44e$-$14 & 22.7 & $-$0.08 & \ion{Fe}{xviii} 14.6160, 14.6566, 14.6887, \ion{O}{viii} 14.6343, 14.6344 \\
\ion{Fe}{xx} & 14.7540 & 7.1 & 7.96e$-$15 & 13.0 & $-$0.24 & \ion{Fe}{xix} 14.7250, \ion{Fe}{xviii} 14.7499, 14.7510, \ion{Fe}{xx} 14.7824 \\
\ion{Fe}{xvi} & 14.9555 & 6.7 & 3.00e$-$14 & 25.7 & $-$0.28 & \ion{O}{viii} 14.8205, 14.8207, \ion{Fe}{xviii} 14.8920, \ion{Fe}{xix} 14.9610, 14.9632 \\
\ion{Fe}{xvii} & 15.0140 & 6.9 & 2.19e$-$13 & 69.4 & $-$0.12 &  \\
\ion{Fe}{xix} & 15.0790 & 7.1 & 3.58e$-$14 & 28.1 & 0.43 &  \\
\ion{O}{viii} & 15.1760 & 6.6 & 5.28e$-$14 & 34.0 & $-$0.05 & \ion{O}{viii} 15.1765, \ion{Fe}{xix} 15.1770, 15.1980 \\
\ion{Fe}{xvii} & 15.2610 & 6.9 & 9.18e$-$14 & 44.7 & 0.02 &  \\
\ion{Fe}{xviii} & 15.3539 & 7.0 & 1.94e$-$14 & 20.2 & 0.08 & \ion{Fe}{xix} 15.3123, 15.3340, \ion{Fe}{xvi} 15.3694, 15.3907, 15.3926, 15.3992 \\
\ion{Fe}{xvii} & 15.4530 & 6.8 & 2.66e$-$14 & 23.9 & 0.03 & \ion{Fe}{xvi} 15.4485, 15.4533, 15.5023, \ion{Fe}{xviii} 15.4940, \ion{Fe}{xx} 15.5170 \\
\ion{Fe}{xviii} & 15.6250 & 7.0 & 2.67e$-$14 & 24.3 & $-$0.10 &  \\
\ion{Fe}{xviii} & 15.7590 & 7.0 & 4.07e$-$15 & 9.5 & $-$0.04 &  \\
\ion{Fe}{xviii} & 15.8240 & 7.0 & 2.24e$-$14 & 21.9 & $-$0.19 & \ion{Fe}{xviii} 15.8700 \\
\ion{O}{viii} & 16.0055 & 6.6 & 1.16e$-$13 & 50.5 & $-$0.11 & \ion{Fe}{xviii} 16.0040, \ion{O}{viii} 16.0067 \\
\ion{Fe}{xviii} & 16.0710 & 7.0 & 5.07e$-$14 & 33.3 & $-$0.08 & \ion{Fe}{xviii} 16.0450 \\
\ion{Fe}{xviii} & 16.1590 & 7.0 & 2.25e$-$14 & 22.1 & $-$0.08 & \ion{Fe}{xix} 16.1100 \\
\ion{Fe}{xix} & 16.2830 & 7.1 & 1.41e$-$14 & 17.5 & $-$0.21 & \ion{Fe}{xvii} 16.2285, 16.3500, \ion{Fe}{xix} 16.2857, \ion{Fe}{xviii} 16.3200 \\
\ion{Fe}{xvii} & 16.7800 & 6.8 & 1.13e$-$13 & 50.0 & $-$0.20 &  \\
\ion{Fe}{xvii} & 17.0510 & 6.8 & 2.85e$-$13 & 79.7 & $-$0.13 & \ion{Fe}{xvii} 17.0960 \\
\ion{O}{vii} & 17.3960 & 6.5 & 1.17e$-$14 & 16.1 & 0.35 & \ion{Fe}{xix} 17.3908 \\
\ion{Fe}{xviii} & 17.6230 & 7.0 & 1.76e$-$14 & 20.0 & $-$0.23 &  \\
\ion{O}{vii} & 17.7680 & 6.5 & 1.09e$-$14 & 15.3 & $-$0.04 & \ion{Ar}{xvi} 17.7371, 17.7468 \\
\ion{O}{vii} & 18.6270 & 6.4 & 1.81e$-$14 & 20.1 & $-$0.40 & \ion{Ca}{xviii} 18.6909 \\
\ion{O}{viii} & 18.9671 & 6.6 & 5.85e$-$13 & 114.0 & $-$0.04 & \ion{O}{viii} 18.9725 \\
\ion{N}{vii} & 20.9095 & 6.4 & 1.76e$-$14 & 18.7 & $-$0.16 & \ion{N}{vii} 20.9106 \\
\ion{O}{vii} & 21.6015 & 6.4 & 2.19e$-$13 & 66.0 & 0.05 &  \\
\ion{O}{vii} & 21.8036 & 6.4 & 4.96e$-$14 & 31.1 & 0.17 &  \\
\ion{O}{vii} & 22.0977 & 6.4 & 1.08e$-$13 & 45.8 & $-$0.08 & \ion{Ca}{xvi} 22.1372 \\
\ion{S}{xiv} & 23.0050 & 6.7 & 8.03e$-$15 & 10.5 & 0.16 &  \\
\ion{N}{vi} & 23.2770 & 6.3 & 5.08e$-$15 & 10.1 & 0.13 & \ion{Ca}{xii} 23.2473, \ion{Ca}{xvi} 23.2608, \ion{Ca}{xiv} 23.2711, \ion{Ca}{xii} 23.3326, 23.3326 \\
\ion{Ar}{xvi} & 23.5063 & 6.8 & 8.04e$-$15 & 13.3 & $-$0.53 & \ion{Ar}{xvi} 23.5465, \ion{Ca}{xvi} 23.5775, \ion{Ca}{xiv} 23.5950,  23.6006 \\
\ion{N}{vi} & 23.7710 & 6.3 & 5.14e$-$15 & 10.5 & 0.38 &  \\
\hline
\end{tabular}
\end{scriptsize}
\end{table*}
\setcounter{table}{1}
\begin{table*}
\caption{(continued) {\em XMM-Newton}/RGS, EUVE, FUSE, and HST/STIS line fluxes of Proxima Centauri\tablefootmark{a}.}
\tabcolsep 3.pt
\begin{scriptsize}
  \begin{tabular}{lrccrcl}
\hline \hline
Ion & $\lambda_{\rm model}$ (\AA) & $\log T_{\rm max}$ & $F_{\rm obs}$ & $S/N$ & Ratio & Blends \\
\hline
\ion{S}{xiv} & 24.2000 & 6.7 & 4.38e$-$15 & 9.8 & 0.16 &  \\
\ion{S}{xiv} & 24.2850 & 6.7 & 6.88e$-$15 & 12.2 & 0.08 &  \\
\ion{S}{xiv} & 24.4180 & 6.7 & 5.82e$-$15 & 11.3 & $-$0.08 & \ion{Ca}{xiv} 24.4046, \ion{Ca}{xv} 24.4120, 24.4282, 24.4442, \ion{S}{xiii} 24.4175 \\
\ion{S}{xiv} & 24.5080 & 6.7 & 1.98e$-$15 & 6.6 & $-$0.04 &  \\
\ion{S}{xiii} & 24.5900 & 6.6 & 5.43e$-$15 & 10.9 & 0.47 &  \\
\ion{N}{vii} & 24.7792 & 6.4 & 9.58e$-$14 & 45.5 & $-$0.29 & \ion{N}{vii} 24.7846 \\
\ion{Ar}{xvi} & 24.9942 & 6.8 & 8.89e$-$15 & 13.6 & $-$0.21 & \ion{Ar}{xvi} 25.0098, \ion{Ca}{xiv} 25.0636 \\
No id. & 25.1640 & 0.0 & 1.93e$-$15 & 6.4 & \ldots &  \\
\ion{Ca}{xi} & 25.3270 & 6.4 & 5.08e$-$15 & 10.3 & 0.00 & \ion{Ca}{xiii} 25.3503 \\
\ion{Ar}{xvi} & 25.5168 & 6.8 & 6.29e$-$15 & 11.3 & $-$0.13 & \ion{Ca}{xi} 25.5170, \ion{Ca}{xiii} 25.5301 \\
\ion{Ar}{xvi} & 25.6844 & 6.8 & 7.50e$-$15 & 12.1 & 0.11 &  \\
\ion{C}{vi} & 26.3572 & 6.3 & 7.86e$-$15 & 12.4 & $-$0.20 & \ion{Ca}{xv} 26.2720, \ion{Ar}{xiv} 26.2730, \ion{C}{vi} 26.3574, \ion{Ca}{xiv} 26.3720, \ion{Ca}{xiii} 26.3762 \\
\ion{Si}{xii} & 26.4560 & 6.5 & 4.05e$-$15 & 8.8 & 0.37 & \ion{Ar}{xiv} 26.4300, \ion{Si}{xii} 26.4590 \\
\ion{C}{vi} & 26.9896 & 6.3 & 2.01e$-$14 & 19.6 & 0.07 & \ion{S}{xiii} 26.9792, \ion{C}{vi} 26.9901, \ion{Ar}{xv} 27.0432 \\
\ion{Ar}{xiv} & 27.4640 & 6.7 & 1.50e$-$14 & 16.6 & 0.07 & \ion{S}{xiii} 27.3918, \ion{Ca}{xii} 27.4131, \ion{Ar}{xiv} 27.6360 \\
\ion{C}{vi} & 28.4652 & 6.3 & 3.40e$-$14 & 23.6 & 0.12 & \ion{Ar}{xv} 28.3464, \ion{C}{vi} 28.4663, \ion{Ca}{xii} 28.4781 \\
\ion{N}{vi} & 28.7870 & 6.3 & 2.48e$-$14 & 19.7 & $-$0.19 &  \\
\ion{N}{vi} & 29.0843 & 6.2 & 1.30e$-$14 & 14.0 & $-$0.14 & \ion{Ca}{xiv} 28.9586, 29.1320, \ion{Ar}{xiii} 29.1420 \\
\ion{S}{xii} & 29.2942 & 6.5 & 8.25e$-$15 & 11.1 & 0.25 & \ion{Ar}{xiv} 29.2060, \ion{Ca}{xiv} 29.2123, 29.2577, \ion{Ar}{xiii} 29.2245 \\
\ion{Si}{xii} & 29.4390 & 6.5 & 1.99e$-$14 & 17.0 & $-$0.22 & \ion{Si}{xii} 29.5090, \ion{N}{vi} 29.5347 \\
\ion{S}{xiv} & 30.4270 & 6.6 & 2.75e$-$14 & 19.2 & $-$0.46 & \ion{Ca}{xi} 30.4480, \ion{S}{xiv} 30.4690 \\
\ion{Si}{xii} & 31.0120 & 6.5 & 5.82e$-$15 & 8.7 & $-$0.28 & \ion{Si}{xii} 31.0230, \ion{S}{xi} 31.0501 \\
\ion{Ar}{xii} & 31.3030 & 6.5 & 3.38e$-$15 & 6.6 & 0.39 &  \\
\ion{S}{xiii} & 32.2391 & 6.6 & 1.05e$-$14 & 11.6 & $-$0.21 & \ion{S}{xiii} 32.1911, \ion{Ca}{xii} 32.2805 \\
\ion{S}{xiv} & 32.4160 & 6.6 & 1.56e$-$14 & 14.1 & $-$0.12 &  \\
\ion{S}{xiv} & 32.5600 & 6.6 & 2.00e$-$14 & 15.7 & $-$0.39 & \ion{S}{xiv} 32.5750, \ion{Ca}{xii} 32.6571 \\
\ion{S}{xiv} & 33.5490 & 6.6 & 1.40e$-$14 & 12.9 & $-$0.11 &  \\
\ion{C}{vi} & 33.7342 & 6.2 & 1.69e$-$13 & 45.0 & 0.23 & \ion{C}{vi} 33.7396 \\
\ion{S}{xiii} & 33.9461 & 6.5 & 1.23e$-$14 & 11.9 & 0.63 & \ion{S}{xiii} 33.9526, \ion{Fe}{xvii} 33.9618 \\
\ion{C}{v} & 34.9728 & 6.1 & 7.26e$-$15 & 9.0 & 0.26 & \ion{Ar}{ix} 35.0240, \ion{Fe}{xvi} 35.1059 \\
\ion{Fe}{xvii} & 35.6844 & 6.9 & 1.86e$-$14 & 14.5 & 0.13 & \ion{Ca}{xi} 35.7370 \\
\ion{Fe}{xviii} & 93.9230 & 7.0 & 1.10e$-$14 & 3.8 & $-$0.80 & \ion{Fe}{xx} 93.7800 \\
\ion{Ne}{viii} & 98.2600 & 5.9 & 2.31e$-$14 & 5.7 & $-$0.04 & \ion{Fe}{xxi} 97.8800, \ion{Ne}{viii} 98.1150, 98.2740 \\
\ion{Fe}{xx} & 121.8300 & 7.1 & 1.63e$-$14 & 4.5 & $-$0.14 &  \\
\ion{Fe}{xxi} & 128.7300 & 7.2 & 1.40e$-$14 & 3.5 & $-$0.37 &  \\
\ion{Fe}{xxiii} & 132.8500 & 7.3 & 4.22e$-$14 & 6.5 & $-$0.34 & \ion{Fe}{xx} 132.8500 \\
\ion{Fe}{xxii} & 135.7800 & 7.2 & 1.45e$-$14 & 3.2 & $-$0.28 & \ion{Fe}{xi} 135.6988 \\
\ion{Fe}{xv} & 284.1630 & 6.4 & 1.53e$-$13 & 4.3 & $-$0.04 &  \\
\ion{Fe}{xvi} & 335.4099 & 6.5 & 1.80e$-$13 & 6.6 & $-$0.07 & \ion{Fe}{xiv} 334.1800 \\
\ion{Mg}{ix} & 368.0577 & 6.1 & 9.96e$-$14 & 4.5 & $-$0.01 & \\
\ion{S}{xiv} & 417.6655 & 6.6 & 1.17e$-$13 & 3.2 & $-$0.06 & \ion{Fe}{xv} 417.2580 \\
\ion{S}{vi} & 933.3788 & 5.4 & 2.28e$-$15 & 3.6 & $-$0.22 &  \\
\ion{O}{vi} & 1031.9121 & 5.6 & 1.42e$-$13 & 18.2 & 0.31 &  \\
\ion{C}{ii} & 1036.3390 & 4.8 & 2.34e$-$15 & 6.6 & $-$0.61 &  \\
\ion{C}{ii} & 1037.0200 & 4.8 & 8.05e$-$15 & 8.7 & $-$0.37 &  \\
\ion{O}{vi} & 1037.6136 & 5.6 & 7.04e$-$14 & 15.5 & 0.31 &  \\
\ion{Si}{iii} & 1108.3610 & 4.9 & 1.22e$-$15 & 4.6 & 0.01 &  \\
\ion{Si}{iii} & 1109.9720 & 4.9 & 1.74e$-$15 & 5.3 & $-$0.20 & \ion{Si}{iii} 1109.9430 \\
\ion{Si}{iv} & 1122.4852 & 5.0 & 7.56e$-$16 & 5.0 & 0.14 &  \\
\ion{Si}{iv} & 1128.3404 & 5.0 & 1.64e$-$15 & 6.4 & 0.23 &  \\
\ion{Ne}{v} & 1145.5959 & 5.5 & 5.90e$-$16 & 5.7 & $-$0.28 &  \\
\ion{C}{iii} & 1176.0000 & 4.8 & 5.81e$-$14 & 7.5 & $-$0.31 &  \\
\ion{C}{iii} & 1176.0000 & 4.8 & 6.20e$-$14 & 20.9 & $-$0.28 &  \\
\ion{S}{iii} & 1200.9610 & 4.9 & 3.18e$-$15 & 42.6 & $-$0.10 &  \\
\ion{O}{v} & 1218.3440 & 5.5 & 1.40e$-$14 & 104.6 & 0.17 &  \\
\ion{N}{v} & 1238.8218 & 5.4 & 4.28e$-$14 & 74.6 & 0.20 &  \\
\ion{N}{v} & 1242.8042 & 5.4 & 2.02e$-$14 & 45.8 & 0.17 &  \\
\ion{C}{iii} & 1247.3830 & 5.1 & 6.29e$-$16 & 18.7 & $-$0.21 &  \\
\ion{Si}{ii} & 1264.7400 & 4.5 & 1.64e$-$15 & 20.6 & $-$0.03 &  \\
\ion{Si}{ii} & 1265.0040 & 4.6 & 1.06e$-$15 & 48.7 & $-$0.08 &  \\
\ion{Si}{iii} & 1294.5480 & 4.9 & 8.71e$-$16 & 12.5 & 0.70 &  \\
\ion{Si}{iii} & 1296.7280 & 4.9 & 9.89e$-$16 & 31.3 & 0.35 &  \\
\ion{Si}{iii} & 1298.9480 & 4.9 & 2.30e$-$15 & 35.0 & 0.65 &  \\
\ion{Si}{iii} & 1301.1510 & 4.9 & 6.32e$-$16 & 15.3 & 0.63 &  \\
\ion{Si}{iii} & 1303.3249 & 4.9 & 8.65e$-$16 & 33.3 & 0.20 &  \\
\ion{Si}{ii} & 1304.3719 & 4.6 & 4.01e$-$16 & 26.3 & $-$0.20 &  \\
\ion{Si}{ii} & 1309.2770 & 4.6 & 7.10e$-$16 & 15.1 & $-$0.25 &  \\
\ion{C}{ii} & 1323.9080 & 4.8 & 4.07e$-$16 & 11.9 & $-$0.55 & \ion{C}{ii} 1323.9540 \\
\ion{Fe}{xix} & 1328.6999 & 7.1 & 4.21e$-$16 & 23.4 & $-$0.44 &  \\
\ion{C}{ii} & 1334.5350 & 4.7 & 1.75e$-$14 & 43.7 & $-$0.66 &  \\
\ion{C}{ii} & 1335.7100 & 4.7 & 4.45e$-$14 & 69.5 & $-$0.05 & \ion{C}{ii} 1335.6650 \\
\ion{Fe}{xii} & 1349.4000 & 6.3 & 4.75e$-$16 & 38.4 & 0.01 &  \\
\ion{O}{v} & 1371.2960 & 5.5 & 2.22e$-$15 & 22.4 & $-$0.03 &  \\
\ion{Si}{iv} & 1393.7552 & 5.0 & 2.34e$-$14 & 70.3 & $-$0.11 &  \\
\ion{O}{iv} & 1397.2309 & 5.3 & 7.47e$-$16 & 24.2 & $-$0.49 &  \\
\ion{Si}{iv} & 1402.7704 & 5.0 & 1.14e$-$14 & 82.1 & $-$0.12 &  \\
\ion{S}{iv} & 1406.0160 & 5.1 & 4.28e$-$15 & 77.5 & $-$0.06 &  \\
\ion{Fe}{xi} & 1467.4230 & 6.2 & 3.28e$-$16 & 12.1 & $-$0.07 &  \\
\ion{Si}{ii} & 1526.7090 & 4.5 & 1.58e$-$15 & 25.5 & 0.04 &  \\
\ion{Si}{ii} & 1533.4320 & 4.5 & 2.60e$-$15 & 25.7 & $-$0.04 &  \\
\hline
\end{tabular}
\end{scriptsize}
\end{table*}
\setcounter{table}{1}
\begin{table*}
\caption{(continued) {\em XMM-Newton}/RGS, EUVE, FUSE, and HST/STIS line fluxes of Proxima Centauri\tablefootmark{a}.}
\tabcolsep 3.pt
\begin{scriptsize}
  \begin{tabular}{lrccrcl}
\hline \hline
Ion & $\lambda_{\rm model}$ (\AA) & $\log T_{\rm max}$ & $F_{\rm obs}$ & $S/N$ & Ratio & Blends \\
\hline
\ion{C}{iv} & 1548.1871 & 5.1 & 1.31e$-$13 & 71.8 & 0.16 &  \\
\ion{C}{iv} & 1550.7723 & 5.1 & 6.60e$-$14 & 40.6 & 0.16 &  \\
\ion{Al}{ii} & 1670.7870 & 4.6 & 1.12e$-$14 & 20.8 & $-$0.01 &  \\
\hline
\end{tabular}
\tablefoot{\tablefoottext{a}{Line fluxes (in erg cm$^{-2}$ s$^{-1}$ units) measured in the spectra, and corrected by the ISM absorption when needed.
  log $T_{\rm max}$ (K) indicates the maximum  temperature of formation of the line (unweighted by the
  EMD). ``Ratio'' is the observed-to-predicted line flux ratio, log($F_{\mathrm {obs}}$/$F_{\mathrm {pred}}$), with predicted flux as calculated with the resulting EMD. 
  ``Blends'' amounting to more than 5\% of the total flux for each line are indicated, with wavelengths in \AA.}}
\end{scriptsize}
\end{table*}

%
\begin{figure*}
  \centering
  \includegraphics[width=0.33\textwidth]{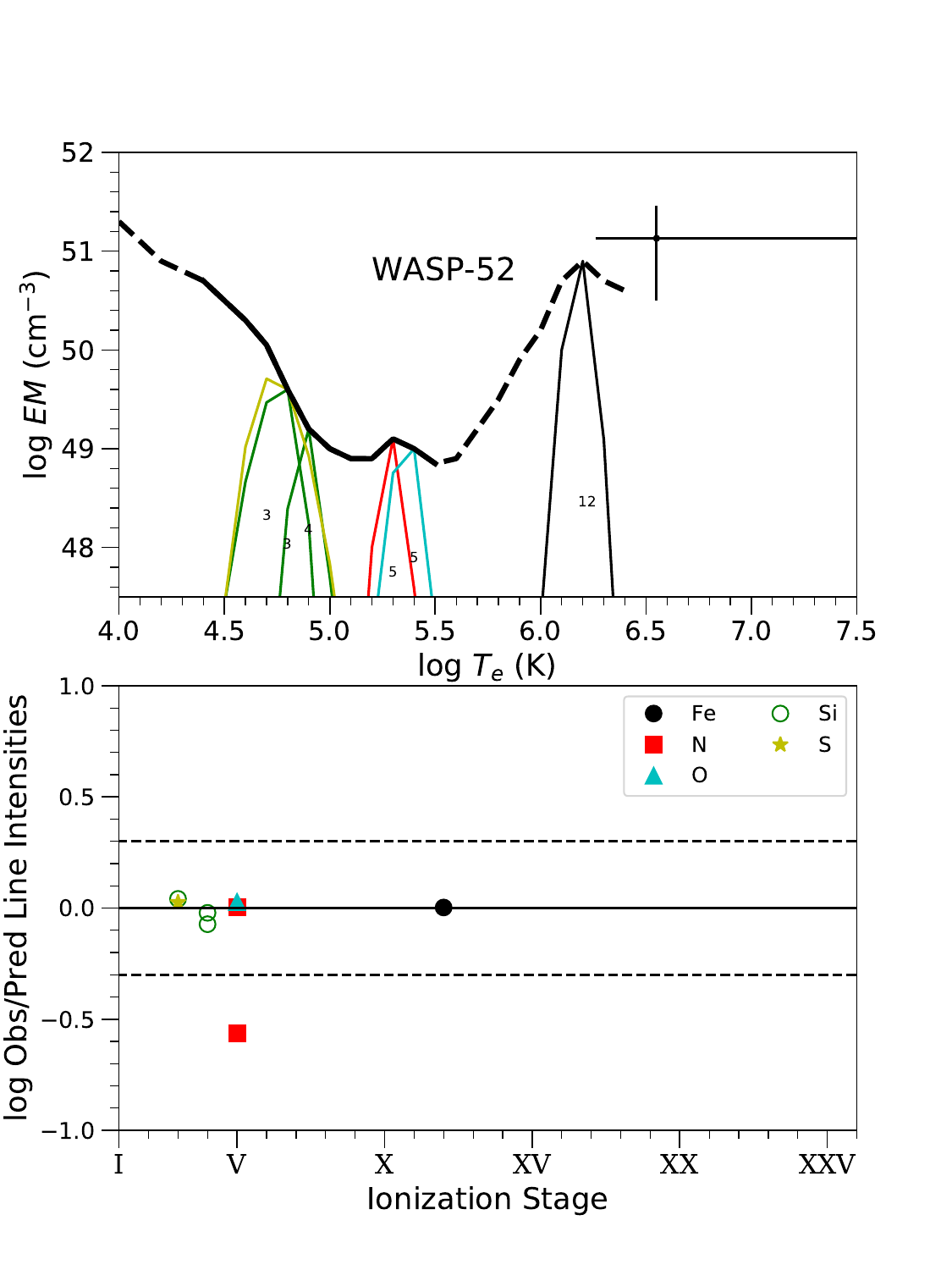}
  \includegraphics[width=0.33\textwidth]{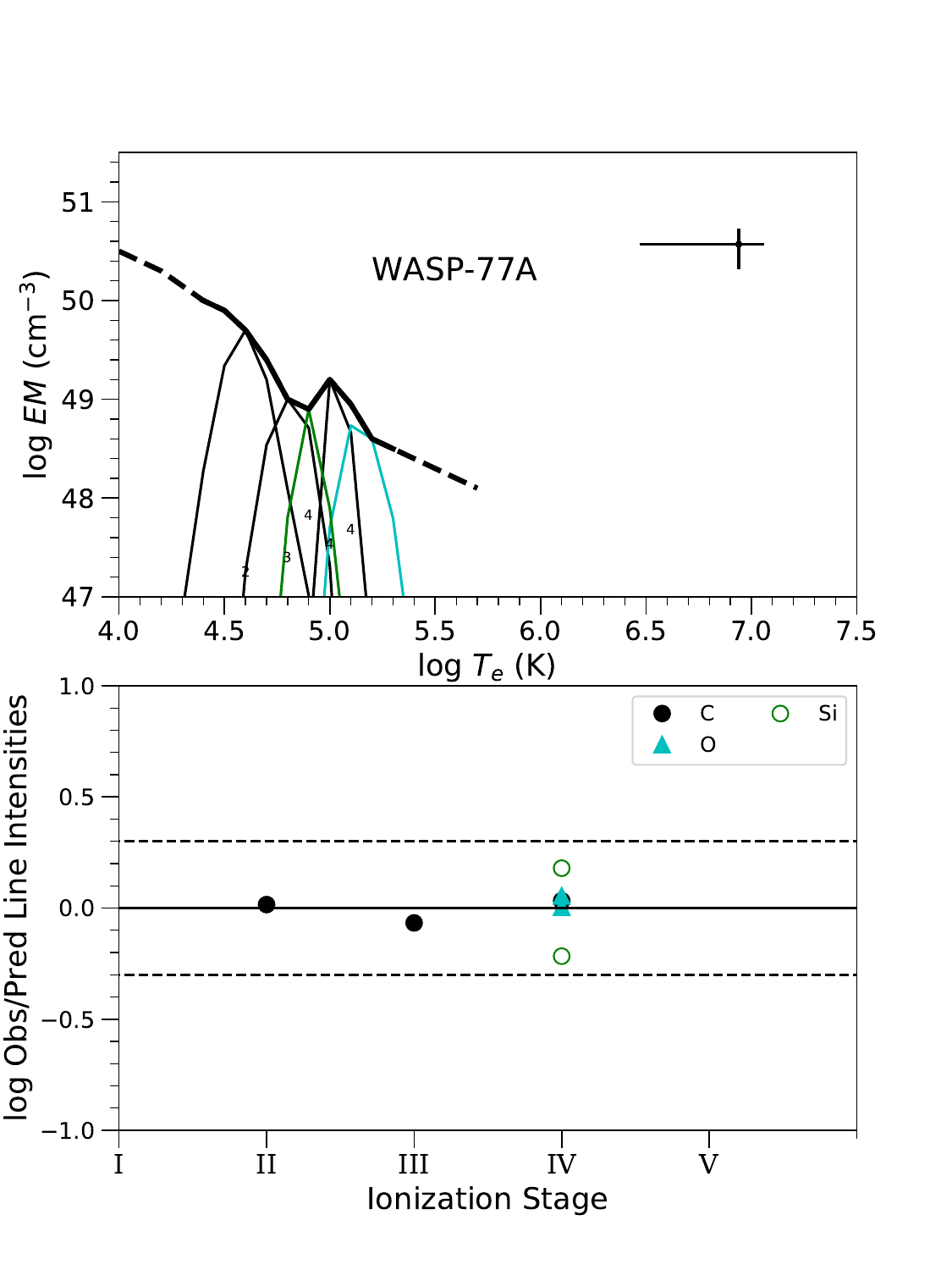}
  \includegraphics[width=0.33\textwidth]{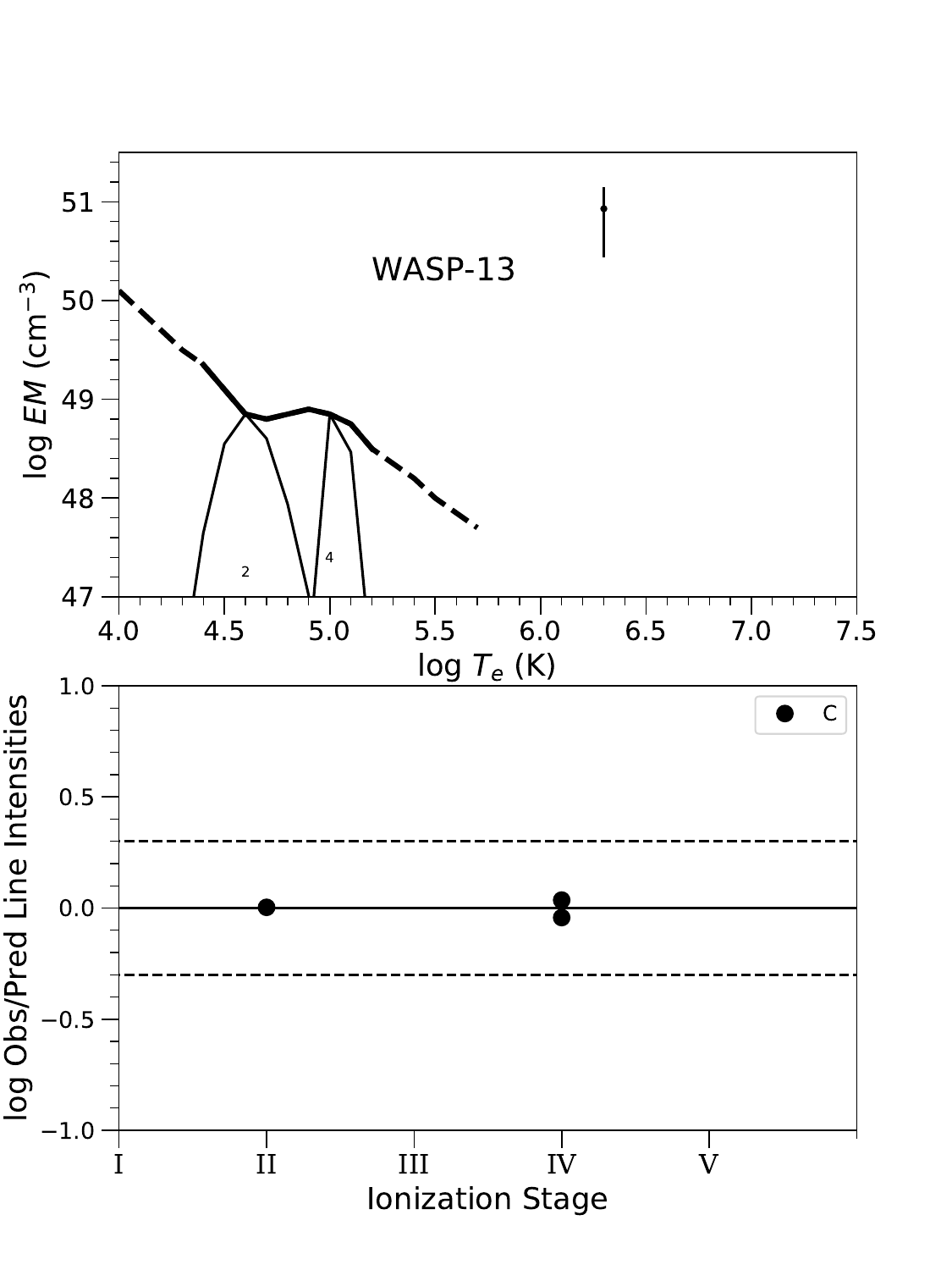}
  \caption{Emission Measure Distributions as in Fig.~\ref{fig:proxcen}, using HST/STIS data for WASP-52 ({\it left}) and WASP-77 ({\it center}), and HST/COS data for WASP-13 ({\it right}). The 1-$T$ global fit to Chandra/ACIS spectra is also indicated with error bars.}\label{fig:wasp52} 
\end{figure*}
%

%
\begin{figure*}
  \centering
  \includegraphics[width=0.33\textwidth]{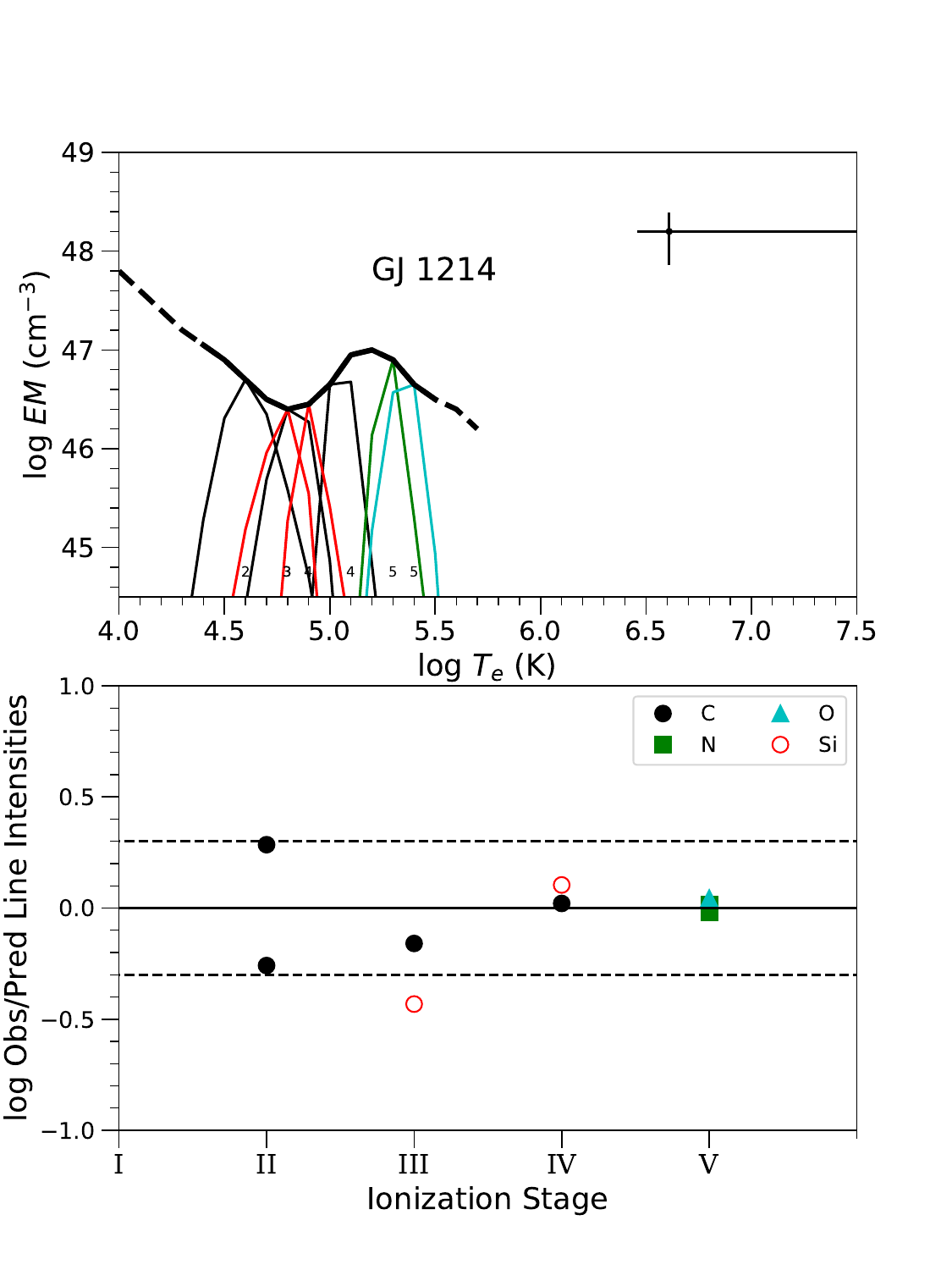}
  \includegraphics[width=0.33\textwidth]{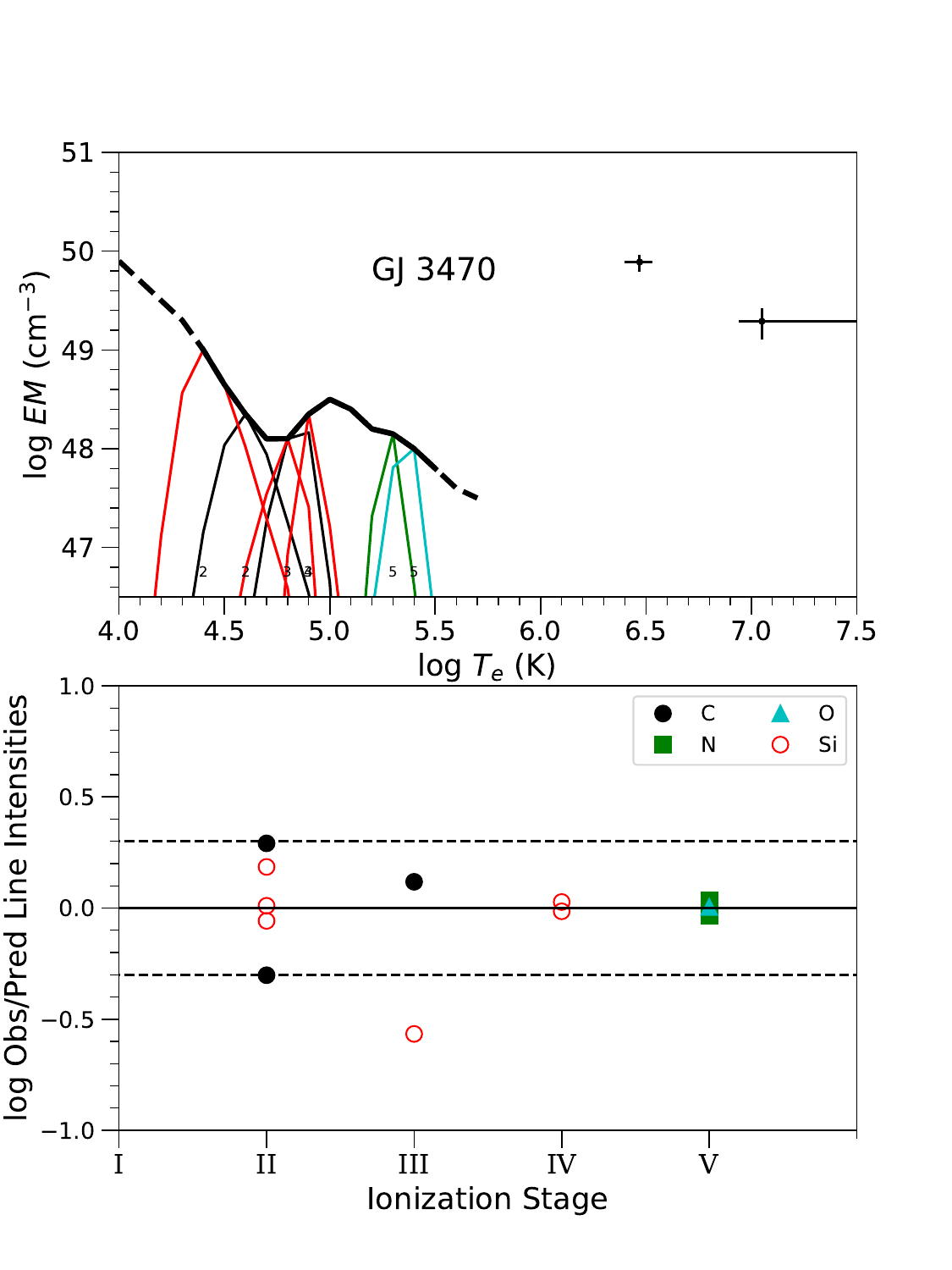}
  \includegraphics[width=0.33\textwidth]{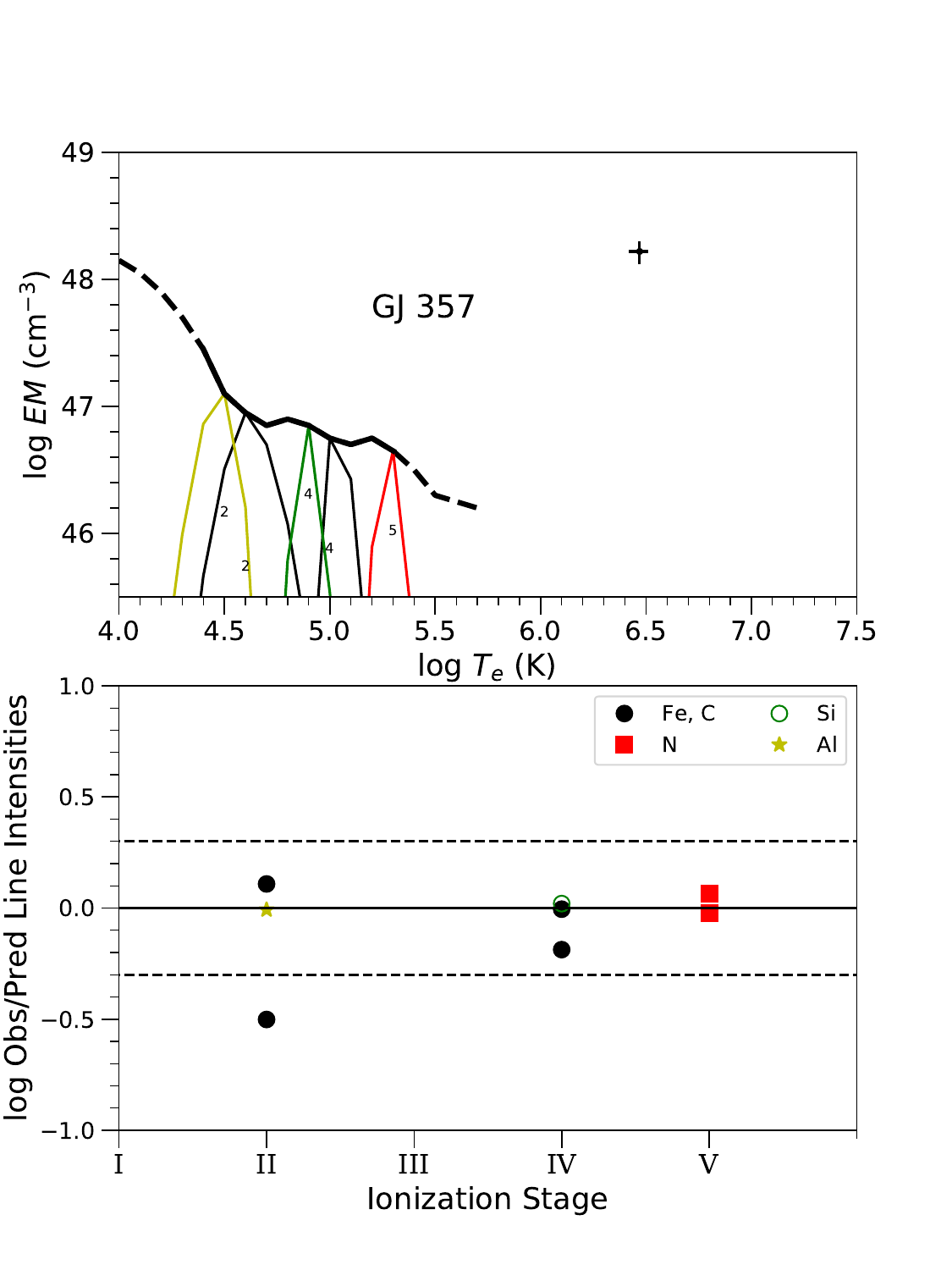}
  \caption{Same as in Fig.~\ref{fig:proxcen} but for GJ~1214, GJ~3470, and GJ~357, using HST/COS data. The 1-$T$ and 2-$T$ global fits to XMM-Newton/EPIC spectra are also indicated with error bars.}\label{fig:gj1214} 
\end{figure*}
%

%
\begin{table*}
\caption[]{{\em XMM-Newton}/RGS, EUVE, FUSE, and HST/STIS line fluxes of AU Mic\tablefootmark{a}.}\label{tab:fluxes2} 
\tabcolsep 3 pt
\begin{scriptsize}
  \begin{tabular}{lrccrcl}
\hline \hline
Ion & $\lambda_{\rm model}$ (\AA) & $\log T_{\rm max}$ & $F_{\rm obs}$ & $S/N$ & Ratio & Blends \\
\hline
\ion{Mg}{xii} & 8.4192 & 7.1 & 2.58e$-$14 & 9.5 & $-$0.43 & \ion{Mg}{xii}  8.4246 \\
\ion{Mg}{xi} & 9.1687 & 6.9 & 5.81e$-$14 & 15.9 & 0.02 & \ion{Fe}{xxi}  9.1944 \\
\ion{Mg}{xi} & 9.2312 & 6.9 & 2.47e$-$14 & 10.5 & $-$0.15 & \ion{Fe}{xxii}  9.2298, \ion{Ne}{x} 9.2461, 9.2462, \ion{Ne}{x}  9.2912, 9.2913 \\
\ion{Mg}{xi} & 9.3143 & 6.9 & 5.40e$-$14 & 15.7 & 0.14 & \ion{Ne}{x}  9.3616 \\
\ion{Ne}{x} & 9.4807 & 6.9 & 4.07e$-$14 & 13.9 & 0.02 & \ion{Ne}{x}  9.4809 \\
\ion{Ne}{x} & 9.7080 & 6.9 & 6.78e$-$14 & 18.5 & $-$0.14 & \ion{Ne}{x}  9.7085 \\
\ion{Ne}{x} & 10.2385 & 6.9 & 1.43e$-$13 & 28.2 & $-$0.23 & \ion{Ne}{x} 10.2396 \\
\ion{Fe}{xviii} & 10.5364 & 7.0 & 3.38e$-$14 & 14.2 & $-$0.04 & \ion{Fe}{xvii} 10.5040, \ion{Fe}{xviii} 10.5382, 10.5640, \ion{Ne}{ix} 10.5406, 10.5650 \\
\ion{Fe}{xix} & 10.6193 & 7.1 & 5.43e$-$14 & 18.2 & $-$0.09 & \ion{Fe}{xix} 10.6001, 10.6116, 10.6840, \ion{Fe}{xxiv} 10.6190, 10.6630, \ion{Ne}{ix} 10.6426, \ion{Fe}{xvii} 10.6570 \\
\ion{Fe}{xvii} & 10.7700 & 6.9 & 2.40e$-$14 & 12.3 & $-$0.10 & \ion{Fe}{xix} 10.7650, \ion{Ne}{ix} 10.7650 \\
\ion{Fe}{xxiii} & 10.9810 & 7.3 & 8.26e$-$14 & 22.8 & $-$0.01 & \ion{Ne}{ix} 11.0010, \ion{Fe}{xxiii} 11.0190, \ion{Fe}{xvii} 11.0260, \ion{Fe}{xxiv} 11.0290 \\
\ion{Fe}{xxiv} & 11.1760 & 7.4 & 3.13e$-$14 & 14.5 & $-$0.11 & \ion{Fe}{xvii} 11.1310 \\
\ion{Fe}{xvii} & 11.2540 & 6.9 & 3.99e$-$14 & 16.4 & 0.08 & \ion{Fe}{xxiv} 11.2680 \\
\ion{Fe}{xviii} & 11.4230 & 7.0 & 3.20e$-$14 & 14.6 & $-$0.20 & \ion{Fe}{xxii} 11.4270, \ion{Fe}{xviii} 11.4274, \ion{Fe}{xxiv} 11.4320, \ion{Fe}{xxiii} 11.4580 \\
\ion{Ne}{ix} & 11.5440 & 6.7 & 1.22e$-$13 & 27.7 & $-$0.04 & \ion{Fe}{xviii} 11.5270 \\
\ion{Fe}{xxii} & 11.7700 & 7.2 & 9.03e$-$14 & 25.5 & $-$0.07 & \ion{Fe}{xxiii} 11.7360 \\
\ion{Ni}{xx} & 11.8320 & 7.1 & 4.61e$-$15 & 5.8 & $-$0.24 & \ion{Ni}{xx} 11.8460 \\
\ion{Fe}{xxii} & 11.9770 & 7.2 & 3.37e$-$14 & 16.0 & $-$0.09 & \ion{Fe}{xxii} 11.8810, 11.9320, 11.9474, \ion{Fe}{xxi} 11.9023, 11.9750 \\
\ion{Ne}{x} & 12.1321 & 6.9 & 9.03e$-$13 & 83.7 & $-$0.25 & \ion{Ne}{x} 12.1375 \\
\ion{Fe}{xxi} & 12.2840 & 7.2 & 1.31e$-$13 & 32.2 & $-$0.08 & \ion{Fe}{xvii} 12.2660 \\
\ion{Ni}{xix} & 12.4350 & 7.0 & 9.12e$-$14 & 27.1 & 0.04 & \ion{Fe}{xxi} 12.3930, 12.4990, \ion{Fe}{xx} 12.4234, 12.4310, 12.5760 \\
\ion{Fe}{xxi} & 12.6490 & 7.2 & 4.33e$-$14 & 18.9 & 0.48 & \ion{Ni}{xix} 12.6560, \ion{Fe}{xvii} 12.6950 \\
\ion{Fe}{xx} & 12.8460 & 7.1 & 1.09e$-$13 & 29.9 & $-$0.14 & \ion{Fe}{xxi} 12.8220, \ion{Fe}{xx} 12.8240, 12.8640 \\
\ion{Fe}{xx} & 12.9120 & 7.1 & 8.79e$-$14 & 26.4 & $-$0.07 & \ion{Fe}{xix} 12.9033, 12.9330, 13.0220, \ion{Fe}{xviii} 12.9494, \ion{Fe}{xxii} 12.9530, \ion{Fe}{xx} 12.9650, 12.9920 \\
\ion{Fe}{xx} & 13.1530 & 7.1 & 3.72e$-$14 & 15.1 & 0.16 & \ion{Fe}{xx} 13.1530, 13.1370,13.1370 \\
\ion{Fe}{xx} & 13.2740 & 7.1 & 5.00e$-$14 & 20.6 & 0.03 & \ion{Fe}{xix} 13.2261, 13.2933, \ion{Fe}{xxii} 13.2360, \ion{Fe}{xx} 13.2453, \ion{Ni}{xx} 13.3090, \ion{Fe}{xviii} 13.3230 \\
\ion{Ne}{ix} & 13.4473 & 6.7 & 8.00e$-$13 & 83.6 & $-$0.04 & \ion{Fe}{xix} 13.4970, 13.5180, \ion{Ne}{ix} 13.5531 \\
\ion{Ne}{ix} & 13.6990 & 6.7 & 4.01e$-$13 & 59.5 & 0.10 & \ion{Fe}{xix} 13.7054 \\
\ion{Fe}{xvii} & 13.8250 & 6.9 & 1.79e$-$13 & 40.2 & 0.13 & \ion{Fe}{xix} 13.7590, 13.7950, 13.8390, \ion{Fe}{xx} 13.7670, \ion{Ni}{xix} 13.7790, \ion{Fe}{xvii} 13.8920 \\
\ion{Fe}{xx} & 13.9620 & 7.1 & 4.74e$-$14 & 20.8 & 0.40 & \ion{Fe}{xviii} 13.9530 \\
\ion{Fe}{xxi} & 14.0080 & 7.2 & 3.97e$-$14 & 19.0 & $-$0.09 & \ion{Fe}{xix} 14.0179, \ion{Cr}{xxi} 14.0339, \ion{Fe}{xix} 14.0340, 14.0388, \ion{Ni}{xix} 14.0430 \\
\ion{Ni}{xix} & 14.0770 & 7.0 & 4.23e$-$14 & 19.7 & 0.42 & \ion{Fe}{xviii} 14.0549, \ion{Fe}{xx} 14.0620, 14.0784 \\
\ion{Fe}{xviii} & 14.2080 & 7.0 & 2.68e$-$13 & 49.5 & $-$0.07 & \ion{Fe}{xviii} 14.2560 \\
\ion{Fe}{xviii} & 14.3730 & 7.0 & 1.40e$-$13 & 35.7 & $-$0.06 & \ion{Fe}{xviii} 14.3430, 14.3990, 14.4250, 14.4555 \\
\ion{Fe}{xviii} & 14.5340 & 7.0 & 1.06e$-$13 & 31.5 & 0.01 & \ion{O}{viii} 14.5242, 14.5243, \ion{Fe}{xviii} 14.5608, 14.5710 \\
\ion{Fe}{xix} & 14.6640 & 7.1 & 7.88e$-$14 & 27.2 & $-$0.17 & \ion{O}{viii} 14.6343, 14.6344, \ion{Fe}{xix} 14.7250, \ion{Fe}{xviii} 14.7499, \ion{Fe}{xx} 14.7540 \\
\ion{O}{viii} & 14.8205 & 6.6 & 5.21e$-$14 & 22.3 & $-$0.03 & \ion{O}{viii} 14.8207 \\
\ion{Fe}{xix} & 14.9610 & 7.1 & 2.41e$-$14 & 15.2 & $-$0.38 & \ion{Fe}{xx} 14.8791, 14.9279, \ion{Fe}{xviii} 14.8920, 14.9717, \ion{Fe}{xix} 14.8995, 14.9170, 14.9632 \\
\ion{Fe}{xvii} & 15.0140 & 6.9 & 3.79e$-$13 & 60.4 & $-$0.09 &  \\
\ion{Fe}{xix} & 15.0790 & 7.1 & 5.52e$-$14 & 23.1 & 0.19 &  \\
\ion{O}{viii} & 15.1760 & 6.6 & 1.47e$-$13 & 37.3 & $-$0.08 & \ion{O}{viii} 15.1765, \ion{Fe}{xix} 15.1770, 15.1980 \\
\ion{Fe}{xvii} & 15.2610 & 6.9 & 1.54e$-$13 & 38.1 & 0.04 &  \\
\ion{Fe}{xviii} & 15.3539 & 7.0 & 3.59e$-$14 & 18.0 & 0.03 & \ion{Fe}{xix} 15.3123, 15.3340, 15.3709, \ion{Fe}{xvi} 15.3907 \\
\ion{Fe}{xvii} & 15.4530 & 6.8 & 4.93e$-$14 & 21.2 & 0.07 & \ion{Fe}{xvi} 15.4533, \ion{Fe}{xviii} 15.4940, \ion{Fe}{xx} 15.5170 \\
\ion{Fe}{xviii} & 15.6250 & 7.0 & 4.65e$-$14 & 21.1 & $-$0.19 &  \\
\ion{Fe}{xviii} & 15.7590 & 7.0 & 1.19e$-$14 & 10.8 & 0.10 &  \\
\ion{Fe}{xviii} & 15.8700 & 7.0 & 4.65e$-$14 & 21.1 & $-$0.20 & \ion{Fe}{xviii} 15.8240 \\
\ion{O}{viii} & 16.0055 & 6.6 & 2.99e$-$13 & 54.4 & $-$0.16 & \ion{Fe}{xviii} 16.0040, \ion{O}{viii} 16.0067 \\
\ion{Fe}{xviii} & 16.0710 & 7.0 & 1.11e$-$13 & 33.1 & $-$0.06 & \ion{Fe}{xviii} 16.0450 \\
\ion{Fe}{xviii} & 16.1590 & 7.0 & 1.79e$-$14 & 13.1 & $-$0.58 & \ion{Fe}{xix} 16.1100 \\
\ion{Fe}{xix} & 16.2830 & 7.1 & 2.92e$-$14 & 16.8 & $-$0.23 & \ion{Fe}{xvii} 16.2285, 16.3500, \ion{Fe}{xix} 16.2857,16.3247, \ion{Fe}{xviii} 16.3200 \\
\ion{Fe}{xvii} & 16.7800 & 6.8 & 2.12e$-$13 & 45.7 & $-$0.12 &  \\
\ion{Fe}{xvii} & 17.0510 & 6.8 & 5.23e$-$13 & 70.5 & $-$0.06 & \ion{Fe}{xvii} 17.0960 \\
\ion{O}{vii} & 17.3960 & 6.5 & 3.61e$-$14 & 18.1 & 0.07 & \ion{Cr}{xviii} 17.4031, \ion{Cr}{xvi} 17.4207, 17.4252 \\
\ion{Fe}{xviii} & 17.6230 & 7.0 & 4.55e$-$14 & 21.0 & $-$0.15 &  \\
\ion{O}{vii} & 17.7680 & 6.5 & 3.50e$-$14 & 17.8 & 0.08 & \ion{Ar}{xvi} 17.7371, 17.7468 \\
\ion{O}{vii} & 18.6270 & 6.4 & 6.85e$-$14 & 26.5 & $-$0.30 & \ion{Ca}{xviii} 18.6909 \\
\ion{O}{viii} & 18.9671 & 6.6 & 1.41e$-$12 & 119.3 & $-$0.16 & \ion{O}{viii} 18.9725 \\
\ion{N}{vii} & 20.9095 & 6.4 & 4.56e$-$14 & 20.4 & 0.26 &  \\
\ion{Ca}{xvi} & 21.4501 & 6.8 & 1.64e$-$14 & 12.6 & $-$0.56 &  \\
\ion{O}{vii} & 21.6015 & 6.4 & 3.59e$-$13 & 59.0 & $-$0.04 &  \\
\ion{O}{vii} & 21.8036 & 6.4 & 7.47e$-$14 & 26.8 & 0.07 &  \\
No id. & 21.9686 & 0.0 & 1.11e$-$14 & 10.3 & \ldots &  \\
\ion{O}{vii} & 22.0977 & 6.4 & 2.65e$-$13 & 50.2 & 0.02 & \ion{Ca}{xvi} 22.1372 \\
\ion{S}{xiv} & 23.0050 & 6.7 & 1.56e$-$14 & 10.2 & $-$0.05 & \ion{S}{xiv} 23.0150, \ion{Ca}{xv} 23.0230, \ion{Ca}{xvi} 23.0433 \\
\ion{N}{vi} & 23.1274 & 6.3 & 1.34e$-$14 & 9.6 & 0.16 & \ion{Ca}{xiv} 23.1518, \ion{Ca}{xvii} 23.1752, \ion{Ca}{xv} 23.1818, \ion{Ca}{xii} 23.1892 \\
\ion{N}{vi} & 23.2770 & 6.3 & 1.64e$-$14 & 12.5 & 0.39 & \ion{Ca}{xii} 23.2473, 23.3326, \ion{Ca}{xvi} 23.2608, \ion{Ca}{xiv} 23.2711 \\
\ion{Ar}{xvi} & 23.5063 & 6.8 & 3.67e$-$14 & 19.3 & 0.05 & \ion{Ca}{xv} 23.5034 \\
\ion{Ar}{xvi} & 23.5465 & 6.8 & 1.38e$-$14 & 11.9 & $-$0.14 & \ion{Ca}{xiv} 23.6006 \\
No id. & 23.7700 & 0.0 & 3.14e$-$14 & 12.6 & \ldots &  \\
\ion{Ca}{xiv} & 24.0860 & 6.7 & 2.16e$-$14 & 15.2 & $-$0.36 & \ion{Ca}{xiv} 24.0335 \\
\ion{S}{xiv} & 24.2000 & 6.7 & 2.48e$-$14 & 16.4 & $-$0.38 & \ion{Ca}{xiv} 24.1331, \ion{Ca}{xvi} 24.2214, \ion{Ca}{xv} 24.2344 \\
\ion{S}{xiv} & 24.2850 & 6.7 & 2.13e$-$14 & 15.2 & 0.12 & \ion{Ca}{xiv} 24.2599, \ion{S}{xiv} 24.2890 \\
\ion{S}{xiv} & 24.4180 & 6.7 & 2.46e$-$14 & 16.3 & 0.16 & \ion{Ca}{xv} 24.3365, 24.4120, 24.4282, 24.4442, \ion{Ca}{xiv} 24.4046 \\
\ion{S}{xiii} & 24.5900 & 6.6 & 2.61e$-$14 & 16.8 & 0.48 & \ion{S}{xiv} 24.5080, \ion{Si}{xiii} 24.5172, \ion{Ca}{xv} 24.5274 \\
\ion{N}{vii} & 24.7792 & 6.4 & 1.97e$-$13 & 45.3 & $-$0.16 & \ion{Ar}{xv} 24.7400, \ion{N}{vii} 24.7846, \ion{Ar}{xvi} 24.8509 \\
\ion{Ar}{xvi} & 24.9942 & 6.8 & 5.24e$-$14 & 23.4 & 0.04 & \ion{Ar}{xvi} 25.0098 \\
No id. & 25.2000 & 0.0 & 1.15e$-$14 & 11.0 & \ldots &  \\
\ion{Ca}{xi} & 25.3270 & 6.4 & 1.55e$-$14 & 12.8 & 0.27 & \ion{Ca}{xiii} 25.3503 \\
\ion{Ar}{xvi} & 25.5168 & 6.8 & 2.21e$-$14 & 15.2 & $-$0.28 & \ion{Ca}{xv} 25.4749, \ion{Ca}{xi} 25.5170, \ion{Ca}{xiii} 25.5301 \\
\ion{Ar}{xvi} & 25.6844 & 6.8 & 7.53e$-$14 & 27.0 & 0.28 & \ion{Si}{xii} 25.6580, \ion{Ca}{xiv} 25.7299, \ion{Ca}{xvi} 25.7330 \\
\ion{C}{vi} & 26.3572 & 6.3 & 3.33e$-$14 & 18.3 & 0.03 & \ion{Ca}{xv} 26.2720, \ion{Ar}{xiv} 26.2730, \ion{C}{vi} 26.3574, \ion{Ca}{xiv} 26.3720, \ion{Ca}{xiii} 26.3762 \\
\ion{Si}{xii} & 26.4560 & 6.5 & 4.21e$-$14 & 20.4 & 0.36 & \ion{Ar}{xiv} 26.4300, \ion{Si}{xii} 26.4590, \ion{Ca}{xiv} 26.5141, \ion{Ca}{xiii} 26.5862 \\
\ion{C}{vi} & 26.9896 & 6.3 & 7.35e$-$14 & 26.8 & 0.13 & \ion{C}{vi} 26.9901, \ion{Si}{xii} 27.0350, \ion{Ar}{xv} 27.0432 \\
\hline
\end{tabular}
\end{scriptsize}
\end{table*}
\setcounter{table}{2}
\begin{table*}
\caption{(continued) {\em XMM-Newton}/RGS, EUVE, FUSE, and HST/STIS line fluxes of AU Mic\tablefootmark{a}.}
\tabcolsep 3.pt
\begin{scriptsize}
  \begin{tabular}{lrccrcl}
\hline \hline
Ion & $\lambda_{\rm model}$ (\AA) & $\log T_{\rm max}$ & $F_{\rm obs}$ & $S/N$ & Ratio & Blends \\
\hline
\ion{Ar}{xiv} & 27.4640 & 6.7 & 4.55e$-$14 & 20.9 & 0.32 & \ion{S}{xiii} 27.3918, \ion{Ca}{xii} 27.4131 \\
\ion{Fe}{xvii} & 27.5097 & 6.9 & 3.14e$-$14 & 16.8 & 0.53 & \ion{S}{xv} 27.5300, 27.5598, \ion{Ar}{xiv} 27.5490, 27.6310, 27.6360, \ion{Ar}{xv} 27.5559 \\
\ion{C}{vi} & 28.4652 & 6.3 & 7.51e$-$14 & 25.8 & $-$0.04 & \ion{Ar}{xv} 28.3464, \ion{C}{vi} 28.4663, \ion{Ca}{xii} 28.4781 \\
\ion{Ca}{xii} & 28.6811 & 6.5 & 1.40e$-$14 & 11.1 & 0.34 & \ion{Ca}{xii} 28.6131, 28.6370 \\
\ion{N}{vi} & 28.7870 & 6.3 & 1.46e$-$14 & 11.3 & $-$0.27 &  \\
\ion{Ca}{xiii} & 28.9161 & 6.6 & 1.37e$-$14 & 11.0 & 0.21 & \ion{Ar}{xiv} 28.8940 \\
\ion{N}{vi} & 29.0843 & 6.2 & 3.69e$-$14 & 17.2 & 0.17 & \ion{Ca}{xiv} 29.1320, \ion{Ar}{xiii} 29.1420 \\
\ion{Ar}{xiii} & 29.2245 & 6.6 & 3.31e$-$14 & 15.9 & 0.47 & \ion{Ar}{xiv} 29.2060, \ion{Ca}{xiv} 29.2123, 29.2577, \ion{Ar}{xv} 29.2210 \\
\ion{N}{vi} & 29.5347 & 6.2 & 4.96e$-$14 & 19.8 & 0.10 & \ion{Si}{xii} 29.4390, 29.5090 \\
\ion{S}{xiv} & 30.4270 & 6.6 & 4.68e$-$14 & 18.6 & $-$0.40 & \ion{Ca}{xi} 30.4480, \ion{S}{xiv} 30.4690 \\
\ion{Ca}{xiii} & 30.5402 & 6.6 & 1.32e$-$14 & 9.8 & 0.64 & \ion{Ca}{x} 30.5033 \\
\ion{Ca}{xi} & 30.8670 & 6.4 & 1.59e$-$14 & 10.9 & 0.26 & \ion{Ar}{xiv} 30.8320, \ion{Ar}{xiii} 30.8854 \\
\ion{Si}{xii} & 31.0120 & 6.5 & 3.50e$-$14 & 15.9 & $-$0.27 & \ion{Si}{xii} 31.0230 \\
\ion{Fe}{xvii} & 31.9515 & 6.9 & 3.04e$-$14 & 15.0 & 0.45 & \ion{Fe}{xvii} 31.8776, \ion{Ar}{xiii} 31.8954, \ion{Ca}{xii} 31.9561, \ion{Ar}{xiv} 32.0140 \\
\ion{Fe}{xvii} & 32.1188 & 6.9 & 1.07e$-$14 & 8.9 & 0.51 & \ion{Fe}{xvii} 32.0941, \ion{Ca}{xii} 32.1057, \ion{Fe}{xvi} 32.1658 \\
\ion{S}{xiii} & 32.2391 & 6.6 & 3.84e$-$14 & 16.8 & 0.18 & \ion{S}{xiii} 32.1911, \ion{Ca}{xii} 32.2805 \\
\ion{S}{xiv} & 32.4160 & 6.6 & 3.38e$-$14 & 15.7 & 0.04 & \ion{Ca}{xii} 32.4184 \\
\ion{S}{xiv} & 32.5600 & 6.6 & 6.38e$-$14 & 21.2 & $-$0.13 & \ion{Ca}{xii} 32.4988, 32.6571, \ion{S}{xiv} 32.5750 \\
\ion{S}{xiv} & 33.3810 & 6.6 & 1.52e$-$14 & 10.2 & 0.10 &  \\
\ion{C}{v} & 33.4257 & 6.1 & 1.13e$-$14 & 8.9 & 0.29 & \ion{Ca}{xiii} 33.4080, \ion{Si}{xiv} 33.4444, \ion{Ar}{xiv} 33.4600 \\
\ion{S}{xiv} & 33.5490 & 6.6 & 3.65e$-$14 & 15.5 & $-$0.05 & \ion{Si}{xiv} 33.5051, \ion{Si}{xi} 33.5301, 33.5731 \\
\ion{C}{vi} & 33.7342 & 6.2 & 2.89e$-$13 & 44.2 & $-$0.08 & \ion{C}{vi} 33.7396 \\
\ion{Fe}{xvii} & 33.8758 & 6.9 & 1.38e$-$14 & 9.5 & 0.33 & \ion{Fe}{xvii} 33.8863, \ion{Si}{xi} 33.9070, \ion{S}{xiii} 33.9461, 33.9526 \\
\ion{C}{v} & 34.9728 & 6.1 & 2.57e$-$14 & 12.9 & 0.47 & \ion{Ar}{xiii} 34.8385, \ion{Fe}{xvi} 34.8569, \ion{Ar}{ix} 35.0240 \\
\ion{Ca}{xi} & 35.2130 & 6.4 & 3.40e$-$14 & 14.7 & $-$0.00 &  \\
\ion{Si}{xi} & 35.4374 & 6.3 & 1.33e$-$14 & 9.4 & 0.40 & \ion{S}{xii} 35.3990, \ion{Ca}{xiii} 35.4444 \\
\ion{Ca}{xi} & 35.5760 & 6.4 & 1.07e$-$14 & 8.5 & $-$0.33 &  \\
\ion{Fe}{xvii} & 35.6844 & 6.9 & 3.39e$-$14 & 14.7 & 0.21 & \ion{Fe}{xxi} 35.6238, \ion{Ca}{xi} 35.7370 \\
\ion{Fe}{xviii} & 93.9230 & 7.0 & 1.04e$-$13 & 9.1 & $-$0.11 & \ion{Fe}{xx} 93.7800 \\
\ion{Ne}{viii} & 98.2600 & 5.9 & 1.02e$-$13 & 9.8 & 0.26 & \ion{Fe}{xxi} 97.8800, \ion{Ne}{viii} 98.1150, 98.2740 \\
\ion{Fe}{xix} & 101.5500 & 7.0 & 4.85e$-$14 & 7.3 & 0.11 &  \\
\ion{Fe}{xxi} & 102.2200 & 7.2 & 1.03e$-$13 & 11.0 & 0.20 & \ion{O}{viii} 102.3476, 102.3552, 102.3919, 102.4897, 102.5497 \\
\ion{Ne}{viii} & 103.0850 & 5.9 & 3.27e$-$14 & 5.6 & 0.30 & \ion{Fe}{ix} 103.5662 \\
\ion{Fe}{xviii} & 103.9370 & 7.0 & 2.42e$-$14 & 4.8 & $-$0.22 &  \\
\ion{Fe}{xix} & 106.3300 & 7.0 & 2.38e$-$14 & 3.9 & 0.38 & \ion{Ni}{xxii} 106.0640, \ion{Fe}{xix} 106.1200, \ion{Ne}{vii} 106.1900 \\
\ion{Fe}{xix} & 108.3700 & 7.0 & 9.00e$-$14 & 9.6 & $-$0.06 & \ion{Fe}{xxi} 108.1200 \\
\ion{Fe}{xix} & 109.9700 & 7.0 & 2.91e$-$14 & 4.7 & 0.21 &  \\
\ion{Fe}{xxii} & 114.4100 & 7.2 & 3.06e$-$14 & 4.9 & 0.26 &  \\
\ion{Fe}{xxii} & 117.1700 & 7.2 & 1.27e$-$13 & 12.5 & 0.10 & \ion{Fe}{xxi} 117.5100 \\
\ion{Ni}{xxii} & 117.9330 & 7.2 & 2.32e$-$14 & 3.9 & 0.15 & \ion{Ni}{xxv} 117.9330, \ion{Si}{v} 117.8540, \ion{Cr}{xx} 117.9580 \\
\ion{Fe}{xx} & 118.6600 & 7.1 & 4.72e$-$14 & 6.7 & 0.14 &  \\
\ion{Fe}{xx} & 121.8300 & 7.1 & 5.56e$-$14 & 7.3 & $-$0.06 &  \\
\ion{Fe}{xxi} & 128.7300 & 7.2 & 1.18e$-$13 & 10.8 & 0.11 &  \\
\ion{Fe}{xxiii} & 132.8500 & 7.3 & 3.76e$-$13 & 20.7 & 0.19 & \ion{Fe}{xx} 132.8500 \\
\ion{Fe}{xxii} & 135.7800 & 7.2 & 8.62e$-$14 & 8.7 & 0.09 &  \\
\ion{Fe}{ix} & 171.0730 & 6.0 & 7.18e$-$14 & 4.7 & $-$0.04 & \ion{O}{v} 172.1690 \\
\ion{Fe}{xi} & 182.1690 & 6.2 & 4.85e$-$14 & 3.7 & 0.59 & \ion{C}{vi} 182.0969, 182.1323, \ion{C}{vi} 182.2304, 182.2900, \ion{Fe}{x} 182.3070 \\
\ion{Fe}{xxiv} & 192.0170 & 7.4 & 2.86e$-$13 & 11.0 & $-$0.29 & \ion{Ca}{xvii} 192.8532 \\
\ion{Fe}{xiii} & 197.4330 & 6.3 & 4.94e$-$14 & 3.4 & 0.70 & \ion{Fe}{xiii} 196.7675, 196.8215, \ion{Fe}{ix} 197.3784 \\
\ion{Ar}{xv} & 221.1356 & 6.7 & 8.41e$-$14 & 4.7 & $-$0.26 & \ion{S}{xii} 221.4005 \\
\ion{S}{xiii} & 256.6852 & 6.5 & 1.50e$-$13 & 6.9 & $-$0.38 & \ion{He}{ii} 256.3183, 256.3194, \ion{Si}{x} 256.3660 \\
\ion{S}{x} & 264.2306 & 6.3 & 5.58e$-$14 & 3.7 & 0.03 & \ion{Fe}{xiv} 264.7900 \\
\ion{Si}{x} & 277.2780 & 6.2 & 3.89e$-$14 & 3.1 & $-$0.18 &  \\
\ion{Fe}{xv} & 284.1630 & 6.4 & 1.96e$-$13 & 8.4 & $-$0.00 &  \\
\ion{Si}{viii} & 314.3560 & 6.0 & 5.45e$-$14 & 3.6 & 0.13 &  \\
\ion{Fe}{xvi} & 335.4099 & 6.5 & 1.55e$-$13 & 9.8 & $-$0.23 & \ion{Fe}{xiv} 334.1800 \\
\ion{Ar}{xvi} & 353.8535 & 6.8 & 7.21e$-$14 & 6.4 & $-$0.26 &  \\
\ion{Fe}{xvi} & 360.7580 & 6.5 & 9.12e$-$14 & 7.3 & $-$0.18 & \ion{Ne}{v} 359.3750 \\
\ion{Mg}{ix} & 368.0577 & 6.1 & 3.78e$-$14 & 4.1 & 0.38 &  \\
\ion{S}{xiv} & 417.6655 & 6.6 & 6.55e$-$14 & 4.0 & $-$0.42 & \ion{Fe}{xv} 417.2580 \\
\ion{Ne}{vi} & 558.6850 & 5.7 & 1.48e$-$13 & 4.1 & $-$0.25 & \ion{O}{iv} 555.2630, \ion{Ca}{x} 557.7640, \ion{Ne}{v} 568.4220 \\
\ion{S}{vi} & 933.3788 & 5.4 & 3.57e$-$15 & 3.3 & $-$0.21 &  \\
\ion{N}{iii} & 991.5770 & 5.0 & 7.91e$-$15 & 4.8 & $-$0.41 &  \\
\ion{Ne}{vi} & 999.1830 & 5.7 & 4.52e$-$15 & 4.0 & 0.45 &  \\
\ion{Ne}{vi} & 1010.2050 & 5.7 & 1.90e$-$15 & 4.2 & 0.48 &  \\
\ion{S}{iii} & 1021.3230 & 4.9 & 2.05e$-$15 & 5.5 & 0.31 &  \\
\ion{O}{vi} & 1031.9121 & 5.6 & 2.29e$-$13 & 16.0 & 0.15 &  \\
\ion{C}{ii} & 1036.3390 & 4.8 & 8.63e$-$15 & 6.2 & $-$0.28 &  \\
\ion{C}{ii} & 1037.0200 & 4.8 & 1.71e$-$14 & 7.2 & $-$0.29 &  \\
\ion{O}{vi} & 1037.6136 & 5.6 & 1.15e$-$13 & 13.7 & 0.15 &  \\
\ion{S}{iv} & 1062.6639 & 5.1 & 2.04e$-$15 & 7.0 & $-$0.73 &  \\
\ion{S}{iv} & 1072.9740 & 5.1 & 5.14e$-$15 & 8.4 & $-$0.14 &  \\
\ion{S}{iii} & 1077.1730 & 4.9 & 1.99e$-$15 & 8.7 & 0.16 &  \\
\ion{N}{ii} & 1085.7030 & 4.7 & 1.16e$-$14 & 3.4 & 0.39 & \ion{N}{ii} 1085.5310, 1085.5480 \\
\ion{Si}{iii} & 1108.3610 & 4.9 & 2.09e$-$15 & 6.4 & $-$0.08 &  \\
\ion{Si}{iii} & 1109.9720 & 4.9 & 5.92e$-$15 & 6.3 & 0.19 &  \\
\ion{Si}{iii} & 1113.2320 & 4.9 & 5.90e$-$15 & 7.3 & $-$0.19 & \ion{Si}{iii} 1113.2061 \\
\ion{Fe}{xix} & 1118.0699 & 7.0 & 5.73e$-$15 & 6.4 & $-$0.19 &  \\
\ion{Si}{iv} & 1122.4852 & 5.0 & 5.01e$-$15 & 7.1 & 0.39 &  \\
\ion{Si}{iv} & 1128.3404 & 5.0 & 4.55e$-$15 & 6.7 & 0.05 & \ion{Si}{iv} 1128.3252 \\
\ion{Ne}{v} & 1145.5959 & 5.5 & 2.01e$-$15 & 6.3 & $-$0.37 &  \\
\ion{S}{iii} & 1190.1990 & 4.9 & 2.80e$-$15 & 8.1 & 0.13 &  \\
\ion{Si}{ii} & 1190.4170 & 4.6 & 1.53e$-$15 & 4.8 & $-$0.07 &  \\
\ion{S}{iii} & 1194.0490 & 4.9 & 4.81e$-$15 & 11.8 & 0.45 &  \\
\hline
\end{tabular}
\end{scriptsize}
\end{table*}
\setcounter{table}{2}
\begin{table*}
\caption{(continued) {\em XMM-Newton}/RGS, EUVE, FUSE, and HST/STIS line fluxes of AU Mic\tablefootmark{a}.}
\begin{scriptsize}
  \begin{tabular}{lrccrcl}
\hline \hline
Ion & $\lambda_{\rm model}$ (\AA) & $\log T_{\rm max}$ & $F_{\rm obs}$ & $S/N$ & Ratio & Blends \\
\hline
\ion{Si}{ii} & 1194.5010 & 4.6 & 4.36e$-$15 & 13.7 & $-$0.36 & \ion{S}{iii} 1194.4430 \\
\ion{S}{iii} & 1200.9610 & 4.9 & 8.55e$-$15 & 23.6 & 0.74 &  \\
\ion{Si}{iii} & 1206.5019 & 4.9 & 1.24e$-$13 & 35.7 & $-$0.68 &  \\
\ion{O}{v} & 1218.3440 & 5.5 & 2.26e$-$14 & 77.0 & $-$0.22 &  \\
\ion{N}{v} & 1238.8218 & 5.4 & 6.78e$-$14 & 37.0 & 0.12 &  \\
\ion{N}{v} & 1242.8042 & 5.4 & 3.30e$-$14 & 23.2 & 0.11 &  \\
\ion{C}{iii} & 1247.3830 & 5.1 & 6.14e$-$15 & 11.5 & 0.40 &  \\
\ion{S}{ii} & 1259.5210 & 4.6 & 3.03e$-$15 & 12.6 & 0.04 &  \\
\ion{Si}{ii} & 1264.7400 & 4.5 & 6.67e$-$15 & 13.4 & $-$0.13 &  \\
\ion{Si}{ii} & 1265.0040 & 4.6 & 3.80e$-$15 & 15.9 & $-$0.20 &  \\
\ion{Si}{iii} & 1296.7280 & 4.9 & 4.88e$-$15 & 17.2 & 0.72 &  \\
\ion{Si}{iii} & 1303.3249 & 4.9 & 4.96e$-$15 & 19.7 & 0.64 &  \\
\ion{Si}{ii} & 1304.3719 & 4.6 & 1.91e$-$15 & 9.3 & $-$0.22 &  \\
\ion{Si}{ii} & 1309.2770 & 4.6 & 4.61e$-$15 & 15.7 & $-$0.14 &  \\
\ion{Si}{iii} & 1312.5930 & 4.9 & 1.80e$-$15 & 9.5 & $-$0.42 &  \\
\ion{C}{ii} & 1323.9080 & 4.8 & 2.88e$-$15 & 8.7 & 0.39 &  \\
\ion{Fe}{xix} & 1328.6999 & 7.1 & 9.39e$-$16 & 6.3 & $-$0.51 &  \\
\ion{C}{ii} & 1334.5350 & 4.7 & 9.08e$-$14 & 56.4 & $-$0.25 &  \\
\ion{C}{ii} & 1335.7100 & 4.7 & 1.96e$-$13 & 59.7 & 0.44 &  \\
\ion{O}{iv} & 1338.6140 & 5.3 & 1.16e$-$15 & 9.0 & 0.49 &  \\
\ion{Fe}{xxi} & 1354.0800 & 7.2 & 7.92e$-$15 & 13.6 & $-$0.27 &  \\
\ion{O}{v} & 1371.2960 & 5.5 & 4.95e$-$15 & 18.6 & $-$0.27 &  \\
\ion{Si}{iv} & 1393.7552 & 5.0 & 1.26e$-$13 & 33.2 & 0.13 &  \\
\ion{Si}{iv} & 1402.7704 & 5.0 & 7.09e$-$14 & 25.3 & 0.18 &  \\
\ion{S}{iv} & 1406.0160 & 5.1 & 9.56e$-$16 & 16.3 & $-$0.59 &  \\
\ion{Cr}{xvi} & 1410.5990 & 6.8 & 4.83e$-$16 & 14.2 & 0.06 &  \\
\ion{S}{v} & 1501.7660 & 5.3 & 1.26e$-$15 & 21.3 & $-$0.04 &  \\
\ion{Si}{ii} & 1526.7090 & 4.5 & 8.62e$-$15 & 19.6 & 0.03 &  \\
\ion{Si}{ii} & 1533.4320 & 4.5 & 1.03e$-$14 & 9.7 & $-$0.19 &  \\
\ion{C}{iv} & 1548.1871 & 5.1 & 4.30e$-$13 & 49.9 & 0.04 &  \\
\ion{C}{iv} & 1550.7723 & 5.1 & 2.24e$-$13 & 35.3 & 0.06 &  \\
\ion{Al}{ii} & 1670.7870 & 4.6 & 2.35e$-$14 & 27.1 & 0.02 &  \\
\hline
\end{tabular}

\tablefoot{\tablefoottext{a}{Line fluxes (in erg cm$^{-2}$ s$^{-1}$ units) measured in the spectra, and corrected by the ISM absorption when needed.
  log $T_{\rm max}$ (K) indicates the maximum  temperature of formation of the line (unweighted by the
  EMD). ``Ratio'' is the log($F_{\mathrm {obs}}$/$F_{\mathrm {pred}}$)   of the line. 
  ``Blends'' amounting to more than 5\% of the total flux for each line are indicated, with wavelengths in \AA.}}
\end{scriptsize}
\end{table*}

%
\begin{figure*}
  \centering
  \includegraphics[width=0.33\textwidth]{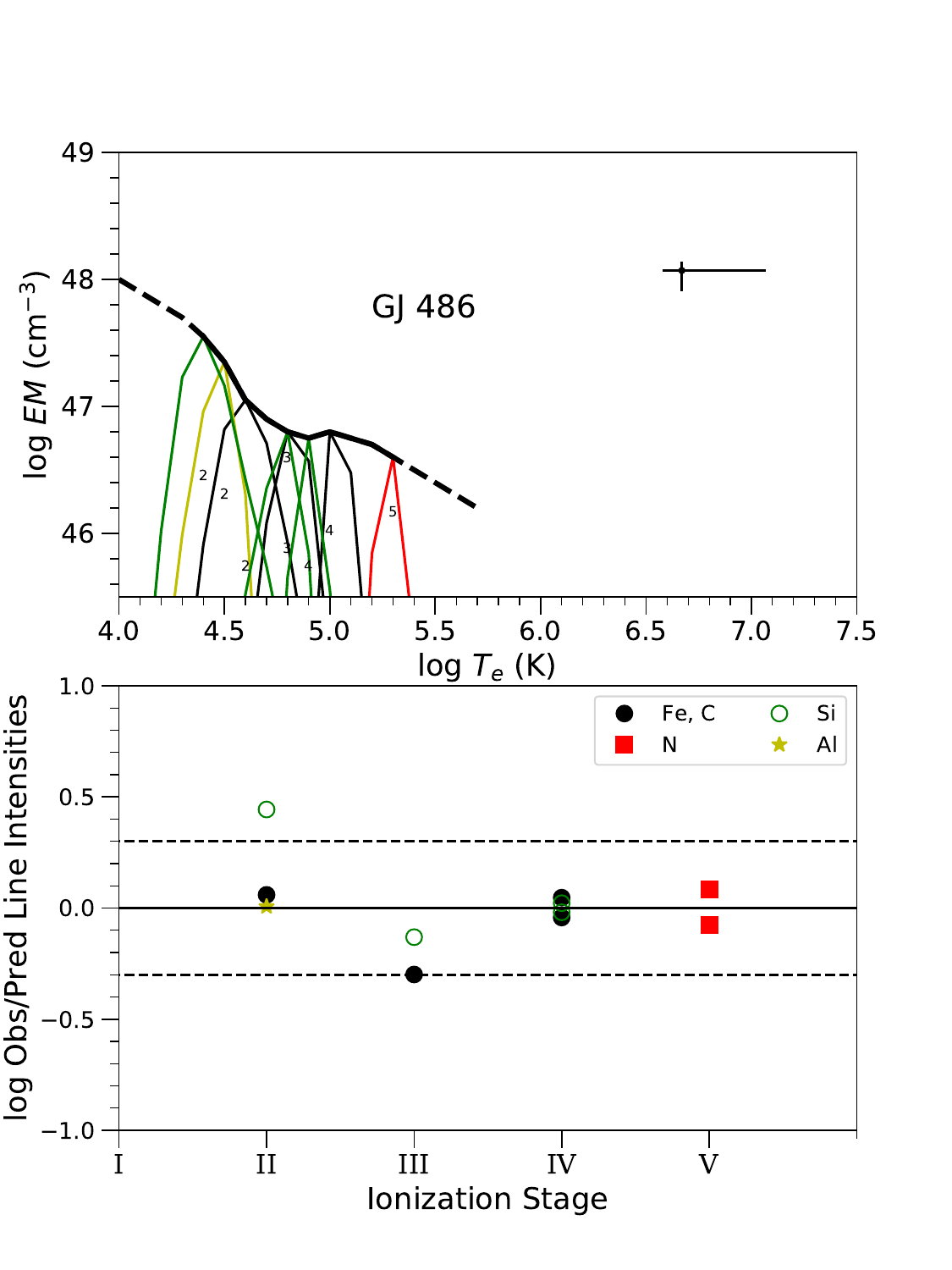}
  \includegraphics[width=0.33\textwidth]{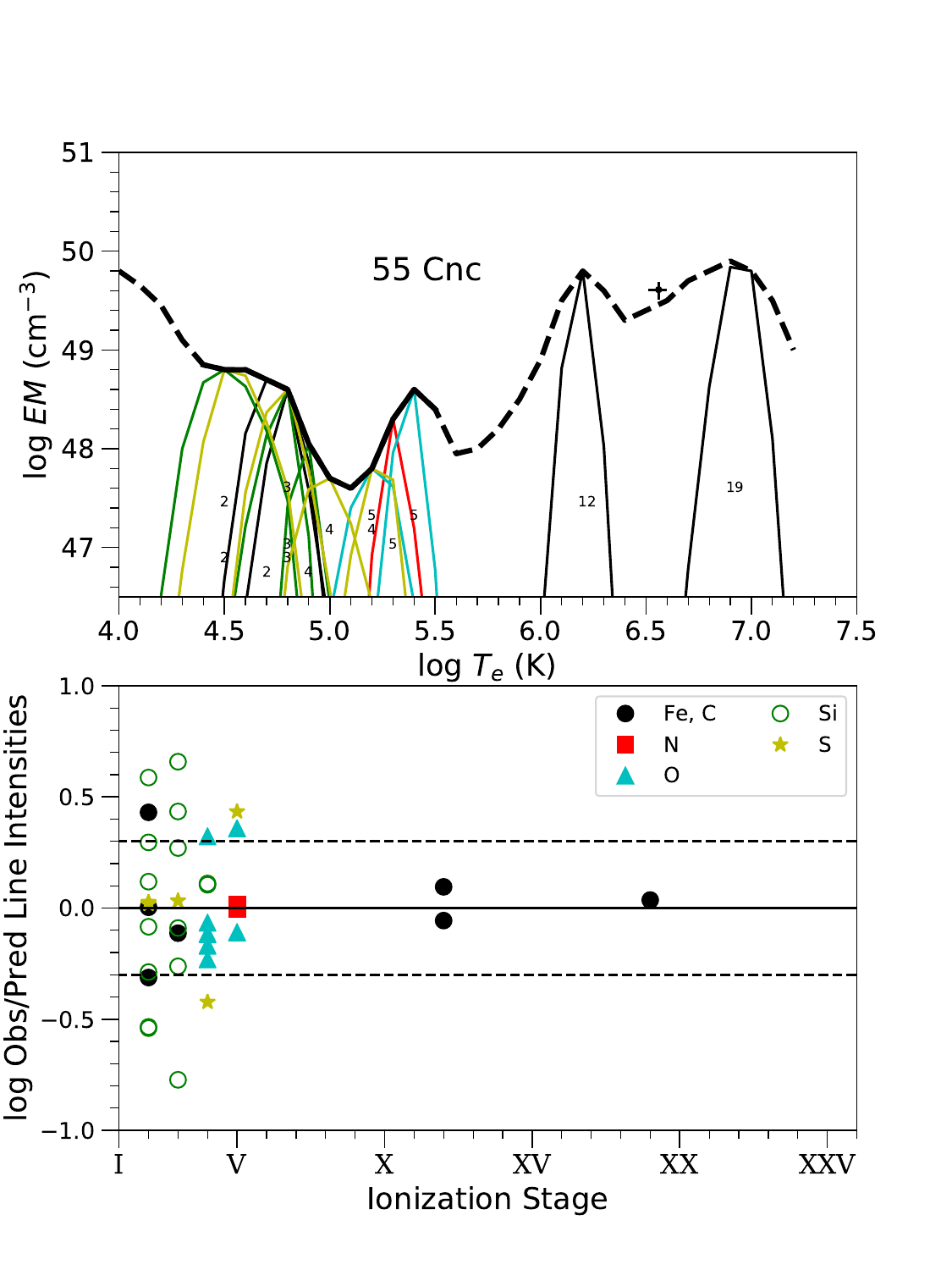}
  \includegraphics[width=0.33\textwidth]{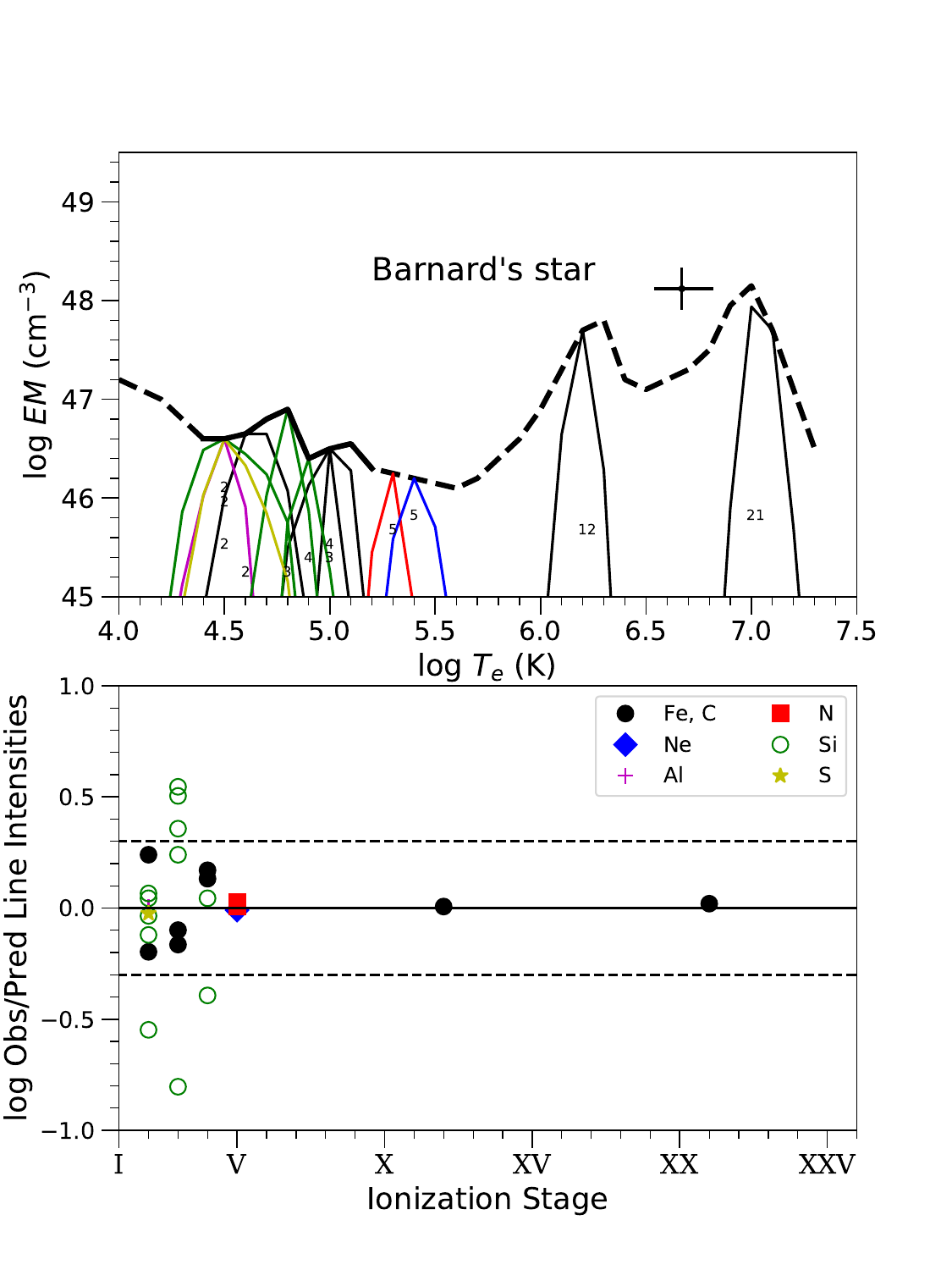}
  \caption{Same as in Fig.~\ref{fig:proxcen} but for GJ~486, 55~Cnc, and Barnard's star (GJ~699), using HST/STIS and COS data. The 1-$T$ global fit to XMM-Newton/EPIC (GJ~486 and 55~Cnc) and Chandra/ACIS-S (Barnard's star) spectra are also indicated with error bars. In 55~Cnc ({\it center}) and Barnard's star ({\it right}), coronal lines in the HST/COS spectra were used to tentatively extend the EMD to coronal temperatures.}\label{fig:gj486} 
\end{figure*}
%

%
\begin{table*}
\caption[]{{\em XMM-Newton}/RGS and HST/COS line fluxes of $\tau$ Boo\tablefootmark{a}.}\label{tab:fluxes5} 
\tabcolsep 3 pt
\begin{scriptsize}
  \begin{tabular}{lrccrcl}
\hline \hline
Ion & $\lambda_{\rm model}$ (\AA) & $\log T_{\rm max}$ & $F_{\rm obs}$ & $S/N$ & Ratio & Blends \\
\hline
\ion{Fe}{xvii} & 11.2540 & 6.9 & 2.68e$-$14 & 7.2 & 0.38 & \ion{Fe}{xviii} 11.3260, 11.3260, 11.3260 \\
\ion{Ne}{x} & 12.1321 & 6.9 & 4.70e$-$14 & 4.1 & $-$0.03 & \ion{Fe}{xvii} 12.1240, \ion{Ne}{x} 12.1375 \\
\ion{Fe}{xvii} & 12.2660 & 6.9 & 1.50e$-$14 & 3.0 & $-$0.10 &  \\
\ion{Ne}{ix} & 13.4473 & 6.7 & 5.02e$-$14 & 7.7 & 0.04 & \ion{Fe}{xix} 13.5180, \ion{Ne}{ix} 13.5531 \\
\ion{Ne}{ix} & 13.6990 & 6.7 & 2.10e$-$14 & 7.3 & 0.00 &  \\
\ion{Fe}{xvii} & 13.8250 & 6.9 & 1.97e$-$14 & 5.0 & $-$0.14 & \ion{Ni}{xix} 13.7790, \ion{Fe}{xvii} 13.8920 \\
\ion{Ni}{xix} & 14.0430 & 7.0 & 2.53e$-$14 & 6.9 & $-$0.01 & \ion{Fe}{xviii} 13.9530, \ion{Ni}{xix} 14.0770 \\
\ion{Fe}{xviii} & 14.2080 & 7.0 & 2.70e$-$14 & 7.9 & $-$0.06 & \ion{Fe}{xviii} 14.2560 \\
\ion{Fe}{xviii} & 14.3730 & 7.0 & 1.15e$-$14 & 4.6 & $-$0.20 & \ion{Fe}{xviii} 14.3430, 14.3430, 14.3990, 14.4250, 14.4250, 14.4555 \\
\ion{Fe}{xviii} & 14.5340 & 7.0 & 5.76e$-$15 & 3.7 & $-$0.26 & \ion{O}{viii} 14.5242, \ion{Fe}{xviii} 14.5608, 14.5710, 14.6160 \\
\ion{Fe}{xvi} & 14.9555 & 6.7 & 2.28e$-$15 & 3.5 & $-$0.63 & \ion{Fe}{xix} 14.9610 \\
\ion{Fe}{xvii} & 15.0140 & 6.9 & 1.48e$-$13 & 28.0 & $-$0.05 & \ion{Fe}{xvi} 15.0496 \\
\ion{O}{viii} & 15.1760 & 6.6 & 2.77e$-$14 & 6.6 & 0.39 & \ion{O}{viii} 15.1765, \ion{Fe}{xix} 15.1980 \\
\ion{Fe}{xvii} & 15.2610 & 6.9 & 5.48e$-$14 & 13.5 & 0.05 &  \\
\ion{Fe}{xvii} & 15.4530 & 6.8 & 3.64e$-$14 & 7.9 & 0.26 & \ion{Fe}{xvi} 15.3907, 15.4485, 15.4533, 15.4955, 15.5023 \\
\ion{Fe}{xviii} & 15.6250 & 7.0 & 1.26e$-$14 & 5.5 & 0.18 &  \\
\ion{Fe}{xviii} & 15.8240 & 7.0 & 8.63e$-$15 & 6.9 & 0.01 & \ion{Fe}{xviii} 15.8700 \\
\ion{O}{viii} & 16.0055 & 6.6 & 3.49e$-$14 & 6.6 & $-$0.07 & \ion{Fe}{xvii} 15.9956, \ion{Fe}{xviii} 16.0040, \ion{O}{viii} 16.0067 \\
\ion{Fe}{xviii} & 16.0710 & 7.0 & 1.96e$-$14 & 5.4 & $-$0.00 & \ion{Fe}{xviii} 16.0450, 16.1590, \ion{Fe}{xix} 16.1100 \\
\ion{Fe}{xvii} & 16.7800 & 6.8 & 9.26e$-$14 & 15.3 & $-$0.03 &  \\
\ion{Fe}{xvii} & 17.0510 & 6.8 & 2.30e$-$13 & 16.6 & $-$0.02 & \ion{Fe}{xvii} 17.0960 \\
\ion{O}{vii} & 18.6270 & 6.4 & 7.02e$-$15 & 3.3 & $-$0.06 &  \\
\ion{O}{viii} & 18.9671 & 6.6 & 1.32e$-$13 & 19.3 & $-$0.14 & \ion{O}{viii} 18.9725 \\
\ion{O}{vii} & 21.6015 & 6.4 & 4.50e$-$14 & 6.6 & $-$0.03 &  \\
\ion{O}{vii} & 22.0977 & 6.4 & 3.71e$-$14 & 5.8 & 0.07 &  \\
\ion{N}{vii} & 24.7792 & 6.4 & 1.28e$-$14 & 4.4 & $-$0.09 & \ion{N}{vii} 24.7846 \\
\ion{C}{vi} & 28.4652 & 6.3 & 7.02e$-$15 & 4.0 & 0.08 & \ion{Ar}{xv} 28.3464, \ion{C}{vi} 28.4663 \\
\ion{C}{vi} & 33.7342 & 6.2 & 2.52e$-$14 & 6.5 & $-$0.11 & \ion{S}{xiv} 33.5490, \ion{C}{vi} 33.7396 \\
\ion{S}{v} & 1199.1360 & 5.3 & 1.68e$-$15 & 12.3 & 0.01 &  \\
\ion{S}{iii} & 1200.9611 & 4.9 & 2.50e$-$15 & 14.1 & 0.02 &  \\
\ion{Si}{iii} & 1206.5019 & 4.9 & 1.20e$-$13 & 43.5 & $-$0.01 &  \\
\ion{O}{v} & 1218.3440 & 5.5 & 1.10e$-$14 & 29.4 & $-$0.18 &  \\
\ion{N}{v} & 1238.8218 & 5.4 & 2.02e$-$14 & 30.2 & 0.05 &  \\
\ion{N}{v} & 1242.8042 & 5.4 & 1.00e$-$14 & 23.0 & 0.04 &  \\
\hline
\end{tabular}
  \tablefoot{\tablefoottext{a}{Line fluxes (in erg cm$^{-2}$ s$^{-1}$) units measured in the spectra, and corrected by the ISM absorption when needed.
    {\it XMM-Newton}/RGS $\tau$~Boo data from \citet{mag11}.  
  $\log T_{\rm max}$ (K) indicates the maximum  temperature of formation of the line (unweighted by the
  EMD). ``Ratio'' is the $\log$ ($F_{\mathrm {obs}}$/$F_{\mathrm {pred}}$)   of the line. 
  ``Blends'' amounting to more than 5\% of the total flux for each line are indicated, with wavelengths in \AA.}}
\end{scriptsize}
\end{table*}

%
\begin{figure*}
  \centering
  \includegraphics[width=0.33\textwidth]{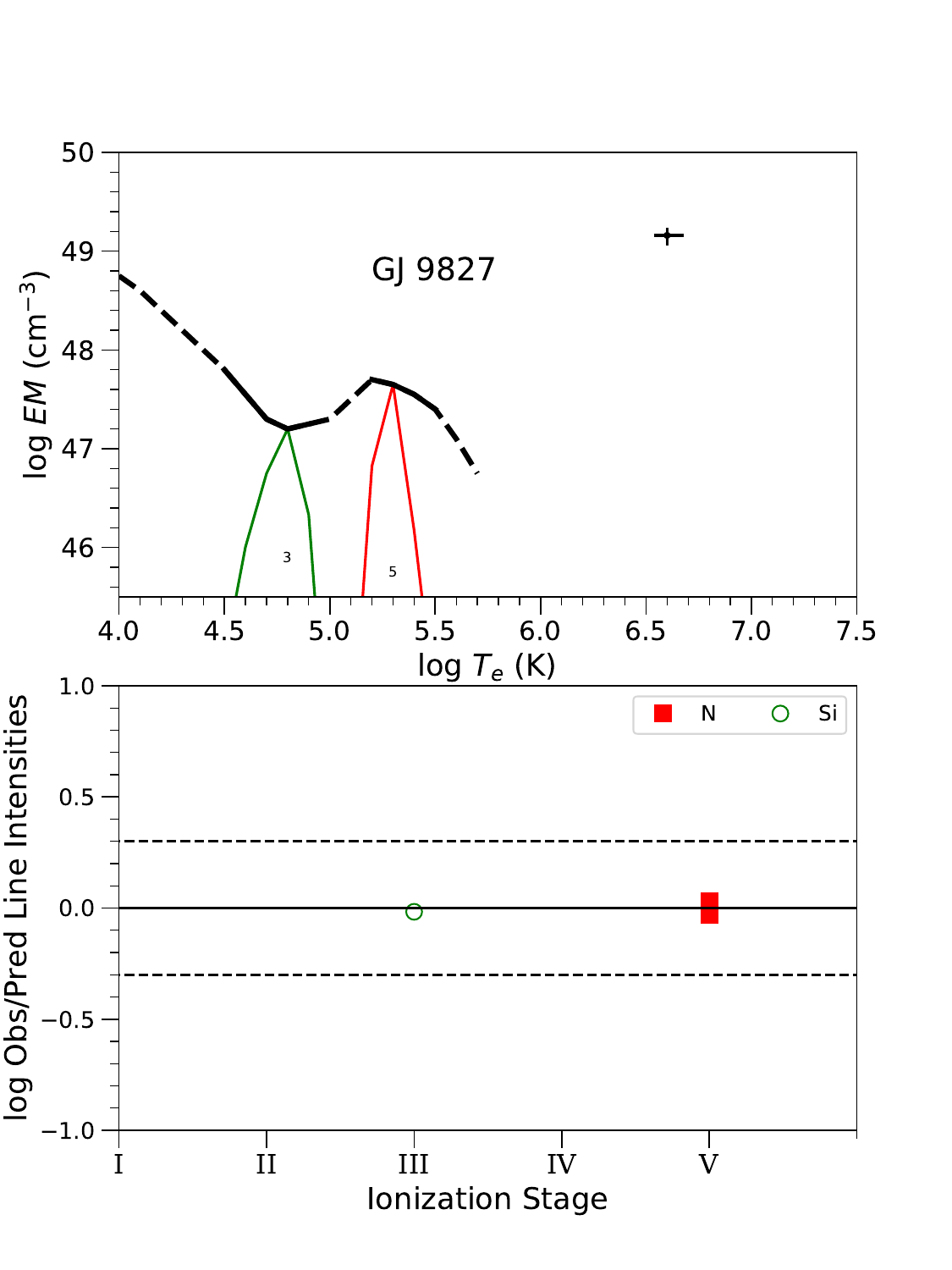}
  \includegraphics[width=0.33\textwidth]{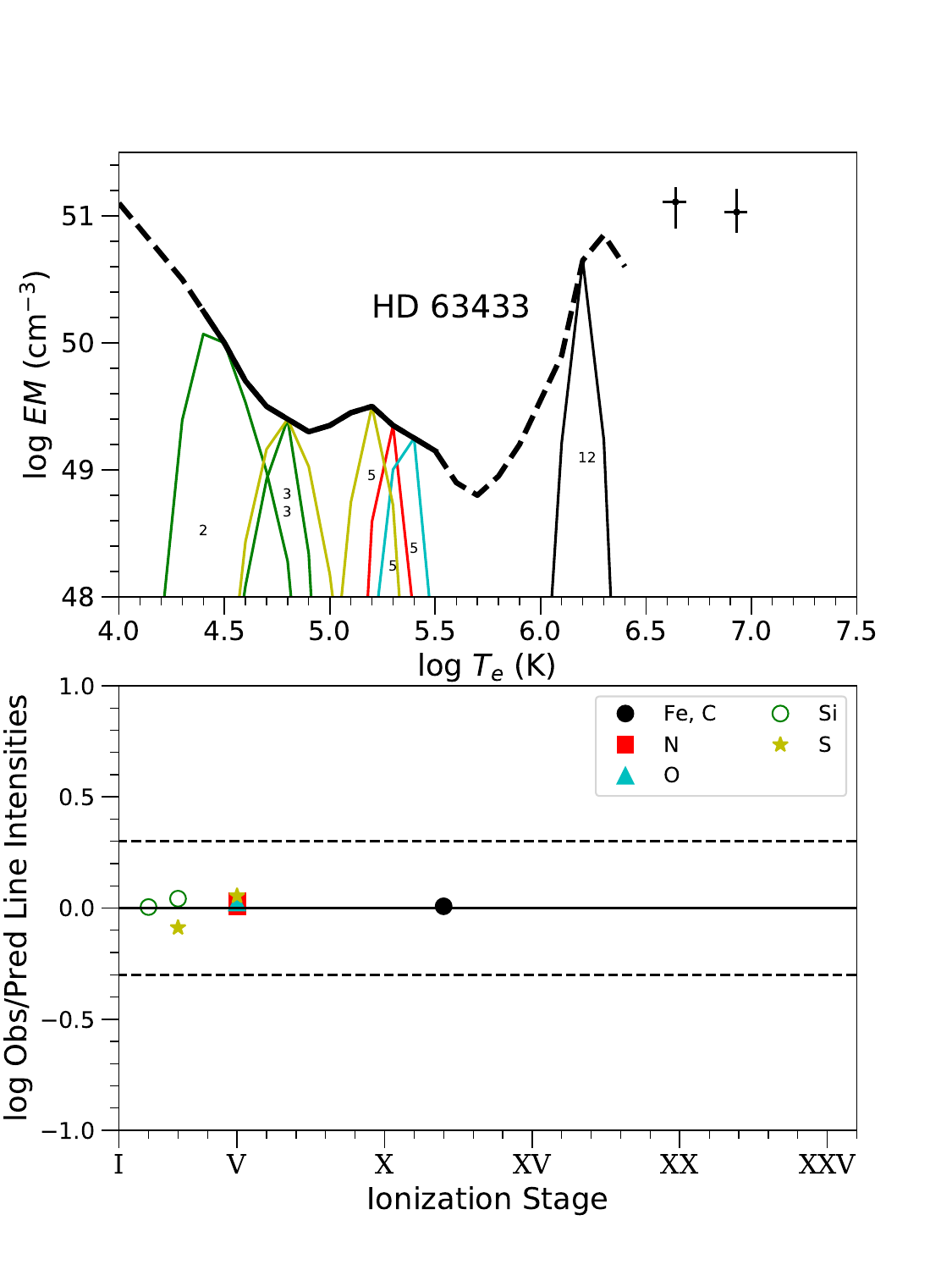}
  \includegraphics[width=0.33\textwidth]{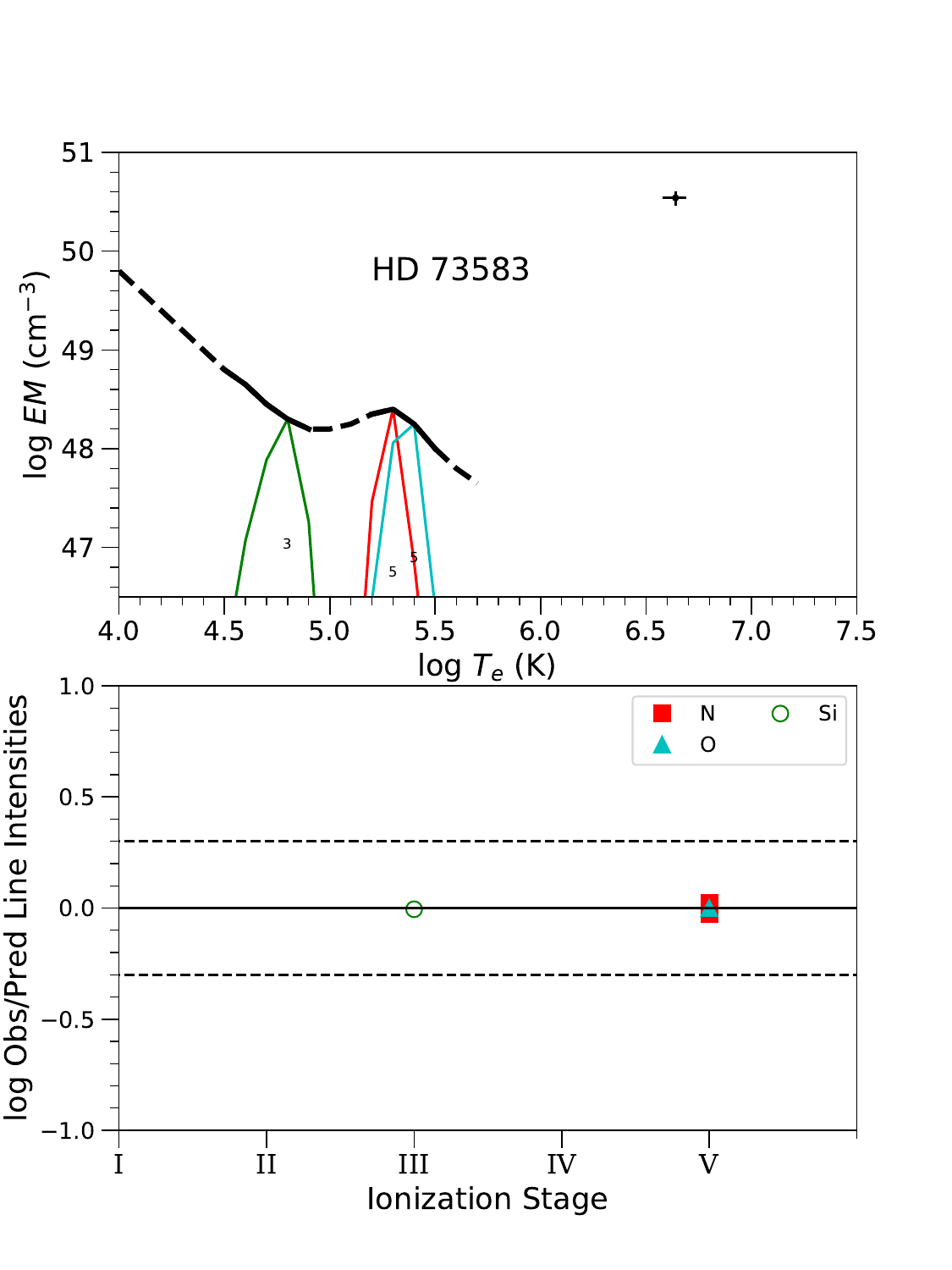}
  \caption{Same as in Fig.~\ref{fig:proxcen} but for GJ~9827, HD~63443 (TOI-1726), and HD~73583 (TOI-560), using HST/STIS data. The 1-$T$ and 2-$T$ global fits to XMM-Newton/EPIC spectra are also indicated with error bars.}\label{fig:gj9827} 
\end{figure*}
%

%
\begin{figure*}
  \centering
  \includegraphics[width=0.33\textwidth]{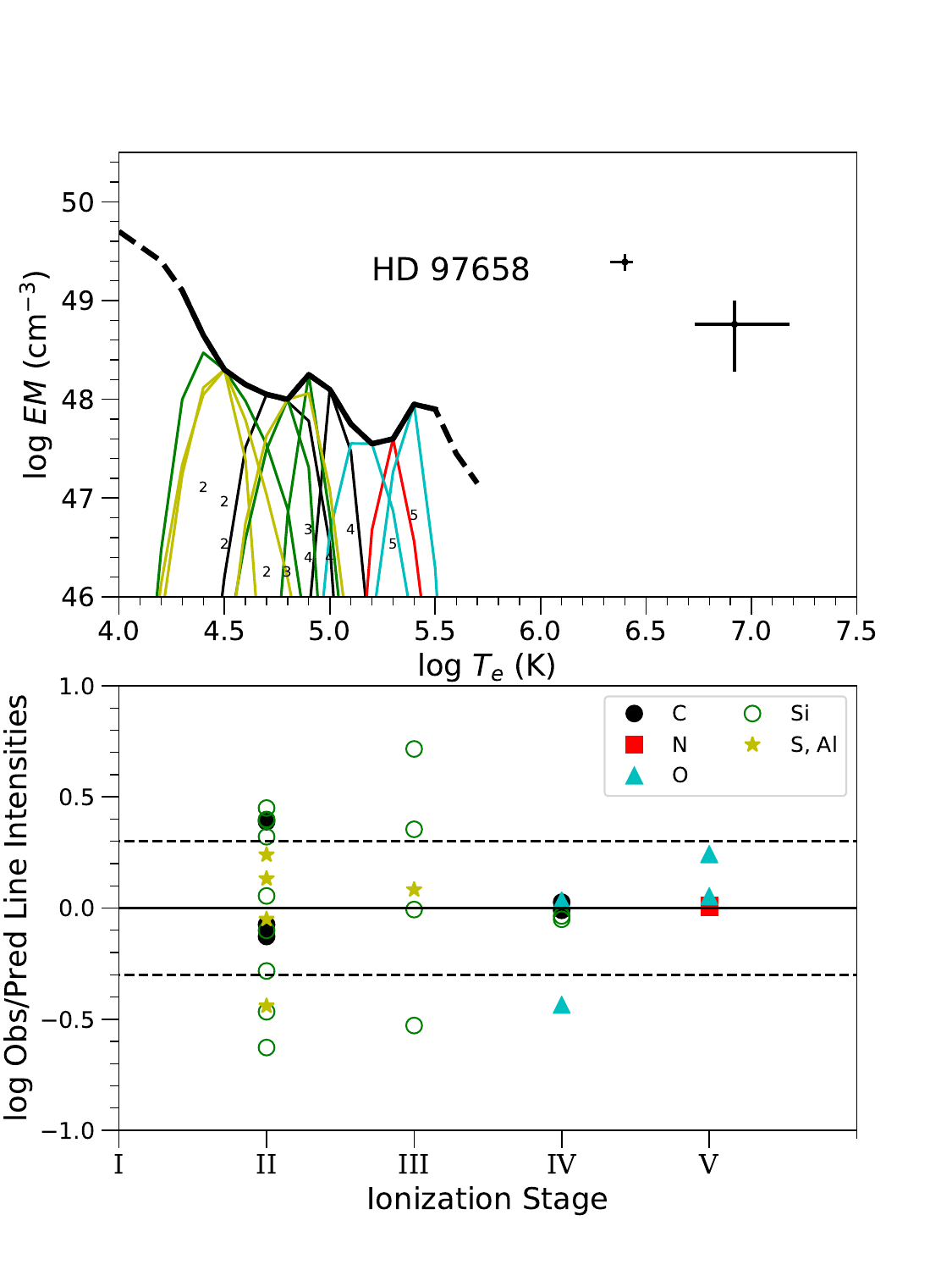}
  \includegraphics[width=0.33\textwidth]{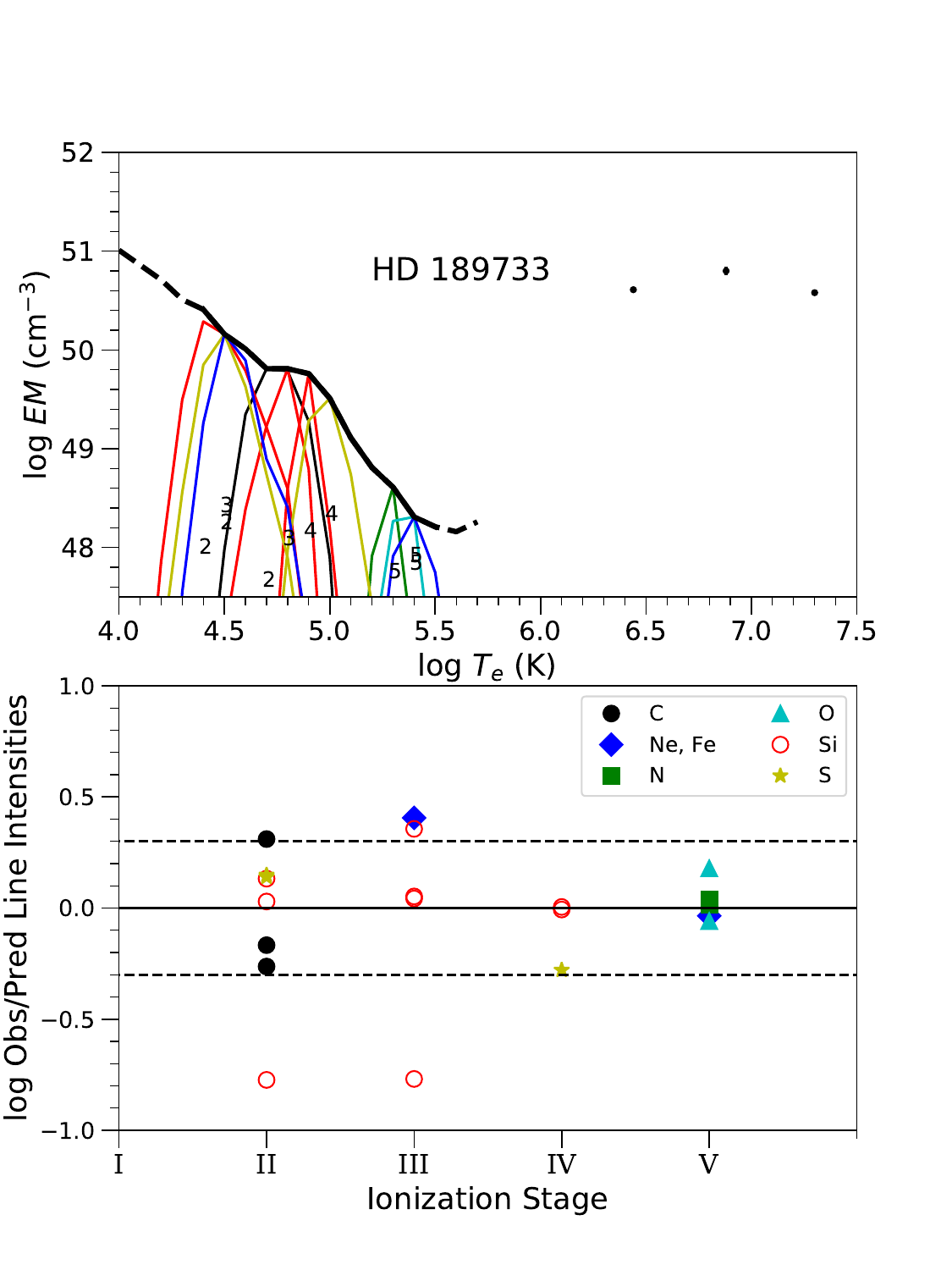}
  \includegraphics[width=0.33\textwidth]{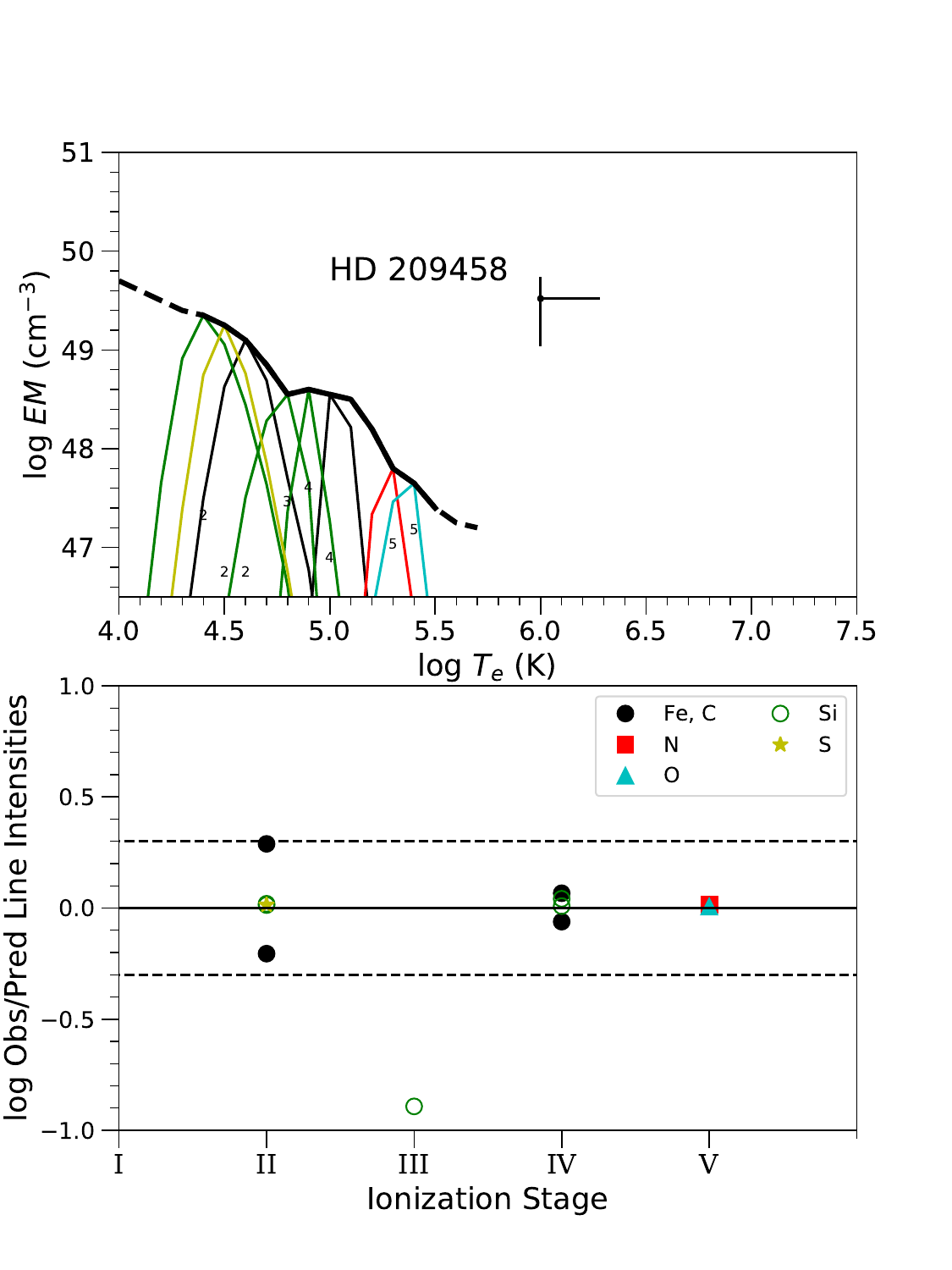}
  \caption{Same as in Fig.~\ref{fig:proxcen} but for HD 97658, HD~189733 and HD~209458, using HST/COS data. HD~209458 COS data taken from \citet{fra10}. The 1 to 3-$T$ global fits to XMM-Newton/EPIC spectra are also indicated with error bars.}\label{fig:hd97658} 
\end{figure*}
%

%
\begin{figure*}
  \centering
  \includegraphics[width=0.33\textwidth]{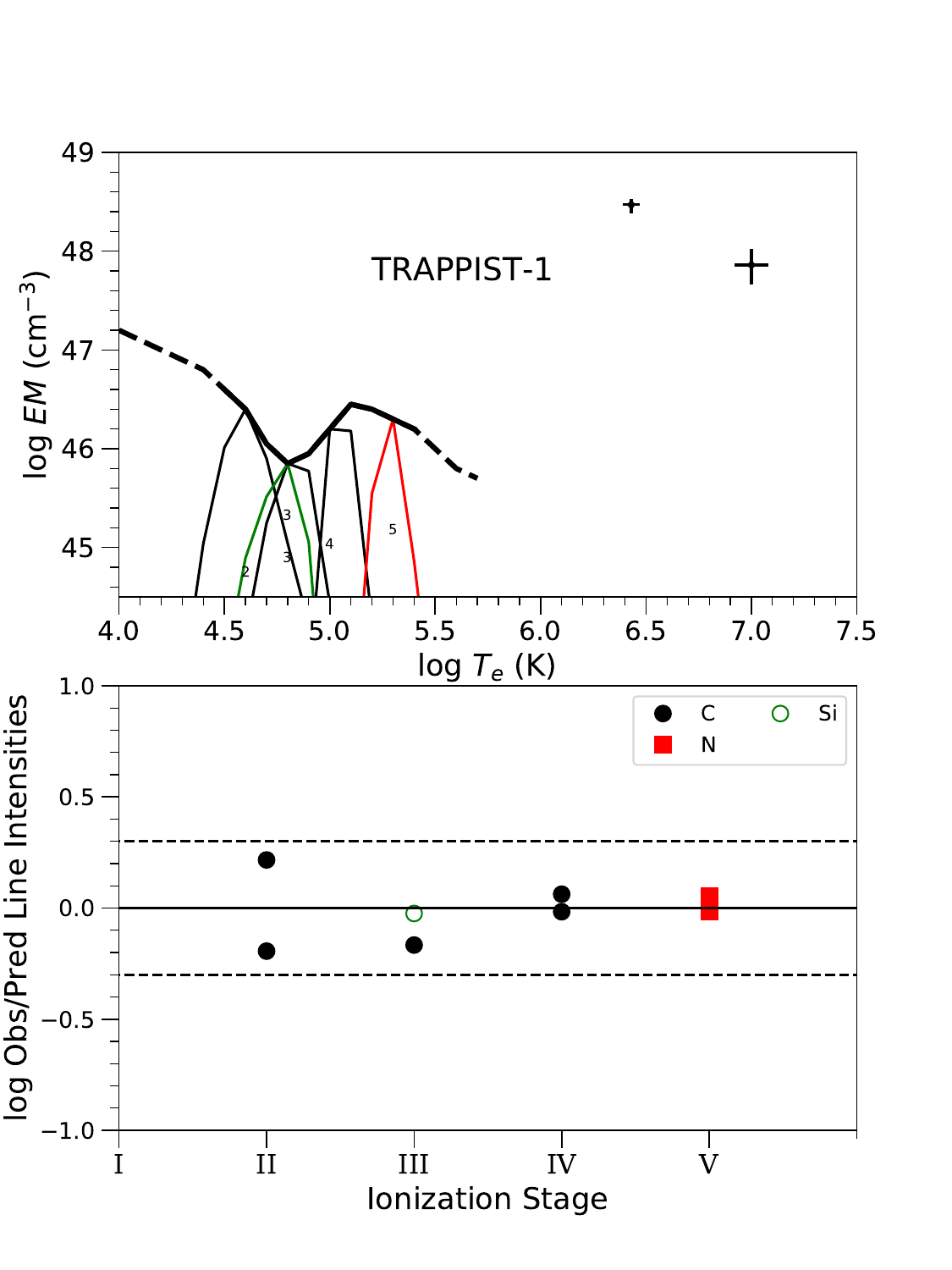}
  \includegraphics[width=0.33\textwidth]{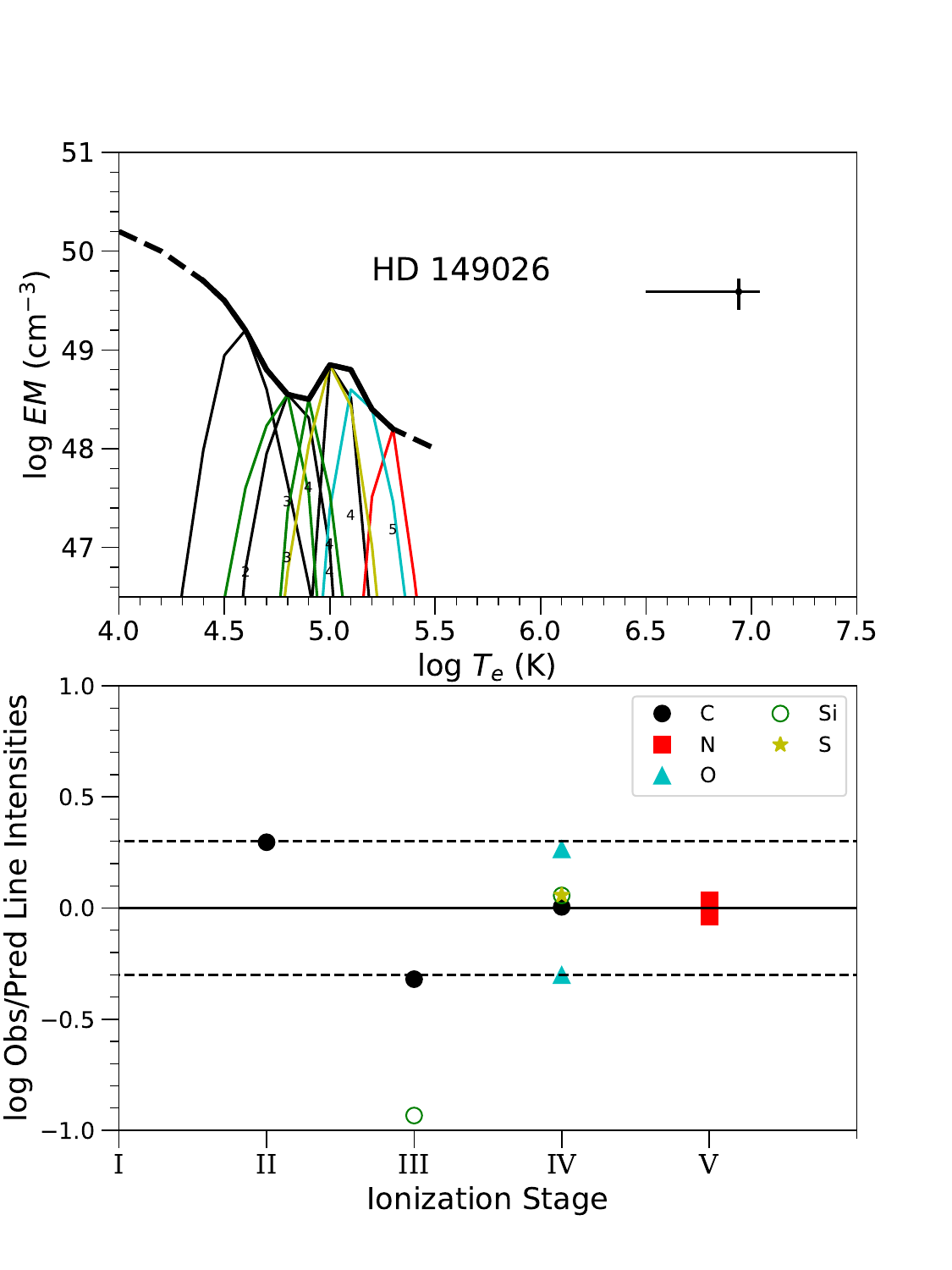}
  \includegraphics[width=0.33\textwidth]{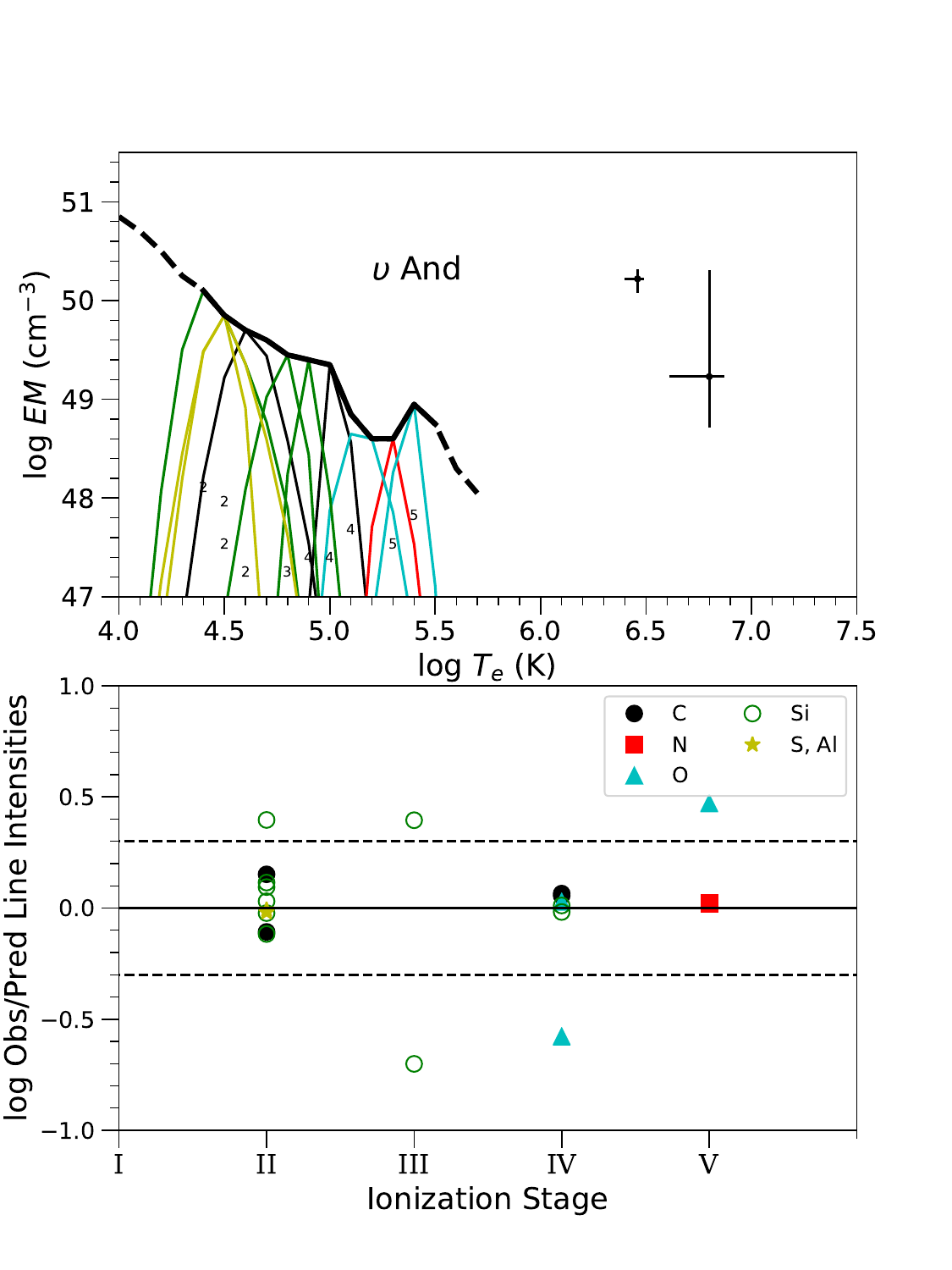}
  \caption{Emission Measure Distributions as in Fig.~\ref{fig:proxcen}, using HST/COS (TRAPPIST-1) and HST/STIS (HD~149026 and $\upsilon$~And) data. The 1-$T$ and 2-$T$ global fits to XMM-Newton/EPIC spectra are also indicated with error bars.}\label{fig:upsand} 
\end{figure*}
%

%
\begin{figure}
  \centering
  \includegraphics[width=0.33\textwidth]{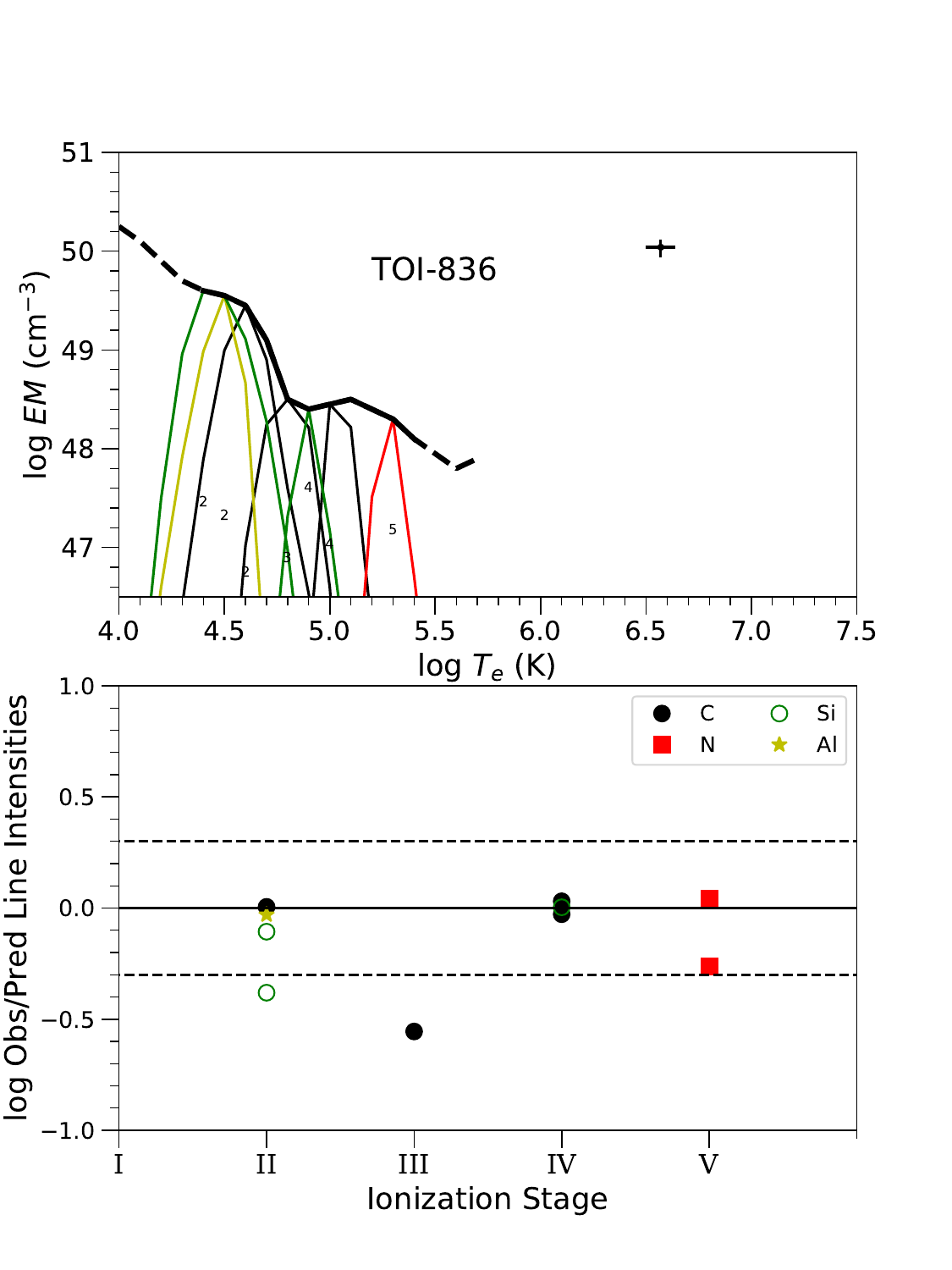}
  \caption{Same as in Fig.~\ref{fig:proxcen} but for TOI-836, using HST/STIS data. The 1-$T$ global fit to XMM-Newton/EPIC spectra is also indicated with error bars.}\label{fig:toi836} 
\end{figure}
%

%
\begin{table*}
  \caption[]{{\em Chandra}/HETG and LETG, {\em EUVE}, and
    {\em HST}/STIS line fluxes of AD Leo\tablefootmark{a}.}\label{tab:fluxes3} 
\tabcolsep 3 pt
\begin{scriptsize}
  \begin{tabular}{lrccrcl}
\hline \hline
Ion & $\lambda_{\rm model}$ (\AA) & $\log T_{\rm max}$ & $F_{\rm obs}$ & $S/N$ & Ratio & Blends \\
\hline
\ion{S}{xv} & 5.0387 & 7.3 & 2.51e$-$14 & 3.1 & $-$0.46 &  \\
\ion{Si}{xiv} & 6.1804 & 7.3 & 6.55e$-$14 & 9.3 & $-$0.29 & \ion{Si}{xiv}  6.1858 \\
\ion{Si}{xiii} & 6.6479 & 7.1 & 9.67e$-$14 & 12.2 & $-$0.39 &  \\
\ion{Si}{xii} & 6.6638 & 7.0 & 1.06e$-$14 & 4.1 & $-$0.01 & \ion{Si}{xii}  6.6627 \\
\ion{Si}{xiii} & 6.6882 & 7.1 & 2.19e$-$14 & 5.8 & $-$0.35 & \ion{Si}{xiii}  6.6850 \\
\ion{Si}{xiii} & 6.7403 & 7.1 & 6.40e$-$14 & 10.7 & $-$0.33 & \ion{Si}{xii}  6.7432 \\
\ion{Mg}{xi} & 7.4730 & 6.9 & 5.61e$-$15 & 3.3 & 0.12 & \ion{Fe}{xxiii}  7.4780 \\
\ion{Al}{xii} & 7.7573 & 7.0 & 4.98e$-$15 & 3.2 & $-$0.16 &  \\
\ion{Mg}{xi} & 7.8503 & 6.9 & 5.57e$-$15 & 3.4 & $-$0.20 &  \\
\ion{Fe}{xxiv} & 7.9857 & 7.4 & 7.43e$-$15 & 4.1 & 0.37 & \ion{Fe}{xxiii}  7.9360, \ion{Fe}{xxiv}  7.9960 \\
\ion{Fe}{xxiv} & 8.3161 & 7.4 & 4.90e$-$15 & 3.3 & $-$0.06 & \ion{Fe}{xxi}  8.3162, \ion{Fe}{xxiii}  8.3038 \\
\ion{Mg}{xii} & 8.4192 & 7.1 & 4.70e$-$14 & 10.4 & $-$0.18 & \ion{Mg}{xii}  8.4246 \\
No id. & 8.4577 & 0.0 & 4.31e$-$15 & 3.1 & \ldots &  \\
\ion{Fe}{xxii} & 8.7140 & 7.2 & 6.13e$-$15 & 3.7 & 0.11 & \ion{Fe}{xxii}  8.7035, 8.7360, \ion{Ni}{xxii}  8.7227, \ion{Ni}{xx}  8.7286, \ion{Ni}{xix}  8.7440 \\
\ion{Fe}{xxii} & 8.9748 & 7.2 & 7.35e$-$15 & 3.9 & 0.24 &  \\
\ion{Fe}{xx} & 9.0659 & 7.1 & 6.48e$-$15 & 3.3 & 0.16 & \ion{Fe}{xxii}  9.0501, 9.0629, \ion{Ni}{xxvi}  9.0603, \ion{Fe}{xx}  9.0647, 9.0683 \\
\ion{Mg}{xi} & 9.1687 & 6.9 & 4.75e$-$14 & 8.0 & $-$0.09 &  \\
\ion{Mg}{xi} & 9.2312 & 6.9 & 1.11e$-$14 & 3.8 & $-$0.07 & \ion{Mg}{xi}  9.2282, \ion{Fe}{xxii}  9.2298 \\
\ion{Ni}{xix} & 9.2540 & 7.0 & 7.66e$-$15 & 3.2 & 0.19 & \ion{Ne}{x}  9.2461,  9.2462 \\
\ion{Mg}{xi} & 9.3143 & 6.9 & 2.35e$-$14 & 5.6 & $-$0.13 &  \\
\ion{Ni}{xx} & 9.3850 & 7.1 & 4.77e$-$15 & 3.4 & 0.51 & \ion{Ni}{xxvi}  9.3853, \ion{Ni}{xxv}  9.3900 \\
No id. & 9.4000 & 0.0 & 5.99e$-$15 & 3.7 & \ldots &  \\
\ion{Ne}{x} & 9.4807 & 6.9 & 1.56e$-$14 & 5.8 & 0.02 & \ion{Ni}{xx}  9.4719, \ion{Ne}{x}  9.4809 \\
\ion{Ne}{x} & 9.7080 & 6.9 & 2.09e$-$14 & 5.4 & $-$0.16 & \ion{Ne}{x}  9.7085 \\
\ion{Ni}{xix} & 9.9770 & 7.0 & 8.01e$-$15 & 3.6 & 0.05 & \ion{Ni}{xxv}  9.9700, \ion{Fe}{xxi}  9.9887 \\
No id. & 10.0100 & 0.0 & 9.21e$-$15 & 3.8 & \ldots &  \\
\ion{Na}{xi} & 10.0232 & 7.0 & 8.56e$-$15 & 3.7 & 0.00 & \ion{Fe}{xx} 10.0235, \ion{Na}{xi} 10.0286 \\
No id. & 10.0500 & 0.0 & 6.03e$-$15 & 3.1 & \ldots &  \\
\ion{Ni}{xix} & 10.1100 & 7.0 & 6.34e$-$15 & 3.2 & 0.15 &  \\
\ion{Fe}{xvii} & 10.1210 & 6.9 & 7.94e$-$15 & 3.5 & 0.54 & \ion{Fe}{xix} 10.1195, 10.1309 \\
\ion{Ne}{x} & 10.2385 & 6.9 & 8.39e$-$14 & 11.0 & $-$0.02 & \ion{Ne}{x} 10.2396 \\
\ion{Fe}{xviii} & 10.4483 & 7.0 & 1.39e$-$14 & 4.4 & 0.26 & \ion{Fe}{xx} 10.4537, \ion{Fe}{xviii} 10.4639, 10.4654, 10.4726 \\
\ion{Fe}{xvii} & 10.5040 & 6.9 & 1.32e$-$14 & 4.2 & 0.34 &  \\
\ion{Fe}{xviii} & 10.5364 & 7.0 & 7.42e$-$15 & 3.1 & $-$0.07 & \ion{Fe}{xviii} 10.5382 \\
\ion{Fe}{xxiv} & 10.6190 & 7.4 & 1.29e$-$14 & 4.2 & $-$0.11 & \ion{Fe}{xix} 10.6193 \\
\ion{Fe}{xvii} & 10.7700 & 6.9 & 2.85e$-$14 & 6.2 & 0.23 & \ion{Fe}{xix} 10.7650, \ion{Ne}{ix} 10.7650 \\
\ion{Fe}{xix} & 10.8160 & 7.1 & 7.02e$-$15 & 3.1 & $-$0.20 & \ion{Fe}{xix} 10.8083 \\
\ion{Fe}{xxiii} & 10.9810 & 7.3 & 1.91e$-$14 & 4.9 & 0.08 & \ion{Fe}{xx} 10.9866 \\
\ion{Ne}{ix} & 11.0010 & 6.7 & 1.26e$-$14 & 4.0 & $-$0.08 & \ion{Fe}{xx} 11.0065 \\
\ion{Fe}{xxiv} & 11.0290 & 7.4 & 1.62e$-$14 & 4.5 & $-$0.06 & \ion{Fe}{xxiii} 11.0190, \ion{Fe}{xvii} 11.0260 \\
\ion{Fe}{xvii} & 11.1310 & 6.9 & 2.13e$-$14 & 5.0 & 0.10 & \ion{Fe}{xx} 11.1415 \\
\ion{Fe}{xxiv} & 11.1760 & 7.4 & 1.43e$-$14 & 3.9 & $-$0.04 & \ion{Fe}{xxiv} 11.1870, \ion{Ni}{xxii} 11.1950 \\
\ion{Fe}{xvii} & 11.2540 & 6.9 & 2.17e$-$14 & 4.9 & $-$0.02 &  \\
\ion{Fe}{xviii} & 11.3260 & 7.0 & 2.40e$-$14 & 5.1 & $-$0.07 & \ion{Fe}{xviii} 11.3260, \ion{Ni}{xxi} 11.3180, \ion{Fe}{xxiii} 11.3360 \\
\ion{Fe}{xviii} & 11.4230 & 7.0 & 1.71e$-$14 & 4.2 & $-$0.27 & \ion{Fe}{xxii} 11.4270, \ion{Fe}{xviii} 11.4274, \ion{Fe}{xxiv} 11.4320 \\
\ion{Fe}{xxiii} & 11.4580 & 7.3 & 9.10e$-$15 & 3.0 & 0.18 & \ion{Fe}{xviii} 11.4445, 11.4471, \ion{Fe}{xxi} 11.4451 \\
\ion{Fe}{xxii} & 11.4900 & 7.2 & 9.37e$-$15 & 3.2 & 0.12 & \ion{Fe}{xxii} 11.4900, 11.5025 \\
\ion{Fe}{xviii} & 11.5270 & 7.0 & 1.79e$-$14 & 4.3 & $-$0.07 & \ion{Fe}{xviii} 11.5270 \\
\ion{Ne}{ix} & 11.5440 & 6.7 & 5.29e$-$14 & 7.5 & 0.10 & \ion{Ni}{xix} 11.5390 \\
No id. & 11.6060 & 0.0 & 1.06e$-$14 & 3.4 & \ldots &  \\
\ion{Fe}{xxiii} & 11.7360 & 7.3 & 3.01e$-$14 & 5.5 & 0.02 &  \\
\ion{Fe}{xxii} & 11.7700 & 7.2 & 1.83e$-$14 & 4.3 & $-$0.25 & \ion{Fe}{xx} 11.7620 \\
\ion{Fe}{xxii} & 11.9320 & 7.2 & 1.36e$-$14 & 3.7 & 0.22 & \ion{Fe}{xxi} 11.9432, \ion{Fe}{xxii} 11.9474 \\
\ion{Ni}{xx} & 12.1120 & 7.1 & 1.03e$-$14 & 3.1 & 0.34 & \ion{Ne}{ix} 12.1110, \ion{Fe}{xxi} 12.1147 \\
\ion{Ne}{x} & 12.1321 & 6.9 & 6.32e$-$13 & 24.4 & 0.03 & \ion{Fe}{xvii} 12.1240, \ion{Ne}{x} 12.1375 \\
\ion{Fe}{xxiii} & 12.1610 & 7.3 & 2.10e$-$14 & 4.4 & 0.17 &  \\
\ion{Fe}{xxii} & 12.2100 & 7.2 & 1.12e$-$14 & 3.2 & 0.11 & \ion{Fe}{xxii} 12.2008, \ion{Fe}{xxi} 12.2040, \ion{Ni}{xxi} 12.2080, \ion{Fe}{xix} 12.2120 \\
\ion{Fe}{xvii} & 12.2660 & 6.9 & 3.58e$-$14 & 5.7 & $-$0.09 &  \\
\ion{Fe}{xxi} & 12.2840 & 7.2 & 3.22e$-$14 & 5.4 & $-$0.25 &  \\
\ion{Ne}{ix} & 12.3560 & 6.8 & 1.30e$-$14 & 3.4 & 0.12 & \ion{Fe}{xxi} 12.3508, 12.3527, 12.3626, 12.3628, 12.3630, \ion{Fe}{xxii} 12.3648 \\
\ion{Fe}{xxi} & 12.3930 & 7.2 & 1.18e$-$14 & 3.2 & 0.07 &  \\
\ion{Ni}{xix} & 12.4350 & 7.0 & 2.16e$-$14 & 4.4 & $-$0.39 & \ion{Fe}{xxi} 12.4220, \ion{Fe}{xx} 12.4234, 12.4310 \\
\ion{Fe}{xx} & 12.4680 & 7.1 & 1.25e$-$14 & 3.3 & 0.05 & \ion{Ni}{xviii} 12.4590, \ion{Fe}{xxi} 12.4625, \ion{Fe}{xxii} 12.4728, \ion{Fe}{xix} 12.4729, 12.4767 \\
\ion{Fe}{xxi} & 12.4990 & 7.2 & 1.16e$-$14 & 3.2 & 0.01 & \ion{Fe}{xxi} 12.5023, 12.5048, \ion{Ni}{xxi} 12.5105 \\
\ion{Fe}{xx} & 12.5760 & 7.1 & 1.40e$-$14 & 3.5 & 0.14 & \\
\ion{Ni}{xix} & 12.6560 & 7.0 & 1.29e$-$14 & 3.3 & 0.02 & \ion{Fe}{xxi} 12.6490 \\
\ion{Ni}{xix} & 12.8120 & 7.0 & 1.49e$-$14 & 3.5 & 0.29 & \ion{Fe}{xxi} 12.8071, 12.8089, 12.8095, 12.8110, \ion{Mn}{xxii} 12.8109 \\
\ion{Fe}{xx} & 12.8240 & 7.1 & 1.02e$-$13 & 9.1 & 0.01 & \ion{Fe}{xxi} 12.8220, \ion{Fe}{xx} 12.8460 \\
\ion{Fe}{xx} & 12.9120 & 7.1 & 1.15e$-$14 & 3.0 & $-$0.22 & \ion{Fe}{xix} 12.9033, 12.9169 \\
\ion{Fe}{xx} & 12.9650 & 7.1 & 2.16e$-$14 & 4.1 & $-$0.15 & \ion{Fe}{xviii} 12.9494, \ion{Fe}{xxii} 12.9530 \\
\ion{Fe}{xix} & 13.0220 & 7.1 & 1.73e$-$14 & 3.7 & $-$0.00 & \ion{Fe}{xviii} 13.0121, \ion{Fe}{xxi} 13.0126, \ion{Fe}{xx} 13.0152, 13.0240 \\
\ion{Fe}{xx} & 13.1000 & 7.1 & 2.49e$-$14 & 4.1 & 0.46 & \ion{Fe}{xix} 13.0810, 13.1011, \ion{Fe}{xxi} 13.1056 \\
\ion{Fe}{xx} & 13.1370 & 7.1 & 1.87e$-$14 & 3.6 & 0.28 & \ion{Fe}{xxi} 13.1223, 13.1338, \ion{Ni}{xx} 13.1350 \\
\ion{Fe}{xx} & 13.1530 & 7.1 & 1.74e$-$14 & 3.5 & 0.07 & \ion{Fe}{xvii} 13.1530, \ion{Fe}{xx} 13.1565, \ion{Ni}{xx} 13.1610 \\
\ion{Fe}{xviii} & 13.3230 & 7.0 & 1.56e$-$14 & 3.3 & 0.02 & \ion{Ni}{xx} 13.3090, 13.3236, \ion{Fe}{xviii} 13.3312 \\
\ion{Fe}{xx} & 13.3850 & 7.1 & 1.68e$-$14 & 3.3 & $-$0.16 & \ion{Fe}{xviii} 13.3550, 13.3807, \ion{Fe}{xx} 13.3752, 13.3953 \\
\ion{Fe}{xix} & 13.4230 & 7.1 & 2.32e$-$14 & 3.8 & 0.42 & \ion{Fe}{xix} 13.4239 \\
\ion{Ne}{ix} & 13.4473 & 6.7 & 4.13e$-$13 & 16.2 & 0.26 &  \\
\ion{Fe}{xix} & 13.4620 & 7.1 & 4.72e$-$14 & 5.6 & 0.18 & \ion{Ne}{viii} 13.4610, 13.4690, 13.4700 \\
\ion{Fe}{xix} & 13.4970 & 7.1 & 2.27e$-$14 & 3.9 & $-$0.24 & \ion{Ne}{viii} 13.4910, \ion{Fe}{xix} 13.4923 \\
\ion{Fe}{xix} & 13.5180 & 7.1 & 7.87e$-$14 & 7.3 & 0.01 &  \\
\hline
\end{tabular}
\end{scriptsize}
\end{table*}
\setcounter{table}{4}
\begin{table*}
\caption{(continued) {\em Chandra}/HETG and LETG, {\em EUVE}, and
    {\em HST}/STIS line fluxes of AD Leo\tablefootmark{a}.}
\tabcolsep 3.pt
\begin{scriptsize}
  \begin{tabular}{lrccrcl}
\hline \hline
Ion & $\lambda_{\rm model}$ (\AA) & $\log T_{\rm max}$ & $F_{\rm obs}$ & $S/N$ & Ratio & Blends \\
\hline
\ion{Ne}{ix} & 13.5531 & 6.7 & 5.60e$-$14 & 6.2 & 0.08 &  \\
\ion{Fe}{xix} & 13.6450 & 7.1 & 2.21e$-$14 & 3.7 & 0.22 & \ion{Fe}{xix} 13.6442, 13.6476 \\
\ion{Ne}{ix} & 13.6990 & 6.7 & 2.50e$-$13 & 12.5 & 0.24 & \ion{Fe}{xix} 13.6878, 13.7054 \\
\ion{Fe}{xix} & 13.7590 & 7.1 & 2.39e$-$14 & 3.6 & 0.42 & \\
\ion{Ni}{xix} & 13.7790 & 7.0 & 2.05e$-$14 & 3.3 & $-$0.07 & \ion{Fe}{xx} 13.7670, \ion{Fe}{xix} 13.8041, 13.8068 \\
\ion{Fe}{xix} & 13.7950 & 7.1 & 2.86e$-$14 & 3.8 & $-$0.04 & \\
\ion{Fe}{xvii} & 13.8250 & 6.9 & 2.03e$-$14 & 3.1 & $-$0.17 &  \\
\ion{Fe}{xix} & 13.8390 & 7.1 & 1.95e$-$14 & 3.0 & 0.27 & \ion{Fe}{xx} 13.8430 \\
\ion{Fe}{xviii} & 13.9530 & 7.0 & 4.66e$-$14 & 4.5 & 0.39 & \ion{Fe}{xix} 13.9263, 13.9330, \ion{Fe}{xx} 13.9620 \\
\ion{Ni}{xix} & 14.0430 & 7.0 & 2.29e$-$14 & 3.5 & $-$0.21 & \ion{Fe}{xix} 14.0179, 14.0340, 14.0388 \\
\ion{Ni}{xix} & 14.0770 & 7.0 & 2.94e$-$14 & 3.9 & 0.13 & \ion{Fe}{xix} 14.0610, \ion{Fe}{xx} 14.0620, 14.0784 \\
No id. & 14.1736 & 0.0 & 1.59e$-$14 & 3.0 & \ldots &  \\
\ion{Fe}{xviii} & 14.2080 & 7.0 & 1.03e$-$13 & 7.8 & $-$0.25 & \\
No id. & 14.2492 & 0.0 & 2.71e$-$14 & 4.0 & \ldots &  \\
\ion{Fe}{xviii} & 14.2560 & 7.0 & 3.51e$-$14 & 4.6 & $-$0.12 & \ion{Fe}{xx} 14.2670 \\
\ion{Fe}{xx} & 14.3259 & 7.1 & 1.53e$-$14 & 3.3 & 0.78 & \ion{Fe}{xix} 14.3275, 14.3391, 14.3397, \ion{Fe}{xx} 14.3321, \ion{Fe}{xxi} 14.3388 \\
\ion{Fe}{xviii} & 14.3430 & 7.0 & 1.89e$-$14 & 3.7 & $-$0.06 & \\
\ion{Fe}{xviii} & 14.3730 & 7.0 & 3.01e$-$14 & 4.7 & $-$0.16 &  \\
\ion{Fe}{xviii} & 14.4250 & 7.0 & 1.68e$-$14 & 3.5 & 0.17 & \ion{Fe}{xix} 14.4106 \\
\ion{Fe}{xviii} & 14.4345 & 7.0 & 1.81e$-$14 & 3.6 & 0.31 & \ion{Fe}{xviii} 14.4432 \\
\ion{Fe}{xx} & 14.4600 & 7.1 & 2.92e$-$14 & 4.6 & 0.14 & \ion{Fe}{xviii} 14.4610, 14.4555, 14.4769, \ion{Fe}{xx} 14.4793 \\
\ion{Fe}{xviii} & 14.5340 & 7.0 & 7.96e$-$14 & 7.4 & 0.03 & \ion{O}{viii} 14.5242, 14.5243, \ion{Fe}{xviii} 14.5608, 14.5710 \\
\ion{O}{viii} & 14.6342 & 6.6 & 1.56e$-$14 & 3.2 & $-$0.12 & \ion{O}{viii} 14.6344 \\
\ion{Fe}{xix} & 14.6640 & 7.1 & 2.12e$-$14 & 3.7 & $-$0.01 &  \\
\ion{Fe}{xx} & 14.7540 & 7.1 & 1.80e$-$14 & 3.5 & 0.12 & \ion{Fe}{xviii} 14.7499, 14.7510 \\
\ion{O}{viii} & 14.8205 & 6.6 & 3.20e$-$14 & 4.5 & $-$0.11 & \ion{O}{viii} 14.8207, \ion{Fe}{xx} 14.8304 \\
\ion{Fe}{xvii} & 15.0140 & 6.9 & 3.77e$-$13 & 14.5 & 0.04 &  \\
\ion{Fe}{xvi} & 15.0496 & 6.7 & 3.54e$-$14 & 4.5 & 0.31 & \ion{Fe}{xx} 15.0470, 15.0470 \\
\ion{Fe}{xix} & 15.0790 & 7.1 & 3.48e$-$14 & 4.5 & 0.12 &  \\
\ion{O}{viii} & 15.1760 & 6.6 & 7.73e$-$14 & 6.4 & $-$0.08 & \ion{O}{viii} 15.1765, \ion{Fe}{xix} 15.1770 \\
\ion{Fe}{xix} & 15.1980 & 7.1 & 3.25e$-$14 & 4.1 & 0.27 & \ion{Fe}{xviii} 15.1898 \\
\ion{Fe}{xvii} & 15.2610 & 6.9 & 1.77e$-$13 & 9.5 & 0.21 &  \\
\ion{Fe}{xvii} & 15.4530 & 6.8 & 2.87e$-$14 & 3.9 & 0.23 &  \\
\ion{Fe}{xviii} & 15.4940 & 7.0 & 1.78e$-$14 & 3.1 & 0.32 & \ion{Fe}{xvi} 15.4955, 15.5023, \ion{Cr}{xviii} 15.5020 \\
\ion{Fe}{xviii} & 15.6250 & 7.0 & 3.51e$-$14 & 4.4 & $-$0.17 &  \\
\ion{Fe}{xviii} & 15.8240 & 7.0 & 3.53e$-$14 & 4.4 & 0.05 &  \\
\ion{Fe}{xviii} & 15.8700 & 7.0 & 2.36e$-$14 & 3.6 & 0.04 & \ion{Fe}{xix} 15.8708 \\
\ion{O}{viii} & 16.0055 & 6.6 & 2.61e$-$13 & 11.6 & $-$0.02 & \ion{Fe}{xviii} 16.0040, \ion{O}{viii} 16.0067 \\
\ion{Fe}{xviii} & 16.0710 & 7.0 & 1.10e$-$13 & 7.5 & 0.09 &  \\
No id. & 16.7276 & 0.0 & 2.07e$-$14 & 3.0 & \ldots &  \\
\ion{Fe}{xvii} & 16.7800 & 6.8 & 2.38e$-$13 & 10.1 & 0.04 &  \\
\ion{Fe}{xvii} & 17.0510 & 6.8 & 3.27e$-$13 & 11.4 & 0.10 &  \\
\ion{Fe}{xvii} & 17.0960 & 6.8 & 2.56e$-$13 & 10.0 & 0.11 &  \\
\ion{Fe}{xviii} & 17.6230 & 7.0 & 2.73e$-$14 & 3.1 & $-$0.24 &  \\
\ion{O}{vii} & 18.6270 & 6.4 & 7.76e$-$14 & 4.9 & 0.03 & \ion{Ar}{xvi} 18.6324 \\
\ion{O}{viii} & 18.9671 & 6.6 & 1.38e$-$12 & 19.0 & 0.01 & \ion{O}{viii} 18.9725 \\
\ion{O}{vii} & 21.6015 & 6.4 & 4.51e$-$13 & 8.1 & $-$0.00 &  \\
\ion{O}{vii} & 21.8036 & 6.4 & 1.12e$-$13 & 3.8 & 0.08 &  \\
\ion{O}{vii} & 22.0977 & 6.4 & 1.93e$-$13 & 4.7 & $-$0.19 &  \\
\ion{N}{vii} & 24.7792 & 6.4 & 1.51e$-$13 & 4.6 & $-$0.25 & \ion{N}{vii} 24.7846 \\
\ion{C}{vi} & 28.4652 & 6.3 & 3.96e$-$14 & 6.5 & 0.18 & \ion{C}{vi} 28.4663 \\
\ion{N}{vi} & 28.7870 & 6.3 & 3.67e$-$14 & 6.2 & $-$0.19 &  \\
\ion{N}{vi} & 29.0843 & 6.2 & 3.27e$-$14 & 5.8 & 0.38 &  \\
\ion{N}{vi} & 29.5347 & 6.2 & 3.86e$-$14 & 6.2 & 0.01 & \ion{Si}{xii} 29.5090 \\
\ion{S}{xiv} & 30.4270 & 6.6 & 3.65e$-$14 & 5.6 & 0.12 &  \\
\ion{S}{xiv} & 30.4690 & 6.6 & 1.52e$-$14 & 3.7 & 0.04 &  \\
\ion{S}{xiii} & 32.2391 & 6.6 & 2.90e$-$14 & 5.5 & 0.39 & \ion{S}{xiii} 32.1911 \\
\ion{S}{xiv} & 32.5600 & 6.6 & 4.60e$-$14 & 6.9 & 0.09 & \ion{S}{xiv} 32.5750 \\
\ion{S}{xiv} & 33.5490 & 6.6 & 4.68e$-$14 & 6.9 & 0.36 & \ion{Si}{xi} 33.5301 \\
\ion{C}{vi} & 33.7342 & 6.2 & 2.61e$-$13 & 16.3 & 0.21 & \ion{C}{vi} 33.7396 \\
\ion{Si}{xi} & 43.7501 & 6.4 & 2.34e$-$14 & 7.8 & 0.03 &  \\
\ion{Si}{xii} & 44.0190 & 6.4 & 3.07e$-$14 & 9.1 & 0.01 &  \\
\ion{Si}{xii} & 44.1650 & 6.4 & 5.73e$-$14 & 12.4 & 0.03 &  \\
\ion{Si}{xii} & 45.5210 & 6.4 & 2.44e$-$14 & 8.1 & 0.24 &  \\
\ion{Si}{xii} & 45.6910 & 6.4 & 2.58e$-$14 & 8.4 & $-$0.03 &  \\
\ion{Fe}{xvi} & 50.3613 & 6.6 & 8.09e$-$15 & 4.3 & 0.00 &  \\
\ion{Fe}{xvi} & 54.1264 & 6.6 & 1.46e$-$14 & 4.2 & 0.12 & \ion{S}{viii} 54.1181, \ion{S}{ix} 54.1750 \\
\ion{Fe}{xvi} & 54.7101 & 6.6 & 1.98e$-$14 & 4.9 & 0.21 & \ion{Fe}{xvi} 54.7472 \\
\ion{Mg}{x} & 57.8760 & 6.2 & 8.36e$-$15 & 4.2 & 0.15 &  \\
\ion{Fe}{xvi} & 63.7106 & 6.6 & 8.70e$-$15 & 3.0 & $-$0.22 &  \\
\ion{Mg}{x} & 65.8450 & 6.2 & 1.59e$-$14 & 5.3 & 0.48 & \ion{Ne}{viii} 65.8940 \\
\ion{Fe}{xvi} & 66.3568 & 6.6 & 4.83e$-$14 & 9.7 & 0.36 & \ion{Fe}{xvi} 66.2489 \\
\ion{Ne}{viii} & 88.0820 & 5.9 & 3.59e$-$14 & 13.2 & 0.24 & \ion{Ne}{viii} 88.1190 \\
\ion{Fe}{xii} & 89.8267 & 6.3 & 1.40e$-$14 & 7.3 & 0.39 & \ion{Fe}{xii} 89.4789, \ion{Fe}{xvii} 89.5977, \ion{Ne}{vii} 89.9472 \\
\ion{Fe}{xix} & 91.0200 & 7.0 & 1.77e$-$14 & 8.4 & 0.06 & \ion{Fe}{xvii} 90.5205, \ion{Fe}{xx} 90.6000, \ion{Fe}{x} 90.6830, \ion{Fe}{xiii} 90.7058, \ion{Fe}{xi} 90.7197 \\
\ion{Fe}{xiv} & 92.4616 & 6.4 & 7.45e$-$15 & 4.3 & 0.17 & \ion{Fe}{xvii} 92.0282, \ion{Fe}{xiv} 92.0572, \ion{Fe}{xi} 92.7554 \\
\ion{Fe}{xviii} & 93.9230 & 7.0 & 7.72e$-$14 & 23.2 & $-$0.07 &  \\
\ion{Ne}{viii} & 98.2600 & 5.9 & 6.45e$-$14 & 20.9 & 0.19 & \ion{Fe}{xxi} 97.8800, \ion{Ne}{viii} 98.1150 \\
\ion{Fe}{xix} & 101.5500 & 7.0 & 1.17e$-$14 & 6.7 & $-$0.37 &  \\
\ion{Fe}{xxi} & 102.2200 & 7.2 & 3.05e$-$14 & 14.0 & $-$0.14 & \ion{Fe}{xiii} 102.0757, \ion{O}{viii} 102.3476, 102.3552, 102.3919, 102.4897, 102.5497 \\
\ion{Ne}{viii} & 103.0850 & 5.9 & 1.57e$-$14 & 8.9 & 0.14 & \ion{Ni}{xxii} 103.3260, \ion{Fe}{xiii} 103.3386 \\
\ion{Fe}{xviii} & 103.9370 & 7.0 & 2.27e$-$14 & 11.3 & $-$0.12 &  \\
\ion{O}{vi} & 104.8130 & 5.6 & 7.03e$-$15 & 5.9 & 0.55 & \ion{O}{vi} 104.8129, \ion{Fe}{xvii} 104.7669 \\
\hline
\end{tabular}
\end{scriptsize}
\end{table*}
\setcounter{table}{4}
\begin{table*}[h!]
\caption{(continued) {\em Chandra}/HETG and LETG, {\em EUVE}, and
    {\em HST}/STIS line fluxes of AD Leo\tablefootmark{a}.}
\tabcolsep 3.pt
\begin{scriptsize}
  \begin{tabular}{lrccrcl}
\hline \hline
Ion & $\lambda_{\rm model}$ (\AA) & $\log T_{\rm max}$ & $F_{\rm obs}$ & $S/N$ & Ratio & Blends \\
\hline
\ion{Ni}{xxii} & 106.0640 & 7.2 & 7.74e$-$15 & 6.0 & $-$0.02 & \ion{Fe}{xii} 106.0264, 106.0341, 106.2027, \ion{Fe}{xix} 106.1200, 106.3300 \\
\ion{Fe}{xix} & 108.3700 & 7.0 & 5.02e$-$14 & 19.4 & $-$0.16 &  \\
\ion{Fe}{xix} & 109.9700 & 7.0 & 6.86e$-$15 & 5.7 & $-$0.27 &  \\
\ion{Fe}{xx} & 110.6300 & 7.1 & 6.43e$-$15 & 4.5 & 0.09 & \ion{Fe}{xx} 110.4464 \\
\ion{Fe}{xix} & 111.7000 & 7.0 & 8.00e$-$15 & 5.2 & $-$0.20 & \ion{Ni}{xxiii} 111.8600 \\
\ion{Fe}{xxii} & 114.4100 & 7.2 & 1.32e$-$14 & 7.8 & 0.10 & \ion{Fe}{xii} 114.5616, 114.9055 \\
\ion{Mn}{xviii} & 115.3653 & 7.0 & 1.04e$-$14 & 7.2 & 0.09 &  \\
\ion{Fe}{xxii} & 117.1700 & 7.2 & 5.58e$-$14 & 19.6 & $-$0.01 & \ion{Fe}{xxi} 117.5100 \\
\ion{Ni}{xxv} & 117.9330 & 7.4 & 7.41e$-$15 & 5.7 & 0.04 & \ion{Ni}{xxii} 117.9330, \ion{Cr}{xx} 117.9580 \\
\ion{Fe}{xx} & 118.6600 & 7.1 & 1.49e$-$14 & 8.5 & $-$0.18 &  \\
\ion{Fe}{xix} & 120.0000 & 7.0 & 1.51e$-$14 & 8.0 & $-$0.22 & \ion{O}{vii} 120.3310, \ion{O}{vii} 120.3327 \\
\ion{Fe}{xx} & 121.8300 & 7.1 & 3.60e$-$14 & 14.1 & $-$0.08 &  \\
\ion{Ne}{vi} & 122.4879 & 5.8 & 6.45e$-$15 & 4.5 & 0.08 & \ion{Ne}{vi} 122.6850, \ion{Cr}{xvii} 122.9718 \\
\ion{Fe}{xxi} & 128.7300 & 7.2 & 5.59e$-$14 & 17.2 & 0.00 &  \\
\ion{Fe}{xxiii} & 132.8500 & 7.3 & 1.65e$-$13 & 30.6 & 0.04 & \ion{Fe}{xx} 132.8500 \\
\ion{Fe}{xxii} & 135.7800 & 7.2 & 4.26e$-$14 & 14.3 & 0.02 &  \\
\ion{Fe}{xxi} & 142.2700 & 7.2 & 1.87e$-$14 & 8.1 & 0.28 & \ion{Fe}{xxi} 142.1600 \\
\ion{Fe}{ix} & 171.0729 & 6.0 & 5.97e$-$14 & 9.9 & $-$0.10 & \ion{Ni}{xiv} 171.3703 \\
\ion{Fe}{x} & 174.5310 & 6.1 & 6.07e$-$14 & 8.6 & $-$0.00 &  \\
\ion{Fe}{xi} & 180.4080 & 6.2 & 2.41e$-$14 & 6.1 & $-$0.69 &  \\
\ion{Fe}{xii} & 193.5090 & 6.3 & 2.02e$-$13 & 19.1 & $-$0.28 & \ion{S}{xi} 191.2664, \ion{Fe}{xxiv} 192.0170, \ion{Fe}{xii} 192.3940, 195.1190 \\
\ion{Fe}{xiii} & 203.8280 & 6.3 & 5.00e$-$14 & 9.2 & 0.23 & \ion{Ar}{xiv} 203.3510, \ion{Fe}{xii} 203.7280, \ion{Mn}{xi} 203.9122, \ion{Fe}{xi} 204.7440, \ion{Fe}{xiii} 204.9450 \\
\ion{Fe}{xiv} & 211.3180 & 6.4 & 6.67e$-$14 & 10.2 & $-$0.28 &  \\
\ion{Ni}{xvii} & 249.1780 & 6.6 & 5.20e$-$14 & 7.1 & 0.30 & \ion{Fe}{xii} 248.7670, 249.3880 \\
\ion{Fe}{xv} & 284.1630 & 6.4 & 2.38e$-$13 & 15.4 & $-$0.13 &  \\
\ion{Fe}{xvi} & 335.4098 & 6.5 & 2.59e$-$13 & 12.7 & $-$0.04 & \ion{Fe}{xiv} 334.1800 \\
\ion{Fe}{xvi} & 360.7581 & 6.5 & 1.09e$-$13 & 6.6 & $-$0.07 & \ion{Fe}{xiii} 359.6420, 359.8420 \\
\ion{C}{iii} & 1176.0000 & 4.8 & 1.93e$-$13 & 20.4 & 0.06 &  \\
\ion{S}{iii} & 1190.1990 & 4.9 & 9.87e$-$16 & 8.2 & $-$0.64 &  \\
\ion{S}{iii} & 1194.0490 & 4.9 & 2.21e$-$15 & 10.7 & $-$0.19 &  \\
\ion{Si}{ii} & 1194.5010 & 4.6 & 3.38e$-$15 & 13.6 & $-$0.27 & \ion{S}{iii} 1194.4430 \\
\ion{Si}{ii} & 1197.3950 & 4.6 & 2.14e$-$15 & 27.6 & 0.00 &  \\
\ion{S}{v} & 1199.1360 & 5.3 & 2.13e$-$15 & 9.1 & $-$0.04 &  \\
\ion{S}{iii} & 1200.9610 & 4.9 & 4.52e$-$15 & 23.4 & 0.16 &  \\
\ion{O}{ii} & 1204.2345 & 4.9 & 5.35e$-$16 & 4.9 & $-$0.19 &  \\
\ion{Si}{iii} & 1206.5019 & 4.9 & 1.29e$-$13 & 56.2 & $-$0.29 &  \\
\ion{O}{v} & 1218.3440 & 5.5 & 2.63e$-$14 & 78.3 & $-$0.13 &  \\
\ion{N}{v} & 1238.8218 & 5.4 & 9.23e$-$14 & 79.2 & 0.12 &  \\
\ion{Fe}{xii} & 1242.0050 & 6.3 & 2.28e$-$15 & 25.8 & 0.38 &  \\
\ion{N}{v} & 1242.8042 & 5.4 & 4.50e$-$14 & 41.2 & 0.11 &  \\
\ion{C}{iii} & 1247.3830 & 5.1 & 1.88e$-$15 & 17.6 & $-$0.24 &  \\
\ion{Ne}{ix} & 1248.1030 & 6.7 & 6.81e$-$16 & 32.1 & 0.03 &  \\
\ion{S}{ii} & 1250.5870 & 4.6 & 1.05e$-$15 & 22.1 & $-$0.05 &  \\
\ion{S}{ii} & 1253.8130 & 4.6 & 1.69e$-$15 & 19.9 & $-$0.15 &  \\
\ion{S}{ii} & 1259.5210 & 4.6 & 2.40e$-$15 & 31.8 & $-$0.22 &  \\
\ion{Si}{ii} & 1264.7400 & 4.5 & 7.76e$-$15 & 24.2 & 0.31 &  \\
\ion{Si}{ii} & 1265.0040 & 4.6 & 3.82e$-$15 & 19.5 & 0.19 &  \\
\ion{Si}{ii} & 1304.3719 & 4.6 & 2.89e$-$15 & 21.5 & 0.33 &  \\
\ion{Si}{ii} & 1309.2770 & 4.6 & 5.77e$-$15 & 35.4 & 0.33 &  \\
\ion{Si}{iii} & 1312.5930 & 4.9 & 8.90e$-$16 & 14.3 & $-$0.47 &  \\
\ion{C}{ii} & 1323.9080 & 4.8 & 2.22e$-$15 & 18.3 & 0.39 &  \\
\ion{Fe}{xix} & 1328.6999 & 7.1 & 1.82e$-$15 & 13.8 & $-$0.08 &  \\
\ion{C}{ii} & 1334.5350 & 4.7 & 5.16e$-$14 & 41.6 & $-$0.29 &  \\
\ion{C}{ii} & 1335.7100 & 4.7 & 1.59e$-$13 & 64.2 & 0.54 &  \\
\ion{O}{iv} & 1338.6140 & 5.3 & 4.70e$-$16 & 13.4 & 0.02 &  \\
\ion{Fe}{xii} & 1349.4000 & 6.3 & 8.71e$-$16 & 16.8 & 0.20 &  \\
\ion{Fe}{xxi} & 1354.0800 & 7.2 & 7.30e$-$15 & 12.5 & $-$0.08 &  \\
\ion{O}{v} & 1371.2960 & 5.5 & 5.97e$-$15 & 25.3 & $-$0.16 &  \\
\ion{Ar}{xi} & 1392.0980 & 6.4 & 1.10e$-$15 & 24.2 & $-$0.02 &  \\
\ion{Si}{iv} & 1393.7552 & 5.0 & 1.09e$-$13 & 52.0 & 0.19 &  \\
\ion{Si}{iv} & 1402.7704 & 5.0 & 6.32e$-$14 & 35.6 & 0.25 &  \\
\ion{Si}{iii} & 1417.2400 & 4.9 & 8.94e$-$16 & 23.1 & 0.34 &  \\
\ion{Si}{viii} & 1445.7371 & 6.0 & 4.82e$-$16 & 9.1 & $-$0.20 &  \\
\ion{S}{v} & 1501.7660 & 5.3 & 1.19e$-$15 & 19.5 & $-$0.19 &  \\
\ion{Si}{ii} & 1526.7090 & 4.5 & 8.05e$-$15 & 21.8 & 0.33 &  \\
\ion{Si}{ii} & 1533.4320 & 4.5 & 1.40e$-$14 & 20.5 & 0.28 &  \\
\ion{C}{iv} & 1548.1871 & 5.1 & 3.54e$-$13 & 64.5 & $-$0.09 &  \\
\ion{C}{iv} & 1550.7723 & 5.1 & 1.93e$-$13 & 48.7 & $-$0.05 &  \\
\ion{Ne}{v} & 1574.7560 & 5.5 & 1.31e$-$15 & 15.2 & 0.36 &  \\
\ion{S}{xi} & 1614.4946 & 6.4 & 1.23e$-$15 & 8.8 & 0.27 &  \\
\ion{O}{vii} & 1623.6672 & 6.4 & 2.41e$-$15 & 12.4 & 0.08 &  \\
\ion{S}{iv} & 1629.1550 & 5.2 & 1.03e$-$15 & 10.7 & 0.33 &  \\
\ion{Al}{ii} & 1670.7870 & 4.6 & 1.52e$-$14 & 17.9 & 0.22 &  \\
\hline
\end{tabular}

\tablefoot{\tablefoottext{a}{Line fluxes (in erg cm$^{-2}$ s$^{-1}$ units) measured in the spectra, and corrected by the ISM absorption when needed.
  log $T_{\rm max}$ (K) indicates the maximum  temperature of formation of the line (unweighted by the
  EMD). ``Ratio'' is the $\log$ ($F_{\mathrm {obs}}$/$F_{\mathrm {pred}}$)   of the line. 
    ``Blends'' amounting to more than 5\% of the total flux for each line are indicated, with wavelengths in \AA.}}
\end{scriptsize}
\end{table*}

%
\begin{table*}
\caption[]{{\em XMM-Newton}/RGS and HST/STIS line fluxes of Lalande 21185\tablefootmark{a}.}\label{tab:fluxes4} 
\tabcolsep 3 pt
\begin{scriptsize}
  \begin{tabular}{lrccrcl}
\hline \hline
Ion & $\lambda_{\rm model}$ (\AA) & $\log T_{\rm max}$ & $F_{\rm obs}$ & $S/N$ & Ratio & Blends \\
\hline
\ion{Si}{xiv} & 6.1804 & 7.3 & 2.47e$-$14 & 3.8 & 0.02 & \ion{Si}{xiv}  6.1858 \\
\ion{Ne}{x} & 10.2385 & 6.9 & 5.38e$-$15 & 3.9 & 0.20 & \ion{Ne}{x} 10.2396 \\
No id. & 10.5040 & 0.0 & 5.56e$-$15 & 4.0 & \ldots &  \\
No id. & 10.7700 & 0.0 & 6.84e$-$15 & 4.6 & \ldots &  \\
\ion{Ne}{ix} & 11.5440 & 6.7 & 4.38e$-$15 & 3.9 & 0.16 & \ion{Fe}{xxii} 11.4900, \ion{Fe}{xviii} 11.5270 \\
\ion{Ne}{x} & 12.1321 & 6.9 & 1.92e$-$14 & 8.5 & $-$0.10 & \ion{Fe}{xvii} 12.1240, \ion{Ne}{x} 12.1375 \\
\ion{Fe}{xvi} & 12.5399 & 6.8 & 2.64e$-$15 & 3.2 & 0.29 & \ion{Fe}{xxi} 12.4990, 12.5048, \ion{Fe}{xx} 12.5260, 12.5760, \ion{Fe}{xvii} 12.5391 \\
\ion{Ne}{ix} & 13.4473 & 6.7 & 3.23e$-$14 & 11.8 & 0.21 & \ion{Fe}{xix} 13.4970, 13.5180, \ion{Ne}{ix} 13.5531 \\
\ion{Ne}{ix} & 13.6990 & 6.7 & 5.52e$-$15 & 4.1 & $-$0.13 & \ion{Fe}{xix} 13.6450, 13.7054 \\
\ion{Fe}{xvii} & 13.8250 & 6.9 & 7.16e$-$15 & 5.6 & 0.26 & \ion{Fe}{xix} 13.7590, 13.7950, 13.8390, \ion{Fe}{xx} 13.7670, \ion{Ni}{xix} 13.7790, \ion{Fe}{xvii} 13.8920 \\
\ion{Fe}{xviii} & 14.2080 & 7.0 & 4.78e$-$15 & 4.7 & $-$0.24 & \ion{Fe}{xviii} 14.2560, \ion{Fe}{xx} 14.2670 \\
\ion{Fe}{xviii} & 14.3730 & 7.0 & 5.79e$-$15 & 5.1 & 0.14 & \ion{Fe}{xviii} 14.3430, 14.3990, 14.4250, 14.4555 \\
\ion{Fe}{xviii} & 14.5340 & 7.0 & 2.17e$-$15 & 3.1 & $-$0.12 & \ion{O}{viii} 14.5242, 14.5243, \ion{Fe}{xviii} 14.5608, 14.5710 \\
\ion{Fe}{xvi} & 14.9555 & 6.7 & 3.70e$-$15 & 4.1 & $-$0.08 & \ion{O}{viii} 14.8205, 14.8207, \ion{Fe}{xix} 14.9610 \\
\ion{Fe}{xvii} & 15.0140 & 6.9 & 1.50e$-$14 & 8.3 & $-$0.02 &  \\
\ion{O}{viii} & 15.1760 & 6.6 & 3.15e$-$15 & 3.8 & $-$0.20 & \ion{O}{viii} 15.1765, \ion{Fe}{xix} 15.1980 \\
\ion{Fe}{xvii} & 15.2610 & 6.9 & 4.38e$-$15 & 4.5 & $-$0.02 &  \\
\ion{Fe}{xvii} & 15.4530 & 6.8 & 3.60e$-$15 & 4.1 & 0.41 & \ion{Fe}{xvi} 15.4485, 15.4533, 15.5023, \ion{Fe}{xviii} 15.4940, \ion{Fe}{xx} 15.5170 \\
\ion{O}{viii} & 16.0055 & 6.6 & 1.60e$-$14 & 8.4 & 0.07 & \ion{Fe}{xviii} 16.0040, \ion{O}{viii} 16.0067 \\
\ion{Fe}{xix} & 16.2830 & 7.1 & 3.24e$-$15 & 3.9 & 0.36 & \ion{Fe}{xvii} 16.2285, 16.3500, \ion{Fe}{xix} 16.2857, 16.3247, \ion{Fe}{xviii} 16.3200 \\
\ion{Fe}{xvii} & 16.7800 & 6.8 & 1.11e$-$14 & 7.2 & 0.11 &  \\
\ion{Fe}{xvii} & 17.0510 & 6.8 & 1.59e$-$14 & 8.5 & $-$0.13 & \ion{Fe}{xvii} 17.0960 \\
\ion{Fe}{xviii} & 17.6230 & 7.0 & 1.86e$-$15 & 3.0 & 0.03 &  \\
\ion{O}{vii} & 18.6270 & 6.4 & 6.27e$-$15 & 5.5 & 0.14 & \ion{Ar}{xvi} 18.6324, \ion{Ca}{xviii} 18.6909 \\
\ion{O}{viii} & 18.9671 & 6.6 & 6.88e$-$14 & 18.1 & $-$0.01 & \ion{O}{viii} 18.9725 \\
\ion{O}{vii} & 21.6015 & 6.4 & 3.61e$-$14 & 12.3 & 0.19 &  \\
\ion{O}{vii} & 21.8036 & 6.4 & 2.41e$-$15 & 3.2 & $-$0.16 &  \\
\ion{O}{vii} & 22.0977 & 6.4 & 2.38e$-$14 & 9.9 & 0.18 &  \\
No id. & 23.0050 & 0.0 & 4.01e$-$15 & 3.6 & \ldots &  \\
\ion{N}{vii} & 24.7792 & 6.4 & 5.49e$-$15 & 5.1 & $-$0.28 & \ion{Ar}{xv} 24.7366, 24.7400, \ion{N}{vii} 24.7846, \ion{Ar}{xvi} 24.8509 \\
\ion{C}{vi} & 26.9896 & 6.3 & 2.88e$-$15 & 3.5 & $-$0.21 & \ion{C}{vi} 26.9901, \ion{Ar}{xv} 27.0432 \\
\ion{Ar}{xiv} & 27.4640 & 6.7 & 3.11e$-$15 & 3.6 & $-$0.17 & \ion{Ar}{xiv} 27.6310, 27.6360 \\
\ion{Ar}{xiii} & 29.1420 & 6.6 & 4.95e$-$15 & 4.2 & 0.10 &  \\
\ion{S}{xiv} & 30.4270 & 6.6 & 3.88e$-$15 & 3.5 & $-$0.09 & \ion{Ca}{xi} 30.4480, \ion{S}{xiv} 30.4690 \\
\ion{S}{xiv} & 32.5600 & 6.6 & 4.02e$-$15 & 3.4 & 0.06 & \ion{S}{xiv} 32.5750, \ion{Ca}{xii} 32.6571 \\
\ion{C}{vi} & 33.7342 & 6.2 & 3.44e$-$14 & 10.0 & 0.23 & \ion{C}{vi} 33.7396 \\
No id. & 35.6844 & 0.0 & 5.70e$-$15 & 3.8 & \ldots &  \\
\ion{O}{v} & 1218.3440 & 5.5 & 2.51e$-$15 & 4.4 & 0.06 &  \\
\ion{N}{v} & 1238.8218 & 5.4 & 2.55e$-$15 & 4.1 & 0.14 &  \\
\ion{N}{v} & 1242.8042 & 5.4 & 1.47e$-$15 & 3.0 & 0.20 &  \\
\ion{Si}{ii} & 1264.7400 & 4.5 & 3.90e$-$16 & 4.2 & 0.06 &  \\
\ion{C}{ii} & 1335.7100 & 4.7 & 4.12e$-$15 & 5.6 & $-$0.22 & \ion{C}{ii} 1335.6650 \\
\ion{O}{v} & 1371.2960 & 5.5 & 2.96e$-$16 & 3.4 & $-$0.28 &  \\
\ion{Si}{iv} & 1393.7552 & 5.0 & 4.73e$-$15 & 5.2 & 0.02 &  \\
\ion{O}{iv} & 1401.1570 & 5.3 & 5.89e$-$16 & 3.0 & $-$0.48 &  \\
\ion{Si}{iv} & 1402.7704 & 5.0 & 1.61e$-$15 & 5.4 & $-$0.15 &  \\
\ion{Si}{ii} & 1533.4320 & 4.5 & 4.96e$-$16 & 3.2 & $-$0.06 &  \\
\ion{C}{iv} & 1548.1871 & 5.1 & 1.66e$-$14 & 8.1 & $-$0.00 &  \\
\ion{C}{iv} & 1550.7723 & 5.1 & 7.85e$-$15 & 5.6 & $-$0.03 &  \\
\ion{Al}{ii} & 1670.7870 & 4.6 & 4.41e$-$15 & 3.1 & $-$0.02 &  \\
\hline
\end{tabular}
\tablefoot{\tablefoottext{a}{Line fluxes (in erg cm$^{-2}$ s$^{-1}$ units) measured in the spectra, and corrected by the ISM absorption when needed.
  log $T_{\rm max}$ (K) indicates the maximum  temperature of formation of the line (unweighted by the
  EMD). ``Ratio'' is the $\log$ ($F_{\mathrm {obs}}$/$F_{\mathrm {pred}}$) of the line. 
    ``Blends'' amounting to more than 5\% of the total flux for each line are indicated, with wavelengths in \AA.}}
\end{scriptsize}
\end{table*}

%
\begin{table*}
\caption[]{{\em XMM-Newton}/RGS, HST/COS, and STIS line fluxes of GJ 674\tablefootmark{a}.}\label{tab:fluxes6} 
\tabcolsep 3.5 pt
\begin{scriptsize}
  \begin{tabular}{lrccrcl}
\hline \hline
Ion & $\lambda_{\rm model}$ (\AA) & $\log T_{\rm max}$ & $F_{\rm obs}$ & $S/N$ & Ratio & Blends \\
\hline
\ion{Fe}{xxiii} & 11.7360 & 7.3 & 2.86e$-$15 & 3.1 & $-$0.39 & \\
\ion{Fe}{xxii} & 11.9770 & 7.2 & 6.49e$-$15 & 4.8 & 0.22 & \ion{Fe}{xxii} 11.8810, 11.9320, 11.9474, \ion{Fe}{xxi} 11.9023, 11.9750, \ion{Ni}{xx} 11.9617 \\
\ion{Ne}{x} & 12.1321 & 6.9 & 2.24e$-$14 & 9.0 & $-$0.11 & \ion{Fe}{xvii} 12.1240, \ion{Ne}{x} 12.1375, \ion{Fe}{xxiii} 12.1610 \\
\ion{Fe}{xxi} & 12.2840 & 7.2 & 2.51e$-$15 & 3.0 & $-$0.79 & \ion{Fe}{xvii} 12.2660 \\
\ion{Ne}{ix} & 13.4473 & 6.7 & 2.15e$-$14 & 9.4 & $-$0.16 & \ion{Fe}{xix} 13.4620, 13.4970, 13.5180, \ion{Fe}{xxi} 13.5070 \\
\ion{Ne}{ix} & 13.6990 & 6.7 & 1.93e$-$14 & 9.0 & 0.41 & \ion{Fe}{xix} 13.6450, 13.6878, 13.7054 \\
\ion{Fe}{xviii} & 14.2080 & 7.0 & 1.01e$-$14 & 6.5 & $-$0.33 & \ion{Fe}{xviii} 14.2560, \ion{Fe}{xx} 14.2670 \\
\ion{Fe}{xviii} & 14.3730 & 7.0 & 3.14e$-$15 & 3.6 & $-$0.55 & \ion{Fe}{xviii} 14.3430, 14.3430, 14.3990, 14.4250, 14.4555, \ion{Fe}{xx} 14.3887 \\
\ion{Fe}{xviii} & 14.5340 & 7.0 & 6.56e$-$15 & 5.3 & 0.02 & \ion{O}{viii} 14.5242, \ion{Fe}{xviii} 14.5608, 14.5710 \\
\ion{Fe}{xvi} & 14.9555 & 6.7 & 4.92e$-$15 & 4.6 & $-$0.26 & \ion{O}{viii} 14.8205, 14.8207, \ion{Fe}{xx} 14.8791, 14.9279, \ion{Fe}{xix} 14.9610, 14.9632 \\
\ion{Fe}{xvii} & 15.0140 & 6.9 & 1.69e$-$14 & 8.6 & $-$0.11 &  \\
\ion{Fe}{xix} & 15.0790 & 7.1 & 7.77e$-$15 & 5.8 & 0.38 &  \\
\ion{O}{viii} & 15.1760 & 6.6 & 5.02e$-$15 & 4.6 & $-$0.17 & \ion{O}{viii} 15.1765, \ion{Fe}{xix} 15.1980 \\
\ion{Fe}{xvii} & 15.2610 & 6.9 & 9.26e$-$15 & 6.3 & 0.16 &  \\
\ion{O}{viii} & 16.0055 & 6.6 & 1.89e$-$14 & 9.2 & $-$0.13 & \ion{Fe}{xviii} 16.0040, 16.0710, \ion{O}{viii} 16.0067 \\
\ion{Fe}{xvii} & 16.7800 & 6.8 & 7.27e$-$15 & 5.7 & $-$0.19 &  \\
\ion{Fe}{xvii} & 17.0510 & 6.8 & 3.35e$-$14 & 12.2 & 0.08 & \ion{Fe}{xvii} 17.0960 \\
\ion{O}{viii} & 18.9671 & 6.6 & 9.82e$-$14 & 20.8 & 0.15 & \ion{O}{viii} 18.9725 \\
\ion{O}{vii} & 21.6015 & 6.4 & 5.02e$-$14 & 14.1 & 0.13 &  \\
\ion{O}{vii} & 21.8036 & 6.4 & 5.34e$-$15 & 4.6 & $-$0.08 &  \\
\ion{O}{vii} & 22.0977 & 6.4 & 2.14e$-$14 & 9.1 & $-$0.12 &  \\
\ion{N}{vii} & 24.7792 & 6.4 & 6.15e$-$15 & 5.1 & $-$0.25 & \ion{N}{vii} 24.7846, \ion{N}{vi} 24.8980 \\
\ion{C}{vi} & 33.7342 & 6.2 & 4.45e$-$14 & 9.9 & 0.48 & \ion{C}{vi} 33.7396 \\
\ion{Si}{iii} & 1108.3610 & 4.9 & 1.13e$-$15 & 3.8 & 0.20 &  \\
\ion{Si}{iii} & 1109.9720 & 4.9 & 2.22e$-$15 & 5.3 & 0.13 & \ion{Si}{iii} 1109.9430 \\
\ion{Si}{iii} & 1113.2321 & 4.9 & 3.00e$-$15 & 7.1 & 0.06 & \ion{Si}{iii} 1113.2061 \\
\ion{Fe}{xix} & 1118.0700 & 7.0 & 1.07e$-$15 & 4.9 & 0.18 &  \\
\ion{Si}{iv} & 1122.4852 & 5.0 & 1.74e$-$15 & 4.9 & 0.53 &  \\
\ion{Si}{iv} & 1128.3405 & 5.0 & 2.74e$-$15 & 6.2 & 0.41 & \ion{Si}{iv} 1128.3252 \\
\ion{Ne}{v} & 1136.5081 & 5.5 & 2.01e$-$16 & 3.6 & 0.02 &  \\
\ion{Fe}{iii} & 1141.7170 & 4.6 & 1.69e$-$16 & 3.9 & 0.45 &  \\
\ion{Ne}{v} & 1145.5959 & 5.5 & 4.22e$-$16 & 4.2 & $-$0.11 &  \\
\ion{C}{iii} & 1176.0000 & 4.8 & 5.24e$-$14 & 21.6 & $-$0.01 &  \\
\ion{C}{iii} & 1176.0000 & 4.8 & 1.74e$-$14 & 4.1 & $-$0.49 &  \\
\ion{S}{iii} & 1190.1990 & 4.9 & 4.99e$-$16 & 8.4 & $-$0.36 &  \\
\ion{Si}{ii} & 1190.4170 & 4.6 & 4.62e$-$16 & 6.5 & 0.30 &  \\
\ion{Si}{ii} & 1193.2910 & 4.6 & 4.15e$-$16 & 6.7 & $-$0.53 &  \\
\ion{S}{iii} & 1194.4430 & 4.9 & 1.65e$-$15 & 8.6 & $-$0.01 & \ion{S}{iii} 1194.0490 \\
\ion{Si}{ii} & 1197.3950 & 4.6 & 3.62e$-$16 & 6.7 & $-$0.29 &  \\
\ion{S}{v} & 1199.1360 & 5.3 & 2.87e$-$16 & 6.6 & $-$0.08 &  \\
\ion{S}{iii} & 1200.9611 & 4.9 & 1.41e$-$15 & 12.8 & 0.22 &  \\
\ion{S}{iii} & 1201.7200 & 4.9 & 3.78e$-$16 & 6.1 & 0.19 &  \\
\ion{O}{ii} & 1204.2345 & 4.9 & 3.66e$-$16 & 9.5 & 0.04 &  \\
\ion{Si}{iii} & 1206.5019 & 4.9 & 3.15e$-$14 & 33.6 & $-$0.68 &  \\
\ion{N}{v} & 1238.8218 & 5.4 & 1.51e$-$14 & 24.4 & 0.09 &  \\
\ion{Fe}{xii} & 1242.0050 & 6.3 & 4.22e$-$16 & 8.1 & 0.18 &  \\
\ion{N}{v} & 1242.8042 & 5.4 & 7.54e$-$15 & 19.5 & 0.09 &  \\
\ion{C}{iii} & 1247.3830 & 5.1 & 1.08e$-$15 & 10.9 & 0.24 &  \\
\ion{S}{ii} & 1250.5870 & 4.6 & 3.57e$-$16 & 4.7 & 0.18 &  \\
\ion{S}{ii} & 1253.8130 & 4.6 & 4.03e$-$16 & 4.8 & $-$0.07 &  \\
\ion{S}{ii} & 1259.5210 & 4.6 & 6.18e$-$16 & 7.2 & $-$0.12 &  \\
\ion{Si}{ii} & 1260.4239 & 4.6 & 1.02e$-$15 & 7.2 & $-$0.59 &  \\
\ion{Si}{ii} & 1264.7400 & 4.5 & 1.59e$-$15 & 10.8 & 0.16 &  \\
\ion{Si}{ii} & 1265.0040 & 4.6 & 7.76e$-$16 & 8.6 & $-$0.00 &  \\
\ion{Si}{iii} & 1303.3250 & 4.9 & 1.76e$-$15 & 13.6 & 0.76 &  \\
\ion{Si}{ii} & 1304.3719 & 4.6 & 4.96e$-$16 & 9.5 & 0.09 &  \\
\ion{Si}{ii} & 1309.2770 & 4.6 & 1.02e$-$15 & 7.9 & 0.11 &  \\
\ion{Si}{iii} & 1312.5930 & 4.9 & 4.14e$-$16 & 5.2 & $-$0.54 &  \\
\ion{C}{ii} & 1323.9080 & 4.8 & 6.49e$-$16 & 6.0 & 0.02 & \ion{C}{ii} 1323.8640, 1323.9540 \\
\ion{Fe}{xix} & 1328.7000 & 7.1 & 2.69e$-$16 & 5.2 & 0.03 &  \\
\ion{C}{ii} & 1334.5350 & 4.7 & 1.83e$-$14 & 21.4 & $-$0.22 &  \\
\ion{C}{ii} & 1335.7100 & 4.7 & 2.98e$-$14 & 27.4 & 0.19 & \ion{C}{ii} 1335.6650 \\
\ion{O}{iv} & 1338.6140 & 5.3 & 1.49e$-$16 & 3.5 & 0.12 &  \\
\ion{Fe}{xxi} & 1354.0800 & 7.2 & 1.22e$-$15 & 5.3 & $-$0.08 &  \\
\ion{Si}{iv} & 1393.7552 & 5.0 & 1.14e$-$14 & 3.3 & $-$0.37 &  \\
\ion{Si}{iv} & 1402.7704 & 5.0 & 7.65e$-$15 & 3.2 & $-$0.24 &  \\
\ion{Si}{ii} & 1526.7090 & 4.5 & 1.51e$-$15 & 8.6 & 0.19 &  \\
\ion{Si}{ii} & 1533.4320 & 4.5 & 1.63e$-$15 & 5.7 & $-$0.08 &  \\
\ion{C}{iv} & 1548.1871 & 5.1 & 3.53e$-$14 & 4.7 & $-$0.20 &  \\
\ion{C}{iv} & 1550.7723 & 5.1 & 2.29e$-$14 & 9.4 & $-$0.08 &  \\
\ion{Al}{ii} & 1670.7870 & 4.6 & 4.14e$-$15 & 11.3 & $-$0.00 &  \\
\hline
\end{tabular}
\tablefoot{\tablefoottext{a}{Line fluxes (in erg cm$^{-2}$ s$^{-1}$) 
  measured in the spectra, and corrected by the ISM absorption when needed. 
  log $T_{\rm max}$ (K) indicates the maximum temperature of formation of the line (unweighted by the
  EMD). ``Ratio'' is the $\log$ ($F_{\mathrm {obs}}$/$F_{\mathrm {pred}}$) of the line. 
    ``Blends'' amounting to more than 5\% of the total flux for each line are indicated, with wavelengths in \AA.}}
\end{scriptsize}
\end{table*}

%
\begin{table*}
\caption{HST/COS and STIS line fluxes of 55 Cnc, HD 97658, Barnard's star, HD 189733, and $\upsilon$ And\tablefootmark{a}.}\label{tab:uvfluxes1}
\tabcolsep 2.0 pt
\begin{scriptsize}
\begin{tabular}{ccccccccccccccccccc}
\hline \hline
Ion & $\lambda_{\rm model}$ & $\log T_{\rm max}$ & \multicolumn{3}{c}{55 Cnc} & \multicolumn{3}{c}{HD 97658} & \multicolumn{3}{c}{Barnard's star} & \multicolumn{3}{c}{HD 189733} & \multicolumn{3}{c}{$\upsilon$ And}  \\
 & (\AA) & (K) & $S/N$ & $F_{\rm obs}$ & Ratio & $S/N$ & $F_{\rm obs}$ & Ratio & $S/N$ & $F_{\rm obs}$ & Ratio & $S/N$ & $F_{\rm obs}$ & Ratio & $S/N$ & $F_{\rm obs}$ & Ratio &  Blends \\
\hline
\ion{Si}{iii} & 1113.2060 & 4.9 & \dots & \dots & \dots & \dots & \dots & \dots & 5.2 & 1.05e$-$16 & $ 0.24$ & \dots & \dots & \dots & \dots & \dots & \dots &  \\
\ion{Ne}{v} & 1145.5959 & 5.5 & \dots & \dots & \dots & \dots & \dots & \dots & 3.5 & 1.92e$-$16 & $-0.01$ & 4.5 & 7.09e$-$17 & $-0.04$ & \dots & \dots & \dots &  \\
\ion{C}{iii} & 1176.0000 & 4.8 & 22.4 & 8.21e$-$15 & $-0.11$ & \dots & \dots & \dots & 6.9 & 5.08e$-$15 & $-0.16$ & \dots & \dots & \dots & \dots & \dots & \dots &  \\
\ion{S}{iii} & 1190.1990 & 4.9 & \dots & \dots & \dots & 40.4 & 5.19e$-$17 & $ 0.08$ & \dots & \dots & \dots & \dots & \dots & \dots & \dots & \dots & \dots & \ion{Si}{ii} 1190.4170 \\
\ion{Si}{ii} & 1193.2910 & 4.6 & 9.5 & 3.78e$-$16 & $-0.54$ & 23.7 & 3.71e$-$17 & $-0.47$ & \dots & \dots & \dots & \dots & \dots & \dots & \dots & \dots & \dots &  \\
\ion{Si}{ii} & 1194.5010 & 4.6 & 16.4 & 8.66e$-$16 & $-0.08$ & 15.7 & 8.23e$-$17 & $-0.10$ & \dots & \dots & \dots & \dots & \dots & \dots & \dots & \dots & \dots & \ion{S}{iii} 1194.4430 \\
\ion{Si}{ii} & 1197.3950 & 4.6 & 18.9 & 3.33e$-$16 & $-0.29$ & 13.2 & 2.79e$-$17 & $-0.28$ & \dots & \dots & \dots & \dots & \dots & \dots & \dots & \dots & \dots &  \\
\ion{S}{v} & 1199.1360 & 5.3 & 5.7 & 2.45e$-$16 & $ 0.43$ & \dots & \dots & \dots & \dots & \dots & \dots & \dots & \dots & \dots & \dots & \dots & \dots &  \\
\ion{S}{iii} & 1200.9611 & 4.9 & 14.8 & 2.62e$-$16 & $ 0.03$ & \dots & \dots & \dots & \dots & \dots & \dots & \dots & \dots & \dots & \dots & \dots & \dots &  \\
\ion{S}{ii} & 1204.3260 & 4.6 & 13.8 & 1.96e$-$16 & $ 1.08$ & \dots & \dots & \dots & \dots & \dots & \dots & \dots & \dots & \dots & \dots & \dots & \dots & \ion{O}{ii} 1204.2345 \\
\ion{Si}{iii} & 1206.5019 & 4.9 & 50.3 & 1.67e$-$14 & $-0.77$ & 18.0 & 1.51e$-$15 & $-0.53$ & 8.0 & 4.00e$-$15 & $-0.80$ & 39.9 & 1.10e$-$14 & $-0.77$ & 11.9 & 7.86e$-$14 & $-0.70$ &  \\
\ion{O}{v} & 1218.3440 & 5.5 & 28.1 & 2.24e$-$15 & $ 0.36$ & 21.2 & 1.81e$-$16 & $ 0.24$ & \dots & \dots & \dots & 39.8 & 2.19e$-$15 & $ 0.18$ & 12.3 & 1.06e$-$14 & $ 0.47$ &  \\
\ion{N}{v} & 1238.8218 & 5.4 & 25.2 & 3.07e$-$15 & $-0.00$ & 14.4 & 2.10e$-$16 & $ 0.01$ & 10.9 & 2.80e$-$15 & $ 0.01$ & 25.9 & 3.32e$-$15 & $ 0.04$ & 5.2 & 9.82e$-$15 & $ 0.02$ &  \\
\ion{Fe}{xii} & 1242.0050 & 6.3 & 18.1 & 3.70e$-$16 & $ 0.10$ & \dots & \dots & \dots & 5.4 & 1.25e$-$16 & $ 0.01$ & \dots & \dots & \dots & \dots & \dots & \dots &  \\
\ion{N}{v} & 1242.8042 & 5.4 & 26.3 & 1.60e$-$15 & $ 0.02$ & 11.4 & 1.03e$-$16 & $ 0.00$ & 8.9 & 1.46e$-$15 & $ 0.03$ & 17.6 & 1.58e$-$15 & $ 0.01$ & 5.9 & 4.87e$-$15 & $ 0.02$ &  \\
\ion{C}{iii} & 1247.3831 & 5.1 & \dots & \dots & \dots & \dots & \dots & \dots & 5.2 & 5.52e$-$17 & $-0.10$ & \dots & \dots & \dots & \dots & \dots & \dots &  \\
\ion{S}{ii} & 1250.5870 & 4.6 & 13.9 & 1.17e$-$16 & $ 0.03$ & 12.7 & 1.82e$-$17 & $ 0.24$ & \dots & \dots & \dots & \dots & \dots & \dots & \dots & \dots & \dots &  \\
\ion{S}{ii} & 1253.8130 & 4.6 & 12.8 & 2.32e$-$16 & $ 0.02$ & 8.5 & 2.87e$-$17 & $ 0.13$ & \dots & \dots & \dots & 9.6 & 1.56e$-$16 & $ 0.14$ & \dots & \dots & \dots &  \\
\ion{S}{ii} & 1259.5210 & 4.6 & 18.1 & 3.94e$-$16 & $ 0.02$ & 9.8 & 1.28e$-$17 & $-0.44$ & 3.8 & 6.93e$-$17 & $-0.02$ & 11.7 & 2.70e$-$16 & $ 0.15$ & 5.5 & 2.07e$-$15 & $-0.02$ &  \\
\ion{Si}{ii} & 1260.4239 & 4.6 & 16.0 & 1.07e$-$15 & $-0.53$ & \dots & \dots & \dots & 5.3 & 1.37e$-$16 & $-0.55$ & \dots & \dots & \dots & \dots & \dots & \dots &  \\
\ion{Si}{ii} & 1260.4240 & 4.6 & \dots & \dots & \dots & 11.0 & 7.44e$-$17 & $-0.63$ & \dots & \dots & \dots & 15.2 & 5.00e$-$16 & $-0.77$ & \dots & \dots & \dots &  \\
\ion{Si}{ii} & 1264.7400 & 4.5 & 27.0 & 3.48e$-$15 & $ 0.30$ & 10.3 & 2.68e$-$16 & $ 0.45$ & 5.8 & 1.48e$-$16 & $ 0.07$ & 18.5 & 1.95e$-$15 & $ 0.13$ & 5.4 & 5.95e$-$15 & $ 0.03$ & \ion{Si}{ii} 1265.0040 \\
\ion{Si}{ii} & 1265.0040 & 4.6 & \dots & \dots & \dots & 8.9 & 1.30e$-$16 & $ 0.32$ & 5.0 & 8.80e$-$17 & $-0.04$ & \dots & \dots & \dots & 4.3 & 2.89e$-$15 & $-0.12$ &  \\
\ion{Si}{iii} & 1294.5480 & 4.9 & 8.8 & 1.15e$-$16 & $ 0.27$ & 4.0 & 1.64e$-$17 & $ 0.72$ & 5.2 & 5.95e$-$17 & $ 0.55$ & \dots & \dots & \dots & \dots & \dots & \dots &  \\
\ion{Si}{iii} & 1296.7280 & 4.9 & 8.8 & 8.62e$-$17 & $-0.26$ & 4.5 & 7.82e$-$18 & $-0.01$ & \dots & \dots & \dots & 6.6 & 1.13e$-$16 & $ 0.05$ & \dots & \dots & \dots &  \\
\ion{Si}{iii} & 1298.9480 & 4.9 & 18.1 & 5.00e$-$16 & $ 0.43$ & 7.7 & 3.45e$-$17 & $ 0.35$ & 5.7 & 1.87e$-$16 & $ 0.36$ & 7.4 & 4.36e$-$16 & $ 0.36$ & 5.3 & 2.92e$-$15 & $ 0.40$ & \ion{Si}{iii} 1296.7280, 1298.8940 \\
\ion{Si}{iii} & 1301.1510 & 4.9 & 14.0 & 2.40e$-$16 & $ 0.66$ & \dots & \dots & \dots & 4.2 & 4.58e$-$17 & $ 0.50$ & \dots & \dots & \dots & \dots & \dots & \dots &  \\
\ion{Si}{iii} & 1303.3249 & 4.9 & 15.9 & 1.58e$-$16 & $-0.09$ & \dots & \dots & \dots & \dots & \dots & \dots & \dots & \dots & \dots & \dots & \dots & \dots &  \\
\ion{Si}{iii} & 1303.3250 & 4.9 & \dots & \dots & \dots & \dots & \dots & \dots & \dots & \dots & \dots & 7.9 & 1.36e$-$16 & $ 0.04$ & \dots & \dots & \dots &  \\
\ion{Si}{ii} & 1304.3719 & 4.6 & 25.5 & 1.45e$-$15 & $ 0.59$ & \dots & \dots & \dots & 4.5 & 3.52e$-$17 & $-0.12$ & \dots & \dots & \dots & 6.3 & 1.90e$-$15 & $-0.02$ &  \\
\ion{Si}{ii} & 1309.2770 & 4.6 & 16.2 & 9.76e$-$16 & $ 0.12$ & 9.7 & 7.66e$-$17 & $ 0.05$ & 5.2 & 1.02e$-$16 & $ 0.04$ & 11.0 & 6.49e$-$16 & $ 0.03$ & 11.5 & 4.92e$-$15 & $ 0.09$ &  \\
\ion{C}{ii} & 1323.9080 & 4.8 & 11.3 & 7.60e$-$17 & $-0.31$ & 5.7 & 1.29e$-$17 & $-0.13$ & \dots & \dots & \dots & 5.9 & 9.90e$-$17 & $-0.26$ & \dots & \dots & \dots & \ion{C}{ii} 1323.8640, 1323.9540 \\
\ion{Fe}{xix} & 1328.7000 & 7.1 & 8.6 & 1.63e$-$16 & $ 0.04$ & \dots & \dots & \dots & \dots & \dots & \dots & \dots & \dots & \dots & \dots & \dots & \dots &  \\
\ion{C}{ii} & 1334.5350 & 4.7 & 47.5 & 1.04e$-$14 & $ 0.00$ & 11.3 & 8.59e$-$16 & $-0.15$ & 12.9 & 2.61e$-$15 & $-0.20$ & 33.6 & 8.96e$-$15 & $-0.17$ & 14.6 & 5.41e$-$14 & $-0.19$ &  \\
\ion{C}{ii} & 1335.7100 & 4.7 & 56.4 & 1.78e$-$14 & $ 0.43$ & 20.9 & 1.94e$-$15 & $ 0.39$ & 16.2 & 4.48e$-$15 & $ 0.24$ & 42.8 & 1.72e$-$14 & $ 0.31$ & 14.5 & 7.56e$-$14 & $ 0.15$ & \ion{C}{ii} 1334.5350, 1335.6650 \\
\ion{Fe}{xii} & 1349.4000 & 6.3 & 20.4 & 1.50e$-$16 & $-0.06$ & \dots & \dots & \dots & \dots & \dots & \dots & \dots & \dots & \dots & \dots & \dots & \dots &  \\
\ion{Fe}{xxi} & 1354.0800 & 7.2 & \dots & \dots & \dots & \dots & \dots & \dots & 7.3 & 4.54e$-$16 & $ 0.02$ & \dots & \dots & \dots & \dots & \dots & \dots &  \\
\ion{Fe}{iii} & 1364.2950 & 4.6 & \dots & \dots & \dots & \dots & \dots & \dots & \dots & \dots & \dots & 9.8 & 1.53e$-$16 & $ 0.41$ & \dots & \dots & \dots &  \\
\ion{O}{v} & 1371.2960 & 5.5 & 20.3 & 2.16e$-$16 & $-0.11$ & 11.1 & 3.28e$-$17 & $ 0.05$ & \dots & \dots & \dots & 11.2 & 2.96e$-$16 & $-0.06$ & \dots & \dots & \dots &  \\
\ion{Si}{iv} & 1393.7552 & 5.0 & 38.5 & 9.08e$-$15 & $ 0.10$ & 14.1 & 7.68e$-$16 & $-0.04$ & 26.6 & 3.38e$-$15 & $ 0.04$ & 25.8 & 6.91e$-$15 & $-0.01$ & 17.3 & 4.03e$-$14 & $-0.02$ &  \\
\ion{O}{iv} & 1397.2310 & 5.3 & 7.3 & 3.52e$-$17 & $-0.17$ & \dots & \dots & \dots & \dots & \dots & \dots & \dots & \dots & \dots & \dots & \dots & \dots &  \\
\ion{O}{iv} & 1399.7800 & 5.3 & 20.7 & 1.26e$-$16 & $-0.07$ & \dots & \dots & \dots & \dots & \dots & \dots & \dots & \dots & \dots & \dots & \dots & \dots &  \\
\ion{O}{iv} & 1401.1570 & 5.3 & 28.5 & 5.94e$-$16 & $ 0.32$ & 6.8 & 7.37e$-$17 & $ 0.03$ & \dots & \dots & \dots & \dots & \dots & \dots & 7.2 & 3.46e$-$15 & $ 0.03$ &  \\
\ion{Si}{iv} & 1402.7704 & 5.0 & 30.1 & 4.61e$-$15 & $ 0.11$ & 11.7 & 3.71e$-$16 & $-0.05$ & 22.2 & 2.42e$-$15 & $-0.39$ & 20.1 & 3.55e$-$15 & $ 0.00$ & 15.3 & 2.16e$-$14 & $ 0.01$ & \ion{O}{iv} 1399.7800, 1401.1570, 1404.8060 \\
\ion{O}{iv} & 1404.8060 & 5.3 & 14.4 & 2.35e$-$16 & $-0.23$ & 24.4 & 3.87e$-$17 & $-0.44$ & \dots & \dots & \dots & \dots & \dots & \dots & 7.2 & 1.31e$-$15 & $-0.58$ & \ion{S}{iv} 1404.8080 \\
\ion{S}{iv} & 1406.0160 & 5.1 & 28.4 & 6.83e$-$17 & $-0.42$ & \dots & \dots & \dots & \dots & \dots & \dots & 5.8 & 6.40e$-$17 & $-0.28$ & \dots & \dots & \dots &  \\
\ion{O}{iv} & 1407.3820 & 5.3 & 22.1 & 1.10e$-$16 & $-0.12$ & \dots & \dots & \dots & \dots & \dots & \dots & \dots & \dots & \dots & \dots & \dots & \dots &  \\
\ion{Si}{ii} & 1526.7090 & 4.5 & \dots & \dots & \dots & 5.8 & 2.36e$-$16 & $ 0.40$ & \dots & \dots & \dots & \dots & \dots & \dots & 11.4 & 1.27e$-$14 & $ 0.40$ &  \\
\ion{Si}{ii} & 1533.4320 & 4.5 & \dots & \dots & \dots & 5.5 & 4.68e$-$16 & $ 0.40$ & \dots & \dots & \dots & \dots & \dots & \dots & 9.6 & 1.30e$-$14 & $ 0.11$ &  \\
\ion{C}{iv} & 1548.1871 & 5.1 & \dots & \dots & \dots & 5.7 & 1.94e$-$15 & $ 0.03$ & 26.9 & 1.14e$-$14 & $ 0.13$ & \dots & \dots & \dots & 11.3 & 8.31e$-$14 & $ 0.06$ &  \\
\ion{C}{iv} & 1550.7723 & 5.1 & \dots & \dots & \dots & 5.6 & 8.93e$-$16 & $-0.01$ & 23.0 & 6.21e$-$15 & $ 0.17$ & \dots & \dots & \dots & 12.3 & 4.24e$-$14 & $ 0.07$ &  \\
\ion{Al}{ii} & 1670.7870 & 4.6 & \dots & \dots & \dots & 6.6 & 1.28e$-$15 & $-0.05$ & 17.0 & 7.69e$-$15 & $ 0.00$ & \dots & \dots & \dots & 8.8 & 3.84e$-$14 & $-0.01$ &  \\
\hline
\end{tabular}
\tablefoot{\tablefoottext{a}{Line fluxes (in erg cm$^{-2}$ s$^{-1}$) 
  measured in the spectra. 
  log $T_{\rm max}$ indicates the maximum temperature of formation of the line (unweighted by the
  EMD). ``Ratio'' is the log($F_{\mathrm {obs}}$/$F_{\mathrm {pred}}$) of the line. 
  ``Blends'', in at least one of the stars, amounting to more than 5\% of the total flux for each line are indicated, with wavelengths in \AA.}}
\end{scriptsize}
\end{table*}

%
\begin{table*}
\caption{HST/COS and STIS line fluxes of GJ 436, HD 209458, GJ 3470, TRAPPIST-1, and GJ 486\tablefootmark{a}.}\label{tab:uvfluxes2}
\tabcolsep 2.5 pt
\begin{scriptsize}
\begin{tabular}{ccccccccccccccccccc}
\hline \hline
Ion & $\lambda_{\rm model}$ & $\log T_{\rm max}$ & \multicolumn{3}{c}{GJ 436} & \multicolumn{3}{c}{HD 209458} & \multicolumn{3}{c}{GJ 3470} & \multicolumn{3}{c}{TRAPPIST-1} & \multicolumn{3}{c}{GJ 486} \\ 
 & (\AA) & (K) & $S/N$ & $F_{\rm obs}$ & Ratio & $S/N$ & $F_{\rm obs}$ & Ratio & $S/N$ & $F_{\rm obs}$ & Ratio & $S/N$ & $F_{\rm obs}$ & Ratio & $S/N$ & $F_{\rm obs}$ & Ratio &  Blends \\
\hline
\ion{C}{iii} & 1176.0000 & 4.8 & \dots & \dots & \dots & \dots & \dots & \dots & 20.0 & 5.41e$-$16 & $ 0.12$ & 4.3 & 2.41e$-$17 & $-0.17$ & 1.0 & 2.58e$-$16 & $-0.30$ &  \\
\ion{Si}{iii} & 1206.5019 & 4.9 & 5.9 & 4.58e$-$16 & $-0.60$ & 36.8 & 1.29e$-$15 & $-0.89$ & 23.5 & 3.40e$-$16 & $-0.57$ & 5.1 & 2.04e$-$17 & $-0.02$ & 3.5 & 9.49e$-$16 & $-0.13$ & \ion{Si}{iii} 1206.5570 \\
\ion{O}{v} & 1218.3440 & 5.5 & \dots & \dots & \dots & 26.5 & 3.97e$-$16 & $ 0.01$ & 23.1 & 1.16e$-$16 & $ 0.00$ & \dots & \dots & \dots & \dots & \dots & \dots &  \\
\ion{N}{v} & 1238.8218 & 5.4 & 11.4 & 5.46e$-$16 & $-0.01$ & \dots & \dots & \dots & 16.9 & 4.02e$-$16 & $ 0.04$ & 4.1 & 3.80e$-$17 & $ 0.05$ & 2.2 & 1.76e$-$16 & $ 0.08$ &  \\
\ion{N}{v} & 1242.8042 & 5.4 & 8.2 & 2.86e$-$16 & $ 0.01$ & 6.5 & 2.60e$-$17 & $ 0.02$ & 21.2 & 1.71e$-$16 & $-0.03$ & 4.5 & 1.62e$-$17 & $-0.02$ & 2.9 & 6.09e$-$17 & $-0.07$ &  \\
\ion{S}{ii} & 1259.5210 & 4.6 & \dots & \dots & \dots & 3.1 & 1.68e$-$16 & $ 0.01$ & \dots & \dots & \dots & \dots & \dots & \dots & \dots & \dots & \dots &  \\
\ion{Si}{ii} & 1264.7400 & 4.5 & 12.5 & 7.57e$-$17 & $ 0.33$ & \dots & \dots & \dots & 13.7 & 4.15e$-$17 & $ 0.19$ & \dots & \dots & \dots & \dots & \dots & \dots &  \\
\ion{Si}{ii} & 1265.0040 & 4.6 & 4.6 & 3.02e$-$17 & $ 0.11$ & \dots & \dots & \dots & 14.0 & 1.83e$-$17 & $ 0.01$ & \dots & \dots & \dots & \dots & \dots & \dots &  \\
\ion{Si}{ii} & 1309.2770 & 4.6 & \dots & \dots & \dots & \dots & \dots & \dots & 14.1 & 1.68e$-$17 & $-0.06$ & \dots & \dots & \dots & \dots & \dots & \dots &  \\
\ion{C}{ii} & 1334.5350 & 4.7 & 7.9 & 2.98e$-$16 & $-0.45$ & 32.2 & 8.04e$-$16 & $-0.21$ & 25.1 & 2.21e$-$16 & $-0.30$ & 5.7 & 2.79e$-$17 & $-0.19$ & \dots & \dots & \dots &  \\
\ion{C}{ii} & 1335.7100 & 4.7 & 9.1 & 8.59e$-$16 & $ 0.20$ & 114.7 & 1.61e$-$15 & $ 0.29$ & 33.3 & 5.61e$-$16 & $ 0.29$ & 4.7 & 4.62e$-$17 & $ 0.22$ & 2.3 & 4.35e$-$16 & $ 0.06$ & \ion{C}{ii} 1335.6650 \\
\ion{Si}{iv} & 1393.7552 & 5.0 & 7.4 & 3.22e$-$16 & $ 0.09$ & 23.7 & 9.47e$-$16 & $ 0.01$ & 24.1 & 2.43e$-$16 & $ 0.03$ & \dots & \dots & \dots & 1.6 & 1.42e$-$16 & $-0.02$ &  \\
\ion{Si}{iv} & 1402.7704 & 5.0 & 7.1 & 1.80e$-$16 & $ 0.14$ & 10.4 & 5.11e$-$16 & $ 0.04$ & 23.6 & 1.10e$-$16 & $-0.01$ & \dots & \dots & \dots & 1.3 & 7.79e$-$17 & $ 0.02$ &  \\
\ion{Si}{ii} & 1526.7090 & 4.5 & \dots & \dots & \dots & 16.5 & 1.19e$-$16 & $ 0.02$ & \dots & \dots & \dots & \dots & \dots & \dots & 1.2 & 5.03e$-$17 & $ 0.44$ &  \\
\ion{Si}{ii} & 1533.4320 & 4.5 & 3.6 & 6.44e$-$17 & $-0.02$ & 16.5 & 2.31e$-$16 & $ 0.01$ & \dots & \dots & \dots & \dots & \dots & \dots & \dots & \dots & \dots &  \\
\ion{C}{iv} & 1548.1871 & 5.1 & 9.3 & 1.40e$-$15 & $-0.01$ & 27.7 & 1.05e$-$15 & $-0.06$ & \dots & \dots & \dots & 3.9 & 1.15e$-$16 & $-0.02$ & 2.8 & 8.90e$-$16 & $ 0.05$ &  \\
\ion{C}{iv} & 1550.7723 & 5.1 & 6.4 & 7.51e$-$16 & $ 0.02$ & 7.4 & 7.07e$-$16 & $ 0.07$ & \dots & \dots & \dots & 3.1 & 6.88e$-$17 & $ 0.06$ & 1.8 & 3.62e$-$16 & $-0.04$ &  \\
\ion{Al}{ii} & 1670.7870 & 4.6 & 6.3 & 3.07e$-$16 & $ 0.00$ & \dots & \dots & \dots & \dots & \dots & \dots & \dots & \dots & \dots & 1.1 & 3.29e$-$16 & $ 0.01$ &  \\
\hline
\end{tabular}
\tablefoot{\tablefoottext{a}{Line fluxes (in erg cm$^{-2}$ s$^{-1}$) measured in the spectra. 
  log $T_{\rm max}$ (K) indicates the maximum temperature of formation of the line (unweighted by the
  EMD). ``Ratio'' is the $\log$ ($F_{\mathrm {obs}}$/$F_{\mathrm {pred}}$) of the line. 
  ``Blends'', in at least one of the stars, amounting to more than 5\% of the total flux for each line are indicated, with wavelengths in \AA.}}
\end{scriptsize}
\end{table*}

%
\begin{table*}
\caption{HST/COS and STIS line fluxes of HD 149026, WASP-52, HD 63433, GJ 357, and TOI-836\tablefootmark{a}.}\label{tab:uvfluxes3}
\tabcolsep 2.5 pt
\begin{scriptsize}
\begin{tabular}{ccccccccccccccccccc}
\hline \hline
Ion & $\lambda_{\rm model}$ & $\log T_{\rm max}$ & \multicolumn{3}{c}{HD 149026} & \multicolumn{3}{c}{WASP 52} & \multicolumn{3}{c}{HD 63433} & \multicolumn{3}{c}{GJ 357} & \multicolumn{3}{c}{TOI-836} \\
 & (\AA) & (K) & $S/N$ & $F_{\rm obs}$ & Ratio & $S/N$ & $F_{\rm obs}$ & Ratio & $S/N$ & $F_{\rm obs}$ & Ratio & $S/N$ & $F_{\rm obs}$ & Ratio & $S/N$ & $F_{\rm obs}$ & Ratio &  Blends \\
\hline
\ion{C}{iii} & 1176.0000 & 4.8 & 1.4 & 2.24e$-$16 & $-0.32$ & \dots & \dots & \dots & \dots & \dots & \dots & \dots & \dots & \dots & 2.3 & 1.24e$-$15 & $-0.56$ &  \\
\ion{Si}{ii} & 1197.3950 & 4.6 & \dots & \dots & \dots & \dots & \dots & \dots & 17.7 & 4.33e$-$16 & $ 0.00$ & \dots & \dots & \dots & \dots & \dots & \dots &  \\
\ion{S}{v} & 1199.1360 & 5.3 & \dots & \dots & \dots & \dots & \dots & \dots & 33.5 & 4.46e$-$16 & $ 0.05$ & \dots & \dots & \dots & \dots & \dots & \dots &  \\
\ion{S}{iii} & 1200.9610 & 4.9 & \dots & \dots & \dots & 5.9 & 2.60e$-$16 & $ 0.03$ & 13.8 & 3.29e$-$16 & $-0.09$ & \dots & \dots & \dots & \dots & \dots & \dots &  \\
\ion{Si}{iii} & 1206.5019 & 4.9 & 1.5 & 4.46e$-$16 & $-0.93$ & 10.5 & 1.88e$-$15 & $ 0.04$ & 68.8 & 3.84e$-$14 & $ 0.04$ & \dots & \dots & \dots & \dots & \dots & \dots &  \\
\ion{O}{v} & 1218.3440 & 5.5 & \dots & \dots & \dots & 2.6 & 3.36e$-$16 & $ 0.03$ & 94.5 & 9.20e$-$15 & $ 0.03$ & \dots & \dots & \dots & \dots & \dots & \dots &  \\
\ion{N}{v} & 1238.8218 & 5.4 & 2.0 & 8.46e$-$17 & $ 0.04$ & 3.4 & 3.08e$-$16 & $ 0.00$ & 36.3 & 4.06e$-$15 & $ 0.01$ & 4.1 & 1.32e$-$16 & $ 0.06$ & 3.2 & 1.33e$-$15 & $ 0.04$ & \\
\ion{Fe}{xii} & 1242.0050 & 6.3 & \dots & \dots & \dots & 1.0 & 2.08e$-$17 & $ 0.00$ & 24.8 & 4.62e$-$16 & $ 0.01$ & \dots & \dots & \dots & \dots & \dots & \dots &  \\
\ion{N}{v} & 1242.8042 & 5.4 & 1.7 & 3.55e$-$17 & $-0.04$ & 1.5 & 4.18e$-$17 & $-0.56$ & 29.7 & 2.14e$-$15 & $ 0.03$ & 4.0 & 5.42e$-$17 & $-0.02$ & 3.0 & 3.29e$-$16 & $-0.26$ & \\
\ion{Si}{ii} & 1264.7400 & 4.5 & \dots & \dots & \dots & \dots & \dots & \dots & \dots & \dots & \dots & \dots & \dots & \dots & 2.7 & 2.03e$-$16 & $-0.38$ & \ion{Si}{ii} 1265.0040 \\
\ion{C}{ii} & 1334.5350 & 4.7 & \dots & \dots & \dots & \dots & \dots & \dots & \dots & \dots & \dots & 4.7 & 1.13e$-$16 & $-0.50$ & \dots & \dots & \dots &  \\
\ion{C}{ii} & 1335.7100 & 4.7 & 7.3 & 1.07e$-$15 & $ 0.30$ & \dots & \dots & \dots & \dots & \dots & \dots & 4.9 & 2.94e$-$16 & $ 0.11$ & 7.1 & 3.38e$-$15 & $ 0.01$ & \ion{C}{ii} 1335.6650 \\
\ion{Si}{iv} & 1393.7552 & 5.0 & 4.7 & 3.89e$-$16 & $ 0.06$ & 2.7 & 5.80e$-$17 & $-0.07$ & \dots & \dots & \dots & \dots & \dots & \dots & 3.8 & 7.19e$-$16 & $-0.08$ &  \\
\ion{O}{iv} & 1401.1570 & 5.3 & 1.4 & 2.69e$-$17 & $-0.30$ & \dots & \dots & \dots & \dots & \dots & \dots & \dots & \dots & \dots & \dots & \dots & \dots &  \\
\ion{Si}{iv} & 1402.7704 & 5.0 & \dots & \dots & \dots & 2.0 & 3.27e$-$17 & $-0.02$ & \dots & \dots & \dots & 1.8 & 5.11e$-$17 & $ 0.02$ & \dots & \dots & \dots &  \\
\ion{O}{iv} & 1404.8060 & 5.3 & 2.4 & 1.37e$-$16 & $ 0.26$ & \dots & \dots & \dots & \dots & \dots & \dots & \dots & \dots & \dots & \dots & \dots & \dots &  \\
\ion{S}{iv} & 1406.0160 & 5.1 & 2.5 & 2.14e$-$16 & $ 0.06$ & \dots & \dots & \dots & \dots & \dots & \dots & \dots & \dots & \dots & \dots & \dots & \dots & \ion{O}{iv} 1407.3820 \\
\ion{Si}{ii} & 1533.4320 & 4.5 & \dots & \dots & \dots & \dots & \dots & \dots & \dots & \dots & \dots & \dots & \dots & \dots & 1.7 & 3.88e$-$16 & $-0.11$ &  \\
\ion{C}{iv} & 1548.1871 & 5.1 & 9.3 & 1.33e$-$15 & $ 0.01$ & \dots & \dots & \dots & \dots & \dots & \dots & 2.8 & 5.55e$-$16 & $-0.01$ & 4.9 & 3.18e$-$15 & $-0.03$ & \ion{C}{iv} 1550.7723 \\
\ion{C}{iv} & 1550.7723 & 5.1 & \dots & \dots & \dots & \dots & \dots & \dots & \dots & \dots & \dots & 2.3 & 1.82e$-$16 & $-0.19$ & 4.6 & 1.82e$-$15 & $ 0.03$ &  \\
\ion{Al}{ii} & 1670.7870 & 4.6 & \dots & \dots & \dots & \dots & \dots & \dots & \dots & \dots & \dots & 3.2 & 1.95e$-$16 & $-0.01$ & 2.2 & 8.05e$-$16 & $-0.03$ &  \\
\hline
\end{tabular}
\tablefoot{\tablefoottext{a}{Line fluxes (in erg cm$^{-2}$ s$^{-1}$) 
  measured in the spectra. 
  log $T_{\rm max}$ (K) indicates the maximum temperature of formation of the line (unweighted by the
  EMD). ``Ratio'' is the log($F_{\mathrm {obs}}$/$F_{\mathrm {pred}}$) of the line. 
  ``Blends'', in at least one of the stars, amounting to more than 5\% of the total flux for each line are indicated, with wavelengths in \AA.}}
\end{scriptsize}
\end{table*}

%
\begin{table*}
\caption{HST/COS and STIS line fluxes of GJ 1214, WASP-77A, HD 73583, WASP-13 and GJ 9827\tablefootmark{a}.}\label{tab:uvfluxes4}
\tabcolsep 2.5 pt
\begin{scriptsize}
\begin{tabular}{ccccccccccccccccccc}
\hline \hline
Ion & $\lambda_{\rm model}$ & $\log T_{\rm max}$ & \multicolumn{3}{c}{GJ 1214} & \multicolumn{3}{c}{WASP-77A} & \multicolumn{3}{c}{HD 73583} & \multicolumn{3}{c}{WASP-13} & \multicolumn{3}{c}{GJ 9827} \\ 
 & (\AA) & (K) & $S/N$ & $F_{\rm obs}$ & Ratio & $S/N$ & $F_{\rm obs}$ & Ratio & $S/N$ & $F_{\rm obs}$ & Ratio & $S/N$ & $F_{\rm obs}$ & Ratio & $S/N$ & $F_{\rm obs}$ & Ratio &  Blends \\
\hline
\ion{C}{iii} & 1176.0000 & 4.8 & 3.7 & 5.10e$-$17 & $-0.16$ & 2.8 & 6.19e$-$16 & $-0.07$ & \dots & \dots & \dots & \dots & \dots & \dots & \dots & \dots & \dots &  \\
\ion{Si}{iii} & 1206.5019 & 4.9 & 3.4 & 5.39e$-$17 & $-0.43$ & \dots & \dots & \dots & 14.6 & 1.52e$-$15 & $-0.01$ & \dots & \dots & \dots & 4.8 & 1.44e$-$16 & $-0.02$ &  \\
\ion{O}{v} & 1218.3440 & 5.5 & 3.6 & 3.77e$-$17 & $ 0.04$ & \dots & \dots & \dots & 10.8 & 5.80e$-$16 & $-0.00$ & \dots & \dots & \dots & \dots & \dots & \dots &  \\
\ion{N}{v} & 1238.8218 & 5.4 & 3.7 & 9.79e$-$17 & $-0.02$ & \dots & \dots & \dots & 11.8 & 5.36e$-$16 & $ 0.02$ & \dots & \dots & \dots & \dots & \dots & \dots &  \\
\ion{N}{v} & 1242.8042 & 5.4 & 3.5 & 5.29e$-$17 & $ 0.02$ & \dots & \dots & \dots & 8.3 & 2.38e$-$16 & $-0.03$ & \dots & \dots & \dots & \dots & \dots & \dots &  \\
\ion{C}{ii} & 1334.5350 & 4.7 & 3.5 & 3.77e$-$17 & $-0.26$ & \dots & \dots & \dots & \dots & \dots & \dots & \dots & \dots & \dots & \dots & \dots & \dots &  \\
\ion{C}{ii} & 1335.7100 & 4.7 & 3.3 & 8.45e$-$17 & $ 0.29$ & 3.2 & 7.45e$-$16 & $ 0.02$ & \dots & \dots & \dots & 4.7 & 8.36e$-$17 & $ 0.00$ & \dots & \dots & \dots & \ion{C}{ii} 1334.5350, 1335.6650 \\
\ion{Si}{iv} & 1393.7552 & 5.0 & 4.3 & 3.27e$-$17 & $ 0.10$ & 4.3 & 5.80e$-$16 & $ 0.18$ & \dots & \dots & \dots & \dots & \dots & \dots & \dots & \dots & \dots &  \\
\ion{O}{iv} & 1399.7800 & 5.3 & \dots & \dots & \dots & 1.7 & 7.85e$-$17 & $ 0.05$ & \dots & \dots & \dots & \dots & \dots & \dots & \dots & \dots & \dots &  \\
\ion{Si}{iv} & 1402.7704 & 5.0 & \dots & \dots & \dots & 4.6 & 1.99e$-$16 & $-0.22$ & \dots & \dots & \dots & \dots & \dots & \dots & \dots & \dots & \dots & \ion{O}{iv} 1401.1570 \\
\ion{O}{iv} & 1404.8060 & 5.3 & \dots & \dots & \dots & 3.4 & 1.90e$-$16 & $ 0.00$ & \dots & \dots & \dots & \dots & \dots & \dots & \dots & \dots & \dots &  \\
\ion{C}{iv} & 1548.1871 & 5.1 & 3.2 & 2.90e$-$16 & $ 0.02$ & 4.0 & 1.41e$-$15 & $ 0.03$ & \dots & \dots & \dots & 4.2 & 1.11e$-$16 & $ 0.04$ & \dots & \dots & \dots & \ion{C}{iv} 1550.7723 \\
\ion{C}{iv} & 1550.7723 & 5.1 & \dots & \dots & \dots & \dots & \dots & \dots & \dots & \dots & \dots & 4.5 & 4.61e$-$17 & $-0.04$ & \dots & \dots & \dots &  \\
\hline
\end{tabular}
\tablefoot{\tablefoottext{a}{Line fluxes (in erg cm$^{-2}$ s$^{-1}$) 
  measured in the spectra. 
  log $T_{\rm max}$ (K) indicates the maximum temperature of formation of the line (unweighted by the
  EMD). ``Ratio'' is the $\log$ ($F_{\mathrm {obs}}$/$F_{\mathrm {pred}}$) of the line. 
  ``Blends'', in at least one of the stars, amounting to more than 5\% of the total flux for each line are indicated, with wavelengths in \AA.}}
\end{scriptsize}
\end{table*}

\begin{table*}
\caption[]{EMDs of stars with UV and X-ray high-resolution spectra$^a$.}\label{tabemd1}
\tabcolsep 3.pt
\vspace{-5mm}
\renewcommand{\arraystretch}{1.3}  
\begin{center}
\begin{small}
\begin{tabular}{ccccccccccc}
\hline \hline
$\log T$ (K) & Proxima Cen & AU Mic & AD Leo & Lalande 21185 & $\tau$ Boo & GJ 674 & WASP 52$^b$ & 55 Cnc$^b$ & Barnard's star$^b$ & HD 63433$^b$ \\
\hline
4.0 & 48.55$^{+0.30}_{-0.20}$ & 50.70: & 50.50: & 47.70: & 50.70: & 49.75: & 51.30: & 49.80: & 47.20: & 51.10: \\
4.1 & 48.30$^{+0.30}_{-0.30}$ & 50.50$^{+0.30}_{-0.30}$ & 50.30$^{+0.50}_{-0.20}$ & 47.60$^{+0.30}_{-0.30}$ & 50.60: & 49.60$^{+0.30}_{-0.20}$ & 51.10$^{+0.20}_{-0.40}$ & 49.65$^{+0.20}_{-0.30}$ & 47.10$^{+0.20}_{-0.30}$ & 50.90$^{+0.30}_{-0.20}$ \\
4.2 & 48.05$^{+0.20}_{-0.30}$ & 50.40$^{+0.30}_{-0.40}$ & 50.10$^{+0.60}_{-0.00}$ & 47.50$^{+0.30}_{-0.30}$ & 50.50: & 49.35$^{+0.40}_{-0.10}$ & 50.90$^{+0.40}_{-0.20}$ & 49.45$^{+0.20}_{-0.30}$ & 47.00$^{+0.10}_{-0.30}$ & 50.70$^{+0.20}_{-0.20}$ \\
4.3 & 47.90$^{+0.20}_{-0.40}$ & 50.20$^{+0.20}_{-0.40}$ & 49.80$^{+0.40}_{-0.20}$ & 47.40$^{+0.20}_{-0.30}$ & 50.40: & 49.10$^{+0.20}_{-0.20}$ & 50.80$^{+0.40}_{-0.30}$ & 49.10$^{+0.20}_{-0.20}$ & 46.80$^{+0.10}_{-0.30}$ & 50.50$^{+0.10}_{-0.30}$ \\
4.4 & 47.80$^{+0.30}_{-0.30}$ & 50.05$^{+0.30}_{-0.30}$ & 49.45$^{+0.00}_{-0.40}$ & 47.30$^{+0.10}_{-0.30}$ & 50.30: & 48.95$^{+0.10}_{-0.30}$ & 50.70$^{+0.30}_{-0.20}$ & 48.85$^{+0.15}_{-0.15}$ & 46.60$^{+0.10}_{-0.40}$ & 50.25$^{+0.10}_{-0.30}$ \\
4.5 & 47.75$^{+0.40}_{-0.20}$ & 49.80$^{+0.40}_{-0.20}$ & 49.00$^{+0.10}_{-0.40}$ & 47.10$^{+0.10}_{-0.30}$ & 50.20$^{+0.30}_{-0.20}$ & 48.75$^{+0.15}_{-0.15}$ & 50.50$^{+0.10}_{-0.30}$ & 48.80$^{+0.20}_{-0.20}$ & 46.60$^{+0.10}_{-0.30}$ & 50.00$^{+0.10}_{-0.40}$ \\
4.6 & 47.75$^{+0.30}_{-0.30}$ & 49.50$^{+0.30}_{-0.40}$ & 48.75$^{+0.20}_{-0.40}$ & 46.90$^{+0.05}_{-0.25}$ & 50.10$^{+0.20}_{-0.30}$ & 48.70$^{+0.10}_{-0.30}$ & 50.30$^{+0.10}_{-0.30}$ & 48.80$^{+0.30}_{-0.30}$ & 46.65$^{+0.10}_{-0.30}$ & 49.70$^{+0.05}_{-0.35}$ \\
4.7 & 47.80$^{+0.20}_{-0.40}$ & 49.40$^{+0.30}_{-0.30}$ & 48.60$^{+0.00}_{-0.40}$ & 46.95$^{+0.10}_{-0.30}$ & 50.00$^{+0.10}_{-0.30}$ & 48.60$^{+0.00}_{-0.30}$ & 50.05$^{+0.10}_{-0.30}$ & 48.70$^{+0.20}_{-0.20}$ & 46.80$^{+0.20}_{-0.20}$ & 49.50$^{+0.05}_{-0.35}$ \\
4.8 & 47.75$^{+0.30}_{-0.30}$ & 49.35$^{+0.20}_{-0.40}$ & 48.70$^{+0.00}_{-0.40}$ & 46.95$^{+0.10}_{-0.30}$ & 49.95$^{+0.05}_{-0.45}$ & 48.75$^{+0.00}_{-0.40}$ & 49.60$^{+0.10}_{-0.40}$ & 48.60$^{+0.20}_{-0.20}$ & 46.90$^{+0.20}_{-0.00}$ & 49.40$^{+0.05}_{-0.35}$ \\
4.9 & 47.55$^{+0.30}_{-0.40}$ & 49.40$^{+0.20}_{-0.40}$ & 49.30$^{+0.10}_{-0.00}$ & 46.80$^{+0.10}_{-0.20}$ & 49.85$^{+0.20}_{-0.30}$ & 48.95$^{+0.40}_{-0.00}$ & 49.20$^{+0.20}_{-0.40}$ & 48.05$^{+0.15}_{-0.15}$ & 46.40$^{+0.00}_{-0.40}$ & 49.30$^{+0.20}_{-0.30}$ \\
5.0 & 47.35$^{+0.40}_{-0.30}$ & 49.55$^{+0.45}_{-0.35}$ & 49.35$^{+0.10}_{-0.10}$ & 46.70$^{+0.10}_{-0.30}$ & 49.80$^{+0.20}_{-0.30}$ & 48.50$^{+0.20}_{-0.20}$ & 49.00$^{+0.30}_{-0.20}$ & 47.70$^{+0.10}_{-0.30}$ & 46.50$^{+0.20}_{-0.10}$ & 49.35$^{+0.20}_{-0.20}$ \\
5.1 & 47.15$^{+0.35}_{-0.45}$ & 49.55$^{+0.30}_{-0.40}$ & 49.00$^{+0.20}_{-0.20}$ & 46.60$^{+0.10}_{-0.30}$ & 49.85$^{+0.10}_{-0.30}$ & 48.35$^{+0.15}_{-0.25}$ & 48.90$^{+0.40}_{-0.20}$ & 47.60$^{+0.10}_{-0.30}$ & 46.55$^{+0.00}_{-0.40}$ & 49.45$^{+0.30}_{-0.30}$ \\
5.2 & 46.95$^{+0.40}_{-0.20}$ & 49.40$^{+0.30}_{-0.30}$ & 48.65$^{+0.10}_{-0.40}$ & 46.65$^{+0.30}_{-0.20}$ & 49.80$^{+0.15}_{-0.15}$ & 48.30$^{+0.15}_{-0.15}$ & 48.90$^{+0.00}_{-0.40}$ & 47.80$^{+0.10}_{-0.40}$ & 46.30$^{+0.00}_{-0.40}$ & 49.50$^{+0.20}_{-0.30}$ \\
5.3 & 46.80$^{+0.25}_{-0.45}$ & 49.20$^{+0.20}_{-0.40}$ & 48.50$^{+0.10}_{-0.20}$ & 46.70$^{+0.10}_{-0.30}$ & 49.55$^{+0.10}_{-0.20}$ & 48.10$^{+0.15}_{-0.25}$ & 49.10$^{+0.00}_{-0.40}$ & 48.30$^{+0.10}_{-0.10}$ & 46.25$^{+0.10}_{-0.30}$ & 49.35$^{+0.05}_{-0.25}$ \\
5.4 & 46.70$^{+0.40}_{-0.30}$ & 49.10$^{+0.20}_{-0.40}$ & 48.35$^{+0.00}_{-0.40}$ & 46.50$^{+0.20}_{-0.30}$ & 49.10$^{+0.05}_{-0.35}$ & 47.70$^{+0.20}_{-0.20}$ & 49.00$^{+0.20}_{-0.40}$ & 48.60$^{+0.20}_{-0.10}$ & 46.20$^{+0.25}_{-0.15}$ & 49.25$^{+0.05}_{-0.35}$ \\
5.5 & 46.60$^{+0.30}_{-0.40}$ & 48.90$^{+0.20}_{-0.40}$ & 48.25$^{+0.50}_{-0.10}$ & 46.35$^{+0.30}_{-0.20}$ & 49.00$^{+0.20}_{-0.30}$ & 47.65$^{+0.15}_{-0.25}$ & 48.85$^{+0.20}_{-0.40}$ & 48.40$^{+0.30}_{-0.30}$ & 46.15$^{+0.35}_{-0.25}$ & 49.15$^{+0.25}_{-0.25}$ \\
5.6 & 46.55$^{+0.40}_{-0.20}$ & 48.70$^{+0.30}_{-0.40}$ & 48.20$^{+0.40}_{-0.20}$ & 46.40$^{+0.30}_{-0.30}$ & 48.90$^{+0.20}_{-0.40}$ & 47.75$^{+0.25}_{-0.25}$ & 48.90$^{+0.40}_{-0.40}$ & 47.95$^{+0.20}_{-0.20}$ & 46.10$^{+0.30}_{-0.30}$ & 48.90$^{+0.30}_{-0.30}$ \\
5.7 & 46.55$^{+0.30}_{-0.30}$ & 48.80$^{+0.40}_{-0.10}$ & 48.25$^{+0.20}_{-0.30}$ & 46.50$^{+0.40}_{-0.20}$ & 48.90$^{+0.20}_{-0.30}$ & 47.90$^{+0.20}_{-0.30}$ & 49.20$^{+0.40}_{-0.20}$ & 48.00$^{+0.20}_{-0.30}$ & 46.20$^{+0.30}_{-0.20}$ & 48.80$^{+0.20}_{-0.30}$ \\
5.8 & 46.75$^{+0.20}_{-0.20}$ & 49.10$^{+0.25}_{-0.15}$ & 48.30$^{+0.20}_{-0.20}$ & 46.70$^{+0.30}_{-0.20}$ & 49.00$^{+0.20}_{-0.30}$ & 48.00$^{+0.20}_{-0.20}$ & 49.50: & 48.20: & 46.40$^{+0.30}_{-0.20}$ & 48.95: \\
5.9 & 47.15$^{+0.20}_{-0.30}$ & 49.35$^{+0.20}_{-0.20}$ & 48.40$^{+0.20}_{-0.30}$ & 47.00$^{+0.30}_{-0.20}$ & 49.10$^{+0.30}_{-0.30}$ & 48.20$^{+0.30}_{-0.20}$ & 49.90: & 48.50: & 46.60$^{+0.30}_{-0.20}$ & 49.20: \\
6.0 & 47.70$^{+0.40}_{-0.20}$ & 49.70$^{+0.25}_{-0.25}$ & 48.60$^{+0.10}_{-0.40}$ & 47.30$^{+0.30}_{-0.20}$ & 49.30$^{+0.30}_{-0.30}$ & 48.70$^{+0.30}_{-0.20}$ & 50.20: & 48.90: & 46.90$^{+0.30}_{-0.30}$ & 49.55: \\
6.1 & 48.25$^{+0.30}_{-0.20}$ & 50.10$^{+0.20}_{-0.20}$ & 49.00$^{+0.10}_{-0.40}$ & 47.70$^{+0.40}_{-0.20}$ & 49.55$^{+0.25}_{-0.25}$ & 49.50$^{+0.40}_{-0.00}$ & 50.70: & 49.50: & 47.30$^{+0.30}_{-0.20}$ & 49.90: \\
6.2 & 48.75$^{+0.20}_{-0.10}$ & 50.40$^{+0.35}_{-0.15}$ & 50.05$^{+0.20}_{-0.10}$ & 48.00$^{+0.40}_{-0.00}$ & 50.05$^{+0.15}_{-0.35}$ & 49.40$^{+0.00}_{-0.30}$ & 50.90: & 49.80: & 47.70$^{+0.00}_{-0.40}$ & 50.65: \\
6.3 & 48.50$^{+0.20}_{-0.20}$ & 50.70$^{+0.30}_{-0.20}$ & 50.35$^{+0.10}_{-0.10}$ & 48.15$^{+0.20}_{-0.20}$ & 50.35$^{+0.00}_{-0.40}$ & 48.80$^{+0.10}_{-0.40}$ & 50.70: & 49.60: & 47.80$^{+0.20}_{-0.30}$ & 50.85: \\
6.4 & 48.30$^{+0.20}_{-0.30}$ & 50.40$^{+0.30}_{-0.20}$ & 49.65$^{+0.20}_{-0.20}$ & 47.90$^{+0.40}_{-0.20}$ & 50.00$^{+0.20}_{-0.30}$ & 48.65$^{+0.10}_{-0.30}$ & 50.60: & 49.30: & 47.20$^{+0.20}_{-0.20}$ & 50.60: \\
6.5 & 48.40$^{+0.20}_{-0.30}$ & 50.75$^{+0.35}_{-0.15}$ & 49.95$^{+0.40}_{-0.20}$ & 48.20$^{+0.20}_{-0.30}$ & 50.30$^{+0.40}_{-0.20}$ & 48.60$^{+0.20}_{-0.30}$ & 50.50: & 49.40: & 47.10$^{+0.20}_{-0.20}$ & 50.80: \\
6.6 & 48.75$^{+0.20}_{-0.30}$ & 51.00$^{+0.45}_{-0.05}$ & 50.20$^{+0.40}_{-0.10}$ & 48.40$^{+0.15}_{-0.35}$ & 51.15$^{+0.15}_{-0.15}$ & 48.70$^{+0.20}_{-0.20}$ & 50.40: & 49.50: & 47.20$^{+0.30}_{-0.30}$ & 51.31: \\
6.7 & 49.20$^{+0.30}_{-0.10}$ & 51.30$^{+0.10}_{-0.40}$ & 50.40$^{+0.35}_{-0.15}$ & 48.50$^{+0.00}_{-0.30}$ & 51.30$^{+0.05}_{-0.15}$ & 48.90$^{+0.20}_{-0.30}$ & 50.20: & 49.70: & 47.30$^{+0.20}_{-0.20}$ & 40.20: \\
6.8 & 49.40$^{+0.20}_{-0.10}$ & 51.50$^{+0.05}_{-0.25}$ & 50.60$^{+0.10}_{-0.20}$ & 48.60$^{+0.00}_{-0.30}$ & 50.40$^{+0.30}_{-0.20}$ & 49.10$^{+0.00}_{-0.40}$ & 49.70: & 49.80: & 47.50$^{+0.20}_{-0.20}$ & 40.20: \\
6.9 & 49.10$^{+0.20}_{-0.30}$ & 51.80$^{+0.10}_{-0.10}$ & 50.85$^{+0.05}_{-0.25}$ & 48.50$^{+0.20}_{-0.30}$ & 49.80$^{+0.30}_{-0.30}$ & 49.50$^{+0.00}_{-0.30}$ & 49.20: & 49.90: & 47.95$^{+0.30}_{-0.30}$ & 51.23: \\
7.0 & 49.05$^{+0.30}_{-0.20}$ & 51.60$^{+0.25}_{-0.25}$ & 50.80$^{+0.10}_{-0.20}$ & 48.70$^{+0.20}_{-0.20}$ & 49.50$^{+0.30}_{-0.20}$ & 50.00$^{+0.10}_{-0.10}$ & 48.50: & 49.80: & 48.15$^{+0.00}_{-0.20}$ & 40.20: \\
7.1 & 49.25$^{+0.30}_{-0.05}$ & 51.80$^{+0.25}_{-0.05}$ & 50.55$^{+0.15}_{-0.25}$ & 48.80$^{+0.10}_{-0.40}$ & 49.30$^{+0.30}_{-0.30}$ & 49.80$^{+0.20}_{-0.10}$ & 44.00: & 49.50: & 47.70$^{+0.20}_{-0.30}$ & 40.20: \\
7.2 & 48.85$^{+0.20}_{-0.20}$ & 51.20$^{+0.30}_{-0.10}$ & 50.40$^{+0.10}_{-0.30}$ & 48.50$^{+0.20}_{-0.30}$ & 49.00$^{+0.20}_{-0.30}$ & 49.30$^{+0.20}_{-0.20}$ & 44.00: & 49.00: & 47.10$^{+0.20}_{-0.30}$ & 40.20: \\
7.3 & 48.45$^{+0.30}_{-0.20}$ & 50.70$^{+0.30}_{-0.10}$ & 50.20$^{+0.30}_{-0.20}$ & 48.10: & 48.70: & 48.80$^{+0.30}_{-0.20}$ & 44.00: & 48.00: & 46.50$^{+0.30}_{-0.30}$ & 40.20: \\
7.4 & 47.95: & 50.20$^{+0.30}_{-0.20}$ & 49.80$^{+0.30}_{-0.20}$ & 47.30: & 48.30: & 48.10$^{+0.30}_{-0.30}$ & 44.00: & 47.40: & 46.00$^{+0.20}_{-0.30}$ & 40.20: \\
7.5 & 46.95: & 49.80: & 49.50: & 46.80: & 48.00: & 47.60: & 44.00: & 46.00: & 45.50: & 40.20: \\
\hline
\end{tabular}
\end{small}
\end{center}
\vspace{-3mm}
 \tablefoot{
 \tablefoottext{a}{Emission measure (EM=log $\int N_{\rm e} N_{\rm H} {\rm d}V$), where $N_{\rm e}$
  and $N_{\rm H}$ are electron and hydrogen densities, in cm$^{-3}$. Errors provided are not independent
between the different temperatures, as explained in \citet{san03}.}
 \tablefoottext{b}{The EM above $\log T$~(K)=5.9 of these stars are listed to show consistency of UV coronal line (i.e. lines formed above that temperature) fluxes. The model used for the SED above that temperature is based on the X-ray low resolution spectra in these cases.}}
\renewcommand{\arraystretch}{1.}
\end{table*}

\begin{table*}
\caption[]{EMDs of stars with no coronal lines available$^a$.}\label{tabemd2}
\tabcolsep 1.pt
\vspace{-5mm}
\renewcommand{\arraystretch}{1.3}  
\begin{center}
\begin{scriptsize}
\begin{tabular}{ccccccccccccccccc}
\hline \hline
$\log T$ & GJ 357 & GJ 436 & GJ 486 & GJ 1214 & GJ 3470 & GJ 9827 & HD 73583 & HD 97658 & HD 149026 & HD 189733 & HD 209458 & TRAPPIST-1 & $\upsilon$ And & WASP-13 & WASP-77A & TOI-836 \\
\hline
4.0 & 48.15: & 48.65: & 48.00: & 47.80: & 49.90: & 48.75: & 49.80: & 49.70: & 50.20: & 51.01: & 49.70: & 47.20: & 50.85: & 50.10: & 50.50: & 50.25: \\
4.1 & 48.05$^{+0.30}_{-0.20}$ & 48.55$^{+0.40}_{-0.20}$ & 47.90$^{+0.40}_{-0.40}$ & 47.60$^{+0.00}_{-0.40}$ & 49.70$^{+0.30}_{-0.20}$ & 48.60$^{+0.00}_{-0.40}$ & 49.60$^{+0.40}_{-0.20}$ & 49.55$^{+0.40}_{-0.10}$ & 50.10$^{+0.30}_{-0.30}$ & 50.86$^{+0.20}_{-0.30}$ & 49.60$^{+0.20}_{-0.20}$ & 47.10$^{+0.30}_{-0.20}$ & 50.70$^{+0.30}_{-0.20}$ & 49.90$^{+0.20}_{-0.40}$ & 50.40$^{+0.20}_{-0.40}$ & 50.10$^{+0.30}_{-0.30}$ \\
4.2 & 47.90$^{+0.30}_{-0.30}$ & 48.40$^{+0.40}_{-0.20}$ & 47.80$^{+0.40}_{-0.40}$ & 47.40$^{+0.40}_{-0.00}$ & 49.50$^{+0.40}_{-0.00}$ & 48.40$^{+0.40}_{-0.00}$ & 49.40$^{+0.20}_{-0.40}$ & 49.40$^{+0.30}_{-0.10}$ & 50.00$^{+0.40}_{-0.20}$ & 50.71$^{+0.10}_{-0.40}$ & 49.50$^{+0.40}_{-0.10}$ & 47.00$^{+0.40}_{-0.30}$ & 50.50$^{+0.30}_{-0.10}$ & 49.70$^{+0.20}_{-0.20}$ & 50.30$^{+0.30}_{-0.40}$ & 49.90$^{+0.30}_{-0.40}$ \\
4.3 & 47.70$^{+0.00}_{-0.40}$ & 48.20$^{+0.20}_{-0.20}$ & 47.70$^{+0.10}_{-0.30}$ & 47.20$^{+0.00}_{-0.40}$ & 49.30$^{+0.30}_{-0.00}$ & 48.20$^{+0.10}_{-0.30}$ & 49.20$^{+0.20}_{-0.40}$ & 49.10$^{+0.10}_{-0.30}$ & 49.85$^{+0.40}_{-0.20}$ & 50.51$^{+0.10}_{-0.30}$ & 49.40$^{+0.40}_{-0.00}$ & 46.90$^{+0.40}_{-0.20}$ & 50.25$^{+0.20}_{-0.20}$ & 49.50$^{+0.20}_{-0.20}$ & 50.15$^{+0.20}_{-0.40}$ & 49.70$^{+0.20}_{-0.40}$ \\
4.4 & 47.45$^{+0.00}_{-0.50}$ & 47.95$^{+0.15}_{-0.35}$ & 47.55$^{+0.10}_{-0.30}$ & 47.05$^{+0.40}_{-0.00}$ & 49.00$^{+0.10}_{-0.40}$ & 48.00$^{+0.10}_{-0.30}$ & 49.00$^{+0.20}_{-0.40}$ & 48.65$^{+0.10}_{-0.30}$ & 49.70$^{+0.40}_{-0.10}$ & 50.41$^{+0.10}_{-0.30}$ & 49.35$^{+0.20}_{-0.00}$ & 46.80$^{+0.30}_{-0.20}$ & 50.10$^{+0.10}_{-0.30}$ & 49.35$^{+0.20}_{-0.20}$ & 50.00$^{+0.10}_{-0.30}$ & 49.60$^{+0.20}_{-0.30}$ \\
4.5 & 47.10$^{+0.00}_{-0.40}$ & 47.50$^{+0.10}_{-0.30}$ & 47.35$^{+0.10}_{-0.30}$ & 46.90$^{+0.00}_{-0.40}$ & 48.65$^{+0.10}_{-0.50}$ & 47.80$^{+0.10}_{-0.30}$ & 48.80$^{+0.20}_{-0.40}$ & 48.30$^{+0.10}_{-0.30}$ & 49.50$^{+0.40}_{-0.10}$ & 50.16$^{+0.30}_{-0.20}$ & 49.25$^{+0.10}_{-0.40}$ & 46.60$^{+0.20}_{-0.30}$ & 49.85$^{+0.10}_{-0.40}$ & 49.10$^{+0.20}_{-0.20}$ & 49.90$^{+0.10}_{-0.30}$ & 49.55$^{+0.00}_{-0.40}$ \\
4.6 & 46.95$^{+0.00}_{-0.40}$ & 47.25$^{+0.10}_{-0.30}$ & 47.05$^{+0.10}_{-0.30}$ & 46.70$^{+0.00}_{-0.40}$ & 48.35$^{+0.10}_{-0.50}$ & 47.55$^{+0.10}_{-0.30}$ & 48.65$^{+0.10}_{-0.30}$ & 48.15$^{+0.10}_{-0.30}$ & 49.20$^{+0.40}_{-0.20}$ & 50.01$^{+0.10}_{-0.40}$ & 49.10$^{+0.10}_{-0.30}$ & 46.40$^{+0.10}_{-0.30}$ & 49.70$^{+0.10}_{-0.30}$ & 48.85$^{+0.20}_{-0.20}$ & 49.70$^{+0.10}_{-0.30}$ & 49.45$^{+0.00}_{-0.40}$ \\
4.7 & 46.85$^{+0.00}_{-0.45}$ & 47.20$^{+0.10}_{-0.30}$ & 46.90$^{+0.10}_{-0.30}$ & 46.50$^{+0.40}_{-0.00}$ & 48.10$^{+0.10}_{-0.40}$ & 47.30$^{+0.10}_{-0.30}$ & 48.45$^{+0.10}_{-0.30}$ & 48.05$^{+0.10}_{-0.20}$ & 48.80$^{+0.40}_{-0.30}$ & 49.81$^{+0.10}_{-0.30}$ & 48.85$^{+0.10}_{-0.50}$ & 46.05$^{+0.10}_{-0.30}$ & 49.60$^{+0.10}_{-0.30}$ & 48.80$^{+0.20}_{-0.20}$ & 49.40$^{+0.20}_{-0.40}$ & 49.10$^{+0.20}_{-0.40}$ \\
4.8 & 46.90$^{+0.00}_{-0.50}$ & 47.15$^{+0.10}_{-0.30}$ & 46.80$^{+0.10}_{-0.30}$ & 46.40$^{+0.00}_{-0.40}$ & 48.10$^{+0.10}_{-0.40}$ & 47.20$^{+0.10}_{-0.30}$ & 48.30$^{+0.10}_{-0.30}$ & 48.00$^{+0.10}_{-0.20}$ & 48.55$^{+0.20}_{-0.30}$ & 49.81$^{+0.20}_{-0.10}$ & 48.55$^{+0.10}_{-0.40}$ & 45.85$^{+0.10}_{-0.30}$ & 49.45$^{+0.05}_{-0.15}$ & 48.85$^{+0.20}_{-0.20}$ & 49.00$^{+0.20}_{-0.30}$ & 48.50$^{+0.10}_{-0.40}$ \\
4.9 & 46.85$^{+0.00}_{-0.40}$ & 47.20$^{+0.10}_{-0.30}$ & 46.75$^{+0.10}_{-0.30}$ & 46.45$^{+0.00}_{-0.40}$ & 48.35$^{+0.10}_{-0.20}$ & 47.25$^{+0.10}_{-0.30}$ & 48.20$^{+0.20}_{-0.20}$ & 48.25$^{+0.10}_{-0.10}$ & 48.50$^{+0.10}_{-0.30}$ & 49.76$^{+0.05}_{-0.25}$ & 48.60$^{+0.20}_{-0.00}$ & 45.95$^{+0.10}_{-0.30}$ & 49.40$^{+0.10}_{-0.10}$ & 48.90$^{+0.20}_{-0.30}$ & 48.90$^{+0.10}_{-0.40}$ & 48.40$^{+0.00}_{-0.40}$ \\
5.0 & 46.75$^{+0.20}_{-0.40}$ & 47.20$^{+0.20}_{-0.20}$ & 46.80$^{+0.40}_{-0.40}$ & 46.65$^{+0.00}_{-0.40}$ & 48.50$^{+0.40}_{-0.20}$ & 47.30$^{+0.20}_{-0.40}$ & 48.20$^{+0.20}_{-0.40}$ & 48.10$^{+0.10}_{-0.20}$ & 48.85$^{+0.00}_{-0.40}$ & 49.51$^{+0.05}_{-0.35}$ & 48.55$^{+0.10}_{-0.10}$ & 46.20$^{+0.10}_{-0.30}$ & 49.35$^{+0.05}_{-0.15}$ & 48.85$^{+0.10}_{-0.30}$ & 49.20$^{+0.15}_{-0.15}$ & 48.45$^{+0.00}_{-0.40}$ \\
5.1 & 46.70$^{+0.40}_{-0.20}$ & 47.15$^{+0.20}_{-0.10}$ & 46.75$^{+0.30}_{-0.10}$ & 46.95$^{+0.40}_{-0.00}$ & 48.40$^{+0.25}_{-0.15}$ & 47.50$^{+0.40}_{-0.40}$ & 48.25$^{+0.40}_{-0.30}$ & 47.75$^{+0.10}_{-0.30}$ & 48.80$^{+0.10}_{-0.20}$ & 49.11$^{+0.10}_{-0.40}$ & 48.50$^{+0.00}_{-0.40}$ & 46.45$^{+0.20}_{-0.30}$ & 48.85$^{+0.10}_{-0.30}$ & 48.75$^{+0.10}_{-0.20}$ & 48.95$^{+0.15}_{-0.25}$ & 48.50$^{+0.10}_{-0.20}$ \\
5.2 & 46.75$^{+0.00}_{-0.40}$ & 47.20$^{+0.20}_{-0.30}$ & 46.70$^{+0.40}_{-0.40}$ & 47.00$^{+0.40}_{-0.40}$ & 48.20$^{+0.00}_{-0.30}$ & 47.70$^{+0.30}_{-0.10}$ & 48.35$^{+0.20}_{-0.40}$ & 47.55$^{+0.10}_{-0.30}$ & 48.40$^{+0.30}_{-0.40}$ & 48.81$^{+0.20}_{-0.20}$ & 48.20$^{+0.00}_{-0.30}$ & 46.40$^{+0.10}_{-0.40}$ & 48.60$^{+0.10}_{-0.30}$ & 48.50$^{+0.40}_{-0.20}$ & 48.60$^{+0.10}_{-0.40}$ & 48.40$^{+0.30}_{-0.20}$ \\
5.3 & 46.65$^{+0.00}_{-0.40}$ & 47.15$^{+0.05}_{-0.35}$ & 46.60$^{+0.00}_{-0.40}$ & 46.90$^{+0.00}_{-0.40}$ & 48.15$^{+0.00}_{-0.30}$ & 47.65$^{+0.30}_{-0.10}$ & 48.40$^{+0.10}_{-0.30}$ & 47.60$^{+0.10}_{-0.30}$ & 48.20$^{+0.40}_{-0.20}$ & 48.61$^{+0.10}_{-0.30}$ & 47.80$^{+0.00}_{-0.20}$ & 46.30$^{+0.00}_{-0.40}$ & 48.60$^{+0.10}_{-0.30}$ & 48.35$^{+0.40}_{-0.40}$ & 48.50$^{+0.10}_{-0.40}$ & 48.30$^{+0.20}_{-0.20}$ \\
5.4 & 46.50$^{+0.20}_{-0.40}$ & 47.00$^{+0.20}_{-0.30}$ & 46.50$^{+0.20}_{-0.00}$ & 46.65$^{+0.40}_{-0.00}$ & 48.00$^{+0.00}_{-0.20}$ & 47.55$^{+0.10}_{-0.40}$ & 48.25$^{+0.10}_{-0.40}$ & 47.95$^{+0.20}_{-0.10}$ & 48.10$^{+0.30}_{-0.20}$ & 48.31$^{+0.10}_{-0.30}$ & 47.65$^{+0.10}_{-0.20}$ & 46.20$^{+0.20}_{-0.40}$ & 48.95$^{+0.30}_{-0.10}$ & 48.20$^{+0.40}_{-0.40}$ & 48.40$^{+0.20}_{-0.40}$ & 48.10$^{+0.40}_{-0.20}$ \\
5.5 & 46.30$^{+0.30}_{-0.20}$ & 46.50$^{+0.30}_{-0.30}$ & 46.40$^{+0.40}_{-0.20}$ & 46.50: & 47.80$^{+0.20}_{-0.10}$ & 47.40$^{+0.30}_{-0.10}$ & 48.00$^{+0.30}_{-0.20}$ & 47.90$^{+0.40}_{-0.20}$ & 48.00$^{+0.20}_{-0.30}$ & 48.21$^{+0.20}_{-0.30}$ & 47.40$^{+0.20}_{-0.20}$ & 46.00$^{+0.20}_{-0.20}$ & 48.75$^{+0.60}_{-0.10}$ & 48.00$^{+0.20}_{-0.40}$ & 48.30$^{+0.30}_{-0.20}$ & 47.95$^{+0.30}_{-0.30}$ \\
5.6 & 46.25$^{+0.30}_{-0.40}$ & 46.20$^{+0.30}_{-0.20}$ & 46.30$^{+0.40}_{-0.40}$ & 46.40: & 47.60$^{+0.20}_{-0.20}$ & 47.10$^{+0.30}_{-0.10}$ & 47.80$^{+0.40}_{-0.20}$ & 47.45$^{+0.20}_{-0.20}$ & 47.90$^{+0.40}_{-0.30}$ & 48.16$^{+0.20}_{-0.30}$ & 47.25$^{+0.20}_{-0.20}$ & 45.80$^{+0.30}_{-0.30}$ & 48.30$^{+0.20}_{-0.20}$ & 47.85: & 48.20$^{+0.30}_{-0.30}$ & 47.80$^{+0.20}_{-0.40}$ \\
5.7 & 46.20$^{+0.30}_{-0.30}$ & 45.90$^{+0.30}_{-0.30}$ & 46.20$^{+0.40}_{-0.00}$ & 46.20$^{+0.00}_{-0.40}$ & 47.50$^{+0.20}_{-0.10}$ & 46.75$^{+0.05}_{-0.15}$ & 47.65$^{+0.20}_{-0.40}$ & 47.15$^{+0.20}_{-0.20}$ & 47.80$^{+0.20}_{-0.30}$ & 48.26$^{+0.30}_{-0.20}$ & 47.20$^{+0.20}_{-0.20}$ & 45.70$^{+0.30}_{-0.30}$ & 48.05$^{+0.20}_{-0.20}$ & 47.70: & 48.10$^{+0.40}_{-0.30}$ & 47.90$^{+0.30}_{-0.30}$ \\
5.8 & 46.10: & 45.70: & 46.00: & 45.80: & 47.70: & 46.60: & 47.60: & 47.00: & 47.70: & 48.51: & 47.00: & 45.60: & 47.80: & 47.85: & 48.00: & 48.10: \\
5.9 & 46.20: & 45.80: & 46.00: & 46.00: & 48.00: & 46.55: & 47.55: & 47.10: & 47.80: & 48.81: & 47.10: & 45.70: & 47.90: & 48.00: & 47.80: & 48.30: \\
\hline
\end{tabular}
\end{scriptsize}
\end{center}
\vspace{-3mm}
 \tablefoot{
 \tablefoottext{a}{The EMD in this table is based on just UV high-resolution spectra, for cases with no lines fromed at coronal temperatures.}}
\renewcommand{\arraystretch}{1.}
\end{table*}

\begin{table*}
\caption[]{Stellar coronal abundances in the EMD temperature range for stars with UV and X-ray high resolution data\tablefootmark{a}.}\label{tababund1}
\tabcolsep 4.5pt
\vspace{-5mm}
\begin{center}
\begin{scriptsize}
  \begin{tabular}{lrcccccccccccccccc}
\hline \hline
Element & FIP & Proxima Cen & AU Mic & AD Leo & Lalande 21185 & $\tau$ Boo & GJ 674 & WASP 52 & 55 Cnc & Barnard's star & HD 63433 \\
\hline
Na & 5.14 & $ 0.75\pm0.06$ & \dots & $-0.15\pm0.27$ & \dots & \dots & \dots & \dots & \dots & \dots & \dots \\
Al & 5.98 & $ 0.20\pm0.05$ & $ 0.15\pm0.06$ & $-0.30\pm0.30$ & $ 1.05\pm0.33$ & \dots & $-0.20\pm0.09$ & \dots & \dots & $ 1.50\pm0.06$ & \dots \\
Ca & 6.11 & $ 0.75\pm0.10$ & $ 0.70\pm0.18$ & \dots & \dots & \dots & \dots & \dots & \dots & \dots & \dots \\
Cr & 6.77 & \dots & $ 0.15\pm0.15$ & \dots & \dots & \dots & \dots & \dots & \dots & \dots & \dots \\
Mn & 7.43 & \dots & \dots & $ 0.65\pm0.24$ & \dots & \dots & \dots & \dots & \dots & \dots & \dots \\
Ni & 7.63 & $-0.15\pm0.08$ & $-0.90\pm0.22$ & $-0.30\pm0.13$ & \dots & $ 0.50\pm0.15$ & \dots & \dots & \dots & \dots & \dots \\
Mg & 7.64 & $-0.20\pm0.09$ & $-0.80\pm0.20$ & $-0.40\pm0.15$ & \dots & \dots & \dots & \dots & \dots & \dots & \dots \\
Fe & 7.87 & $-0.50\pm0.05$ & $-0.80\pm0.05$ & $-0.60\pm0.04$ & $-0.50\pm0.12$ & $-0.30\pm0.10$ & $-0.60\pm0.12$ & $ 0.00\pm1.04$ & $ 0.00\pm0.09$ & $ 0.00\pm0.12$ & $-0.20\pm0.04$ \\
Si & 8.15 & $ 0.15\pm0.10$ & $ 0.55\pm0.10$ & $ 0.10\pm0.09$ & $ 0.70\pm0.13$ & $-0.25\pm0.04$ & $ 0.00\pm0.12$ & $ 0.00\pm0.22$ & $ 0.85\pm0.14$ & $ 0.45\pm0.16$ & $-0.05\pm0.05$ \\
S & 10.36 & $ 0.30\pm0.10$ & $-0.10\pm0.10$ & $ 0.05\pm0.11$ & $ 0.35\pm0.24$ & $ 0.00\pm0.06$ & $-0.30\pm0.13$ & $ 1.20\pm0.18$ & $ 0.25\pm0.13$ & $ 0.00\pm0.27$ & $ 0.00\pm0.12$ \\
C & 11.26 & $-0.10\pm0.12$ & $ 0.10\pm0.10$ & $-0.10\pm0.13$ & $ 0.40\pm0.16$ & $-0.15\pm0.20$ & $-0.40\pm0.14$ & \dots & $ 0.05\pm0.21$ & $ 0.00\pm0.14$ & \dots \\
O & 13.61 & $-0.10\pm0.10$ & $-0.10\pm0.08$ & $ 0.00\pm0.08$ & $ 0.10\pm0.13$ & $-0.30\pm0.13$ & $-0.15\pm0.10$ & $ 0.50\pm0.39$ & $-0.70\pm0.14$ & \dots & $-0.15\pm0.06$ \\
N & 14.53 & $ 0.35\pm0.14$ & $ 0.05\pm0.13$ & $ 0.30\pm0.16$ & $ 0.05\pm0.27$ & $-0.30\pm0.11$ & $-0.10\pm0.18$ & $ 0.50\pm0.51$ & $ 0.00\pm0.03$ & $ 0.30\pm0.08$ & $-0.45\pm0.04$ \\
Ar & 15.76 & $ 0.25\pm0.18$ & $ 0.20\pm0.15$ & $ 0.45\pm0.06$ & $ 1.00\pm0.26$ & \dots & \dots & \dots & \dots & \dots & \dots \\
Ne & 21.56 & $ 0.15\pm0.10$ & $ 0.20\pm0.12$ & $ 0.10\pm0.08$ & $ 0.15\pm0.17$ & $-0.35\pm0.11$ & $-0.20\pm0.17$ & \dots & \dots & $ 0.25\pm0.29$ & \dots \\
\hline
\end{tabular}
  \tablefoot{\tablefoottext{a}{Element abundances and First Ionization Potential (FIP) based on solar photospheric values by \citet{and89}}}
\end{scriptsize}
\end{center}
\end{table*}

\begin{table*}
\caption[]{Stellar coronal abundances in the EMD temperature range for stars with no coronal lines\tablefootmark{a}.}\label{tababund2}
\vspace{-5mm}
\begin{center}
\begin{scriptsize}
  \begin{tabular}{lccccccc}
\hline \hline
Element & Al & Si & S & C & O & N & Ne \\
FIP (eV) & 5.98 & 8.15 & 10.36 & 11.26 & 13.61 & 14.53 & 21.56 \\
\hline
GJ 357 & $ 0.65\pm0.31$ & $ 0.15\pm0.54$ & \dots & $ 0.00\pm0.26$ & \dots & $ 0.00\pm0.19$ & \dots \\
GJ 436 & $ 0.40\pm0.16$ & $ 0.25\pm0.17$ & \dots & $ 0.00\pm0.18$ & \dots & $ 0.25\pm0.08$ & \dots \\
GJ 486 & $ 0.65\pm0.91$ & $ 0.25\pm0.38$ & \dots & $ 0.00\pm0.34$ & \dots & $ 0.00\pm0.31$ & \dots \\
GJ 1214 & \dots & $ 0.20\pm0.38$ & \dots & $ 0.00\pm0.23$ & $-0.40\pm0.30$ & $ 0.10\pm0.20$ & \dots \\
GJ 3470 & \dots & $ 0.00\pm0.14$ & \dots & $-0.40\pm0.25$ & $-0.55\pm0.04$ & $ 0.00\pm0.07$ & \dots \\
GJ 9827 & \dots & $ 0.00\pm0.21$ & \dots & \dots & \dots & $ 0.00\pm0.30$ & \dots \\
HD 73583 & \dots & $ 0.00\pm0.07$ & \dots & \dots & $ 0.00\pm0.09$ & $ 0.01\pm0.08$ & \dots \\
HD 97658 & $ 0.90\pm0.19$ & $ 0.50\pm0.14$ & $ 0.00\pm0.21$ & $ 0.00\pm0.15$ & $-0.60\pm0.20$ & $-0.10\pm0.06$ & \dots \\
HD 149026 & \dots & $ 0.70\pm0.57$ & $ 1.10\pm0.41$ & $ 0.00\pm0.34$ & $-0.40\pm0.54$ & $ 0.00\pm0.39$ & \dots \\
HD 189733 & \dots & $-0.21\pm0.15$ & $-1.01\pm0.22$ & $-0.76\pm0.26$ & $-0.26\pm0.19$ & $ 0.04\pm0.06$ & $-0.31\pm0.23$ \\
HD 209458 & \dots & $ 0.80\pm0.18$ & $ 0.70\pm0.32$ & $ 0.00\pm0.17$ & $ 0.65\pm0.04$ & $-0.30\pm0.16$ & \dots \\
TRAPPIST-1 & \dots & $-0.35\pm0.20$ & \dots & $ 0.00\pm0.17$ & \dots & $ 0.05\pm0.18$ & \dots \\
$\upsilon$ And & $ 0.60\pm0.12$ & $ 0.55\pm0.13$ & $ 0.00\pm0.19$ & $ 0.00\pm0.11$ & $-0.45\pm0.34$ & $ 0.15\pm0.13$ & \dots \\
WASP-13 & \dots & \dots & \dots & $ 0.00\pm0.14$ & \dots & \dots & \dots \\
WASP-77A & \dots & $ 0.65\pm0.30$ & \dots & $ 0.00\pm0.19$ & $ 0.05\pm0.33$ & \dots & \dots \\
TOI-836 & $ 0.00\pm0.47$ & $ 0.35\pm0.32$ & \dots & $ 0.00\pm0.23$ & \dots & $ 0.00\pm0.33$ & \dots \\
\hline
\end{tabular}
  \tablefoot{\tablefoottext{a}{Element abundances and FIP based on solar photospheric values by \citet{and89}.}}
\end{scriptsize}
\end{center}
\end{table*}

\end{appendix}
\end{document}